\documentclass[aps,prd,showpacs,superscriptaddress,preprintnumbers,floatfix]{revtex4}
\usepackage{graphics}
\usepackage{latexsym,amssymb,amsmath}
\usepackage{amscd,array,dsfont,graphicx,kbordermatrix,mathtools,multirow,rotating,subfig,tensor,times,verbatim,youngtab}

\usepackage{epsfig}
\usepackage{amsmath}
\usepackage{amssymb}
\usepackage{calc}
\usepackage[bookmarksnumbered,bookmarksopen=true,bookmarksopenlevel=1,linktocpage,pdfstartview=FitH]{hyperref}
\usepackage[all]{hypcap}

\newcolumntype{M}[1]{>{$}{#1}<{$}}

\newcommand{\beq}{\begin{equation}}
\newcommand{\eeq}{\end{equation}}
\newcommand{\bea}{\begin{eqnarray}}
\newcommand{\be}{\begin{equation}}
\newcommand{\ee}{\end{equation}}
\newcommand{\eea}{\end{eqnarray}}
\newcommand{\ba}{\begin{array}}
\newcommand{\ea}{\end{array}}
\newcommand{\bit}{\begin{itemize}}
\newcommand{\eit}{\end{itemize}}
\newcommand{\ben}{\begin{enumerate}}
\newcommand{\een}{\end{enumerate}}

\def\b{\beta}
\def\g{\gamma}

\def\d{\delta}
\def\e{\epsilon}

\allowdisplaybreaks[1]

\DeclareMathOperator{\id}{id}
\DeclareMathOperator{\Aut}{Aut}

\DeclareMathOperator{\tr}{tr}
\DeclareMathOperator{\Tr}{Tr} % for partial traces only
\DeclareMathOperator{\Det}{Det}

\DeclareMathOperator{\Iso}{Iso}

\DeclareMathOperator{\Or}{O}
\DeclareMathOperator{\SO}{SO}
\DeclareMathOperator{\USp}{USp}
\DeclareMathOperator{\SL}{SL}
\DeclareMathOperator{\GL}{GL}
\DeclareMathOperator{\SU}{SU}

\DeclareMathOperator{\U}{U}
\DeclareMathOperator{\Sym}{Sym}

\DeclareMathOperator{\Sp}{Sp}

\DeclareMathOperator{\Orth}{O}

\newcommand{\F}{\mathds{F}}
\newcommand{\R}{\mathds{R}}
\newcommand{\C}{\mathds{C}}

\newcommand{\Z}{\mathds{Z}}
\newcommand{\J}{\mathfrak{J}}
\newcommand{\FTS}{\mathfrak{F}}

\newcommand{\AutF}[1]{\Aut(#1)}

\newcommand{\half}{\ensuremath{\tfrac{1}{2}}}

\newcommand{\SUSY}{\ensuremath{\mathcal{N}}}

\newcommand{\field}[1]{\ensuremath{\mathds{#1}}}
\newcommand{\rep}[1]{\ensuremath{\mathbf{#1}}}
\newcommand{\tyoung}{\tiny\young}
\newcommand{\braket}[1]{\langle#1|#1\rangle}
\newcommand{\bra}[1]{\langle #1|}
\newcommand{\ket}[1]{|#1\rangle}

\newtheorem{theorem}{Theorem}

\newtheorem{definition}[theorem]{Definition}

\newcommand{\sfx}{\textsf{x}}
\newcommand{\sfo}{\textsf{o}}

\newcolumntype{D}[1]{>{$\displaystyle}{#1}<{$}}

\newcolumntype{C}[1]{>{\centering} m{#1}}
\newcolumntype{X}[1]{>{\centering $} m{#1}<{$}}

\begin{document}
\title{The black-hole/qubit correspondence: an up-to-date review}
\author{L. Borsten}
\email[]{leron.borsten@imperial.ac.uk}
\author{M. J. Duff}
\email[]{m.duff@imperial.ac.uk}
\affiliation{Theoretical Physics, Blackett Laboratory, Imperial College London, London SW7 2AZ, United Kingdom}
\author{P. L\'evay}
\email[]{levay@phy.bme.hu}
\affiliation{Department of Theoretical Physics, Institute of
Physics, Budapest University of Technology, H-1521 Budapest,
Hungary}

\date{\today}
\begin{abstract}
{We give a review of the black-hole/qubit correspondence that incorporates not only the earlier results  on black hole entropy and entanglement measures, seven qubits and the Fano plane, wrapped branes as qubits and the attractor mechanism as a distillation procedure, but also newer material including error-correcting codes, Mermin squares, Freudenthal triples and 4-qubit entanglement classification.}
\end{abstract}

\preprint{Imperial/TP/2012/mjd/3}

\maketitle{}

\renewcommand\thesection{\arabic{section}}
\renewcommand\thesubsection{\arabic{subsection}}

\numberwithin{subsection}{section}
\numberwithin{subsubsection}{subsection}

\numberwithin{equation}{section}

\tableofcontents

\section{Introduction}

%\subsection{Proba subsection}
%\subsubsection{Proba subsubsection}

Six years have passed since the realization
  \cite{Duff:2006uz,Kallosh:2006zs,Levay:2006kf} that the apparently
separate fields of quantum information   \cite{Nielsen:2000} and string theory  can be
related. When two different branches of theoretical physics share many of the same features, it frequently allows
knowledge on one side to provide new insights on the other. This has certainly proved to be the case with string theory and quantum information as we hope to show in this review.  The original
correspondence was between the structure
of the Bekenstein-Hawking entropy  \cite{Bekenstein:1973ur,Hawking:1974sw}
formulae of certain black hole solutions in string theory, and certain multipartite entanglement
measures  \cite{Plenio:2007} in quantum information. However, many more striking structural similarities between these fields have since been uncovered forming what has become known as
the ``black-hole/qubit correspondence" (BHQC).

As far as the BHQC is concerned the main reason for the occurrence of
these striking coincidences seems to be the presence of similar symmetry structures.
Namely, within the field of stringy black holes there are the
 U-duality groups  \cite{Cremmer:1978ds,Cremmer:1979up,Duff:1990hn,Hull:1994ys}, for a review see \cite{Obers:1998fb}, and in the field of
multipartite entanglement the groups of admissible local manipulations
preserving entanglement type  \cite{Bennett:1999,Dur:2000}.
While in the former case the groups in question are real
(in the supergravity approximation) and in the latter they are complex, in particular instances a suitable
complex extension places these stringy dualities in the realm of
local multipartite entanglement transformations.
Then the U-duality invariants can be mapped to invariants
(entanglement measures) under the local group of admissible manipulations.
Then it is not so surprising that the  most obvious invariants
that can show up in these different scenarios are U-duality invariant
formulae of the black hole entropy.

This realization paved the way for studying the structure of
black-hole entropy formulae via the techniques of entanglement
measures  \cite{Duff:2006uz,Kallosh:2006zs,Levay:2006kf, Duff:2006ue,Levay:2006pt,Duff:2007wa} and
three-qubit Pauli operators  \cite{Levay:2008mi,Levay:2009bp} or, conversely,
getting hints for constructing new and physically interesting
measures from the structure of entropy
formulae  \cite{levay-2008,Levay:2009}. Another useful aspect of this
correspondence is that the classification problem of certain
types of black hole can be mapped to the classification problem
of entanglement types of qubit systems   \cite{Kallosh:2006zs,Levay:2006kf,Levay:2009,Borsten:2010db,Borsten:2011is}
.

Apart from clarifying the structure of black-hole entropy
formulae, there has been some progress in understanding the
dynamical aspects of the moduli, in particular the famous
attractor mechanism, in entanglement terms  \cite{Ferrara:1995ih,Strominger:1996kf,Ferrara:1996dd}. It is
known that in general the entropy of a black hole can depend on
the values of moduli or scalar fields, having their origin in the
compactification of extra dimensions. However, for the special
case of extremal black holes, having zero Hawking temperature, the
values of such scalar fields are fixed on the black hole horizon
in terms of the charges. The crucial point is that the fixed
values are independent of the asymptotic values of such moduli.
The radial evolution of the scalar fields starting from the
asymptotically Minkowski region to the horizon shows a fixed point
behaviour. In the special case of the so called
$STU$-model  \cite{Duff:1995sm,Bellucci:2008sv} it has been shown that such
moduli stabilisation can be recast in the language of quantum
information as a radial evolution of a three-qubit state depending
on the charges, the  moduli and the warp factor resulting in a
distillation  procedure of GHZ (Greenberger-Horne-Zeilinger) like states on the
horizon  \cite{Levay:2006kf,Levay:2007nm,Levay:2010qp,Levay:2010yh,Levay:2010ua}.

The BHQC has shed new light on well-known results in quantum entanglement  \cite{Borsten:2009yb} and has
clarified previously controversial issues  \cite{Borsten:2010db,Borsten:2011is}. The techniques
employed in such cases have originated from the string theoretical
side. For example, although Freudenthal triple
systems  (FTS) \cite{Freudenthal:1954,McCrimmon:2004,Krutelevich:2004} have been
well-known to the supergravity community  \cite{Gunaydin:1983bi,Gunaydin:1983rk,Gunaydin:1984ak}, their relevance to special entangled
systems  \cite{levay-2008,Borsten:2008,Levay:2009} has only recently been realised within
the framework of the BHQC. This, in turn, inspired new applications of the Freudenthal triple system to black holes in the form of  
Freudenthal and Jordan dualities   \cite{Borsten:2009zy, Ferrara:2011gv}.

In some special cases multiqubit entanglement measures have turned
out to be useful for classifying multicenter black hole
solutions  \cite{Ferrara:2010ug,Levay:2011bq}. These studies connected the
structure of four-qubit invariants  \cite{Luque:2002,2006JPhA...39.9533L} to the
structure of elliptic curves and the $j$ invariant  \cite{Levay:2011bq}.
The idea that such objects might play some role in four-qubit
systems and the BHQC was first suggested in   \cite{Gibbs:2010uz} and some
related discussion also appeared in the
supergravity  \cite{Bellucci:2010zd} and quantum
entanglement  \cite{2011PhRvL.106r0502R} literature.

The BHQC attempted to clarify the possible microscopic origin of
qubits (and qutrits) in this entanglement based approach. It has
been suggested that in the case of toroidal compactifications the
appearance of qubits can be traced back to wrapping configurations
of membranes  \cite{Borsten:2008ur} on noncontractible loops of the
extra dimensions. This heuristic picture has been generalised and
made more precise  \cite{Levay:2011ph} by identifying the Hilbert
space where these qubits reside within the cohomology of the extra
dimensions. As a bonus it was also
shown  \cite{Levay:2011ph} that in this special case the phenomenon of
flux compactification  \cite{Gukov:1999ya} can also be included within the
realm of the BHQC.

The BHQC combined with the methods of finite geometry have
provided a new conceptual framework for understanding the role of
incidence geometries in quantum theory. It has been shown that the
structure of certain black hole entropy formulae are encapsulated
in the incidence structure of geometric hyperplanes  \cite{Ronan:1987:EHD:38247.38256} of
finite geometries based on the two, three and four-qubit Pauli
groups  \cite{Levay:2008mi,Levay:2009bp}. Such groups have already made their debut
to quantum error correcting codes  \cite{1996PhRvA..54.1862G}, objects that
have also been shown to play some role in the black hole attractor
mechanism  \cite{Levay:2006pt,Levay:2007nm,2010SPPhy.134...85L}. A surprising result is
that Mermin squares  \cite{Mermin} as geometric hyperplanes show up
naturally in noncommutative parametrizations of incidence
geometries characterising the structure of black hole entropy
formulae. Such results initiated a further study of automorphism
groups of finite geometric structures related to special subgroups
of the U-dualities  \cite{Levay:2008mi,Levay:2009bp,Vrana:2009ph,2010JMP....51l2203C}, and a
systematic study of the Veldkamp space of geometric hyperplanes
for multiple qubits  \cite{Vrana:2009ph}.

Though the BHQC is still at its infancy,  it has repeatedly proved  useful for obtaining interesting results on both sides of the correspondence
by employing the techniques and methods of the other \cite{Levay:2006kf,Duff:2006ue,Levay:2006pt,Duff:2007wa,Levay:2007nm,Borsten:2008ur,Borsten:2008,Levay:2008mi,Borsten:2008wd,Levay:2009bp,Borsten:2009ae,Levay:2010qp,Levay:2010ua,
Borsten:2010db,Borsten:2011is,Rios:2011fa,Levay:2011bq,Levay:2011ph}.
Over the past six years
joint efforts of two groups culminated in establishing a precise dictionary between the two sides of the
correspondence  \cite{Borsten:2008wd,2010SPPhy.134...85L}.
The aim of the present review is to give an account of these efforts.

\section{Cayley's hyperdeterminant and black hole entropy}

\subsection{Entanglement and three-qubit systems}
\label{ent3qb}

Since entanglement  may be used in the course of a quantum computation \cite{Nielsen:2000}, characterising the ``amount'' of entanglement possessed by a given state is an important problem.  There are several criteria for good measures of
entanglement  \cite{Plenio:2007}. In particular, since entanglement is a global
 phenomenon of a quantum nature (in the sense that it leads to correlations between spatially separated systems that admit no classical explanation),    any good measure should  be
monotonically decreasing under  local operations (LO) on the constituent systems supplemented by classical communication  (CC) between them \cite{Bennett:1999,Dur:2000,Plenio:2007}.  LOCC operations cannot create entanglement. Hence, two states that may be stochastically (S) interrelated  by an LOCC protocol have the same entanglement under any good measure \footnote{The condition that a measure be
monotonically decreasing \emph{on average} under LOCC is actually a stronger condition.}. Two states of a $k$-constituent composite system with Hilbert space $\mathcal{H}_1\otimes\cdots\otimes\mathcal{H}_k$, $\dim \mathcal{H}_i=n_i$, are SLOCC-equivalent if and only if they are related by the subset of invertible local
operation in SLOCC,  i.e. elements of $\GL(n_1,{\mathds C})\times\cdots\times\GL(n_k,{\mathds C})$ \cite{Dur:2000}. This SLOCC-equivalence group (which we will often refer to loosely as  simply SLOCC) partitions the state space to entanglement classes.  Any relative invariant of SLOCC is a good entanglement measure \cite{Verstraete:2003}.
 
For three qubits we have three two-state systems  each with Hilbert
space ${\mathds C}^2$: ${\cal H}_A$,
${\cal H}_B$ and ${\cal H}_C$ where the labels refer to Alice, Bob
and Charlie.  The Hilbert space of the total system is ${\cal
H}={\mathds C}^2\otimes {\mathds C}^2\otimes {\mathds C}^2$. A
three-qubit state of general form can be represented as \beq
\vert\psi\rangle=\sum_{ABC=0,1}\psi_{ABC}\vert ABC\rangle,\qquad
\vert ABC\rangle =\vert A\rangle \otimes\vert B\rangle\otimes\vert
C\rangle\in {\cal H}_A\otimes {\cal H}_B\otimes {\cal H}_C.
\label{3qbitstate} \eeq \noindent Under SLOCC transformations our
state transforms as a $\mathbf{(2,2,2)}$ namely \beq
\vert\psi\rangle\mapsto ({\cal A}\otimes {\cal B}\otimes {\cal
C})\vert\psi\rangle, \qquad \psi_{ABC}\mapsto {{\cal
A}_A}^{A^{\prime}}{{\cal B}_B}^{B^{\prime}}{{\cal
C}_C}^{C^{\prime}}\psi_{A^{\prime} B^{\prime} C^{\prime}}, \qquad
{\cal A},{\cal B},{\cal C}\in \GL(2,{\mathds C}). \label{slocc}
\eeq \noindent Now one can define the quantity (Cayley's
hyperdeterminant)  \cite{Cayley:1845,Gelfand:1994}

\begin{eqnarray}
\label{Cayley}
D(\psi)&=&
[\psi_0\psi_7-\psi_1\psi_6-\psi_2\psi_5-\psi_3\psi_4]^2-
4[(\psi_1\psi_6)(\psi_2\psi_5)+(\psi_2\psi_5)(\psi_3\psi_4)\\\nonumber&+&
(\psi_3\psi_4)(\psi_1\psi_6)]+
4\psi_1\psi_2\psi_4\psi_7+4\psi_0\psi_3\psi_5\psi_6 
\end{eqnarray}
where $(\psi_0,\psi_1,\dots,\psi_7)\equiv
(\psi_{000},\psi_{001},\dots,\psi_{111})$, which gives rise to a
famous entanglement measure called the {\it
three-tangle}  \cite{Coffman:1999jd} which for normalised states satisfies \beq
0\leq {\tau}_{ABC}=4\vert D(\psi)\vert\leq 1. \label{threetangle}
\eeq \noindent Under SLOCC transformations $D(\psi)$ transforms as
\beq D(\psi)\mapsto ({\rm Det}{\cal A})^2 ({\rm Det}{\cal
B})^2({\rm Det}{\cal C})^2D(\psi) \eeq \noindent hence this
polynomial is a relative invariant. Notice that the expression of
the three-tangle is invariant under permutations (triality) and
the subgroup $[\SL(2,{\mathds C})]^{\otimes 3}$ of SLOCC
transformations. The physical meaning of the three-tangle is the
residual distributed entanglement not contained in either the pure
state or the mixed state entanglement of any
bipartite-singlepartite split  \cite{Coffman:1999jd}. Considerations of
distributed entanglement have also been used in connection with
attractors of $STU$ black holes \cite{Levay:2010yh}.

In this formalism the classification problem of entanglement types amounts to
finding the $[\GL(2,{\mathds C})]^{\otimes 3}$ orbits of a particular $\vert\psi\rangle$. This problem has been solved by mathematicians  \cite{Gelfand:1994} and later rediscovered by physicists  \cite{Dur:2000}.
The result is that apart from the trivial class with $\vert\psi\rangle =0$
we have six SLOCC classes.
The four classes that represent states with some degree of separability
are the totally separable states with representative $\vert 000\rangle$,
the biseparable states with (unnormalized) representatives:  $\vert 0\rangle\otimes(\vert 00\rangle +\vert 11\rangle)$ and two similar states with the qubits cyclically permuted.
Three qubits can be entangled in two inequivalent ways  \cite{Dur:2000}, the unnormalized representatives of these classes are the so-called $\vert GHZ\rangle$ and $\vert W\rangle$ states with the form
\beq
\vert GHZ\rangle =\vert 000\rangle +\vert 111\rangle,\qquad \vert W\rangle =\vert 001\rangle+\vert 010\rangle +\vert 100\rangle.
\label{GHZW}
\eeq
\noindent
The important point is that these two classes can be separated from the rest as follows.
The GHZ-class is characterised by $D(GHZ)\neq 0$, i.e. this state has nonvanishing three-tangle.
On the other hand, it can be shown  \cite{Borsten:2009yb} that one can introduce a dual three-qubit state $\vert\tilde{\psi}\rangle$ which also transforms as a $\mathbf{(2,2,2)}$ of $[\GL(2,{\mathds C})]^{\otimes 3}$. The dual state is cubic in the original amplitudes of $\vert \psi\rangle$ and its explicit expression  \cite{Borsten:2009yb} is connected to the so-called trilinear form of the corresponding Freudenthal triple system  \cite{Krutelevich:2004,Borsten:2009yb}.
Then one can show that the $W$-class is characterized by the conditions
$D(W)=0$, $\vert\tilde{\psi}\rangle\neq 0$.
States that are having $D=0$ and $\vert\tilde{\psi}\rangle=0$ are either separable or biseparable.
There is a nice geometric characterisation of these entanglement classes  \cite{Levay:2006kf,Levay:2004,Brody:2007}
in terms of twistors.
Notice also that for states in the GHZ-class one can define the new state
$\vert\hat{\psi}\rangle\equiv\vert\tilde{\psi}\rangle/\sqrt{\vert D(\psi)\vert}$which is a special case of the Freudenthal dual state which plays an important role in the physics of black holes admitting a Freudenthal dual  \cite{Borsten:2009zy}.

\subsection{Three-qubit entanglement and Black Hole Entropy}

An interesting subsector of string compactification to four dimensions is provided by the  $STU$ model.
This model has a low energy limit which is described by $\mathcal{N}=2$
supergravity coupled to three vector multiplets  \cite{Duff:1995sm,Bellucci:2008sv,Sen:1995ff,Gregori:1999ns}.
This model can be obtained as a consistent truncation of different string theories in a number of ways.
One possibility is to take the type IIA string theory compactified on a Calabi-Yau manifold and then consider a convenient truncation of the $\mathcal{N}=2$ theory arising as the low energy limit.
As an alternative possibility one can start with the heterotic string on the six torus $T^6$. Then the $STU$ model  arises as a truncation of the resulting $\mathcal{N}=4$ theory.
The $STU$ model  got its name from the names of the three complex scalar fields  ($S,T$ and $U$), which play different roles in the different interpretations \cite{Duff:1995sm}.
The $STU$ model admits extremal black hole solutions carrying four electric and four magnetic charges.
Within the framework of this model the macroscopic black hole entropy can be calculated  \cite{Behrndt:1996hu}.

The starting point of the black-hole/qubit correspondence was the observation that if we organize the eight charges of the solution into a $2\times 2\times 2$ array, i.e. a hypermatrix, for BPS (Bogomolnyi-Prasad-Sommerfield) solutions the macroscopic black hole entropy can be expressed as the negative of the square root of Cayley's hyperdeterminant  \cite{Duff:2006uz}.
The four electric $(q_0,q_1,q_2,q_3)$
 and four magnetic $(p^0,p^1,p^2,p^3)$ charges, and the amplitudes of an {\it unnormalized}
three-qubit state as follows
\beq
(p^0,p^1,p^2,p^3,q_0,q_1,q_2,q_3)\leftrightarrow (\psi_0,\psi_1,\psi_2,\psi_4,-\psi_7,\psi_6,\psi_5,\psi_3).
\label{corres}
\eeq
\noindent
Then the macroscopic entropy is
\beq
S=\pi\sqrt{-D(\psi)}=\frac{\pi}{2}\sqrt{\tau_{ABC}(\psi)}.
\label{entropy}
\eeq
\noindent
From this expression we see that the entropy of such black holes can be related to a tripartite entanglement measure, namely the three-tangle.
Notice, however, these are not  three-qubit states, in the conventional sense.
First of all, the amplitudes are {\it real} and unnormalized. Moreover, the charges should be also quantized, hence the amplitudes should be {\it integer}. To cap all this, for BPS solutions, for which half of the supersymmetry is conserved, $D(\psi)$  is negative  \cite{Bellucci:2008sv}.

However, the apparent issue of unnormalized states is not  serious since SLOCC transformations  do not  preserve the norm in any case.
As far as the reality of the amplitudes is concerned, one can regard the three-qubit states as real versions of the usual qubits called rebits  \cite{Caves:2000}.
In this case the SLOCC group should be modified accordingly to three copies of $\GL(2,{\mathds R})$.
Restricting to determinant one transformations what we get is precisely the symmetry group of the $STU$ model at the classical level, ie. $\SL(2,{\mathds R})^{\otimes 3}$.
After implementing the (quantum) constraint coming from the usual Dirac-Zwanziger charge quantization the group we get is the U-duality group of the model
namely $\SL(2,{\mathds Z})^{\otimes 3}$.

Now let us have a look at the constraint $D(\psi)<0$.
It can be shown that we can relax this constraint as well, provided we are willing to embark in the rich field of non-BPS black hole solutions  \cite{Tripathy:2005qp,Kallosh:2006bt} for which $D(\psi)>0$.
Moreover,  as discussed in the paper of Kallosh and Linde  \cite{Kallosh:2006zs} there are also solutions
for which $D(\psi)=0$. These are called small black holes, referring to the fact that though they have vanishing Bekenstein-Hawking entropy,  they can develop a nonvanishing entropy via higher-order and quantum corrections \cite{Sen:1995in}.
The final result of these considerations is that the classification of entanglement types of three rebits under the SLOCC group $[\SL(2,{\mathds R})]^{\otimes 3}$ can be mapped to the classification of  different types of black holes solutions
in the $STU$ model  \cite{Kallosh:2006zs,Levay:2006kf,Borsten:2008wd} and vice versa.

In summary: the formula for the macroscopic black hole entropy
in the $STU$ model can be expressed in terms of the three-tangle which is a
triality and U-duality  invariant tripartite measure of entanglement as
\beq
S=\frac{\pi}{2}\sqrt{\tau_{ABC}(\psi)}=\pi\sqrt{\vert D(\psi)\vert}
\label{threetanglebh}
\eeq
\noindent
where $\vert\psi\rangle$ is an unnormalized three-rebit state with amplitudes being the eight quantized charges. Cayley's hyperdeterminant is negative for BPS, and positive for non-BPS large black holes.
These have nonzero horizon area and nonzero semiclassical Bekenstein-Hawking entropy. For small black holes we have $D(\psi)=0$.

The question left to be answered is whether such rebits can somehow  be embedded into
the realm of genuine {\it complex} three-qubit states.
Within the framework of conventional quantum information theory this problem has already been discussed  \cite{Acin:2001}.
In order to do this also in our black hole context we clearly have to see how other ingredients of the $STU$ model (namely the complex scalar fields $S$, $T$ and $U$) can be incorporated into the formalism. The introduction of such  structures will be discussed later.

\section{$E_7$ and the tripartite entanglement of seven qubits}

\subsection{Embedding the $STU$ model}

Having discussed the $STU$ black hole entropy and its connection to three-qubit entanglement, the question now is whether we can extend our considerations
to more general charge configurations.
The extremal spherically symmetric black hole solutions in $\mathcal{N}=8$ supergravity \cite{Cremmer:1978ds,Cremmer:1979up} are defined by $28+28$ electric/magnetic charges and the entropy formula is given by the square root of the quartic Cartan-Cremmer-Julia $E_{7(7)}$ invariant  \cite{Cartan,Cremmer:1979up,Kallosh:1996uy}
\beq
S=\pi\sqrt{\vert I_4\vert}
\eeq
\noindent
where the Cartan form
\beq
I_4=-{\rm Tr}(xy)^2+\frac{1}{4}\left({\rm Tr}xy\right)^2-4\left({\rm Pf}x+{\rm Pf}y\right)
\label{Cartan}
\eeq
\noindent
depends on the $8\times 8$ antisymmetric quantized charge matrices $x$ and $y$. The $28$ independent components of $x$ and $y$ correspond to electric and magnetic charges, respectively.
From an M-theoretic point of view these charges  originate from wrapping configurations of membranes on the extra dimensions, as discussed in  \autoref{sec:qfed}.

An alternative (the Cremmer-Julia) form of this invariant is given in terms of the
$8\times 8$ complex central charge matrix ${\cal Z}$ \beq I_4={\rm
Tr}({\cal Z}\overline{{\cal Z}})^2-\frac{1}{4}({\rm Tr}{\cal
Z}\overline{{\cal Z}})^2+4({\rm Pf}{\cal Z}+{\rm
Pf}\overline{{\cal Z}}) \label{cremmerform} \eeq \noindent where
the overbars refer to complex conjugation. The definition of the
Pfaffian is \beq {\rm Pf}{\cal Z}=\frac{1}{2^4\cdot
4!}{\epsilon}^{ABCDEFGH}{\cal Z}_{AB}{\cal Z}_{CD}{\cal Z}_{EF}{\cal Z}_{GH}. \label{Pfaffian}
\eeq \noindent

The relation between the Cremmer-Julia and Cartan forms can be
established by using the relation \beq {\cal
Z}_{AB}=-\frac{1}{4\sqrt{2}}(x^{IJ}+iy_{IJ})({\Gamma}^{IJ})_{AB}
\label{relation} \eeq \noindent where summation through the
indices $A,B$ is implied only for $A<B$. Here
$({\Gamma}^{IJ})_{AB}$ are the generators of the $\SO(8)$ algebra,
where $(IJ)$ are the vector indices ($I,J=0,1,\dots, 7$) and
$(AB)$ are the spinor ones ($A,B=0,1,\dots, 7$). Triality of
$\SO(8)$ ensures that we can transform between its vector and
spinor representations. A consequence of this is that we can also
invert the relation of \,(\ref{relation}) and express
$x^{IJ}+iy_{IJ}$ in terms of the   central charge matrix ${\cal
Z}_{AB}$.

Let 
\begin{eqnarray}
\label{cimkestu} x^{01}+iy_{01}&=&-\psi_7-i\psi_0,\qquad
x^{34}+iy_{34}=\psi_1+i\psi_6\\\nonumber
x^{26}+iy_{26}&=&\psi_2+i\psi_5,\qquad
x^{57}+iy_{57}=\psi_4+i\psi_3
\end{eqnarray}
\noindent where the remaining components of $x$ and $y$ are set to
zero. Then a calculation shows that \beq I_4=-D(\psi) \eeq
\noindent where $D(\psi)$ is Cayley's hyperdeterminant of 
(\ref{Cayley}). This result suggests that we should be able to obtain
the three-qubit interpretation of the $STU$ model as a consistent
truncation of  a larger entangled system living within our $\mathcal{N}=8$,
$D=4$ supergravity theory.

By a transformation of the form
${\cal Z}\mapsto U^{t}{\cal Z}U$, where $U\in SU(8)$, ${\cal Z}$
can be brought to the form \beq {\cal
Z}_{canonical}=\begin{pmatrix}z_1&0&0&0\\0&z_1&0&0\\0&0&z_3&0\\0&0&0&z_4
\end{pmatrix}
\otimes
{\varepsilon}, \label{kanonikus} \eeq \noindent where $\varepsilon$ is the antisymmetric $2\times 2$ matrix ${\varepsilon}^{12}=1$ and all four $z_j$ can be
chosen to have the same phase, or three of the $z_i$ can be
chosen to be real   \cite{2007stmt.book.....B}.
Our choice of (\ref{cimkestu}) can then be related to this canonical form as
\beq
z_1=\frac{1}{\sqrt{8}}(-\psi_7+\psi_1+\psi_2+\psi_4+i(-\psi_0+\psi_6+\psi_5+\psi_3))
,\quad
z_2=\frac{1}{\sqrt{8}}(-\psi_7-\psi_1+\psi_2-\psi_4+i(-\psi_0-\psi_6+\psi_5-\psi_3))
\nonumber
\eeq
\noindent
where $z_3$ and $z_4$ is obtained from $z_2$ by a cyclic permutation of the $+$ sign. 
As a result of these considerations it can be shown that the $STU$ truncation is a natural one related to the canonical form of the central charge matrix  \cite{Ferrara:2006em}.
Can we interpret this truncation as a one arising from some larger entangled system?

\subsection{$E_7$ in the cyclic representation}

\begin{figure}[pth!]
\centerline{\includegraphics[width=14.0truecm,clip=]{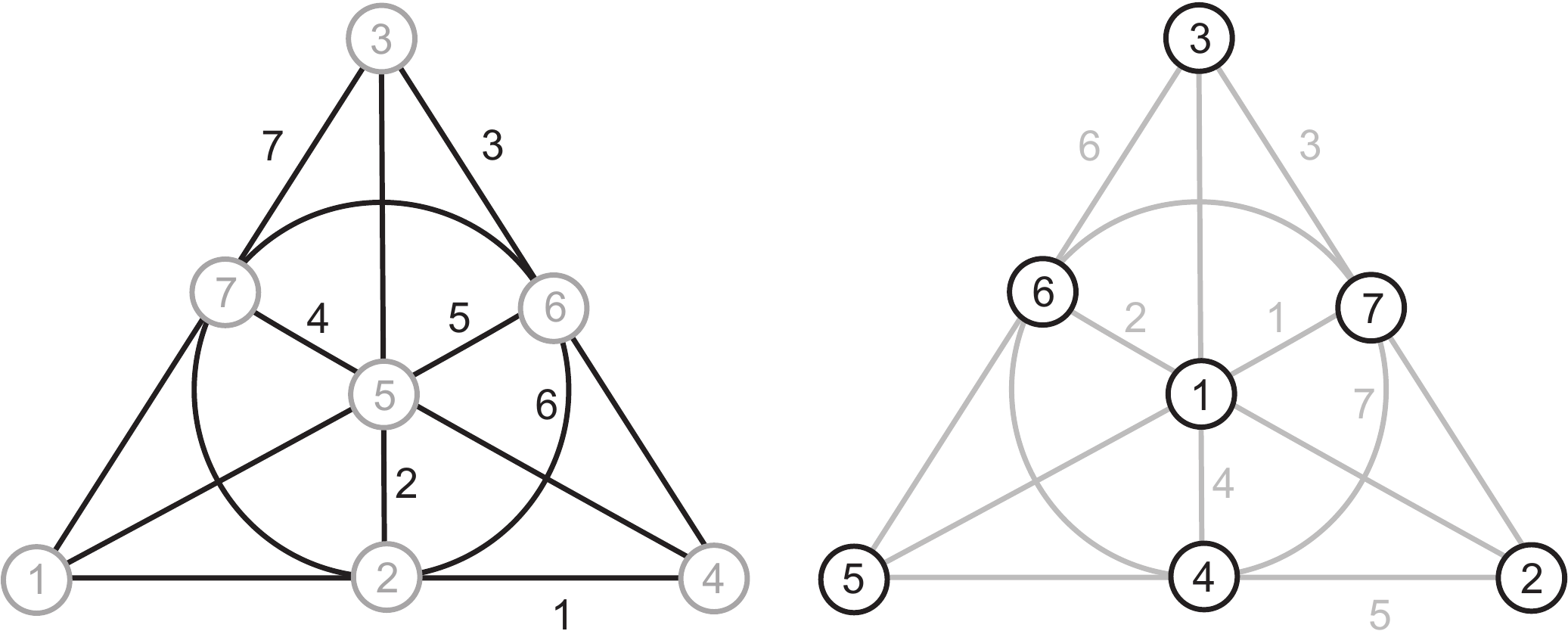}}
\caption{\label{TwoFanos} The Fano plane and its dual. }
\end{figure}

Our success with the three-qubit interpretation of the $STU$ model                is clearly related to the underlying $[\SL(2,{\mathds R})]^{\otimes 3}$ symmetry group of the corresponding $\mathcal{N}=2$ supergravity which can be related to
\emph{real states}
 or \emph{rebits} which  also transform  as the $\mathbf{(2,2,2)}$ of the  complex SLOCC subgroup $[\SL(2,{\mathds C})]^{\otimes 3}$ of a three-qubit system. However, in the $\mathcal{N}=8$ context the symmetry group in question is $E_{7(7)}$ which
is not of the product form hence a qubit interpretation seems to be impossible. However, we know that the $56$ charges of the $\mathcal{N}=8$ model  transform  as the fundamental $56$-dimensional representation of $E_{7(7)}$.
We can try to arrange these $56$ charges as the integer-valued amplitudes
of a reference state.
However, $56$ is not a power of $2$ so the entanglement of this reference state
if it exists at all should be of unusual kind.
A trivial observation is that $56=7\times 8$ hence the direct sum of seven copies of three-qubit state spaces produces the right count.
Moreover, a multiqubit description is possible if the complexification $E_7({\mathds C})$ contains the product of some number of copies of the SLOCC subgroup $\SL(2, {\mathds C})$.
Since the rank of $E_7$ is seven we expect that it should contain seven copies
of $\SL(2,{\mathds C})$ groups. Hence this $56$-dimensional representation space
might be constructed as some combination of tripartite states of seven qubits.
This construction is indeed possible  \cite{Duff:2006ue,Levay:2006pt}.
The relevant decomposition of the $\mathbf{56}$ of $E_7({\mathds C})$ with respect to the $[\SL(2,{\mathds C})]^{\otimes 7}$ subgroup is  \cite{Duff:2006ue}
\begin{eqnarray}
\label{dek}
{\bf 56}&\to& ({\bf 2},{\bf 2},{\bf 1},
{\bf 2},{\bf 1},{\bf 1},{\bf 1})+
({\bf 1},{\bf 2},{\bf 2},{\bf 1},{\bf 2},{\bf 1},{\bf 1})+
({\bf 1},{\bf 1},{\bf 2},{\bf 2},{\bf 1},{\bf 2},{\bf 1})\nonumber\\&+&
({\bf 1},{\bf 1},{\bf 1},{\bf 2},{\bf 2},{\bf 1},{\bf 2})+
({\bf 2},{\bf 1},{\bf 1},{\bf 1},{\bf 2},{\bf 2},{\bf 1})\\&+&
({\bf 1},{\bf 2},{\bf 1},{\bf 1},{\bf 1},{\bf 2},{\bf 2})+
({\bf 2},{\bf 1},{\bf2},{\bf 1},{\bf 1},{\bf 1},{\bf 2})\nonumber.
\end{eqnarray}
\noindent
While this is clearly not a subspace of the 7-qubit Hilbert space, it is in fact a subspace of seven qutrits closed under $\SL(2,{\mathds C})^{\otimes 7}\subset\SL(3,{\mathds C})^{\otimes 7}$ \cite{Duff:2006ue}, and so admits a conventional interpretation despite the appearance of the direct sum. Let us now formally replace  the $\mathbf{2}$'s with $1$'s,
and the $\mathbf{1}$'s with $0$'s, and form a $7\times 7$
 matrix by regarding the seven vectors obtained in this way as its rows.
Let the rows correspond to lines and the
columns to points, and the location of a ``1'' in the corresponding slot correspond to incidence.
Then this correspondence results in the incidence matrix of the Fano plane
in the cyclic, or Paley  \cite{2010SPPhy.134...85L},  realization.
Changing the roles of rows and columns we obtain the incidence structure of the dual Fano plane.
Hence the multiqubit state we are searching for is
 a state associated with the incidence geometry of the Fano plane 
 (see Figure (\ref{TwoFanos})).

Let us reproduce here this incidence matrix with the following labelling for the
rows (r) and columns (c)

\beq
\begin{pmatrix}r/c&A&B&C&D&E&F&G\\
                  a&1&1&0&1&0&0&0\\
              b&0&1&1&0&1&0&0\\
              c&0&0&1&1&0&1&0\\
              d&0&0&0&1&1&0&1\\
              e&1&0&0&0&1&1&0\\
              f&0&1&0&0&0&1&1\\
              g&1&0&1&0&0&0&1\end{pmatrix}\mapsto
\begin{pmatrix}a_{ABD}\\ b_{BCE}\\c_{CDF}\\d_{DEG}\\e_{EFA}\\f_{FGB}\\g_{GAC}
\end{pmatrix}
\label{konvencio1}
\eeq
\noindent
where we also displayed the important fact that this labelling automatically    defines
the index structure for the amplitudes
 of seven three-qubit states formed
  out of seven qubits $A,B,C,D,E,F,G$ (Alice, Bob, Charlie, Daisy, Emma, Fred and
George). If we introduce the notation $V_{ijk}\equiv V_i\otimes
V_j\otimes V_k$ where $i,j,k\in \{A,B,C,D,E,F,G\}$ then the ${\bf
56}$ of $E_7$ denoted by ${\cal H}$ decomposes as \beq {\cal
H}=V_{ABD}\oplus V_{BCE}\oplus V_{CDF}\oplus V_{DEG}\oplus
V_{EFA}\oplus V_{FGB}\oplus V_{GAC}. \label{decompcycle} \eeq
\noindent Clearly this structure encompasses an unusual type of
entanglement; entanglement is usually associated with tensor
products, however here we also encounter {\it direct sums}. One
can regard the seven tripartite sectors as seven superselection
sectors corresponding to seven different $STU$
truncations  \cite{Duff:2006ue,Levay:2006pt}. This structure  is usually
referred to in the literature as the tripartite entanglement of
seven qubits. When the amplitudes are reinterpreted as quantized
charges the elements of ${\cal H}$ are states associated to the incidence geometry
of the Fano plane. In order to understand which amplitudes of the
seven three-qubit states correspond to electric, and magnetic
charges we need to relate the amplitudes of the correspondence of
 (\ref{konvencio1}) to the matrices $x^{IJ}$ and $y_{IJ}$ of
the Cartan form of (\ref{Cartan}). Using the decimal labelling
we obtain the so-called Cartan-Fano dictionary  \cite{Borsten:2008wd}.

\beq x^{IJ}=\begin{pmatrix}
0&-a_7&-b_7&-c_7&-d_7&-e_7&-f_7&-g_7\\
a_7&0&f_1&d_4&-c_2&g_2&-b_4&-e_1\\
b_7&-f_1&0&g_1&e_4&-d_2&a_2&-c_4\\
c_7&-d_4&-g_1&0&a_1&f_4&-e_2&b_2\\
d_7&c_2&-e_4&-a_1&0&b_1&g_4&-f_2\\
e_7&-g_2&d_2&-f_4&-b_1&0&c_1&a_4\\
f_7&b_4&-a_2&e_2&-g_4&-c_1&0&d_1\\
g_7&e_1&c_4&-b_2&f_2&-a_4&-d_1&0
\end{pmatrix}
\label{x}
\eeq
\noindent

\beq
y^{IJ}=\begin{pmatrix}
0&-a_0&-b_0&-c_0&-d_0&-e_0&-f_0&-g_0\\
a_0&0&f_6&d_3&-c_5&g_5&-b_3&-e_6\\
b_0&-f_6&0&g_6&e_3&-d_5&a_5&-c_3\\
c_0&-d_3&-g_6&0&a_6&f_3&-e_5&b_5\\
d_0&c_5&-e_3&-a_6&0&b_6&g_3&-f_5\\
e_0&-g_5&d_5&-f_3&-b_6&0&c_6&a_3\\
f_0&b_3&-a_5&e_5&-g_3&-c_6&0&d_6\\
g_0&e_6&c_3&-b_5&f_5&-a_3&-d_6&0
\end{pmatrix}.
\label{y} \eeq \noindent As explained elsewhere the structure of
these matrices is encoded into the structure constants of the dual
Fano plane  (see the second of  Figure \ref{TwoFanos}) and the structure constants of the
octonions  \cite{Borsten:2008wd}. 
One can now
see that the choice of (\ref{cimkestu}) corresponds to
identification of $\vert\psi\rangle$ with the three-qubit state
$\vert a\rangle$ with amplitudes $a_{ABD}$, $A,B,D=0,1$ built from
the qubits of Alice, Bob and Daisy, and the remaining $48$
amplitudes are zero. This $STU$ truncation is just one of seven possibilities
 corresponding to the three-qubit states $\vert
b\rangle,\dots \vert g\rangle$. The quartic invariant $I_4$
truncates to $-D(a),\dots -D(g)$ in these seven possible cases.
This relates the black hole entropies of the seven possible $STU$
sectors to the corresponding three-tangles of the relevant charge
states.

Concerning the seven possible $STU$ truncations it is important to realise
that there is an automorphism $\alpha$ of order seven which  transforms cyclically
the amplitudes $a,b,\dots g$ of the relevant three qubit states into each other.
Note, $\alpha$  transforms cyclically the points $1,2,\dots 7$ of the dual Fano plane of Figure 1.
One can find an $8\times 8$ orthogonal matrix representation ${\cal D}(\alpha)$
acting on the central charge as ${\cal Z}\mapsto {\cal D}(\alpha){\cal Z}{\cal D}^T(\alpha)$.
It can be expressed  \cite{Levay:2008mi} in terms of the ``controlled not" (CNOT) operators  \cite{Nielsen:2000} as
\beq
{\cal D}(\alpha)=(C_{12}C_{21})(C_{12}C_{31})C_{23}(C_{12}C_{31}).
\label{fifth} \eeq \noindent
It can be shown  \cite{Levay:2008mi} that using ${\cal Z}$ as given by (\ref{relation}) with a convenient representation for
the gamma matrices the effect of ${\cal D}(\alpha)$ is to rotate the seven groups of three-qubit amplitudes showing up in (\ref{x})-(\ref{y}) cyclically.

This representation for the automorphism of order seven can be generalized  \cite{Levay:2008mi,Borsten:2008wd}
to  one for the full automorphism group of the Fano plane which is $PSL_2(7)$.
Moreover, it turns out that $PSL_2(7)$ can also be represented on the $28$ charges regarded as composites of electric and magnetic ones with their incidence geometry corresponding to the Coxeter graph  \cite{Levay:2008mi}.
This configuration shows up as a subgeometry of an object called the split-Cayley hexagon related to the incidence geometry of the real three-qubit Pauli group  \cite{Levay:2008mi,Vrana:2009ph}.
Now $PSL_2(7)$ can be embedded into the Weyl group $W(E_7)$, which is a
subgroup of the full U-duality group $E_7({\mathds Z})$ implementing
electric-magnetic duality  \cite{Lu:1996ge}.
The fact that the Weyl-group of $E_7$ is naturally connected to three-qubit quantum gates was first emphasized by Planat and Kibler  \cite{2008arXiv0807.3650P}.
For a recent elaboration on this connection with a description of $W(E_7)$
and three qubits in terms of symplectic transvections  \cite{Vrana:2009ph} see the paper of Cherchiai and van Geemen  \cite{2010JMP....51l2203C}.

\subsection{$E_7$ and the Hamming code}

Let us now see yet another realization of the tripartite
entanglement of seven qubits living inside $E_7$. This realization
is  related to a famous error correcting code: the Hamming code.
As a starting point let us consider the matrix of the three-qubit
discrete Fourier transformation i.e. the tensor product 
$H\otimes H\otimes H$ of three Hadamard gates where
\beq
H\equiv\frac{1}{\sqrt{2}}\begin{pmatrix}1&1\\1&-1\end{pmatrix}.
                                  \label{Hadamard3}
\eeq \noindent
Delete now the first column of the matrix 
$H\otimes H\otimes H$ and replace the $-1$s with $0$s in the remaining $8\times
 7$ matrix.
 Alternatively we can replace the $+1$s with $0$s and the $-1$s with $1$s.
 Then we obtain the following matrices which are complements of each other

\beq
\label{Hamming}
\begin{pmatrix}1&1&1&1&1&1&1\\0&1&0&1&0&1&0\\
               1&0&0&1&1&0&0\\
               0&0&1&1&0&0&1\\
               1&1&1&0&0&0&0\\
               0&1&0&0&1&0&1\\
               1&0&0&0&0&1&1\\
               0&0&1&0&1&1&0\end{pmatrix},\qquad\qquad\qquad
  \begin{pmatrix}0&0&0&0&0&0&0\\1&0&1&0&1&0&1\\
                 0&1&1&0&0&1&1\\
                1&1&0&0&1&1&0\\
               0&0&0&1&1&1&1\\
               1&0&1&1&0&1&0\\
               0&1&1&1&1&0&0\\
              1&1&0&1&0&0&1\end{pmatrix}.
\eeq
\noindent
One can then regard the rows of these matrices as seven binary digit codewords
encoding messages of \emph{four} digits.
For this purpose let us now regard the \emph{first}, \emph{second} and \emph{fourth} digits as \emph{check digits}. The remaining ones are the \emph{message digits}. Hence for example the codeword $({\bf 0},{\bf 1},0,{\bf 1},0,1,0)$ encodes the message $0010$ and the check digits are ${\bf 011}$.
If we would like to send four message bits through a noisy channel
we can encode our $16$  possible $4$ digit message bits into our $16$ seven digit long codewords as discussed above.
Let us suppose that the noisy channel has the effect of flipping just
one of the seven bits. The recipient  would like to
know whether the seven bit sequence  has been corrupted or not. Moreover, if it is corrupted she would like
to correct it unambiguously.
In order to see that she can perform this task just notice that all of our codewords  differ from each other in \emph{at least  three digits}.
If we define the Hamming distance between two codewords as the number of places in which the
codewords differ we see that all pairs of our codewords have distance
at least three. Now if one error is made in the
transmission then the received binary sequence will still be closer
to the origial one than to any other.
As a result the received sequence can be unambiguously
corrected by chosing the codeword from the list which is the closest to it.

\begin{figure}
\centerline{\resizebox{5.0cm}{!}{\includegraphics{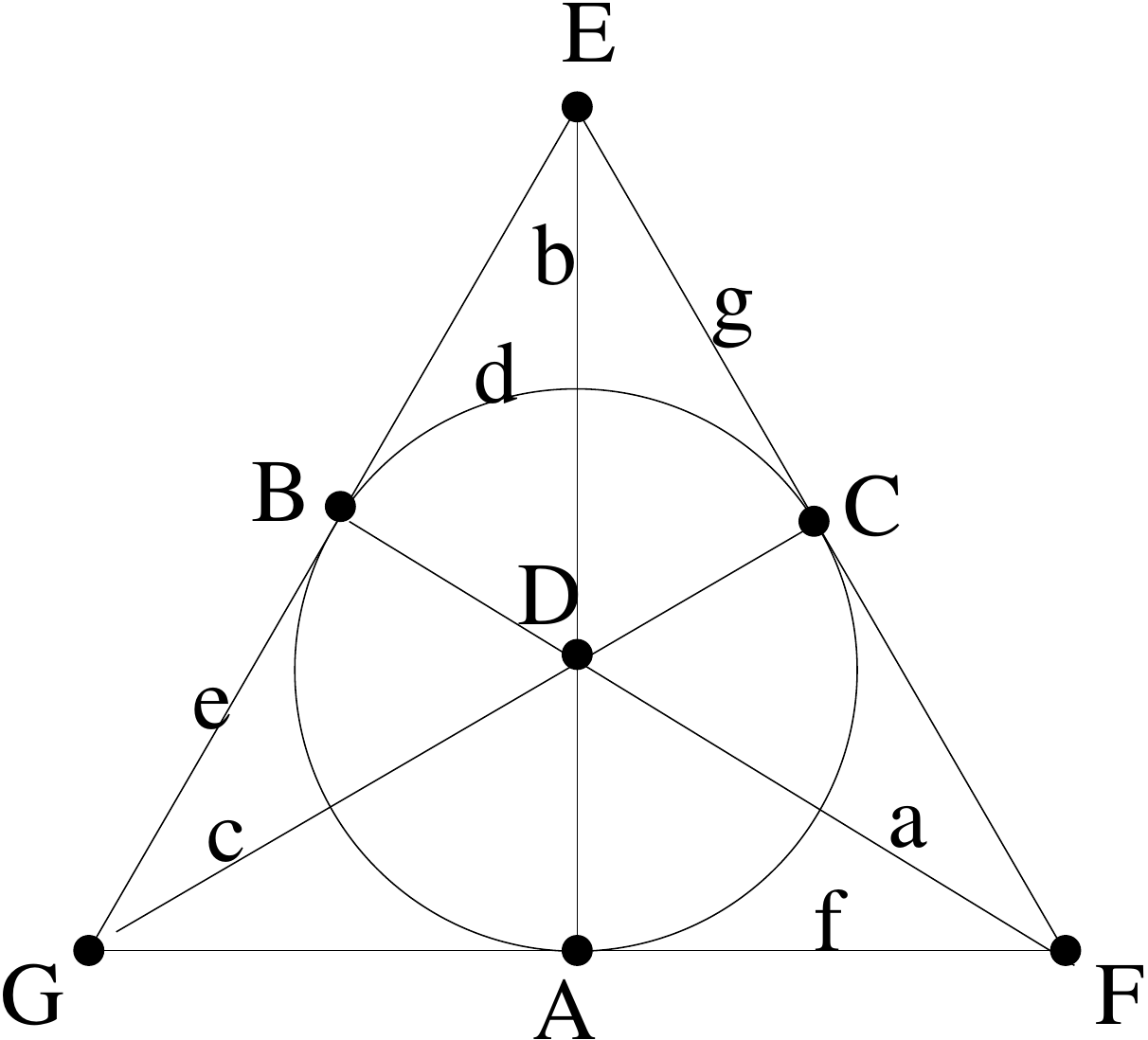}}}
\caption{\label{fig2}
The Hadamard labelling convention for the points and lines of the Fano plane.}
 \end{figure}

Now our aim is to demonstrate that the two matrices of (\ref{Hamming})
related to the codewords of the Hamming code encode another version of the tripartite entanglement of seven qubits and the structure of the Lie-algebra of $E_7$.
Let us first use the first of the two matrices of (\ref{Hamming}) as the
incidence matrix of yet another copy of the Fano plane in the Hadamard
parametrization  \cite{2010SPPhy.134...85L}. For this purpose write the
incidence matrix with the following labelling for the rows (r) and
columns (c)
\beq\begin{pmatrix}r/c&A&B&C&D&E&F&G\\
a&0&1&0&1&0&1&0\\b&1&0&0&1&1&0&0\\
c&0&0&1&1&0&0&1\\d&1&1&1&0&0&0&0\\
e&0&1&0&0&1&0&1\\f&1&0&0&0&0&1&1\\
g&0&0&1&0&1&1&0\end{pmatrix}\mapsto
\begin{pmatrix}a_{BDF}\\ b_{ADE}\\c_{CDG}\\d_{ABC}\\e_{BEG}\\f_{AFG}\\g_{CEF}
\end{pmatrix}
\label{konvencio} \eeq \noindent where 
this labelling automatically defines the index
structure for the amplitudes of seven three-qubit states formed
out of seven qubits $A,B,C,D,E,F,G$. This convention also fixes
the labelling of lines and points of the Fano plane, see Figure
\ref{fig2}.

To the points again we associate qubits 
and to the lines three-qubit systems
with vector spaces $V_{BDF}, V_{ADE},\dots V_{CEF}$.
A list of these three-qubit Hilbert spaces ${\cal H}_{\sigma}$ ,
$\sigma\in {\mathds Z}^3_2-(000)$ is given by the correspondence
\beq
\begin{pmatrix}{\cal H}_{001}&
{\cal H}_{010}&
{\cal H}_{011}&
{\cal H}_{100}&
{\cal H}_{101}&
{\cal H}_{110}&
{\cal H}_{111}\end{pmatrix}
\leftrightarrow
\begin{pmatrix}V_{BDF}&V_{ADE}&V_{CDG}&V_{ABC}&V_{BEG}&V_{AFG}&V_{CEF}\end{pmatrix}.
\nonumber
\label{forreversed}
\eeq
\noindent
Now we switch to a new
ordering of the spaces ${\cal H}_{\sigma}$  according to the rule
$
(1,2,3,4,5,6,7)\to ((100),(010),(110),(001),(101),(011),(111))$
which is the reverse binary labelling.
This yields our definition
for the representation space of the $\mathbf{56}$ of $E_7$
in terms of the spaces ${\cal H}_{\sigma}$ as
\begin{eqnarray}
\label{deco}
{\cal H}=&V_{ABC}&\oplus V_{ADE}\oplus V_{AFG}\oplus V_{BDF}\oplus V_{BEG}\oplus V_{CDG}\oplus V_{CEF}.
\end{eqnarray}
\noindent

In order to shed some light
on the possibility of describing also the structure of the Lie-algebra of $E_7$
in terms of data provided by the Hamming code
let us consider the second matrix of (\ref{Hamming}).

The Lie-algebra of $E_7$ has $133$ dimensions.
We clearly have $\mathfrak{sl}(2)^{\oplus 7}$
as a subalgebra of dimension $7\times 3=21$. These $21$ generators
 act on ${\cal H}$  of (\ref{deco}) via the
well known action of the SLOCC subgroup.
To define the remaining $112$ generators and their action on ${\cal H}$
we consider the complements of the lines of the Fano plane of Figure 
\ref{fig2}.
These seven sets of four points form seven {\it quadrangles}.
Since we have already attached to the points of the Fano plane qubits, and this assignment automatically defined our three-qubit states corresponding to the lines, it then follows that the quadrangles define seven four-qubit states.
They  form the $112$ dimensional  complex vector space
\beq
 W\equiv V_{DEFG}\oplus V_{BCFG}\oplus V_{BCDE}\oplus V_{ACEG}\oplus V_{ACDF}\oplus V_{ABEF}\oplus V_{ABDG}
 \eeq
which we can use as the space of $E_7$ generators not belonging to the SLOCC subalgebra.
Notice that since the complements of the quadrangles are lines that can be associated to seven three-qubit states one can label each of these $16$ dimensional spaces as ${\cal W}_{001},{\cal W}_{010},\dots {\cal W}_{111}$.

Let us denote the basis vectors of the corresponding four-qubit spaces in the computational base as
$(T_{ACEG},\dots,T_{ABDG})$.
A clear indication that we are on the right track for defining the $\mathfrak{e}_7$ algebra via four-qubit states comes from the possibility of defining the Lie-bracket on $W$ using,
\beq
\label{komm}
[T_{ACEG},T_{BC^{\prime}FG^{\prime}}]=\Phi(ACEG,BC^{\prime}FG^{\prime}){\varepsilon}_{CC^{\prime}}{\varepsilon}_{GG^{\prime}}T_{ABEF}
\eeq
\noindent
where in this example the pair $CG$  is common to both  quadrangles, $ACEG$ and $BCFG$.
It can be shown that the explicit form for ${\Phi}$ is arising
from the octonionic multiplication rule that is in turn also encoded into the
Hamming code via the structure of the Fano plane.
Adding also the $21$ generators of the SLOCC group as an extra  vector space
${\cal W}_{000}$  one can show that
after introducing the $133$ dimensional vector space
${\cal W}\equiv {\cal W}_{000}\oplus W$ the Lie bracket can be extended using the obvious commutators.
Denoting this extended bracket by $[\cdot,\cdot]$ one can show  \cite{Manivel:2005,Elduque:2005,2010SPPhy.134...85L}
\beq
e_7=({\cal W},[\quad]).
\eeq
\noindent
As we see ${\cal W}$ has a deep connection with 
the division algebra of octonions.
 In technical terms $\mathfrak{e}_7$, as a vector space, has an octonionic grading \cite{Manivel:2005}.

Using this formalism based on the Hamming code one can show that the generators of $\mathfrak{e}_7$
can be written as combinations of tripartite entanglement transformations  \cite{Levay:2006pt,2010SPPhy.134...85L}.
Some of them are of SLOCC form (those operating in the diagonal blocks), while
 others  generate correlations between the different tripartite sectors. One can also show that the representation theoretic details are entirely encoded in
a so-called $(7,3,1)$ design and its complementary $(7,4,2)$ one  \cite{2010SPPhy.134...85L}, which correspond to the two matrices of (\ref{Hamming}) and are related to
lines and quadrangles of the smallest finite projective plane: the Fano plane.
Moreover, these designs are described in a unified form via the nontrivial
codewords of the Hamming code  of (\ref{Hamming}).
The Hamming code in turn is clearly related to the Hadamard matrix 
(\ref{Hadamard3})
which is the discrete Fourier transform on three-qubits.
We will see in later sections that such Hadamard transformations on three-qubits also play a role in obtaining
a nice characterisation of BPS and non-BPS solutions of the $STU$ truncation. This
suggests that black hole solutions of more general
type might be understood
in a framework related to error correcting codes.

\subsection{The structure of the $E_7$ symmetric black hole entropy formula}

We have already discussed  Cartan's quartic invariant (\ref{Cartan}) well-known from studies concerning $\SO(8)$ supergravity  \cite{Cartan,Cremmer:1979up,Kallosh:1996uy}.
$I_4$ is the singlet in the tensor product representation $\mathbf{56}\times \mathbf{56}\times \mathbf{56}\times \mathbf{56}$.
Its explicit form in connection with stringy black holes with their $E_{7(7)}$ symmetric area form  \cite{Kallosh:1996uy} is given either in the Cremmer-Julia form  \cite{Cremmer:1979up} in terms of the complex $8\times 8$
central charge matrix $Z$ or in the Cartan form  \cite{Cartan} in terms of two real $8\times 8$ ones ${x}$ and ${y}$ containing the quantized electric and magnetic charges of the black hole.
Let us now present its  new form in terms of the $56$ amplitudes of our  seven qubits  \cite{Duff:2006ue}.
In the Hadamard representation of (\ref{deco}) its  new expression is

\begin {eqnarray}
\label{tagok}
I_4=\frac{1}{2}( a^4&+&b^4+c^4+d^4+e^4+f^4+g^4)+\nonumber\\
2[a^2b^2&+&b^2c^2+c^2d^2+d^2e^2+e^2f^2+f^2g^2+g^2a^2+\nonumber\\
a^2c^2&+&b^2d^2+c^2e^2+d^2f^2+e^2g^2+f^2a^2+g^2b^2+\nonumber\\
a^2d^2&+&b^2e^2+c^2f^2+d^2g^2+e^2a^2+f^2b^2+g^2c^2]\nonumber\\
+8[aceg&+&bcfg+abef+defg+acdf+bcde+abdg].
\end {eqnarray}
\noindent
Here we have for example
\beq
bcde={\varepsilon}^{A_1A_3}{\varepsilon}^{B_3B_4}{\varepsilon}^{C_2C_3}{\varepsilon}^{D_1D_2}{\varepsilon}^{E_1E_4}{\varepsilon}^{G_2G_4}b_{A_1D_1E_1}c_{C_2D_2G_2}d_{A_3B_3C_3}e_{B_4E_4G_4}.
\nonumber
\eeq
\noindent

The remaining terms can be described in a unified manner by employing the following definition.
Let us consider for example the two three-qubit states $\vert d\rangle$ and $\vert b\rangle$ with amplitudes $d_{ABC}$ and
$b_{ADE}$ made of five different qubits $A,B,C,D,E$ with
qubit A as the common one.
For this situation we define
\beq
d^2b^2\equiv {\cal Q}(d,b)=\\
{\varepsilon}^{A_1A_3}{\varepsilon}^{B_1B_2}{\varepsilon}^{C
_1C_2}{\varepsilon}^{A_2A_4}{\varepsilon}^{D_3D_4}{\varepsilon}^{E_3E_4}
{d}_{A_1B_1C_1}{d}_{A_2B_2C_2}{b}_{A_3D_3E_3}{b}_{A_4D_4E_4}.
\nonumber
\eeq
\noindent
In this notation
\beq
\label{j42}
{d}^4\equiv{\cal Q}(d,d)=-2D(d)
\nonumber
\eeq
i.e. $-2$ times the usual expression for Cayley's hyperdeterminant.

\begin{figure}
\centerline{\resizebox{5.0cm}{!}{\includegraphics{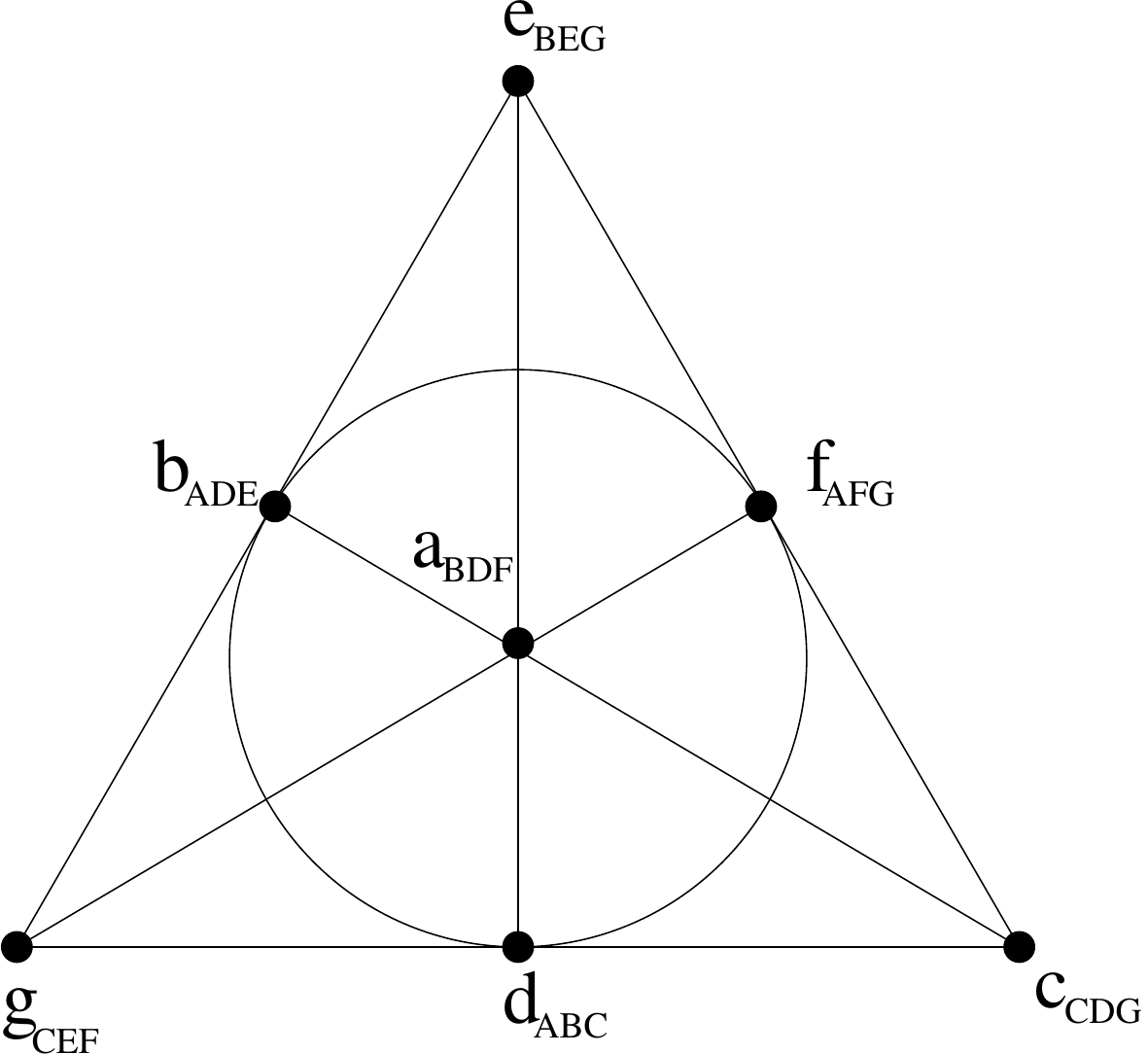}}}
\caption{\label{fig5} \label{dualfan}
The dual Fano plane. To its points now we attached three qubit states with the
representative amplitudes indicated.
To the lines we associate the common qubits these tripartite states share.}
\end{figure}

Another important observation is that the terms occurring in 
(\ref{tagok}) can be understood using the
{\it dual} Fano plane. To see this note that the Fano plane is a
projective plane hence we can use projective duality to exchange
the role of lines and planes. Originally we attached qubits to the
points, and tripartite sysems to the lines of the Fano plane. Now
we take the dual perspective, and attach the tripartite states to
the points and qubits to the lines  of the dual Fano plane see
Figure \ref{fig5}. In the ordinary Fano plane the fact that three
lines  intersect in a unique point correspondeds to the fact
that any three entangled tripartite systems share a unique qubit.
In the dual perspective this entanglement property corresponds to
the geometric one that three points are always lying on a unique
line. For example let us consider the three points corresponding
to the tripartite states with amplitudes $d$, $b$, and $f$.
Looking at Figure \ref{dualfan}
 these amplitudes define the corresponding
points lying on the line $dbf$. This line is defined by the common
qubit these tripartite states share namely qubit $A$.

In the dual Fano plane we have seven points,
with  seven tripartite states attached
to them. The corresponding entanglement
measures are proportional to seven copies
of Cayley's hyperdeterminant, then in $I_4$ we have the terms  $a^4, b^4,
c^4, d^4, e^4, f^4$ and $g^4$.
We also have seven lines with three tripartite states on each of them.
We can group the $21$ terms of the
form $a^2b^2$ etc. into seven groups associated to such lines.
They  describe the pairwise
entanglement between the three different
tripartite systems (sharing a common qubit).
For example for the line $dbf$ we have the terms $b^2d^2$, $d^2f^2$ and $b^2f^2$
 describing such pairwise entanglements.
 Finally we have seven quadrangles (as complements to the lines) with four entangled tripartite systems
 giving rise to the last seven terms in $I_4$.

Apart from immediately identifying the seven different $STU$ truncations, the form
of (\ref{tagok}) has many other virtues.
One of them is that we can easily understand some of the nontrivial truncations
and as an extra bonus we can also quickly  realize their finite geometric meaning.
As an example let us consider
the decomposition of the ${\bf 56}$ of $E_7$
\beq
{\bf 56}\to ({\bf 2},{\bf 12})\oplus ({\bf 1}, {\bf 32})
\eeq
\noindent
with respect to the maximal subgroup $\SL(2,{\mathds C})\times \SO(12,{\mathds C})$.
Notice that the $({\bf 2},{\bf 12})$ part of the representation space ${\cal H}$
consists of the amplitudes of the form
\beq
\begin{pmatrix}d_{ABC}\\b_{ADE}\\f_{AFG}\end{pmatrix}\in
{\cal H}_{({\bf 2},{\bf 12})}\equiv
V_{ABC}\oplus V_{ADE}\oplus V_{AFG}=V_A\otimes (V_{BC}\oplus V_{DE}\oplus V_{FG})
\eeq
\noindent
and the $({\bf 1},{\bf 32})$ part of the ones
\beq
\begin{pmatrix}a_{BDF}\\e_{BEG}\\c_{CDG}\\g_{CEF}\end{pmatrix}\in{\cal H}_{(
{\bf 1}
,{\bf 32})}\equiv V_{BDF}\oplus V_{BEG}\oplus V_{CDG}\oplus
V_{CEF}.
\eeq
\noindent
We see that the $({\bf 2},{\bf 12})$ space consists of all the amplitudes
sharing qubit $A$ in common, and the $({\bf 1},{\bf 32})$ all those
{\it excluding} qubit $A$.

It is also clear that by writing our representation space as
\beq
V_{ADE}\oplus V_A\otimes(V_{BC}\oplus V_{FG})\oplus V_D\otimes(V_{BF}\oplus V_{C
G})\oplus V_E\otimes(V_{BG}\oplus V_{CF}).
\eeq
\noindent
one can easily understand the decomposition
\beq
({\bf 2},{\bf 12})\oplus ({\bf 1}, {\bf 32})\to ({\bf 2},{\bf 2},{\bf 2},{\bf 1})\oplus ({\bf 2},{\bf 1},{\bf 1},{\bf 8_v})\oplus ({\bf 1},{\bf 2},{\bf 1}, {\bf 8_s})\oplus ({\bf 1}, {\bf 1}, {\bf 2}, {\bf 8_s})
\label{stu222}
\eeq
\noindent
with respect to the inclusion $\SL(2)\times \SL(2)\times \SL(2)\times \SO(4,4)\subset \SL(2)\times \SO(6,6)$.

Let us discuss the meaning of the $(\mathbf{2,12})$ truncation in the black hole context.
In this case the corresponding groups are real, hence in the supergravity approximation we have $\SL(2,{\mathds R})\times \SO(6,6)$.
We have in this case $2\times 12$ charges. In the quantum theory they are integer valued and the  U-duality group is broken to $\SL(2,{\mathds Z})\times \SO(6,6,{\mathds Z})$.
The  groups $\SL(2,{\mathds Z})$ and  $\SO(6,6,{\mathds Z})$ are  the $S$ and $T$ duality transformations of toroidally compactified type IIA string theory \cite{Hull:1994ys}.
For the black hole solutions in the corresponding models one can obtain
entropy formulae that are truncations of (\ref{tagok})
with integer amplitudes.
Since we have seven lines in the dual Fano plane such truncations can be obtained
in seven different ways.
As a particular truncation one can take for example the line $dbf$ in the dual Fano plane.
The relevant truncation of $I_4$ interpreted as a measure of pure state entanglement one can take

\beq
\label{3measure}
{\tau}_3^{(3)}=2\vert b^4+d^4+f^4+2(b^2d^2+d^2f^2+b^2f^2)\vert
\eeq
\noindent
where the notation ${\tau}_3^{(3)}$ indicates that now we have three tripartite
states.
Now we write the state corresponding to the line $dbf$ in the form
\beq
\vert\psi\rangle =\sum_{ABCDEFG=0,1}\vert A\rangle\otimes (d_{ABC}\vert BC
\rangle
 +b_{ADE}\vert DE\rangle+f_{AFG}\vert FG\rangle).
\eeq
\noindent
This notation clearly displays that  qubit $A$ is entangled 
with the remaining pairs $(BC)(DE)(FG)$.
Recalling that  this state  transforms as the $\mathbf{(2,12)}$  of $\SL(2)\times \SO(6,6)$          we can write,                                                           \beq                                                                            \label{292}                                                                     \vert\psi\rangle=\sum_{A\mu}{\psi}_{A\mu}\vert A\rangle\otimes\vert \mu\rangle,\quad A=0,1,\quad \mu=1,2,\dots 12.                                              \eeq                                                                            \noindent
Introducing the notation
\beq
p^{\mu}\equiv{\psi}_{0\mu}=
\begin {pmatrix}d_{0BC}\\b_{0DE}\\f_{0FG}\end {pmatrix},\quad
q^{\mu}\equiv{\psi}_{1\mu}=
\begin {pmatrix}d_{1BC}\\b_{1DE}\\f_{1FG}\end {pmatrix}
.
\eeq
\noindent
In these variables for
 ${\tau}_3^{(3)}$ we get the following expression
\beq
\label{pluckeres}
{\tau}_3^{(3)}=4\vert({\bf pp})({\bf qq})-({\bf pq})^2\vert
\eeq
\noindent
where the scalar products above are defined
with respect to
the $12\times 12$ block-diagonal matrix 
containing three copies of 
${\varepsilon}
\otimes {\varepsilon}$.
Now in the black hole analogy $p^{\mu}$ and $q^{\mu}$ are
integer and
the measure of entanglement in (\ref{3measure}) can be related to the black
 hole entropy  \cite{Cvetic:1995bj}
\beq
S=\frac{\pi}{2}\sqrt{\tau_3^{(3)}}
\eeq
\noindent
coming from the truncation of the $\mathcal{N}=8$ theory with $E_{7(7)}({\mathds Z})$ symmetry to
 $\mathcal{N}=4$ supergravity coupled to 6 vector multiplets  with  $\SL({2,{\mathds Z}})\times \SO(6,6,{\mathds Z})$ U-duality.
From the string theoretical point of view this sector describes the  NS-NS charges.
Note, that in the cyclic representation of (\ref{decompcycle}) the formula above can also be reinterpreted as Cayley's hyperdeterminant over the imaginary quaternions  \cite{Borsten:2008wd}.
For a similar discussion of the $(\mathbf{1,32})$ truncation of $I_4$ and its interpretation as an entanglement measure featuring the $32$ so called R-R
charges we refer the reader to the literature  \cite{Borsten:2008wd}.

\section{$E_6$ and the bipartite entanglement of three qutrits}

\subsection{The Octonions and the cubic invariant}

In quantum information one can consider entangled states
representing quantum systems with more than two states (qubits).
Apart from qubits the simplest objects to consider are entangled
three-state systems called {\it qutrits}. A qutrit is an element
of ${\mathds C}^3$ and the corresponding SLOCC group acting on it
is $\GL(3,{\mathds C})$. The entanglement measures for such systems
should come from relative invariants under the action of the
multipartite local SLOCC group. For example as the simplest
relative SLOCC invariant for a two-qutrit state of the form \beq
\vert\psi\rangle= \sum_{A,B=0,1,2}\psi_{AB}\vert A\rangle\otimes
\vert B\rangle \eeq \noindent one can take the determinant ${\rm
Det}(\psi)$ of the $3\times 3$ matrix $\psi_{AB}$. This quantity
is obviously a relative invariant which is invariant under the the
$\SL(3,{\mathds C})\times \SL(3,{\mathds C})$ subgroup of two-qutrit
SLOCC transformations. The classification of two-qutrit states is
very simple. Different SLOCC classes are labelled by the rank of
the $3\times 3$ matrix $\psi_{AB}$.

Based on our experience of relating special entangled systems
built from few {\it real} {\it qubits} (or rebits) to the structure
of black hole entropy formulae for $D=4$ supergravity theories
with $\mathcal{N}=8,4,2$ supersymmetries the question now is the following:
can we find entangled systems of real {\it qutrits} that can be
related to black hole entropy formulae of other kind? This
generalization is indeed possible, provided we consider black
hole, and black sring solutions in five dimensions  \cite{Duff:2007wa}.
Moreover, we can then also use these structures in the complex
domain, as new entanglement measures of some hypothetical
entangled system.

In order to see how these structures arise let us recall that
magic $\mathcal{N}=2$, $D=5$ supergravities   \cite{Gunaydin:1983bi,Gunaydin:1983rk,Gunaydin:1984ak,Borsten:2008wd}
coupled to $5,8,14$ and $26$ vector multiplets with symmetries
$\SL(3,{\mathds R})$, $\SL(3,{\mathds C})$, $\SU^{\star}(6)$ and
$E_{6(-26)}$ can be described by Jordan algebras of $3 \times 3$
Hermitian matrices with entries taken from the reals, complexes, quaternions and octonions. It is also known that in these
cases we have black hole solutions that have {\it cubic
invariants} whose square roots yield the corresponding black hole
entropy   \cite{Ferrara:2006yb}. Moreover, we can also replace in these
Jordan algebras the division algebras by their split versions. For
example,  in the case of split octonions we arrive at
the $\mathcal{N}=8$, $D=5$ supergravity   \cite{Maldacena:1999bp} with $27$ Abelian
gauge fields transforming in the fundamental of $E_{6(6)}$. In
this theory the corresponding black hole solutions have an $E_{6(6)}(\Z)$ symmetric entropy
formula  
  \cite{Ferrara:1996um,Andrianopoli:1997hb,Borsten:2008wd}. It is also important to note that
the magic $\mathcal{N}=2$ supergravities associated
 with the reals, complexes and  quaternions can be
obtained as consistent reductions of the $\mathcal{N}=8$ theory   \cite{Ferrara:2006yb} based on the split octonions. On the other hand, the $\mathcal{N}=2$
supergravity based on the division algebra of the octonions is
exceptional
 since it cannot be obtained from the split
octonionic $\mathcal{N}=8$ theory by truncation.

Since in all these cases the black hole entropy is given in terms
of a cubic invariant,  to relate them to entangled
systems of some kind we need to first understand  the structure of these
invariants. 

 An element of a magic cubic Jordan algebra can be represented
as a $3\times 3$ Hermitian matrix with entries taken from a
division algebra ${\mathds A}$, i.\,e. ${\mathds R}$, ${\mathds
C}$, ${\mathds H}$ or ${\mathds O}$. Explicitly, we have
\beq J_3(Q)=\begin{pmatrix}q_1&Q^v&\overline{Q^s}\\
\overline{Q^v}&q_2&Q^c\\ Q^s&\overline{Q^c}&q_3\end{pmatrix}\qquad
q_i\in {\mathds R},\qquad Q^{v,s,c}\in {\mathds A} \label{Jordan}\eeq
\noindent where an overbar refers to conjugation in ${\mathds A}$.
These charge configurations describe electric black holes of the
$\mathcal{N}=2$, $D=5$ magic supergravities
  \cite{Gunaydin:1983bi,Gunaydin:1983rk,Gunaydin:1984ak,Borsten:2008wd}. In the octonionic case the
superscripts of $Q$ refer to the fact that the fundamental $27$-dimensional representation of the U-duality group $E_{6(-26)}$
decomposes under the subgroup $\SO(8)$ to three
$8$-dimensional representations (vector, spinor and conjugate
spinor connected by triality) plus three singlets corresponding
to the $q_i, i=1,2,3$. A general element in this case is
of the form $Q=Q_0+Q_1e_1+\dots +Q_7e_7$, where the imaginary
units $e_1,e_2,\dots,e_7$ satisfy the rules of the octonionic
multiplication table. The norm of an octonion is
$Q\overline{Q}=(Q_0)^2+ \dots +(Q_7)^2$. The real part of an
octonion is defined as ${\rm Re}(Q)=\frac{1}{2}(Q+\overline{Q})$.
The magnetic analogue of $J_3(Q)$ is obtained by replacing $Q$
with $P$ referring now to the magnetic charges. $J_3(P)$   describes black strings related to the previous case by  electric-magnetic duality.

The black hole entropy is given by the cubic invariant \beq
I_3(Q)=q_1q_2q_3-(q_1Q^c\overline{Q^c}+q_2Q^s\overline{Q^s}+q_3Q^v\overline{Q^v})+2{\rm
Re}(Q^vQ^sQ^c) \label{i3} \eeq \noindent as \beq
S=\pi\sqrt{I_3(Q)} \label{entropycubic} \eeq \noindent and for
the black string we get a similar formula with $I_3(Q)$ replaced
by $I_3(P)$.

 In the spit octonionic case ${\mathds O}_s$ the
norm is defined as \beq
Q\overline{Q}=(Q_0)^2+(Q_1)^2+(Q_2)^2+(Q_3)^2-(Q_4)^2-(Q_5)^2-(Q_6)^2-(Q_7)^2
\label{splitnorm} \eeq \noindent and the group preserving the
cubic invariant of the corresponding Jordan algebra is $E_{6(6)}$,
which decomposes similarly under $\SO(4,4)$. This is the case of
$\mathcal{N}=8$ supergravity with duality group $E_{6(6)}$   \cite{Ferrara:1997uz}. In the quantum
theory the black hole/string charges become integer-valued and the
relevant $3 \times 3$ matrices are defined over the {\it integral}
octonions and {\it integral} split octonions, respectively. The U-duality groups are in this case broken to
$E_{6(-26)}({\mathds Z})$ and $E_{6(6)}({\mathds Z})$.
In all these cases the entropy formula is given by
\,(\ref{i3})--(\ref{entropycubic}), with the norm given by
either the usual one or its split analogue  (\ref{splitnorm}).

\subsection{Qutrits and the cubic invariant}

Since  all the $\mathcal{N}=2$ magic, aside from the octonionic case,
supergravities can be obtained as consistent truncations of the
$\mathcal{N}=8$ split-octonionic case, let us consider the cubic invariant
$I_3$ of \,(\ref{i3}) with the U-duality group $E_{6(6)}$.
Let us also consider the decomposition of the $27$-dimensional
fundamental representation of $E_{6(6)}$ with respect to its
$[\SL(3,{\mathds R})]^{\otimes 3}$ subgroup,
 \beq E_{6(6)}\supset \SL(3,{\mathds R})_A\times
\SL(3,{\mathds R})_B\times \SL(3,{\mathds R})_C \label{griddecomp}
\eeq \noindent under which \beq {\bf 27}\to ({\bf 3}^{\prime},{\bf
3},{\bf 1})\oplus ({\bf 1},{\bf 3}^{\prime},{\bf
3}^{\prime})\oplus ({\bf 3},{\bf 1},{\bf 3}). \eeq \noindent The
above-given decomposition gives  the bipartite entanglement of
three-qutrits interpretation  \cite{Duff:2007wa,Borsten:2008wd} of the ${\bf
27}$ of $E_{6(6)}$. Just as the tripartite entanglement of seven qubits was not a subspace of seven qubit, clearly this is not a subspace of the three qutrits. However,  it is  a subspace of three 7-dits closed under $[\SL(3,{\mathds C})]^{\otimes 3}\subset[\SL(7,{\mathds C})]^{\otimes 7}$ \cite{Duff:2007wa}, and so once again admits a conventional interpretation despite the appearance of the direct sum.  Neglecting the details, all we need is three $3
\times 3$ real matrices $a,b$ and $c$ with the index structure
\beq {a^A}_B,\qquad {b^{BC}},\qquad {c_{CA}},\qquad A,B,C=0,1,2
\label{ampl} \eeq \noindent where the upper indices are
transformed according to the (contragredient) ${\bf 3}^{\prime}$
and the lower ones by ${\bf 3}$. The explicit dictionary between
the qutrit amplitudes $a,b$ and $c$ above and the components of
the Jordan algebra as given in (\ref{Jordan}) can be found in
the literature  \cite{Borsten:2008wd,Levay:2009bp}.

Now the new expression for the cubic invariant $I_3$ of
\,(\ref{i3}) is  \cite{Duff:2007wa} \beq I_3={\rm
Det}J_3(Q)=a^3+b^3+c^3+6abc. \label{i3v} \eeq \noindent Here \beq
a^3=\frac{1}{6}{\varepsilon}_{A_1A_2A_3}{\varepsilon}^{B_1B_2B_3}{a^{A_1}}_{B_1}
{a^{A_2}}_{B_2}{a^{A_3}}_{B_3}, \quad
abc=\frac{1}{6}{a^{A}}_Bb^{BC}c_{CA} \eeq \noindent 
with similar expressions for $b^3$ and $c^3$. Notice that
the terms like $a^3$ produce just the determinant of the
corresponding $3 \times 3$ matrix. Since each determinant
contributes $6$ terms, altogether we have $18$ terms from the
first three terms in \,(\ref{i3v}).
The fourth term contains $27$ terms hence altogether $I_3$
contains precisely $45$ terms. This observation will be of
importance for setting up a finite geometric
interpretation  \cite{Levay:2009bp} of the structure of $I_3$.

The qutrits giving rise to this nice interpretation are again
real. After quantization the amplitudes $a,b$ and $c$ are
integer, and the cubic invariant $I_3$ is an
$E_{6(6)}({\mathds Z})$ invariant. 

\subsection{The qubic invariant, qutrits and generalized quadrangles}

A {\it finite generalized quadrangle} of order $(s, t)$, usually
denoted GQ($s, t$), is an incidence structure $S = (P, B, {\rm
I})$, where $P$ and $B$ are disjoint (non-empty) sets of objects,
called respectively points and lines, and where I is a symmetric
point-line incidence relation satisfying the following axioms
  \cite{678649}: (i) each point is incident with $1 + t$ lines ($t
\geq 1$) and two distinct points are incident with at most one
line; (ii) each line is incident with $1 + s$ points ($s \geq 1$)
and two distinct lines are incident with at most one common point;  and
(iii) if $x$ is a point and $L$ is a line not incident with $x$,
then there exists a unique pair $(y, M) \in  P \times B$ for which
$x {\rm I} M {\rm I} y {\rm I} L$; from these axioms it readily
follows that $|P| = (s+1)(st+1)$ and $|B| = (t+1)(st+1)$.

Given two points $x$ and $y$ of $S$ one writes $x \sim y$ and says
that $x$ and $y$ are collinear if there exists a line $L$ of $S$
incident with both. For any $x \in P$ denote
$x^{\perp} = \{y \in P | y \sim x \}$
 and note that $x \in x^{\perp}$;  obviously, $|x^{\perp}|
= 1+s+st$. Given an arbitrary subset $A$ of $P$, the {\it
perp}(-set) of $A$, $A ^{\perp}$, is defined as $A^{\perp} =
\bigcap \{x^{\perp} | x \in A\}$.

Here we shall be concerned with generalized
quadrangles having lines of size {\it three}, GQ$(2,t)$. From the
above-given restrictions one readily
sees that these are of three distinct kinds, namely GQ$(2,1)$,
GQ$(2,2)$ and $GQ(2,4)$. GQ$(2,1)$ is a grid of 9
points on 6 lines. GQ$(2,2)$ is the smallest thick generalized
quadrangle, also known as the ``doily." It is the pentagon-like
object shown within Figure 4. The pairs of numbers clearly show
its duad construction. This quadrangle is endowed with 15
points/lines, with each line containing 3 points and, dually, each
point being on 3 lines. The last case in the hierarchy is
$GQ(2,4)$, which possesses 27 points and 45 lines, with lines of
size 3 and 5 lines through a point.
 One of its
constructions goes as follows. One starts with the duad
construction of GQ$(2,2)$, adds 12 more points labelled simply as
$1,2,3,4,5,6,1',2',3',4',5',6'$ and defines 30 additional lines as
the three-sets $\{a,b',\{a,b\}\}$ of points, where $a,b \in
\{1,2,3,4,5,6\}$ and $a \neq b$. This process is diagrammatically
illustrated, after Polster   \cite{Polster}, in Figure 4.

\begin{figure}[t]
\centerline{\includegraphics[width=7.5truecm,clip=]{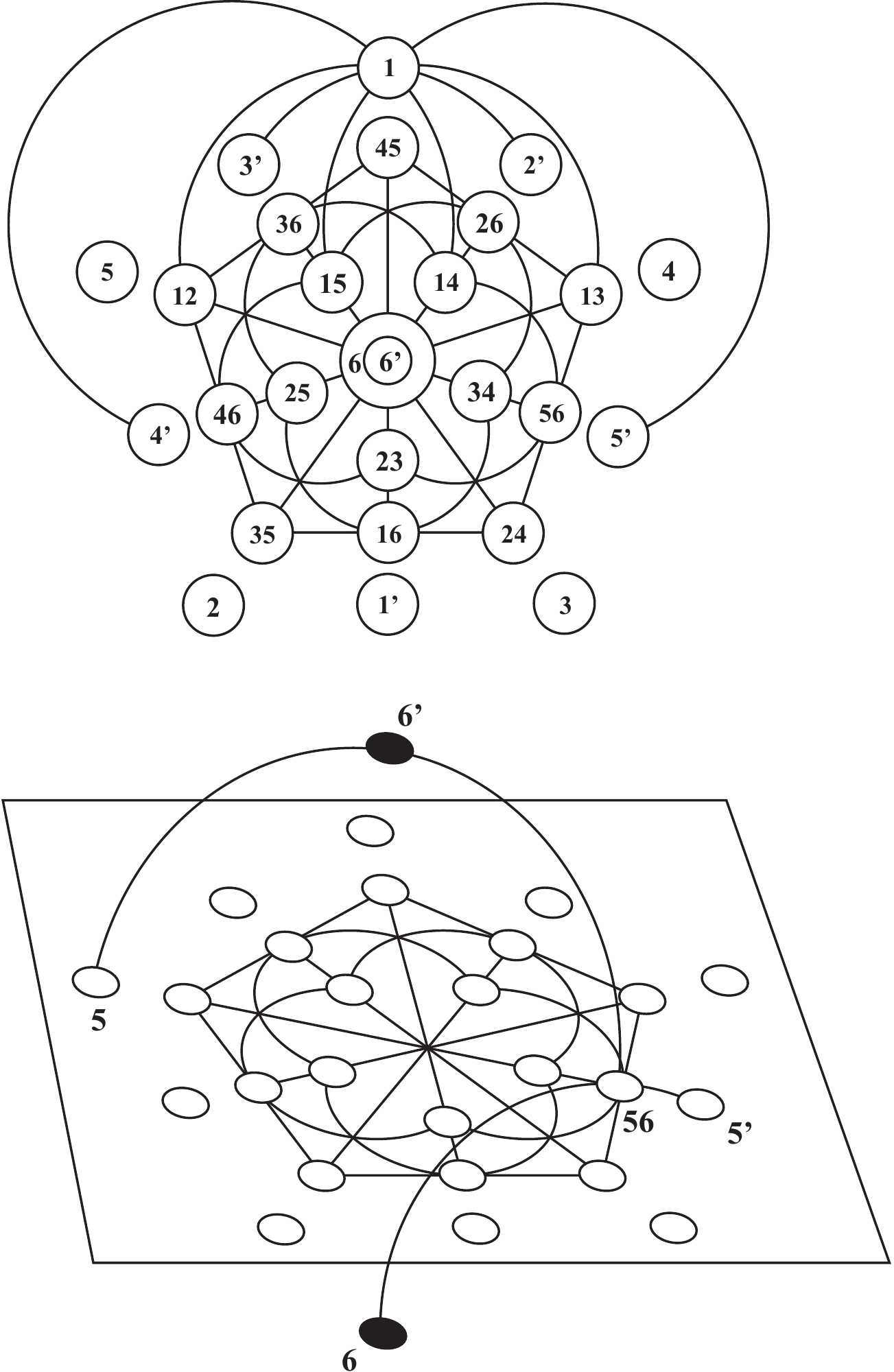}}
\vspace*{0.5cm} \caption{\label{duad}A diagrammatic illustration
of the structure of the generalized quadrangle GQ($2,4)$ after
Polster   \cite{Polster}. In both the figures, each picture depicts
all 27 points (circles). The top picture shows only 19 lines (line
segments and arcs of circles) of $GQ(2,4)$, with the two points
located in the middle of the doily being regarded as lying one
above and the other below the plane the doily is drawn in. 16 out
of the missing 26 lines can be obtained by successive rotations of
the figure through 72 degrees around the center of the pentagon.
The bottom picture shows a couple of lines which go off the
doily's plane; the remaining 8 lines of this kind are again got by
rotating the figure through 72 degrees around the center of the
pentagon.}
\end{figure}
 
   The
structure of the generalized quadrangle $GQ(2,4)$  nicely
encapsulates the structure of the cubic invariant $I_3$ {\it up
to signs}  \cite{Levay:2009bp}. Recall that the number
of lines (45)  matches the number of terms in the explicit
expression of the cubic invariant of (\ref{i3v}).  Writing
out explicitly $I_3$ one can  deduce the  labelling for
$GQ(2,4)$ described in Figure 5.

\begin{figure}[t]
\centerline{\includegraphics[width=10truecm,clip=]{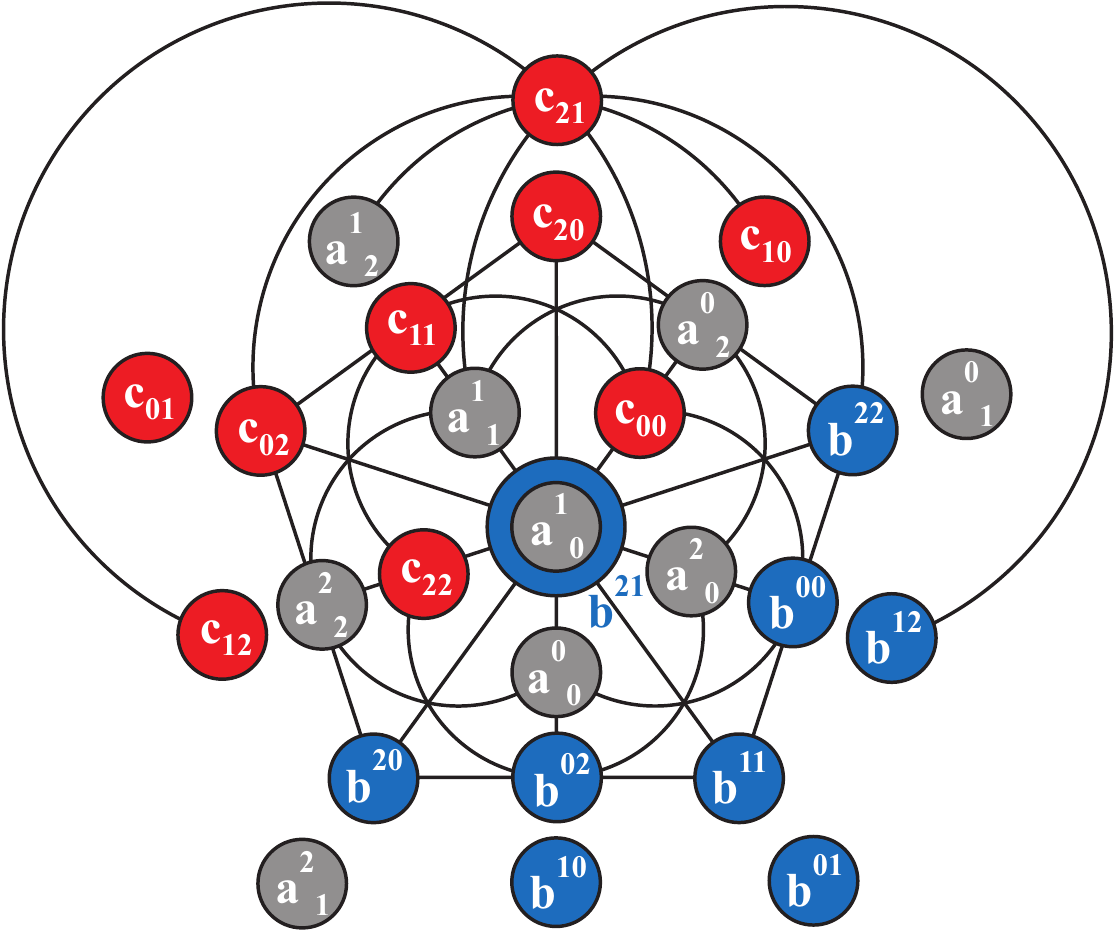}}
\vspace*{0.5cm} \caption{\label{qutritlab} A qutrit labelling of
the points of $GQ(2,4)$. Three different colours (online only) are
used to illustrate a triple of grids partitioning the point set.}
\end{figure}

 Notice that the three two-qutrit states of
\,(\ref{ampl}) partition the $27$ points of $GQ(2,4)$ to $3$
disjoint grids, i.\,e. $GQ(2,1)$s. 
The $27$
lines corresponding to the terms of ${\rm Tr}(abc)$ of
\,(\ref{i3v}) are of the type like the one
${a^1}_2b^{22}c_{21}$, and the $3\times 6=18$ terms are coming
from the three $3\times 3$ determinants $a^3,b^3,c^3$. These terms
are of the form  $b^{20}b^{02}b^{11}$.
One can check that each of $45$ lines of $GQ(2,4)$ correspond to
exactly one monomial of \,(\ref{i3v}).

It
is well-known that the automorphism group of the generalized
quadrangle $GQ(2,4)$ is the Weyl group  \cite{678649}  $W(E_6)$ of
order 51840. 
This group is a subgroup of the U-duality group $E_{6(6)}({\mathds Z})$.
For an explicit realization of this subgroup in a quantum information theoretic setting see  \cite{Levay:2009bp}.
The cubic invariant is  also connected to
the geometry of smooth (non-singular)) cubic  surfaces 
\cite{Manivel:2005}. It was Elie Cartan who first realized
  \cite{Cartan2} that the $45$ monomials of our cubic form
stabilized by $W(E_6)$ are in correspondence with the tritangent
planes of the cubic.

\subsection{Geometric hyperplanes and truncations}

A {\it geometric hyperplane} $H$ of a point-line geometry $\Gamma
(P,B)$ is a   proper subset of $P$ such that each line of $\Gamma$
meets $H$ in one or all points   \cite{Ronan:1987:EHD:38247.38256}. The only type of
hyperplanes featured in $GQ(2,4)$ are doilies (we have 36 of them)
and perp sets (their number is 27). Moreover, $GQ(2,4)$ also
contains $3 \times 40 = 120$ grids. However, these are {\it not}
its geometric hyperplanes   \cite{Saniga:2009ik}. (This is quite different
from the $GQ(2,2)$ case, where grids are geometric hyperplanes.)
Though they are not hyperplanes, they have an important property: there exits $40$ triples of them, each partitioning the point
set of $GQ(2,4)$.

A decomposition of $E_{6(6)}$ directly related to a doily as a
geometric hyperplane sitting inside $GQ(2,4)$ is the following one
  \cite{Borsten:2008wd,Baez:2001dm} \beq E_{6(6)}\supset \SL(2, \R)\times
\SL(6, \R) \eeq \noindent under which \beq {\bf 27}\rightarrow ({\bf
2},{\bf 6})\oplus ({\bf 1},{\bf 15}). \eeq \noindent One can show
that under this decomposition $I_3$ schematically factors as \beq
I_3={\rm Pf}(A)+u^TAv, \label{Pfaffdecomp} \eeq \noindent where
$u$ and $v$ are two six-component vectors and ${\rm Pf}(A)$ is the
Pfaffian of an antisymmetric $6\times 6$ matrix with $15$
independent components. Clearly the decomposition above featuring
the doily is nicely mapped to the duad construction of $GQ(2,4)$,
see Figure 4.

The next important type of subconfiguration of $GQ(2,4)$ is the
grid. The decomposition underlying this type of subconfiguration
is the one given by \,(\ref{griddecomp}). It is also obvious
that the $40$ triples of pairwise disjoint grids are intimately
connected to the $40$ different ways we can obtain a qutrit
description of $I_3$. Note that there are $10$ grids which are
geometric hyperplanes of a particular copy {\it of the doily} of
$GQ(2,4)$. This is related to the fact that the quaternionic magic
case with $15$ charges can be truncated to the 9 charge complex  case.

The second type of hyperplanes we should consider are perp-sets.
Perp-sets are obtained by selecting an arbitrary point and
considering all the points collinear with it. Since we have five
lines through a point, any perp set has $1+10=11$ points. A
decomposition which corresponds to perp-sets is thus of the form
  \cite{Borsten:2008wd} \beq E_{6(6)}\supset \SO(5,5)\times \SO(1,1) \eeq
\noindent under which \beq {\bf 27}\rightarrow {\bf 16}_{\bf
1}\oplus {\bf 10}_{\bf -2}\oplus {\bf 1}_{\bf 4}. \eeq \noindent
This is the usual decomposition of the U-duality group into the
$T$ duality and $S$ duality. It is interesting to see that the
last term (i.\,e. the one corresponding to the fixed/central point
in a perp-set) describes the $NS$ five-brane charge. Notice that
we have five lines going through this fixed point of a perp-set.
These correspond to the $T^5$ used to compactify    type II
string theory to five dimensions. The two remaining points on each of these $5$
lines correspond to $2\times 5=10$ charges.  They correspond to
the $5$ directions of $KK$ momentum and the $5$ directions of
fundamental string winding. In this picture the $16$ charges {\it
not belonging to} the perp-set correspond to the $16$ D-brane
charges. Notice that we can get $27$ similar truncations based on
the $27$ possible central points of the perp-set.

\subsection{Three-qubit operators and the cubic invariant}

In the previous subsections we managed to understand the structure
of the cubic invariant giving rise to the black hole entropy of
five dimensional black holes and black strings in terms of
qutrits. Now we show that interestingly there is a dual way of
understanding $I_3$ using the {\it real three-qubit Pauli
group}  \cite{Levay:2009bp}. This way of looking at $I_3$
provides a geometric framework for understanding the connection
between the $d=4$ and $d=5$ duality groups i.e. $E_{7(7)}({\mathds
Z})$ and $E_{6(6)}({\mathds Z})$.

Let us define the {\it real} three qubit Pauli operators by
introducing the notation   \cite{Levay:2008mi} $X\equiv {\sigma}_1$,
$Y=i{\sigma}_2$ ,and $Z\equiv {\sigma}_3$; here, ${\sigma}_j,
j=1,2,3$ are the usual $2 \times 2$ Pauli matrices. Then we can
define the real operators of the three-qubit Pauli
group  \cite{Nielsen:2000,1996PhRvA..54.1862G,2010JMP....51l2203C} by forming the tensor
products of the form $ABC \equiv A \otimes B \otimes C$ that are
$8 \times 8$ matrices. All possible combinations of these
operators of the form $\pm ABC$ make up the real Pauli
group, a set of 128 matrices endowed with the usual matrix
multiplication.
Notice that operators containing an
even number of $Y$s are symmetric and those containing an
odd number of $Y$s are  antisymmetric. From the $64$ possible
combinations of the form $ABC$ we have $36$ symmetric matrices and
$28$ antisymmetric.

From the set of antisymmetric matrices let us chose the seven
element subset \beq (g_1,g_2, g_3,g_4,g_5,g_6,g_7
)=(IIY,ZYX,YIX,YZZ,XYX,IYZ,YXZ) \label{cliff} \eeq \noindent
satisfying the relation $\{g_a,g_b\}=-2{\delta}_{ab},\quad
a,b=1,2,\dots 7$. These operators form the generators of a
seven-dimensional Clifford algebra. The remaining $21$
antisymmetric operators are of the form
$\frac{1}{2}[g_a,g_b]$. They generate
an $\mathfrak{so}(7)$ algebra. We relate these matrices to the generators of
$\SO(8)$ using \beq
-{\Gamma}^{0a}=g_{0a}\equiv g_a,\qquad
-{\Gamma}^{ab}=g_{ab}\equiv\frac{1}{2}[g_a,g_b]. \label{so8} \eeq
\noindent We can make use of these three-qubit operators for
expanding the $\mathcal{N}=8$ central charge ${\cal Z}_{AB}$ as in
(\ref{relation}).

Note that the decomposition \beq E_{7(7)}\supset E_{6(6)}\times
\SO(1,1) \label{45decomp} \eeq \noindent under which \beq {\bf
56}\rightarrow {\bf 1}\oplus{\bf 27}\oplus{\bf
27}^{\prime}\oplus{\bf 1}^{\prime} \eeq \noindent describes the
relation between the $D=4$ and $D=5$ duality groups
  \cite{Ferrara:1996um,Kallosh:1996uy,Cvetic:1996zq,Balasubramanian:1997az,Bertolini:2000ei,Andrianopoli:1997wi}. In order to connect the qutrit and
three-qubit operator pictures we assign to one of the three-qubit
operators a special status \beq
{\Omega}=IIY=g_1. \label{omega} \eeq
\noindent Now we use the $\mathcal{N}=8$ central charge parametrized as in
\,(\ref{relation}) and look at the structure of the cubic
invariant. It can be also be written in the alternative form
  \cite{2007stmt.book.....B} \beq I_3=-\frac{1}{48}{\rm Tr}(\Omega {\cal Z}\Omega
{\cal Z}\Omega {\cal Z}). \label{nicecube} \eeq \noindent In order
to get the correct number of components, we impose the 
constraints   \cite{Ferrara:2006yb} \beq {\rm Tr}(\Omega {\cal Z})=0,\qquad
\overline{\cal Z}=\Omega {\cal Z}{\Omega}^T. \label{cubeconstr}
\eeq \noindent The first of these 
restricts the number of antisymmetric matrices to be considered in
the expansion of ${\cal Z}$ from $28$ to $27$. The second
constraint is the usual reality condition which restricts the $27$
{\it complex} expansion coefficients to $27$ {\it real} ones.
The group theoretical
meaning of these constraints is the expansion of the $\mathcal{N}=8$ central
charge in an $\USp(8)$ basis, which is appropriate since $\USp(8)$
is the automorphism group of the $\mathcal{N}=8$, $D=5$ supersymmetry
algebra.

It is easy to see that the reality constraint yields 
for $\Omega{\cal Z}$ the form \beq
\Omega{\cal Z}={\cal S}+i{\cal A}\equiv
x^{jk}g_{1jk}+i(y_{0j}g_{1j}-y_{1j}g_j). \label{hudefontos}\eeq
\noindent 
Performing standard manipulations, we get \beq
I_3=-\frac{1}{48}\left( {\rm Tr}({\cal SSS})-3{\rm Tr}({\cal
SAA})\right). \label{i3vv} \eeq \noindent Hence, with the notation
\beq A^{jk}\equiv x^{j+1k+1}, \qquad u_j\equiv y_{1j+1},\qquad
v_j\equiv y_{0j+1},\qquad j,k=1,2,\dots,6, \eeq \noindent the
terms of \,(\ref{i3vv}) give rise to the form of
\,(\ref{Pfaffdecomp}). Notice also  that the parametrization
\beq
u^T=\begin{pmatrix}-c_{21},&-{a^{2}}_1,&-b^{01},&-{a^0}_1,&c_{01},&b^{21}\end{pmatrix},
\quad
v^T=\begin{pmatrix}b^{10},&-c_{10},&{a^1}_2,&c_{12},&b^{12},&{a^1}_0\end{pmatrix}
\eeq \noindent \beq
A=\begin{pmatrix}0&c_{02}&b^{22}&-c_{00}&{a^1}_1&b^{02}\\
c_{02}&0&{a^0}_0&b^{11}&c_{22}&{-a^0}_2\\
-b^{22}&-{a^0}_0&0&{a^2}_0&b^{20}&c_{11}\\
c_{00}&-b^{11}&-{a^2}_0&0&c_{20}&{a^2}_2\\
-{a^1}_1&-c_{22}&-b^{20}&-c_{20}&0&-b^{00}\\
-b^{02}&{a^0}_2&-c_{11}&-{a^2}_2&b^{00}&0\end{pmatrix} \eeq
\noindent yields for $I_3$ its qutrit version of \,(\ref{i3v}).

\subsection{Mermin squares}

At this point it is instructive to have a look again at the finite
geometric structure of $I_3$. The careful reader might have
noticed that there is one important issue we have not clarified
yet.  We have established a connection between the qutrit
interpretation and the structure of the generalized quadrangle
$GQ(2,4)$. However, our labelling of the points of $GQ(2,4)$ by
the real $3\times 3$ matrices $a,b$ and $c$ serving as qutrit
amplitudes did not manage to take care of the {\it signs} of the
$45$ terms showing up in $I_3$. It is easy to see that no
distribution of charges for these amplitudes is available matching
the structure of $I_3$ and the incidence structure of $GQ(2,4)$ at
the same time.

The reason for this is very simple. According to Figure
\ref{qutritlab} the points of $GQ(2,4)$ can be split into three
grids. Moreover, according to (\ref{i3v}) the
relevant part of $I_3$ answering a particular grid is just the
$3\times 3$ determinant of the corresponding two-qutrit state. The
structure of this determinant is encapsulated in the structure of
the corresponding grid. We can try to arrange the $9$ amplitudes
in a way that the $3$ plus signs for the determinant should occur
along the rows and the $3$ minus signs along the columns. But this
is impossible since multiplying all of the nine signs  ``row-wise"
yields a plus sign, but ``column-wise" yields a minus sign.

Readers familiar with Bell-Kochen-Specker type theorems ruling out
noncontextual hidden variable theories may immediately suggest
that if we have failed to associate signs with the points of the
grid, what about trying to use noncommutative objects instead?
More precisely, we can try to associate objects that are generally
noncommuting but that are pairwise commuting along the lines of
the grid. This is exactly what is achieved by using Mermin squares
  \cite{Mermin,Mermin2,1991JPhA...24L.175P}. Mermin squares are obtained by
assigning pairwise commuting  {\it two-qubit Pauli matrices} to
the lines of the grid in such a way that the naive sign assignment
does not work, but we get the identity operators with the correct
signs by multiplying the operators row- and column-wise.

However, we have merely $16$ real two-qubit Pauli operators up to
sign, which is simply not enough to label the 27 points of our
$GQ(2,4)$. Hence we are forced to try the next item in the line:
namely some subset of the real three-qubit Pauli group.
Let us recall the duad
labelling of $GQ(2,4)$ as discussed in Figure \ref{duad}.
According to this a natural {\it noncommutative labelling} for the
27 points of $GQ(2,4)$ is the following. Let us remove the special
operator $\Omega$ from the $28$ antisymmetric ones. Then set up
the correspondence between the points of Figure \ref{duad} and the
remaining operators as
 \beq
\{1^{\prime},2^{\prime},3^{\prime},4^{\prime},5^{\prime},6^{\prime}\}\leftrightarrow
\{g_2,g_3,g_4,g_5,g_6,g_7\},\quad
\{1,2,3,4,5,6\}\leftrightarrow\{g_{12},g_{13},g_{14},g_{15},g_{16},g_{17}\}\label{six2}
\eeq \noindent \beq \{12,13,14,15,16,23,24,25,26\}\leftrightarrow
\{g_{23},g_{24},g_{25},g_{26},g_{27},g_{34},g_{35},g_{36},g_{37}\}
\label{doily1}\eeq \noindent \beq
\{34,35,36,45,46,56\}\leftrightarrow
\{g_{45},g_{46},g_{47},g_{56},g_{57},g_{67}\} \label{doily2}\eeq
\noindent i.\,e., by shifting all the indices of $g_{IJ}$ not
containing $0$ or $1$ by $-1$ we get the duad labels.

However, in order for this noncommutative labelling of $GQ(2,4)$
to represent a generalization of a Mermin square:  (i) the
operators on each line should be pairwise commuting and (ii)
at the same time their products (not depending on the order)
should produce the identity operator up to sign. It is easy to
check that the noncommutative labelling above fails to satisfy
these criteria  \cite{Levay:2009bp}.

\begin{figure}[t]
\centerline{\includegraphics[width=10truecm,clip=]{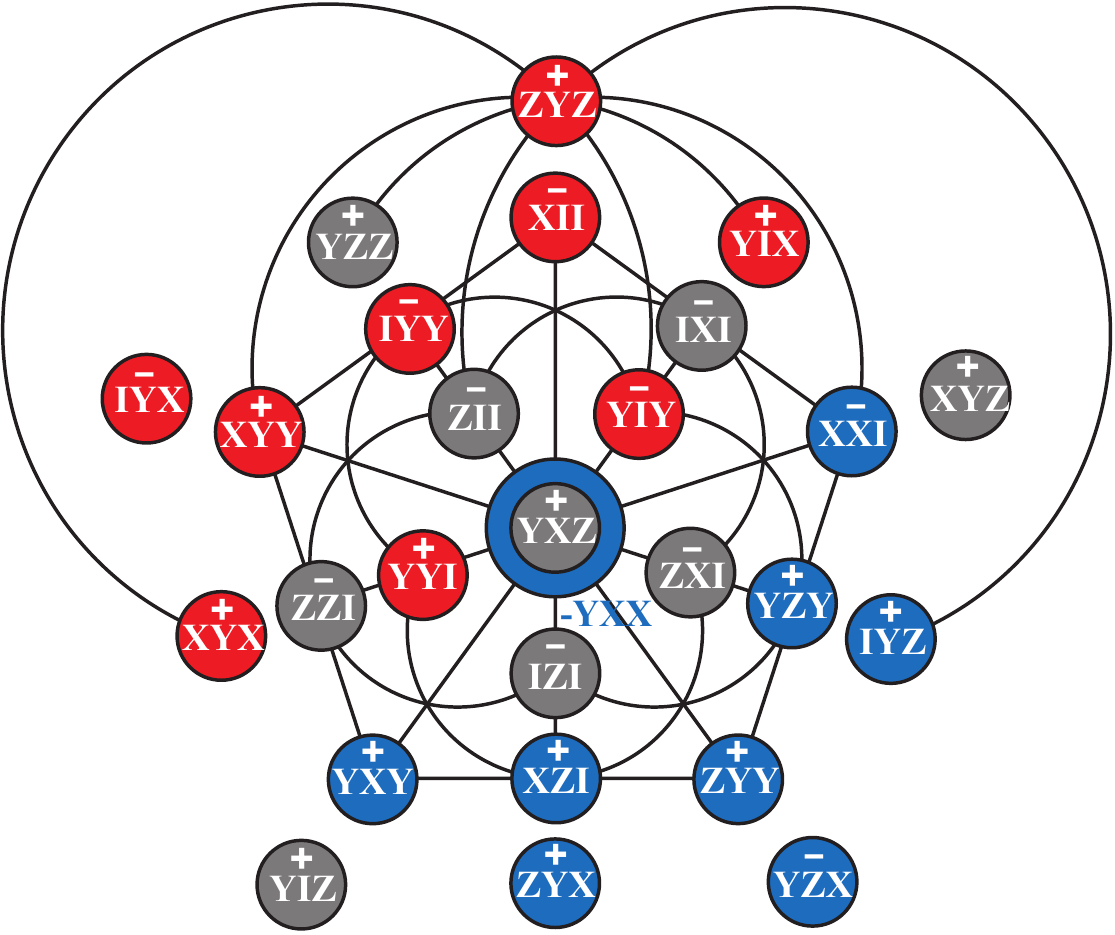}}
\vspace*{0.5cm} \caption{\label{noncomm}An illustration of the
non-commutative labelling of the points of $GQ(2,4)$. For better
readability of the figure, the sign of an operator is placed above
the latter.}
\end{figure}

Luckily this is  easily remedied. Notice that our special
operator ${\Omega}$ of (\ref{omega}) commutes with all of
the operators  in (\ref{doily1})-(\ref{doily2}).
Hence we can multiply the operators of
(\ref{doily1})-(\ref{doily2}) by $\Omega$ from either side.
One can then check that the resulting labelling, with $12$
antisymmetric and $15$ symmetric operators, now satisfies the
criteria required by a genuine generalization of a Mermin square. In summary for a Mermin-like noncommutative labelling for
$GQ(2,4)$: use (\ref{six2}), and for the
remaining points the new labels \beq
\{12,13,14,15,16,23,24,25,26\}=
\{g_{123},g_{124},g_{125},g_{126},g_{127},g_{134},g_{135},g_{136},g_{137}\},
\label{3v} \eeq \noindent \beq \{34,35,36,45,46,56\}=
\{g_{145},g_{146},g_{147},g_{156},g_{157},g_{167}\}. \label{4v}
\eeq \noindent Using the explicit form of the $8 \times 8$
matrices $g_a$ of \,(\ref{cliff}), we  get
three-qubit operators with a natural choice of signs as
non-commutative labels for the points of $GQ(2,4)$. This
is displayed in Figure \ref{noncomm}.

Let us now recall (\ref{nicecube}), (\ref{hudefontos}) and
(\ref{entropycubic}). These formulae relate our considerations
based on Mermin squares to the structure of the black hole
entropy. The expression in (\ref{hudefontos}) clearly shows
that the charges are expansion coefficients of $\Omega{\cal Z}$
with respect to basis vectors that are precisely our
noncommutative labels for $GQ(2,4)$. Hence employing the simple
criteria  (i) and (ii) for constructing Mermin square-like
configurations for the generalized quadrangle $GQ(2,4)$ lead us
naturally to a finite geometric understanding of the structure of
the black hole entropy formula. Recall that for
$GQ(2,1)$ (the grid) we have an ordinary Mermin square with
entropy formula related to the determinant as and entanglement
measure for a two qutrit system, for $GQ(2,4)$ (the doily) we have
a Mermin square-like configuration with entropy related to the
Pfaffian (see (\ref{Pfaffdecomp})). These observations put our
considerations on the structure of the $D=5$ semiclassical black
hole entropy into a nice unified picture based on ``Mermin-squares"
associated to generalized quadrangles of type $GQ(2,t)$.
We note in closing that there are other interesting subconfigurations of finite geometries called ovoids
that can be associated to Mermin pentagrams  \cite{Mermin2,2012EL.....9750006S}.
Their possible interpretation within the framework of the BHQC is unclear.

\section{$STU$ Black Holes and attractors.}

\subsection{The $STU$ model}

In the previous section we saw how the structure of the macroscopic Bekenstein-Hawking entropy is encoded into entanglement measures of several qubits and qutrits.
Apart from the BHQC contributing to our understanding of structural issues concerning black-hole entropy in quantum information theoretic terms, the desire for an entanglement based understanding for issues of dynamics also arose.
In this section we would like to discuss results connected to the dynamics of the well-known attractor mechanism \cite{Ferrara:1995ih,Strominger:1996kf,Ferrara:1996dd} in the special case of the $STU$ model \cite{Duff:1995sm,Bellucci:2008sv}.

Theories such as the $STU$ model arise in string theory, whose low energy limit is described by two derivative supergravity theories involving massless fields of spins $\leq 2$. We consider the effective action for these fields to leading order in the string coupling constant and the inverse tension. Since these string theories (and M-theory)  live in ten (eleven) spacetime dimensions we have to deduce the four dimensional massless
spectrum by compactification of the extra dimensions. This process is effected by decomposing the $10$-dimensional fields according to harmonic forms determined by the cohomology of the extra dimensions. Thus, the geometric data of these spaces gives rise to extra fields in the low energy effective four dimensional theory. Among these fields especially important are the so-called moduli, massless scalar fields describing the fluctuation of the shape and size of the extra dimensions. The space of  deformation parameters of ``size and shape'' is called the moduli space.  The scalar fields in the four-dimensional spacetime manifold  take values in this space.

The massless spectrum of string theories  also contains Maxwell-like fields described by differential forms.
Like the familiar Maxwell field, which is a one-form coupled to the world-line of point-like objects, these $(p+1)$-form fields  couple to the world-volumes of extended objects called $p$-branes.
In the low energy effective four-dimensional picture such objects also give rise to Maxwell fields ($\U(1)$ gauge fields) with their couplings depending on the scalar (moduli) fields.

As we have already mentioned the low energy four-dimensional actions are supergravity theories implying that accompanying   the {\it bosonic} fields, namely the metric, the scalar fields and the Maxwell fields, are their {\it fermionic} counterparts. Supersymmetry transformations mix the fermionic and bosonic fields. Since our main concern here is finding the classical black hole solutions
we can restrict our attention merely to the bosonic part of the Lagrangian by setting the fermionic fields to zero. Nevertheless we shall be primarily concerned with solutions that preserve some fraction of the supersymmetry. In other words they admit covariantly constant (Killing) spinors. These solutions will  be called {\it supersymmetric} or BPS (Bogolmolny, Prasad, Summerfield).

The $STU$ model is a rigid $\mathcal{N}=2$ supergravity model in $D=4$, coupled to three vector multiplets.   The $D=4, \mathcal{N}=2$  supergravity multiplet contains the metric (graviton), two spin-3/2 fermions (gravitini)  and one spin-1 gauge potential $A_\mu$,  the so-called ``graviphoton''. The vector multiplets each consist of one gauge potential, two spin-1/2 fermions (gaugini) and a single complex scalar field. 
In summary, the bosonic sector is precisely of the form we discussed above. Namely it contains the four-dimensional spacetime metric $g_{\mu\nu}$, three complex scalar fields $z^j, j=1,2,3$, and $4$ $\U(1)$ vector fields ${\cal A}_{\mu}^I$  with field strengths ${\cal F}_{\mu\nu}^I$ where
$I=0,1,2,3$. Sometimes the three complex scalars are denoted by the letters $S,T$ and $U$,
hence the name of the model.

There are a number of different ways of obtaining the  $STU$ model
from string or M-theory compactifications \cite{Duff:1995sm}.
For example, when type IIA string theory is compactified on a six torus $T^6$
(or equivalently when M-theory is compactified on a $T^7$) one recovers $\mathcal{N}=8$ supergravity in $D=4$ with $28$ vectors and $70$ scalars. The moduli space is the coset space $E_{7(7)}/\SU(8)$. This theory with an on shell $U$-duality
symmetry $E_{7(7)}$ is already familiar from our considerations of the tripartite entanglement of seven qubits.
There we  saw that the $STU$ model is a consistent $\mathcal{N}=2$ truncation of this $\mathcal{N}=8$ model. We have seven equivalent $STU$ truncations corresponding to the seven points of the dual Fano plane. One may also obtain the $STU$ model directly by orbifold compactification \cite{Sen:1995ff}. This version comes with an additional four hypermultiplets and is the one obtained by truncating the Fano plane from 7 lines to one. These hypermultiplets will not play a role in the present paper, however.

It is therefore not surprising that the study of $STU$ black hole solutions is very important.
For example,  the  single-center 1/2-BPS solutions of the $STU$ model with non-zero Bekenstein-Hawking entropy may be embedded in the 1/8-BPS solutions with non-zero Bekenstein-Hawking entropy of the $\mathcal{N}=8$ model.
This implies that in order to generate the most general solution one has to act with and $\SU(8)$ transformation rotating the $4+4$ charges of the STU model associated with the $4$ $\U(1)$ gauge fields
to the $28+28$ charges of the $\mathcal{N}=8$ theory.
In the language of group theory this process is encapsulated in (\ref{stu222}).
The charges of the $STU$ model correspond to a singlet of $\SO(4,4)$ and transform
as a three-qubit state, i.e. a $({\bf 2},{\bf 2},{\bf 2})$ under $[\SL(2,{\mathds Z})]^{\otimes 3}$.
Although in the $STU$ model we have just  $6$ real ($3$ complex) scalars
one can generate generic values of the scalars via applying an $E_{7(7)}$ transformation.
Thus in many ways, the $STU$ model serves as a basic building block.

\subsection{$STU$ black holes as four-qubit systems}
\subsubsection{Timelike dimensional reduction of the $STU$
Lagrangian}

The bosonic part of the action of the $STU$ model is
\begin{eqnarray}
\label{action11}
{\cal S}&=&\frac{1}{16\pi}\int d^4x\sqrt{\vert g\vert }\{-\frac{R}{2}+G_{i\overline{j}}{\partial}_{\mu}z^i{\partial}_{\nu}{\overline{z}}^{\overline{j}}g^{\mu\nu}\nonumber\\&+&({\rm Im}{\cal N}_{IJ}{\cal F}^I\cdot{\cal F}^J+{\rm Re}{\cal N}_{IJ}{\cal F}^I\cdot{\star{\cal F}^J})\}
\end{eqnarray}
\noindent
Here $\star{\cal F}$ refers to the Hodge-dual of the two-form ${\cal F}$ and  ${\cal F}\cdot{\cal F}={\cal F}_{\mu\nu}{\cal F}^{\mu\nu}$.
The manifold of the scalar fields for the $STU$ model is
$[\SL(2, \mathds R)/\U(1)]^{\otimes 3}$.
In the following we will denote the three complex scalar fields as
\beq
z^j\equiv x^j-iy^j,\qquad j=1,2,3,\qquad y^j>0.
\label{moduli}
\eeq
\noindent
With these definitions the metric on the scalar manifold (moduli space) is
\beq
G_{i\overline{j}}=\frac{\delta_{i\overline{j}}}{(2y^i)^2}\qquad
.
\label{targetmetric}
\eeq
\noindent
The metric above can be derived from the K\"ahler potential
\beq
K= -\log(8y_1y_2y_3)
\label{Kahler}
\eeq
\noindent
as $G_{i\overline{j}}={\partial}_i{\partial}_{\overline{j}}K$.
For the $STU$ model the scalar dependent vector couplings ${\rm Re}{\cal N}_{IJ}$
and ${\rm Im}{\cal N}_{IJ}$ take the following form
\beq
\nu_{IJ}\equiv{\rm Re}{\cal N}_{IJ}=\begin{pmatrix}2x_1x_2x_3&-x_2x_3&-x_1x_3&-x_1x_2\\-x_2x_3&0&x_3&x_2\\-x_1x_3&x_3&0&x_1\\-x_1x_2&x_2&x_1&0\end{pmatrix}
\label{valos}
\eeq
\noindent
\beq
{\mu}_{IJ}\equiv{\rm Im}{\cal N}_{IJ}=-y_1y_2y_3\begin{pmatrix}1+{\left(\frac{x_1}{y_1}\right)}^
2+{\left(\frac{x_2}{y_2}\right)}^2+{\left(\frac{x_3}{y_3}\right)}^2&-\frac{x_1}{
y_1^2}&-\frac{x_2}{y_2^2}&-\frac{x_3}{y_3^2}\\-\frac{x_1}{y_1^2}&\frac{1}{y_1^2}
&0&0\\-\frac{x_2}{y_2^2}&0&\frac{1}{y_2^2}&0\\-\frac{x_3}{y_3^2}&0&0&\frac{1}{y_
3^2}\end{pmatrix}
\label{kepzetes}
\eeq
\noindent
Our aim is to describe stationary solutions of the Euler-Lagrange equations arising from the  $STU$  action (\ref{action11})   in an entanglement based language.

It is well-known that the most general ansatz for stationary solutions in four dimensions is  \cite{Breitenlohner:1987dg}
\beq
ds^2=-e^{2U}(dt+\omega)^2+e^{-2U}h_{ab}dx^adx^b
\label{ans1}
\eeq
\noindent
\beq
{\cal F}^{I}=d{\cal A}^{I}=d(\xi^I(dt+\omega)+ A^I)
\label{ans2}
\eeq
\noindent
where $a,b=1,2,3$ correspond to the spacial directions. The quantities $U$,     $\xi^I$, $A^I_a$, $\omega_a$ and $h_{ab}$ are regarded as $3D$ fields, i.e.
the ansatz above corresponds to dimensional reduction to $D=3$ along the timelike direction.
In achieving this we have chosen the gauge such that the Lie-derivative of ${\cal A}^I$ with respect to the timelike Killing vector vanishes, and have chosen coordinates such that the isometry corresponding to this Killing vector is just a (time) translation. In this case the quantities in (\ref{ans1}-\ref{ans2}) depend only on $x^a$, $a=1,2,3$.

After performing the dimensional reduction to $D=3$ our starting Lagrangian of
 (\ref{action11}) takes the following form  \cite{Breitenlohner:1987dg,Gaiotto:2007ag}
\beq
{\cal L}={\cal L}_1+{\cal L}_2+{\cal L}_3
\eeq
\noindent
where
\beq{\cal L}_1=-\frac{1}{2}\sqrt{h}R[h]+dU\wedge\star dU+\frac{1}{4}e^{-4U}
(d\sigma
+\tilde{\xi}_Id \xi^I
-\xi^Id\tilde{\xi}_I
)\wedge\star(d\sigma+\tilde{\xi}_Jd \xi^J
-\xi^Jd \tilde{\xi}_J
)
\label{L1}
\eeq
\noindent
\beq
{\cal L}_2=G_{i\overline{j}}dz^i\wedge\star d\overline{z}^{\overline{j}}
\label{L2}
\eeq
\noindent
\beq
{\cal L}_3=\frac{1}{2}e^{-2U}\mu_{IJ}d\xi^I\wedge\star d\xi^J+\frac{1}{2}e^{-2U}\mu^{IJ}(d\tilde{\xi}_I-\nu_{IK}d\xi^K)\wedge\star(d\tilde{\xi}_J-\nu_{JL}d\xi^L).
\label{L3}
\eeq
\noindent
\noindent
Here the new (axionic) scalars $\sigma$ and $\tilde{\xi}_I$ come from dualizing $\omega$ and $A^I$ by  \cite{Breitenlohner:1987dg}
\beq
d\tilde{\xi}_I\equiv
\nu_{IJ}d\xi^J
-e^{2U}\mu_{IJ}\star(dA^J+\xi^Jd\omega)
\eeq
\noindent
\beq
d\sigma\equiv e^{4U}\star d\omega
+\xi^I d\tilde{\xi}_I
-\tilde{\xi}_Id\xi^I
.
\label{sigmaegy}
\eeq
\noindent
Note also that here  the exterior derivative is understood on the spatial slice with local coordinates $x^a$, $a=1,2,3$.

The dimensionally reduced Lagrangian ${\cal L}$ can be written in
the nice form of 3-dimensional  gravity coupled to a nonlinear sigma model
defined on the spatial slice with target manifold  \cite{Bossard:2009we}
${\cal M}_3=\SO(4,4)/[\SL(2,{\mathds R})]^{\otimes 4}$ with the
Lagrangian \beq {\cal
L}=-\frac{1}{2}\sqrt{h}R[h]+g_{mn}{\partial}_a{\Phi}^m{\partial}^a{\Phi}^n\label{teljeslag}
\eeq \noindent where ${\Phi}^m, m=1,2,\dots 16$ refers to the
scalar fields: $U,\sigma,\xi^I,\tilde{\xi}_I, z^j,
\overline{z}^{\overline{j}}$ with $I=0,1,2,3$ and $j=1,2,3$. Here
the line element on ${\cal M}_3$ defines $g_{mn}$ as $ds^2_{{\cal
M}_3}= g_{mn}d{\Phi}^md{\Phi}^n$ with the explicit form
\begin{eqnarray}
\frac{1}{4}ds^2_{{\cal M}_3}&=& G_{i\overline{j}}(z,\overline{z})dz^id\overline{
z}^{\overline{j}}+dU^2+
\frac{1}{4}e^{-4U}(d\sigma+\tilde{\xi}_Id \xi^I
-\xi^Id\tilde{\xi}_I
)^2\nonumber\\&+&
\frac{1}{2}e^{-2U}\left[\mu_{IJ}d\xi^Id\xi^J+
\mu^{IJ}(d\tilde{\xi}_I-\nu_{IK}d\xi^K)(d\tilde{\xi}_J-\nu_{J
L}d\xi^L)\right].
\label{modulimetrika}
\end{eqnarray}
\noindent

\subsubsection{The line element as a four-qubit measure}

We have seen that in the $3D$ picture the moduli space is the
coset ${\cal M}_3=\SO(4,4)/\SL(2,{\mathds R})^{\otimes 4}$. Due to
the presence of $\SL(2,{\mathds R})^{\otimes 4}$ which is a subgroup
of the real SLOCC group $\GL(2,{\mathds R})^{\otimes 4}$ one should
be tempted to try a four-qubit reformulation incorporating all the
quantities of the $STU$ model.

 In order to do this note that our  coset can be locally parametrized by $16$ independent quantities.
These are the $6$ quantities $(x_j,y_j), j=1,2,3$ coming from the scalar fields
of (\ref{moduli}), the $8$ potentials
${\xi}^I,\tilde{\xi}_I$, the NUT potential  \cite{1963JMP.....4..915N} $\sigma$ defined by 
(\ref{sigmaegy}) and the warp factor $U$ showing up in the metric ansatz of (\ref{ans1}).
We introduce new quantities
\beq                                                                            \zeta^I\equiv\sqrt{2}\xi^I,\qquad \tilde{\zeta}_I=\sqrt{2}\tilde{\xi}_I,\qquad
x_0\equiv \sigma,\qquad y_0\equiv e^{\phi}=e^{2U}.
\label{ujkoord}
\eeq
\noindent
Then in the Iwasawa parametrization  \cite{Levay:2010ua} we can
describe our coset by the matrix $\mathcal{V}$
\beq
\mathcal{V}=
\begin{pmatrix}M_3\otimes M_2&0\\0&M_1\otimes M_0\end{pmatrix}
\begin{pmatrix}{\bf 1}&-\zeta g\\\zeta^Tg&{\bf 1}+\frac{1}{2}\Delta\end{pmatrix}.
\label{coset} \eeq \noindent Here \beq
M_{\alpha}\equiv\frac{1}{\sqrt{y_{\alpha}}}\begin{pmatrix}1&-x_{\alpha}\\0&
y_{\alpha}\end{pmatrix} \qquad \alpha=0,1,2,3 \label{mmatrix}
\eeq \noindent \beq
\Delta=\begin{pmatrix}{\zeta}^{(0)}\cdot{\zeta}^{(0)}&
{\zeta}^{(0)}\cdot{\zeta}^{(1)}\\{\zeta}^{(0)}\cdot{\zeta}^{(1)}&{\zeta}^{(1)}\cdot{\zeta}^{(1)}\end{pmatrix}\varepsilon\otimes
\begin{pmatrix}0&1\\0&0\end{pmatrix}. \label{cayleys} \eeq
\noindent Here the $\SL(2)\times \SL(2)$ invariant $\cdot$ product
of two four component vectors is defined with respect to the
$4\times 4$ matrix  $g=\varepsilon\otimes\varepsilon$, with
$\varepsilon$ defined as in (\ref{kanonikus}). The
$4$-component vectors $\zeta^{(0)}$ and $\zeta^{(1)}$ are just the
first and third columns of the matrix $\zeta$ defined as \beq
\zeta\equiv
\begin{pmatrix}{\zeta}_{0000}&{\zeta}_{0001}&{\zeta}_{0010}&{\zeta}_{0011}\\
{\zeta}_{0100}&{\zeta}_{0101}&{\zeta}_{0110}&{\zeta}_{0111}\\
{\zeta}_{1000}&{\zeta}_{1001}&{\zeta}_{1010}&{\zeta}_{1011}\\
{\zeta}_{1100}&{\zeta}_{1101}&{\zeta}_{1110}&{\zeta}_{1111}\end{pmatrix}=
\begin{pmatrix}-\tilde{\zeta}_0&0&\tilde{\zeta}_1&0\\
                            \tilde{\zeta}_2&0&{\zeta}^3&0\\
                          \tilde{\zeta}_3&0&{\zeta}^2&0\\
              {\zeta}^1&0&{\zeta}^0&0\end{pmatrix}
    \label{zetaparam}
    \eeq
   \noindent
and ${\bf 1}$ is the $4\times 4$ identity matrix.

Using the coset representative $\mathcal{V}$ the line element on ${\cal
M}_3$ is given by the formula  \cite{Bossard:2009we,Levay:2010ua} \beq ds^2={\rm
Tr}(P)^2
\label{line}
\eeq
\noindent
where
\beq
 P\equiv\frac{1}{2}(d\mathcal{V}\mathcal{V}^{-1}+\eta (d\mathcal{V}\mathcal{V}^{-1})^T{\eta})
\label{calp} \eeq
\noindent and the involution compatible with our conventions is
\beq \eta=\begin{pmatrix}I\otimes I&0\\0&-I\otimes I\end{pmatrix}.
\eeq \noindent

Let us introduce a four-qubit state which is a differential form
on the symplectic torus determined by the Wilson lines \beq \vert
\Psi\rangle=(M_3\otimes M_2\otimes M_1\otimes M_0)\vert
d\zeta\rangle. \label{state} \eeq \noindent Using   this we obtain
for the line element on ${\cal M}_3$ the following form \beq
ds_{{\cal
M}_3}^2=\sum_{j=1}^3\frac{dx_j^2+dy_j^2}{y_j^2}+\frac{(dx_0-w)^2+dy_0^2}{y_0^2}-\vert\vert
\Psi\vert\vert^2\label{line1} \eeq \noindent where $\vert\vert
\Psi\vert\vert^2\equiv\langle \Psi\vert \Psi\rangle$, and \beq
w=\frac{1}{2}({\zeta}^Id\tilde{\zeta}_I-\tilde{\zeta}_Id\zeta^I).
\eeq \noindent

Looking at the expression at (\ref{state}) we see that $\vert
\Psi\rangle$ is on the {\it real} SLOCC orbit of $\vert
d\zeta\rangle$ which determines the orbit type. It is useful to
embed this real state in a {\it complex} state,  using
the Hadamard gate  appearing in (\ref{Hadamard3}) and the phase
gate \beq P=\begin{pmatrix}i&0\\0&1\end{pmatrix} \label{hadphase},
\eeq \noindent  by defining \beq \vert \hat {\Psi}\rangle
=(H\otimes H\otimes H\otimes H)(P\otimes P\otimes P\otimes
 P)(M_3\otimes M_2\otimes M_1\otimes M_0)\vert d\zeta\rangle.
\label{fourierstate} \eeq \noindent Note, this new $4$-qubit
state is now on the SLOCC, i.e. $[\GL(2,\mathds C)]^{\otimes 4}$, orbit.
It can be shown  \cite{Levay:2010ua} that the amplitudes of this state depend only on the following four quantities and their
conjugates \beq {\cal E}_0=\sqrt{2}e^{\frac{K}{2}-U}X^I({\cal
N}_{IJ}d\zeta^J-d\tilde{\zeta}_I),\qquad {\cal
E}_j=2i\sqrt{2}y_je^{-U}f^I_j(\overline{\cal
N}_{IJ}d{\zeta}^J-d\tilde{\zeta}_I) \label{nahalisten} \eeq
\noindent well known from special K\"ahler geometry  \cite{Ferrara:1995ih,Strominger:1996kf,Craps1997565}. Here
\beq
f_1^I=e^{\frac{K}{2}}D_1X^I=e^{\frac{K}{2}}({\partial}_1+({\partial}_1K))X^I=e^{\frac{K}{2}}\frac{1}{\overline{z}_1-z_1}
\begin{pmatrix}1\\\overline{z}_1\\z_2\\z_3\end{pmatrix},
\label{specialgeo}
\eeq
\noindent
where $X^I=(1,z_1,z_2,z_3)^T$, $K=-\log(y_1y_2y_3)$ and ${\cal N}_{IJ}$ is defined by (\ref{valos}-\ref{kepzetes}).
The line element is then given by,
\beq
ds^2_{{\cal M}_3}=\sum_{\alpha=0}^3(\overline{e}_{\alpha}e_{\alpha}-
\overline{\cal E}
_{\alpha}{\cal E}_{\alpha})
=\sum_{\alpha=0}^3\frac{d\overline{z}_{\alpha}z_{\alpha}}{y_{\alpha}^2}-\vert
\vert \hat{\Psi}\vert\vert^2.
\label{line2}
\eeq
\noindent
Here $e_{\alpha}=\frac{-i}{y_{\alpha}}dz_{\alpha}$ are the right invariant one forms with $dz_0=(dx_0-w)-idy_0$ and $dz_j=dx_j-idy_j$.

Notice that according to (\ref{zetaparam}) the four-qubit
state $\vert   d\zeta\rangle$ which determines the orbit type of
$\vert \hat{\Psi}\rangle$ is very special. In particular, though written
in a four-qubit form, it contains merely $8$ nonzero amplitudes
reminiscent of a three-qubit state. This special structure is due
to the special status of the fourth $\SL(2,{\mathds R})$, the so
called Ehlers group  \cite{1955ZPhy..143..239E}, associated with the fourth qubit.
Moreover, the only quantities which play any role in
$\vert \hat{\Psi}\rangle$ are given by
(\ref{nahalisten}). In order to  incorporate the
information contained in the right-invariant forms $e_{\alpha}$
we introduce yet another
four-qubit state which already contains all  $16$ {\it real}
quantities associated to our coset. Neglecting the
details  \cite{Levay:2010ua} this state is given by \beq
\vert\Lambda\rangle
=\sum_{a_3,a_2,a_1,a_0=0,1}{\Lambda}_{a_3a_2a_1a_0}\vert
a_3a_2a_1a_0\rangle \label{lambdastate} \eeq \noindent with
amplitudes \beq \Lambda=
\begin {pmatrix}{\Lambda}_{0000}&{\Lambda}_{0001}&{\Lambda}_{0010}&{\Lambda}_{00
11}\\
{\Lambda}_{0100}&{\Lambda}_{0101}&{\Lambda}_{0110}&{\Lambda}_{0111}\\
{\Lambda}_{1000}&{\Lambda}_{1001}&{\Lambda}_{1010}&{\Lambda}_{1011}\\
{\Lambda}_{1100}&{\Lambda}_{1101}&{\Lambda}_{1110}&{\Lambda}_{1111}\end{pmatrix}
\equiv
\begin{pmatrix}-{\cal E}_0&-{e}_0&-{e}_1&-\overline{\cal E}_1\\{e}_2&
\overline{\cal E}_2
&{\cal E}_3&\overline{e}_3\\
{e}_3&\overline{\cal E}_3&{\cal E}_2&\overline{e}_2\\-{\cal
E}_1&-\overline{e}_1 &-\overline{e}_0&-\overline{\cal
E}_0\end{pmatrix}. \label{ittalenyeg} \eeq \noindent This state is
of central importance for the considerations of the following
sections. It is a {\it complex} four-qubit state satisfying the
reality condition \beq \overline{\vert\Lambda\rangle}
=(\sigma_1\otimes \sigma_1\otimes \sigma_1\otimes
\sigma_1)\vert\Lambda\rangle,\qquad
\sigma_1=\begin{pmatrix}0&1\\1&0\end{pmatrix} \label{realagain}
\eeq \noindent where $\sigma_1$ is the bit flip gate of quantum information theory.
It is straightforward to check that the
subgroup of  $[\SL(2,\mathds C)]^{\otimes
4}$ preserving the reality condition is $[\SU(1,1)]^{\otimes
4}$. Hence the admissible transformations are of the form \beq
\vert\Lambda\rangle\mapsto (S\otimes S_2\otimes S_1\otimes
S_0)\vert\Lambda\rangle,\qquad S_3,S_2,S_1,S_0\in \SU(1,1). \eeq
\noindent

 The number of algebraically independent
$[\SL(2,{\mathds C})]^{\otimes 4}$ invariants is four  \cite{Luque:2002}. We
have a quadratic, two quartic, and one sextic invariant. The
structure and geometry of these invariants have been
investigated  in \cite{2006JPhA...39.9533L}. Now  we observe that the quadratic
four-qubit invariant  \cite{Luque:2002} for our state
$\vert\Lambda\rangle$ is precisely the line element $ds^2_{{\cal
M}_3}$ \beq ds^2_{{\cal
M}_3}=-\frac{1}{2}{\varepsilon}^{a_3a_3^{\prime}}
{\varepsilon}^{a_2a_2^{\prime}}{\varepsilon}^{a_1a_1^{\prime}}{\varepsilon}^{a_0a_0^{\prime}}{\Lambda}_{a_3a_2a_1a_0}{\Lambda}_{a_3^{\prime}a_2^{\prime}a_1^{\prime}a_0^{\prime}}=\sum_{\alpha=0}^3(\overline{e}_{\alpha}e_{\alpha}-\overline{\cal
E}_{\alpha}{\cal E}_{\alpha}). \label{lineasinv} \eeq \noindent
This quadratic invariant is also a permutation invariant. However,
the special role we have attached to the first qubit (associated
with the Ehlers group) obviously breaks this permutation
invariance.

\subsubsection{Conserved quantities. A three qubit reformulation.}

Looking at the Lagrangian  (\ref{teljeslag}) we see that the second term
  describes  geodesic motion on the target space
${\cal M}_3=\SO(4,4)/[\SL(2,{\mathds R
})]^{\otimes 4}$ with line element given by the quadratic four-qubit invariant  (\ref{lambdastate}).
For pseudo-Riemann symmetric target spaces, such as
${\cal M}_3=\SO(4,4)/[\SL(2,{\mathds R
})]^{\otimes 4}$,
stationary spherically symmetric black hole solutions can be obtained as geodesic curves on this target space. See, for example \cite{Bergshoeff:2008be} and the references therein.
Such geodesic curves are classified in terms of the Noether charges of the solutions.
Combining these results we can relate different black hole solutions to the different SLOCC entanglement classes of four-qubits  \cite{Levay:2010ua,Borsten:2010db,Borsten:2011is}.
In order to set the stage for reviewing these results
lets look at the conserved quantities related to the Noether charge ${\cal Q}$.

The $3$-dimensiional U-duality group $\SO(4,4)$ of the $STU$ model acts
isometrically on our coset ${\cal M}_3$ by right multiplication
and yields a conserved Noether
charge  \cite{Bossard:2009we,Bossard:2009at,Bergshoeff:2008be} \beq Q=\mathcal{V}^{-1}P\mathcal{V}
\label{Noether} \eeq \noindent where $P$ and $\mathcal{V}$ are
defined by (\ref{coset}) and (\ref{calp}). An
analysis  \cite{Bossard:2009we,Levay:2010ua} of the relevant parts of ${\cal Q}$
shows that we have the following conseved quantities.

First of all we have the NUT charge
\beq
k=p_{\sigma}=\frac{dx_0-w}{2y_0^2}.
\label{1963JMP.....4..915N}
\eeq
\noindent
Here the notation $p_{\sigma}$ refers to the fact that when using the relevant part of the Lagrangian this quantity is canonically conjugated to $x_0\equiv\sigma$.
We also have $8$ conserved quantities arranged within a conserved four-qubit state $\vert\Gamma\rangle$ defined as
\beq
\vert \Gamma\rangle =\frac{1}{2}e^{-2U}({\cal N}\otimes I)\vert d\zeta\rangle -{p_{\sigma}}
(\epsilon\otimes I)\vert\zeta\rangle
\label{ezagamma}
\eeq
\noindent
with
\beq
{\cal N}\equiv N_3\otimes N_2\otimes N_1,\qquad \epsilon\equiv \varepsilon\otimes\varepsilon\otimes\varepsilon,\qquad N_{\alpha}=M_{\alpha}^TM_{\alpha}.
\eeq
\noindent
Here we also displayed the special role of the qubit corresponding to the Ehlers group, facilitating an effective three-qubit picture.
The $8$ conserved components as amplitudes of a four-qubit state are arranged as
\beq
\Gamma =\frac{1}{\sqrt{2}}\begin{pmatrix}p^0&0&-p^1&0\\
-p^2&0&q_3&0\\-p^3&0&q_2&0\\q_1&0&q_0&0\end{pmatrix}
\label{electricmagnetic}
\eeq
\noindent
and are related to the usual charges of the $STU$ model.
Comparing with (\ref{zetaparam})
we see that only the ${\Gamma}_{lkj0}$ amplitudes are nonzero.

The momenta canonically conjugate to $\zeta^I$ and $\tilde{\zeta}_I$  \cite{Bossard:2009we}  suggest that it is rewarding to introduce the new conserved quantity
\beq
\vert\hat{\Gamma}\rangle\equiv\vert\Gamma\rangle +
{p_{\sigma}}(\epsilon\otimes I)\vert\zeta\rangle
\label{bigpexpl}
\eeq
\noindent
 One can then show that the Hamiltonian governing the dynamics of our $16$ fields  depending on the conserved charges is
\beq
H=\sum_{\alpha =0}^3y_{\alpha}^2(p_{x_{\alpha}}^2+p_{y_{\alpha}}^2)-
e^{2U}\langle \hat{\Gamma}\vert {\cal N}^{-1}\otimes
I\vert \hat{\Gamma}\rangle
\label{3hami}
\eeq
\noindent
where ${\cal N}=N_1\otimes N_2\otimes N_3$.

For vanishing NUT charge $k=p_{\sigma}=0$ the second term is \beq
e^{2U}V_{BH} =e^{2U}\langle\Gamma\vert {\cal N}^{-1}\otimes
I\vert\Gamma\rangle
=e^{2U}\frac{1}{2}\begin{pmatrix}p^I&q_I\end{pmatrix}\begin{pmatrix}(\mu+\nu{\mu}^{-1}\nu)_{IJ}&-(\nu{\mu}^{-1})^J_I\\-({\mu}^{-1}\nu)^I_J&({\mu}^{-1})^{IJ}\end{pmatrix}\begin{pmatrix}p^J\\q_J\end{pmatrix}
\label{potential} \eeq \noindent which gives the usual expression
for the Black Hole Potential   $V_{BH}$.

Now using (\ref{ezagamma}) and (\ref{bigpexpl}) one can express
$\vert d\zeta\rangle$ hence an explicit formula for $\vert \hat{\Psi}\rangle$ the discrete Fourier transformed state (\ref{fourierstate}) can be derived.
Using $HPM^{T-1}=VS\sigma_3$
where
\beq
{ V}\equiv\frac{1}{\sqrt{2}}\begin{pmatrix}i&-1\\i&1\end{pmatrix},\qquad
S_j\equiv\frac{1}{\sqrt{y_j}}\begin{pmatrix}y_j&0\\-x_j&1
\end{pmatrix}
,\qquad j=1,2,3 \label{Smatrix} \eeq the $3$-qubit part of $\vert
\hat{\Psi}\rangle$ can be written as
$\vert\hat{\Psi}\rangle_3\equiv i\sqrt{2}\vert\hat{\chi}\rangle$
where \beq \vert\hat{\chi}\rangle =e^{U}({ V}\otimes { V}
\otimes { V})(S_3\otimes S_2\otimes
S_1)\vert\hat{\gamma}\rangle.
 \label{ezittaklassz}
\eeq                                                                            \noindent
Here by virtue of (\ref{bigpexpl})
\beq
\vert\hat{\gamma}\rangle=(\sigma_3\otimes \sigma_3\otimes \sigma_3)            \vert\Gamma\rangle+{p_{\sigma}}(\sigma_1\otimes\sigma_1\otimes\sigma_1)
\vert\zeta\rangle.
\label{naeztkapdki}                                                             \eeq

Equations (\ref{Smatrix})-(\ref{naeztkapdki}) constitute the final
result of our investigations. These expressions show that after
performing the timelike dimensional reduction of our starting
Lagrangian, stationary black hole solutions can be characterized
by a {\it complex} three-qubit state $\vert\hat{\chi}\rangle$
depending on the charges (electric, magnetic and NUT), the warp
factor, the moduli and the potentials $\zeta^I$ and
$\tilde{\zeta}_I$. For nonvanishing NUT charge the SLOCC class of
this state  depends on the class of $\vert\hat{\gamma}\rangle$
of (\ref{naeztkapdki}). If we assume also spherical symmetry
this class is a
function of  the
 radial coordinate. However, for vanishing NUT charge the SLOCC
 class is entirely determined by the constant electric and magnetic charges.
Moreover, in this special case a calculation shows that
\beq
\vert\vert\hat{\chi}\vert\vert^2=e^{2U}V_{BH}
\label{normsquared}
\eeq
\noindent
i.e. for vanishing NUT charge the Black Hole Potential is given by the norm of the corresponding three-qubit state $\vert\psi\rangle$ obtained from $\vert\chi\rangle$ after removing the warp factor and putting $p_{\sigma}=0$ in (\ref{naeztkapdki}).
Notice that though our state $\vert\hat{\chi}\rangle$ is complex  it satisfies the {\it reality condition}
\beq
\hat{\chi}_{111}=-\hat{\overline{\chi}}_{000},\quad
\hat{\chi}_{001}=-\hat{\overline{\chi}}_{110},\quad
\hat{\chi}_{010}=-\hat{\overline{\chi}}_{101},\quad
\hat{\chi}_{100}=-\hat{\overline{\chi}}_{011}.
\label{chireality}
\eeq
\noindent

\subsection{Static spherically symmetric extremal solutions}\label{sec:timelikereduction}

In the next sections we would like to present an entanglement
based understanding  of weakly extremal solutions of the $STU$
model. These are black hole solutions for which the spacial slices
provided by the metric $h_{ab}$ of (\ref{ans1}) are
flat  \cite{Bossard:2009we,Gaiotto:2007ag}. Single centered black holes with
spherical symmetry are of this type. In this case the dynamics of
the moduli are decoupled from the $3D$ gravity and the metric
ansatz can be chosen to be of the form \beq
ds^2=-e^{2U}(dt+\omega)+e^{-2U}(dr^2+r^2(d\theta^2+{\sin}^2\theta
d{\varphi})\label{metricans} \eeq \noindent with the warp factor depending merely
on $r$. It can be shown that now the equations of motion are
equivalent to light-like geodesic motion on ${\cal M}_3$ with the
affine parameter $\tau=\frac{1}{r}$. We have seen that due to the
fact that ${\cal M}_3$ is a symmetric space there is a number of
conserved Noether charges associated with this geodesic motion.
The most important ones are the electric and magnetic charges
$p^I$ and $q_I$ and the NUT charge
$k$  \cite{Bossard:2009we,Bossard:2009at,Bergshoeff:2008be}. Static solutions are
characterized by the vanishing of the NUT charge i.e. $k=0$. In
this case the dynamics is described by the Lagrangian (or
equivalently by the Hamiltonian of (\ref{3hami})) of
a fiducial particle in the black-hole potential $V_{BH}$ of
(\ref{potential})

\beq {\cal L}(U(\tau),
z^a(\tau),\overline{z}^{\overline{a}}(\tau))=\left(\frac{dU}
{d\tau}\right)^2+G_{a\overline{a}}\frac{dz^a}{d\tau}\frac{d\overline{z}^{\overline{a}}}{d\tau}+e^{2U}V_{BH}(z,\overline{z},p,q)
\label{Lagrange} \eeq \noindent with the constraint \beq \left(
\frac{d{\cal
U}}{d\tau}\right)^2+G_{a\overline{a}}\frac{dz^a}{d\tau}
\frac{d\overline{z}^{\overline{a}}}{d\tau}-e^{2U}V_{BH}(z,\overline{z},p,
q)=0. \label{constraint} \eeq \noindent Notice that the latter is
just about the vanishing of the Hamiltonian of
(\ref{3hami}). Equivalently,  the line element
 (\ref{line2}) is vanishing. This is just another way of
saying that our black hole solutions give rise to a
light-like geodesic motion on ${\cal M}_3$. According to
(\ref{lineasinv})
our constraint is also
equivalent to the vanishing of an entanglement measure for the
four-qubit state of (\ref{lambdastate}).

Later we will need an alternative expression for $V_{BH}$ that can be given in terms of
the central charge  of $\mathcal{N}=2$ supergravity \cite{Gibbons:1996af,Andrianopoli:1996ve,Ferrara:1997tw} \beq
V_{BH}=Z\overline{Z}+G^{i\overline{j}}(D_iZ)({\overline{D}}_{\overline{j}}\overline{Z})
\label{alternativepot}\eeq \noindent where for the $STU$ model  \cite{Behrndt:1996hu} \beq
Z=e^{K/2}W=e^{K/2}(q_0+z_1q_1+z_2q_2+z_3q_3+z_1z_2z_3p^0-z_2z_3p^1-z_1z_3p^2-z_1z_2p^3),
\label{central} \eeq  $D_a$ is the K\"ahler covariant
derivative \beq D_iZ=({\partial}_i+\frac{1}{2}{\partial}_iK)Z
\label{kovika}\eeq \noindent  and $W$ is referred to as the {\it
superpotential}.

Extremization of the effective Lagrangian (\ref{Lagrange}) with
respect to the warp factor and the scalar fields yields the
Euler-Lagrange equations \beq \ddot{U}=e^{2U}V_{BH},\qquad
\ddot{z}^{i}+\Gamma^i_{jk}\dot{z}^j\dot{z}^k=e^{2U}{\partial}^iV_{BH}.
\label{Euler} \eeq \noindent In these equations the dots denote
derivatives with respect to $\tau=\frac{1}{r}$. These radial
evolution equations taken together with the constraint
(\ref{constraint}) determine the structure of static,
spherically symmetric, extremal black hole solutions in the $STU$
model. For the more general stationary case with nonvanishing NUT
charge ($k\neq 0$)
 the motion along $\zeta^I$, $\tilde{\zeta}_I$ and $\sigma$ does not decouple from
   $U$ and $z^j$.
In this case we obtain a generalization of (\ref{Euler}).
We will not consider solutions of
such kind hence we will not give the corresponding equations here.

We  conclude that the radial evolution associated
to stationary spherical symmetric black hole solutions of the
$D=4$ $STU$ model can be described  as geodesic
motion in the moduli space ${\cal M}_3$ of the time-like  reduced
$D=3$ theory. The
four-qubit picture hinges on the enlargement of the $D=4$ symmetry
 $[\SL(2,{\mathds R})]^{\otimes 3}$ to the $D=3$ symmetry
$\SO(4,4)$ containing $[\SL(2,{\mathds R})]^{\otimes 4}$ as a subgroup.
We are now in a position to see how the entanglement encoded in our state
$\vert\Lambda\rangle$ of (\ref{lambdastate}) helps to classify
static spherically symmetric extremal single-centre black hole
solutions in the $STU$ model.

\subsection{Black hole solutions as entangled systems.}
\label{blacksol}

\subsubsection{BPS
solutions} \label{G} Let us consider the four-qubit state
$\vert\Lambda\rangle$ of (\ref{lambdastate}). In this
subsection we will be interested in the sufficient
 and necessary condition for the separability of its {\it first qubit},
 labelled by $a_0$ in (\ref{lambdastate}). From
our previous considerations it is clear that this qubit has a special status. In quantum information theoretic terms separability of this qubit is {\it equivalent} to the condition that the
(unnormalized) $2\times 2$ reduced density matrix
${\varrho}_1\equiv{\rm Tr}_1\vert
\Lambda\rangle\langle\Lambda\vert$ represents a pure state. This
density matrix is of the form \beq
{\varrho}_1=\begin{pmatrix}\langle\Lambda_0\vert\Lambda_0\rangle&
\langle\Lambda_0\vert\Lambda_1\rangle\\
\langle\Lambda_1\vert\Lambda_0\rangle&\langle\Lambda_1\vert\Lambda_1\rangle\end{pmatrix},\qquad
\langle\Lambda_{a_0}\vert\Lambda_{a^{\prime}_0}\rangle\equiv
\sum_{a_3,a_2,a_1=0,1}
\overline{\Lambda}_{a_3a_2a_1a_0}\Lambda_{a_3a_2a_1a^{\prime}_0}.
\eeq \noindent This is a pure state if and only
if ${\rm Det}\varrho_1=0$. Equivalently 
this condition is satisfied iff 
$\Lambda_{a_3a_2a_10}=\lambda{\Lambda}_{a_3a_2a_11}$. By virtue of
the reality condition of (\ref{realagain}) we  also have the
constraint $\vert\lambda\vert =1$. Using the definitions in
(\ref{ittalenyeg}) this means that \beq {\cal E}_0=\lambda
e_0,\qquad {\cal E}_j=\lambda\overline{e}_j,\qquad
\vert\lambda\vert=1. \label{bps} \eeq \noindent Clearly now the
constraint of (\ref{constraint}) is satisfied, equivalently the
quadratic four-qubit invariant is vanishing. A calculation also
shows that actually all of the four-qubit invariants are
vanishing  \cite{Levay:2010ua}. Such states are called nilpotent. It can
be shown that such states gives rise to a Noether charge $Q$ of
(\ref{Noether}) which is a nilpotent matrix.

In order to link these considerations to the usual BPS black hole
solutions we choose $\lambda$ as \beq
\lambda=-i\sqrt{\frac{Z}{\overline{Z}}}. \label{bps2} \eeq
\noindent 
In the language of supergravity the above 
condition on separability correspondes to the existence
of Killing spinors   characterizing supersymmetric
solutions \cite{Bossard:2009we}. These considerations give rise to the well-known
attractor flow equations   \cite{Ferrara:1995ih,Strominger:1996kf,Ferrara:1996dd,Bossard:2009we} \beq
\dot{U}=-e^{U}\vert Z\vert,\qquad
\dot{z}^j=-2e^{U}G^{j\overline{k}}{\partial}_{\overline{k}}\vert
Z\vert. \label{bpsexpl} \eeq \noindent These
first order equations imply that the corresponding second order
equations of (\ref{Euler}) also hold.

 Note, for weakly extremal solutions to be also extremal we also
have to ensure that the solutions are smooth. To ensure this
one must fine tune the boundary conditions at spatial infinity so
that the fiducial particle reaches the top of the potential hill
defined by $V_{BH}$ in infinite proper time and with zero
velocity. In our case an analysis of the first order equations
(\ref{bpsexpl}) shows that this indeed can be
achieved  \cite{Moore:1998pn}.

From this analysis we have learnt that the condition of
separability for the first qubit for the four-qubit state
$\vert\Lambda\rangle$ taken together with the special choice of
(\ref{bps2}) yields the first order attractor flow equations.
Moreover, in this case $\vert\Lambda\rangle$ is
a {\it nilpotent} state. This property of $\vert\Lambda\rangle$ is
related to the well-known nilpotency of the Noether
charge  $Q$  \cite{Bossard:2009we,Bergshoeff:2008be,Bossard:2009at}.
An analysis of the explicit form of these solutions will be given in Section
5.6.2.

\subsubsection{Non-BPS solutions}
\label{F} As our first example of non-supersymmetric solutions let us
discuss the separability properties of $\vert\Lambda\rangle$
associated with the remaining three qubits not playing a distinguished
role. Here we chose to consider separability of the fourth qubit.
An argument similar to the one as given in the previous subsection
shows that the sufficient and necessary condition of separability
for this qubit is that the first row of (\ref{ittalenyeg}) is
proportional to the third and the second is proportional to the
fourth. Due to the reality condition we again have
$\vert\lambda\vert=1$ and we get \beq {\cal E}_0=-\lambda
e_3,\qquad \overline{\cal E}_1=-\lambda\overline{e}_2, \qquad
\overline{\cal E}_2=-\lambda\overline{e}_2,\qquad {\cal E}_3=
-\lambda\overline{e_0}. \label{furafelt} \eeq \noindent Using the
definitions of (\ref{nahalisten}) these conditions take the
explicit form \beq
\frac{\dot{z}_0}{y_0}=\overline{\lambda}e^UZ_3,\qquad
\frac{\dot{z}_1}{y_1}=\lambda
e^UZ_2,\qquad\frac{\dot{z}_2}{y_2}=\lambda e^UZ_1,\qquad
\frac{\dot{z}_3}{y_3}=-\overline{\lambda}e^U Z, \label{veryexpl}
\eeq \noindent where $Z_j\equiv D_jZ$ as given by
(\ref{kovika}). Now for static solutions we again have no twist
potential, $x_0=0$, hence by choosing \beq \lambda
=-i\sqrt{\frac{Z_3}{\overline{Z_3}}} \label{lambdika} \eeq
\noindent we get \beq \dot{U}=-e^U\vert Z_3\vert,\qquad
\dot{z}^j=-2e^UG^{j\overline{k}}{\partial}_{\overline{k}}\vert
Z_3\vert. \label{vanish4} \eeq \noindent These expressions show
that demanding separability for the fourth qubit taken together
with the choice of (\ref{lambdika}) yields the first order
equations characterizing attractors with vanishing central
charge  \cite{Bellucci:2007zi}.

Clearly similar considerations apply for issues of separability
for the second and third qubits. The result will be similar sets
of equations with $\vert Z_3\vert$ replaced by $\vert Z_1\vert$
and $\vert Z_2\vert$. This amounts to taking different forms for
the so-called fake superpotential  \cite{Bossard:2009we}. Calculations
again show that the four algebraically independent four-qubit
invariants are zero, hence our considerations on the nilpotency of
$\vert\Lambda\rangle$ familiar from the previous subsection still
apply.

Let us now discuss a non-BPS solution with non-vanishing central
charge. Obviously the
vanishing of the quadratic four-qubit invariant i.e. the (\ref{constraint}) constraint 
can be satisfied in a number of different
ways. Explicitly, \beq
\sum_{\alpha=0}^3\overline{\cal E}_{\alpha}{\cal
E}_{\alpha}=\sum_{\alpha=0}^3 \overline{e}_{\alpha}e_{\alpha}.
\label{hmm} \eeq \noindent For static solutions we have already
remarked that $\overline{e}_0=e_0$, hence for BPS solutions 
(\ref{bps})-(\ref{bps2}) can be written in the form ${\cal
E}_{\alpha}=\lambda\overline{e}_{\alpha}$, i.e. ${\cal
E}_{\alpha}$ is related to $\overline{e}_{\alpha}$ via a special
element of $\U(4)$ containing only phase factors $\lambda$ in its
diagonal. In the case of non-BPS solutions with vanishing central
charge these elements of $\U(4)$ are just permutation matrices
combined with similar phase factors and their conjugates. This
structure is related to the {\it separability} of one of the
qubits in the state $\vert\Lambda\rangle$.

In order to obtain states $\vert\Lambda\rangle$ which are {\it
entangled} and at the same time give rise to static spherically
symmetric non-BPS black hole solutions with non-vanishing central
charge let us consider the following choice \beq
\begin{pmatrix}{\cal E}_0\\{\cal E}_1\\{\cal E}_2\\{\cal E}_3\end{pmatrix}=\frac{1}{2i}
\begin{pmatrix}1&1&1&1\\1&1&-1&-1\\
1&-1&1&-1\\1&-1&-1&1\end{pmatrix}\begin{pmatrix}e_0\\e_1\\e
_2\\e_3\end{pmatrix}. \label{nonbpsunitary} \eeq \noindent Due to
the unitarity of the relevant matrix 
(\ref{hmm}) is satisfied moreover, one can show that none of
the qubits can be separated from the rest. However, we still have
to satisfy the equations of motion (\ref{Euler}).
 Let us
illustrate that the choice of (\ref{nonbpsunitary}) indeed
gives rise to a solution of the latter equations. This solution is
 the non-BPS seed solution  \cite{Gimon:2007mh}. First recall the
definition of the three-qubit state $\vert\hat{\chi}\rangle$ of
(\ref{ezittaklassz}) and denote the state with $k=p_{\sigma}=0$
by $\vert\chi\rangle$. Then employ a discrete Fourier
transformation, \beq \vert\tilde{\chi}(\tau)\rangle =(H\otimes
H\otimes H)\vert\chi(\tau)\rangle
.
\label{hadi} \eeq \noindent Now one can check that the amplitudes
of
 $\tilde{\chi}$ can be related to the derivatives of the moduli
 as  \cite{Levay:2010ua} \beq \tilde{\chi}_{000}=\frac{i}{2}\frac{\dot{x_0}}{y_0},\qquad
\tilde{\chi}_{110}=\frac{i}{2}\frac{\dot{x_1}}{y_1},\qquad
\tilde{\chi}_{101}=\frac{i}{2}\frac{\dot{x_2}}{y_2},\qquad
\tilde{\chi}_{011}=\frac{i}{2}\frac{\dot{x_3}}{y_3},
\label{elsofele} \eeq \noindent \beq
\tilde{\chi}_{111}=\frac{1}{4}\left(\frac{\dot{y_0}}{y_0}-
\frac{\dot{y_1}}{y_1}- \frac{\dot{y_2}}{y_2}-
\frac{\dot{y_3}}{y_3}\right),\qquad
\tilde{\chi}_{001}=\frac{1}{4}\left(-\frac{\dot{y_0}}{y_0}+
\frac{\dot{y_1}}{y_1}- \frac{\dot{y_2}}{y_2}-
\frac{\dot{y_3}}{y_3}\right) \label{masodikfele1} \eeq \noindent
where the remaining amplitudes are given by a cyclic shift of the $+$ sign.
For static solutions we have vanishing NUT charge and $x_0=0$
hence the first of these equations reads 
$\tilde{\chi}_{000}=0$. Writing out explicitly the amplitudes
$\tilde{\chi}_{jkl}$ in terms of the moduli, warp factor and the
charges this constraint implies $p^0=0$. Hence our candidate
for a non-BPS solution should have only seven nonvanishing Fourier
amplitudes and vanishing $p^0$ charge.

Let us now introduce the notation \beq y_0=e^{\phi_0},\qquad
y_{j}=e^{\phi_{j}},\qquad \beta\equiv
U-\frac{1}{2}(\phi_1+\phi_2+\phi_3), \qquad \alpha_j\equiv
U+\frac{1}{2}\phi_j \label{phidefi} \eeq \noindent with and
$j=1,2,3$. Now our equations take the form \beq
\tilde{\chi}_{111}=\frac{1}{2}\dot{\beta},\qquad
\tilde{\chi}_{110}=\frac{i}{2}e^{-\phi_1}\dot{x}_1,\qquad
\tilde{\chi}_{101}=\frac{i}{2}e^{-\phi_2}\dot{x}_2,\qquad
\tilde{\chi}_{011}=\frac{i}{2}e^{-\phi_3}\dot{x}_3 \label{gim1}
\eeq \noindent \beq
\tilde{\chi}_{001}=\frac{1}{2}(\dot{\alpha}_1-\dot{\alpha}_2-\dot{\alpha}_3),
\qquad
\tilde{\chi}_{010}=\frac{1}{2}(\dot{\alpha}_2-\dot{\alpha}_3-\dot{\alpha}_1),
\qquad
\tilde{\chi}_{100}=\frac{1}{2}(\dot{\alpha}_3-\dot{\alpha}_1-\dot{\alpha}_2).
\label{gim2} \eeq \noindent With the further charge constraints
$q_j=0$, $q_0<0$, and $p^1,p^2,p^3>0$ one can see that the
equations above are precisely the ones found in the paper of Gimon
et.al.  \cite{Gimon:2007mh} characterizing the seed solutions for the
so-called $D0-D4$ system. We remark in closing that one can also
verify by an explicit calculation that all of the four
algebraically independent four-qubit invariants are again
vanishing. This means that the corresponding matrix $Q$ of
conserved charges is again nilpotent.

Let us give a brief summary of our results. The
central object of our considerations was the {\it complex}
$4$-qubit state $\vert\Lambda\rangle$,  satisfying a reality condition.
The amplitudes of this state of {\it odd parity} contain the right
invariant one-forms $e_{\alpha}$, $\alpha=0,1,2,3$. On the other
hand the $8$ amplitudes of {\it even parity} are related to the
$8$ amplitudes of  a $3$-qubit state $\vert\hat{\chi}\rangle$. We
have shown that the state $\vert\Lambda\rangle$ is connected to
the line element on ${\cal M}_3$. We also realized that this
expression for the line element is just the quadratic $4$-qubit
$\SL(2,\mathds C)$ invariant. After expressing the $8$ amplitudes
of the embedded $3$-qubit state in terms of the conserved
electric, magnetic and NUT charges this invariant also has the
physical interpretation as the {\it extremality} parameter.

Note that one of the qubits of the state $\vert\Lambda\rangle$ was
special. The separability properties of this special qubit are
related to the solution being BPS or non-BPS. We demonstrated
within our formalism that static, extremal BPS and
non-BPS-solutions with vanishing central charge correspond to
states for which one of the qubits is separable from the rest. On
the other hand using the non-BPS seed solution for nonvanishing
central charge we have shown that $\vert\Lambda\rangle$ in this
case is entangled. We revealed a connection between the
classification of {\it nilpotent states} within the realm of
quantum information theory and the similar classification of
nilpotent orbits. The details of this connection will be explored
further in the next section.

\subsection{Four-qubit entanglement from string theory}

In the proceeding section it was shown how the time-like reduced $STU$ model may be naturally related to a four-qubit system,  rephrasing various important features of the black hole solutions in  quantum information theoretic terms. In particular, certain BPS and non-BPS spherically symmetric black hole solutions were related  to (partially and totally) entangled four-qubit states. Here, developing this correspondence,  we will describe how the  classification of all  extremal, both single-centre and multi-centre, black hole solutions provides a complete characterisation of the four-qubit entanglement classes.

The extremal black hole solutions are determined by the nilpotent orbits of the 3-dimensional U-duality group \cite{Bergshoeff:2008be,Bossard:2009at,Bossard:2009my,Bossard:2009we,Bossard:2011kz}. The nilpotent orbits are then related to the four-qubit entanglement classes through the Kostant-Sekiguchi correspondence \cite{Sekiguchi:1987,Collingwood:1993,Levay:2010ua, Borsten:2010db, Borsten:2011is}. Using these tools we find that there are 31 four-qubit entanglement families, which reduce to 9 under permutations of Alice, Bob, Charlie and Dave in agreement with the quantum information and mathematical literature \cite{Verstraete:2002, Chterental:2007}.

An interesting new feature, first treated in \cite{Rubens:2011phd, Borsten:2012wr}, of the four-qubit correspondence, which goes beyond the three-qubit case, is the appearance of interacting multi-centre black hole solutions as will be described below.

Although two and three
qubit entanglement is well-understood (see \textit{e.g.}
\cite{Dur:2000}), the literature on four qubits can be confusing and
seemingly contradictory, as illustrated in \autoref{tab:authors}.
This is due in part to genuine calculational disagreements, and in
part to the use of distinct (but in principle consistent and
complementary) perspectives on the criteria for classification.

\begin{table*}[th]
\caption[Classification of $D=4,\ensuremath{\mathcal{N}}=2$
\ensuremath{STU}{} black holes.]{Various results on four-qubit
entanglement.} \label{tab:authors}
\begin{tabular*}{\textwidth}{@{\extracolsep{\fill}}clllr@{,}rr@{,}r}
\hline Paradigm & Author & Year & Ref & \multicolumn{2}{c}{result
mod perms} & \multicolumn{2}{c}{result incl. perms} \\ \hline
\multirow{5}{*}{classes} & Wallach & 2004 & \cite{Wallach:2008} &
\multicolumn{2}{c}{?} & \multicolumn{2}{c}{90} \\
& Lamata et al & 2006 & \cite{Lamata:2006b} & 8 genuine & 5
degenerate & 16
genuine & 18 degenerate \\
& Cao et al & 2007 & \cite{Cao:2007} & 8 genuine & 4 degenerate & 8
genuine
& 15 degenerate \\
& Li et al & 2007 & \cite{Li:2007c} & \multicolumn{2}{c}{?} &
$\geq31$
genuine & 18 degenerate \\
& Akhtarshenas et al & 2010 & \cite{Akhtarshenas:2010} &
\multicolumn{2}{c}{? } & 11 genuine & 6 degenerate \\
& Buniy et al & 2010 & \cite{Buniy:2010a} & 21 genuine & 5
degenerate & 64
genuine & 18 degenerate \\
\hline
\multirow{3}{*}{families} & Verstraete et al & 2002 &
\cite{Verstraete:2002}
& \multicolumn{2}{c}{9} & \multicolumn{2}{c}{?} \\
& Chterental et al & 2007 & \cite{Chterental:2007} &
\multicolumn{2}{c}{9} &
\multicolumn{2}{c}{?} \\
& String theory & 2010 & \cite{Borsten:2010db} &
\multicolumn{2}{c}{9} & \multicolumn{2}{c}{31} \\ \hline
\end{tabular*}
\end{table*}

On the one hand, there is the ``covariant'' approach which
distinguishes the SLOCC orbits by the vanishing or not of
$[\SL(2,\mathds{C})]^{\otimes n}$ covariants/invariants. This
philosophy is adopted for the 3-qubit case in
\cite{Dur:2000,Borsten:2009yb}. The analogous
4-qubit case was treated, with partial results, in
\cite{Briand:2003a}. 

On the other hand, there is the ``normal form'' approach which
considers ``families'' of orbits. An arbitrary state may be
transformed into one of a finite number of normal forms. If the
normal form depends on some of the algebraically independent SLOCC
invariants it constitutes a family of orbits parametrised by these
invariants. On the other hand, a parameter-independent family
contains a single orbit. This philosophy is adopted for the 4-qubit
case  in \cite
{Verstraete:2002,Chterental:2007}. There are four algebraically independent SLOCC
invariants \cite{Briand:2003a}. Up to permutation of the four
qubits,
these authors found 6 parameter-dependent families called $G_{abcd}$, $%
L_{abc_{2}}$, $L_{a_{2}b_{2}}$, $L_{a_{2}0_{3\oplus \bar{1}}}$, $L_{ab_{3}}$%
, $L_{a_{4}}$ and 3 parameter-independent families called
$L_{0_{3\oplus \bar{1}}0_{3\oplus \bar{1}}}$, $L_{0_{5\oplus
\bar{3}}}$, $L_{0_{7\oplus \bar{1}}}$, see \autoref{tab:four_qubit_reps}.
\begin{table}\caption{The 9 ways of entangling four qubits }\label{tab:four_qubit_reps}
\begin{tabular*}{\textwidth}{@{\extracolsep{\fill}}>{\centering$}m{1.5cm}<{$}>{\centering$}m{1.5cm}<{$}>{\centering$}m{\textwidth-4cm}<{$}c}
\toprule\\
 & $Verstraete/Chterental Family$ & \text{Representative state} & \\\\
\hline\\
 & G_{abcd} & \frac{a+d}{2}\left(\ket{0000} + \ket{1111}\right) + \frac{a-d}{2}\left(\ket{0011} + \ket{0011}\right) + \frac{b+c}{2}\left(\ket{0101} + \ket{1010}\right)  + \frac{b-c}{2}\left(\ket{0110} + \ket{1001}\right) &\\ \\
&  L_{abc_2} & \frac{a+b}{2}\left(\ket{0000} + \ket{1111}\right) + \frac{a-b}{2}\left(\ket{0011} + \ket{0011}\right) + c\left(\ket{0101} + \ket{1010}\right)  + \ket{0110} &\\ \\
 & L_{a_2b_2} & a\left(\ket{0000} + \ket{1111}\right) + b\left(\ket{0101} + \ket{1010}\right)  + \ket{0110} + \ket{0011} & \\ \\
&  L_{ab_3} & a\left(\ket{0000} + \ket{1111}\right) + \frac{a+b}{2}\left(\ket{0101} + \ket{1010}\right)  + \frac{a-b}{2} \left(\ket{0110} + \ket{1001}\right)  + \frac{i}{\sqrt{2}}\left(\ket{0001} + \ket{0010} + \ket{0111} + \ket{1011}\right) &\\ \\
&  L_{a_4} & a\left(\ket{0000} +\ket{0101} + \ket{1010}+ \ket{1111}\right)  + i \ket{0001} + \ket{0110} - i\ket{1011} &\\ \\
& L_{a_20_{3\oplus\bar{1}}} & a\left(\ket{0000} + \ket{1111}\right) + \ket{	0011} + 	\ket{0101} + \ket{0110} & \\ \\
& L_{0_{5\oplus\bar{3}}} & \ket{0000} + \ket{0101} + \ket{1000} + \ket{1110} &\\ \\ 
& L_{0_{7\oplus\bar{1}}} & \ket{0000} + \ket{1011} + \ket{1101} + \ket{1110} &\\ \\ 
& L_{0_{3\oplus\bar{1}}0_{3\oplus\bar{1}}} & \ket{0000} + \ket{0111} & \\\\
\hline
\end{tabular*}
\end{table}

To illustrate the difference between these two approaches, consider
the separable EPR-EPR state $(|00\rangle +|11\rangle )\otimes
(|00\rangle
+|11\rangle )$. Since this is obtained by setting $b=c=d=0$ in %
\autoref{tab:four_qubit_reps}, it belongs to the $G_{abcd}$ family, whereas in
the
covariant approach it forms its own class. Similarly, a totally separable $A$%
-$B$-$C$-$D$ state, such as $|0000\rangle $, for which all
covariants/invariants vanish, belongs to the family $L_{abc_{2}}$,
which however also contains genuine four-way entangled states. These
interpretational differences were also noted in \cite{Lamata:2006b}. As we shall see, our black hole perspective lends itself naturally to the
``normal form'' framework.

In order to relate the extremal BH solutions to the entanglement
classes of four qubits, we invoke the aforementioned
Kostant-Sekiguchi theorem \cite
{Sekiguchi:1987,Collingwood:1993}. By
applying the Kostant-Sekiguchi correspondence to the Cartan decomposition,
\be
\mathfrak{so}(4,4)=4 \mathfrak{sl}(2)\oplus (\mathbf{2,2,2,2})
\ee
 one can state that the nilpotent orbits of
$\SO_{0}(4,4)$ acting on its adjoint representation, which classify the extremal black hole solutions, are in
one-to-one correspondence with the nilpotent orbits of
$[\SL(2,\ensuremath{\mathds{C}})]^{4}$ acting on its fundamental
$\ensuremath{\mathbf{(2,2,2,2)}}$ representation and, hence, with
the classification of 4-qubit entanglement. Note furthermore that it
is the complex qubits that appear automatically, thereby relaxing
the restriction to real qubits (sometimes called rebits) that
featured in earlier versions of the BH/qubit correspondence.

It follows that there are 31 nilpotent orbits for four qubits under
SLOCC \cite{Borsten:2010db}. For each nilpotent orbit there is
precisely one family of SLOCC orbits since each family contains one
nilpotent orbit on setting all invariants to zero. The nilpotent
orbits and their associated families are summarized in
\autoref{tab:realcosets} \cite{Borsten:2010db}, which is split into
upper and lower sections according as the nilpotent orbits belong to
parameter-dependent or parameter-independent families.

\subsubsection{Extremal black hole solutions and nilpotent orbits}

In \autoref{sec:timelikereduction} we saw how one can study the stationary black hole solutions of the $STU$ model by performing a time-like dimensional reduction. The intuition is that since a strictly stationary spacetime by definition admits an everywhere time-like Killing vector it can be dimensionally reduced to a 3-dimensional field theory. Solutions of the 3-dimensional theory then up-lift to stationary solutions of its 4-dimensional parent theory, providing the ideal framework for the study of stationary 4-dimensional black holes, as first demonstrated to great effect in \cite{Breitenlohner:1987dg}. 

For 4-dimensional supergravity theories with scalars $\phi$ living in a symmetric space $\mathcal{M}_4 = G_4/H_4$, where $G_4$ is the U-duality group and $H_4$ its maximal compact subgroup we can  exploit these group-theoretic structures to systematically characterise the various classes of stationary black holes. Note, this includes all supergravity theories with $\mathcal{N}>2$ and a large class of $\mathcal{N}=2$ theories, including the Einstein-Maxwell theories such as the $STU$ model. 

Performing a time-like reduction of such models leads to a theory of 3-dimensional gravity coupled to the 3-dimensional scalars $\Phi$ through a nonlinear sigma model described by the Lagrangian,
\be\label{eq:3dgravity}
\mathcal{L}=-\frac{1}{2}\sqrt{h}R[h]+g_{mn}\partial^a \Phi^m \partial^b \Phi^n h_{ab}.
\ee
The scalars $\Phi$ come not only directly from  the 4-dimensional scalars, but also the metric and the  3-dimensional   gauge fields after dualization, which in turn originate from both the metric and the 4-dimensional gauge fields. This procedure is described in detail for the $STU$ model in \autoref{sec:timelikereduction}.
Einstein's equations and equations of motion for the scalar fields derived from \eqref{eq:3dgravity} are given respectively by,
\be\label{eq:3dEE}
R_{ab}-\frac{1}{2}h_{ab}R=g_{mn}\partial_a \Phi^m \partial_b \Phi^n-\frac{1}{2}h_{ab}g_{mn}\partial_c \Phi^m \partial^c \Phi^n 
\ee
and
\be
\nabla_a\nabla^a\Phi^m+\Gamma^{m}_{np}\partial_a\Phi^n\partial^a\Phi^p=0.
\ee
The 3-dimensional scalars $\Phi$ parametrize a pseudo-Riemannian symmetric space $\mathcal{M}_3=G_3/H_{3}^{*}$, with line element given by the scalar manifold metric $g_{mn}$.  Here, $G_3$ is the 3-dimensional U-duality group and $H^{*}_{3}$ is the maximally non-compact real form of $H_3(\mathds{C})$, the complexification of the maximal compact subgroup $H_3\subset G_3$.  If we had perform instead a space-like reduction we would have obtained the Riemannian symmetric space  $\mathcal{M}_3=G_3/H_{3}$. 
Such a model may be described in terms of a coset representative
$\mathcal{V}\in G_3/H_{3}^{*}$,
\be
\mathcal{V}\mapsto g\mathcal{V}, \quad g \in G_3,\qquad
\mathcal{V}\mapsto \mathcal{V}h(\Phi), \quad h \in H^{*}_{3},
\ee
  containing all the scalar degrees of freedom, as well as 
the three-dimensional metric $h_{ab}$, which carries no physical degrees of freedom
in three dimensions.

The Maurer-Cartan form $\mathcal{V}^{-1}d\mathcal{V}$ decomposes as,
\be
\mathcal{V}^{-1}d\mathcal{V}=P+B, \qquad P=P_{a}dx^a\in \mathfrak{p}, \qquad B=B_{a}dx^a\in \mathfrak{h}^{*}_{3},
\ee
where
\be
\mathfrak{g}_3=\mathfrak{h}^{*}_{3}\oplus \mathfrak{p},
\ee
and $\mathfrak{g}_3, \mathfrak{h}^{*}_{3}$ denote the Lie algebras of $G_3$, $H^{*}_{3}$, respectively. 

In this language the Einstein equation and the scalar equations of motion are respectively given by,
\be
R_{ab}-\frac{1}{2}h_{ab}R=\Tr P_aP_b, \qquad
d\star P+[B, \star P]=0
\ee
and the Bianchi identity gives 
\be
dB + B^2=-P^2, \qquad dP+[B,P]=0.
\ee
A few lines using the above shows
\be
d\star \mathcal{V}P\mathcal{V}^{-1}=0,
\ee
so we  have a $\mathfrak{g}_3$ valued Noether current $ \star\mathcal{V}P\mathcal{V}^{-1}$, which for a  2-cycle $\Sigma$ defines a $\mathfrak{g}_3$ valued Noether charge matrix,
\be\label{eq:noethercharge}
Q=\frac{1}{4\pi}\int_{\Sigma} \star \mathcal{V}P\mathcal{V}^{-1}.
\ee

 Restricting our attention to weakly extremal solutions, i.e. assuming  the 3-dimensional spatial slice to be flat, from  \eqref{eq:3dEE} we have,
\be\label{eq:exscalar}
g_{mn}\partial_a \Phi^m \partial_b \Phi^n=0,
\ee
or equivalently,
\be
\Tr P_aP_b =0,
\ee
so the moduli decouple from gravity. With these restrictions the  solutions correspond to harmonics maps from $\R^3/\{x_A\}$ to $\mathcal{M}_3$, where the  $x_A, A=1,\ldots, k$ correspond to a finite set of removed points.

Further imposing spherical symmetry,   focusing on single-centre solutions, the scalars will depend only on the radial coordinate and solutions to \eqref{eq:exscalar} are  given by null geodesics in   $\mathcal{M}_3$ \cite{Breitenlohner:1987dg}.
In this case the Noether charge $Q\in\mathfrak{g}_3$ given in \eqref{eq:noethercharge} will be \emph{nilpotent}. In general, an element  $X\in\textrm{End}(V)$, where $V$  is a finite dimensional complex vector space, is said to be \emph{semisimple} if its eigenvectors form a basis for $V$, in which case it is diagonalizable. An element  $X$ is said to be nilpotent if $X^n=0$ for some finite $n$. This definition naturally applies to the adjoint representation of Lie algebras,
\begin{equation}
	 \textrm{ad}: \mathfrak{g} \mapsto \textrm{End}(\mathfrak{g}); \qquad X\mapsto \textrm{ad}_X \qquad\text{where}\qquad\textrm{ad}_X Y := [X,Y].
\end{equation}
An element $X\in\mathfrak{g}$ is semisimple (resp. nilpotent) if $\textrm{ad}_X$ is a semisimple (resp. nilpotent) endomorphism. Then the Jordan Decomposition theorem
	says for any $X\in\textrm{End}(V)$ there exits a unique commuting pair $X_s, X_n \in  \textrm{End}(V)$ such that $X = X_s + X_n$, where $X_s$ is semisimple and $X_n$ is nilpotent.
	
	A solution corresponding to a  given null geodesic is precisely determined by the nilpotent conserved charge $Q$ \cite{Bergshoeff:2008be, Bossard:2009at, Bossard:2009my}. Since $\mathcal{V}\mapsto g\mathcal{V}h$ and $P\mapsto h^{-1}Ph$ under U-duality, the Noether charge transforms according as $Q\mapsto gQg^{-1}$.  Consequently, the physically distinct weakly extremal  black hole solutions are classified in terms of the adjoint orbits,
	  \begin{equation}
	\mathcal{O}_Q = G_3\cdot Q := \left\{ g Q g^{-1} \; \big\lvert \; g \in G_3 \right\},
\end{equation}
for  $Q$  nilpotent \cite{Bossard:2009at, Bossard:2009my}. We will refer to these simply as   the \emph{nilpotent orbits of} $G_3$. Moreover the scalar momentum $P$ will lie in the intersection of  $\mathcal{O}_Q$ with the Lie algebra component $\mathfrak{p}$. 

To summarise, the single-centre extremal black hole solutions are in one-to-one correspondence with the nilpotent orbits of $G_3$. Well not quite, one must still pay due diligence to the regularity of the solutions.    A weakly extremal solution is not necessarily a smooth extremal solution.  In particular, it was recognised in \cite{Gunaydin:2005mx} that regularity places a constraint on the degree $n$ of nilpotency of $Q$, i.e. the smallest $n$ such that $[\textrm{ad}_Q]^n=0$. This is a subtle matter in general, but for our case of interest, the $STU$ model, it turns out we require $Q^3=0$ \cite{Bossard:2009at}, where $Q$  lies strictly in $\mathfrak{p}$, which is the case for asymptotically flat solutions since $\mathcal{V}$ goes to  identity in the asymptotic limit.

Dropping the assumption of spherical symmetry we return to the case in which the stationary solutions correspond to harmonics maps from $\R^3/\{x_A\}$ to $\mathcal{M}_3$ and the scalars depend on generically on $\R^3$.  In order to study the extremal multi-centre solutions  it was assumed in \cite{Bossard:2011kz} that the cycles $\Sigma$ defining the Noether charges are characterised by the black hole centres $x_A$ they enclose.  In particular, the Noether charge $Q_A$ associated to a cycle enclosing a  single centre at $x_A$  is characterised
by the pole of $\star \mathcal{V}P\mathcal{V}^{-1}$ at $x=x_A$. Since the presence of other centres modifies the  horizon geometry of a given centre  by subleading corrections, regularity requires the associated Noether charge to satisfy the single-centre constraint \cite{Bossard:2011kz}, that is $Q_{A}^3=0$ for all $A$.

However, the crucial observation made in  \cite{Bossard:2011kz} is that, while the individual centres must satisfy $Q_{A}^3=0$, the Noether charge $Q$ of a cycle $\Sigma_I$ enclosing a set of centres $x_A, A\in I$, which is given by,
\be
Q=\sum_A Q_A,
\ee
need not be of nilpotency degree 3. It is then reasonable to anticipate that the nilpotent orbits which have nilpotency degree too high to accommodation regular single-centre black holes  do support \emph{regular multi-centre solutions}, where each centre individually falls into one of the well-defined single-centre orbits. For the $STU$ model this is explicitly shown to be the case in  \cite{Bossard:2011kz} and has recently been extended to $\mathcal{N}=8$ supergravity in \cite{Bossard:2012ge}. We will briefly describe some the features of the $STU$ solutions in the following sections.

\subsubsection{Nilpotent orbits of $\SO(4,4)$}

As described in \autoref{sec:timelikereduction} the $STU$ model has ${\cal M}_3=\SO(4,4)/\SL(2,{\mathds R})^{\otimes 4}$. Here we summarise the structure of the nilpotent orbits of the identity component $\SO_0(4,4)$.
 
 The classification of nilpotent orbits is a rather elegant and well developed subject. We will not describe here the general theory of the orbit classification,  but interested reader can refer to \cite{Collingwood:1993, Djokovic:2000}. We will merely present the results and their labelling for the relevant case of $\Orth(p,q)$, where $p+q=n$. 

The nilpotent orbits of $\Orth(p,q)$ may be labelled by ``signed Young tableaux'' (sometimes referred to as $ab$-diagrams in the mathematical literature). A signed Young tableau for $\Orth(p,q)$ is an $n$ box Young tableau whose boxes filled with signs $+/-$ such that:
\begin{enumerate}
 \item The signs alternate along the rows. 
 \item The total number of $+$'s must be $p$. 
 \item Rows of even length must come in pairs such that if one row starts with ``$+$''  the other row starts with ``$-$''. 
\end{enumerate}
Two such diagrams are equivalent if they may be related by row permutation. There is precisely one nilpotent orbit of $\Orth(p,q)$ on $\mathfrak{so}(p, q)$ for each equivalence class of signed Young tableaux.  

Since the identity component $\SO_0(p, q)$ has index 4 in $\Or(p, q)$, for each nilpotent $\Or(p, q)$ orbit there may be either 1, 2 or 4 nilpotent $\SO_0(p, q)$ orbits. This number is also determined by the corresponding signed Young tableau. If the middle sign of every odd length row is  ``$-$'' (``$+$'') there are 2 orbits and we label the diagram to its left (right) with a $I$ or a $II$.  If it only has even length rows there are 4 orbits and we label the diagram to both its left and right with a $I$ or a $II$. If it is none of these it is said to be stable and there is only one orbit.

Following these rules for our case of $\SO_0(4,4)$ we find 31 labelled signed Young tableaux as given in \autoref{tab:realcosets}. The corresponding to the 31 nilpotent orbits are presented in the same table. The closure ordering of the orbits is given in \autoref{fig:hasse}. 

The Kostant-Sekiguchi correspondence \cite{Sekiguchi:1987,Collingwood:1993} then implies that the nilpotent orbits of $\SO_0(4,4)$ acting on the adjoint representation \rep{28}  are in one-to-one correspondence with the nilpotent orbits of $[\SL(2, \field{C})]^4$ acting on the fundamental representation  $\rep{(2,2,2,2)}$ and hence with the classification of four-qubit entanglement \footnote{Conventionally,  the Kostant-Sekiguchi theorem is formulated in terms of the maximal compact subgroup of $G$, $\SO(4)\times\SO(4)$ here, but it also applies to the maximally split case  $H_{\mathds{R}}^{*}$. The details are in the first appendix of \cite{Bossard:2009we}}.

\subsubsection{Extremal black holes and entanglement classes.} 

\begin{figure}[pth!]
\centerline{\includegraphics[width=\textwidth]{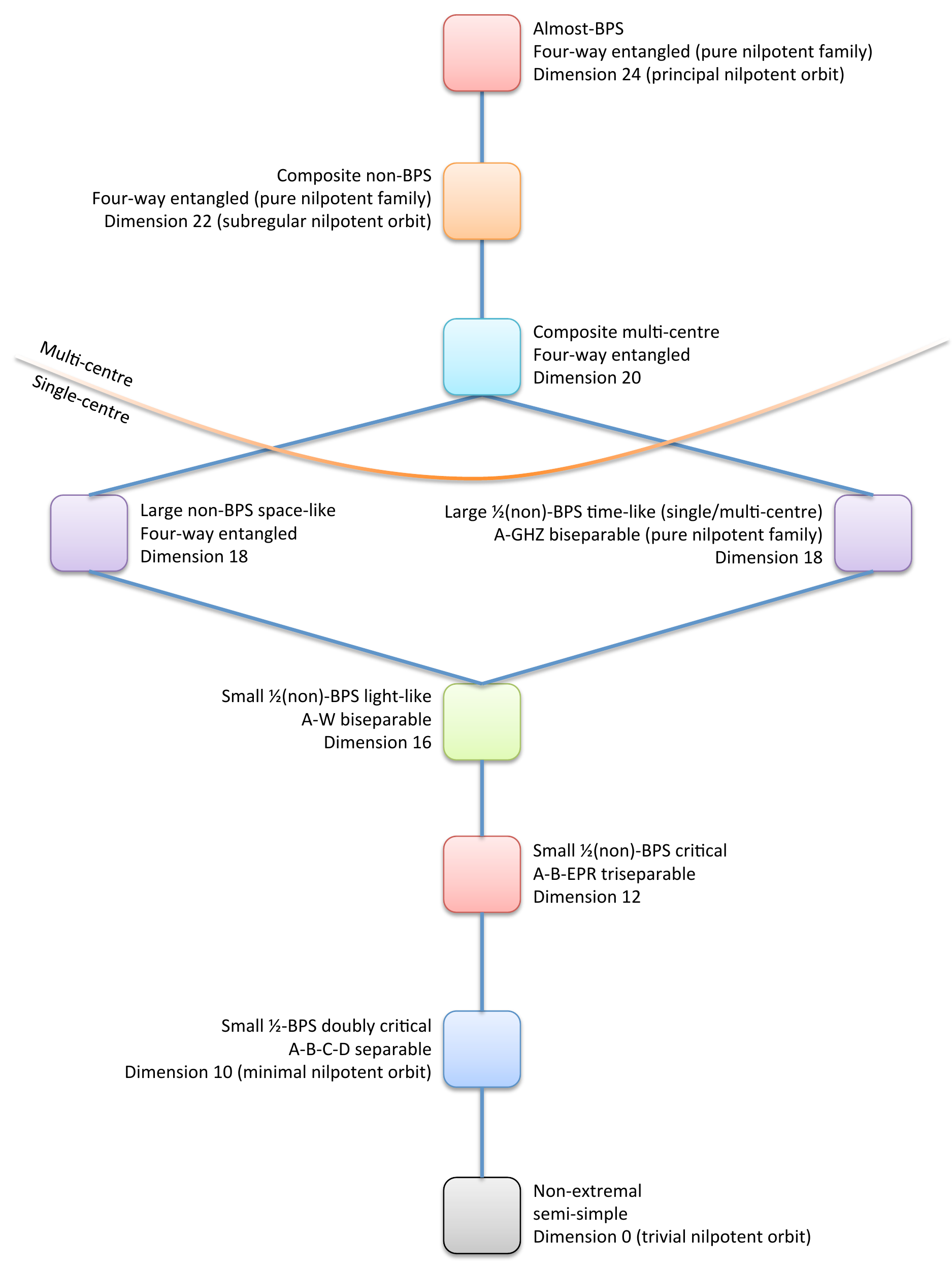}}
%\vspace*{7.2cm}
\caption{\label{fig:pres} Basic structure of the orbit classification with orbit dimension, black hole class and entanglement properties. Above the horizontal line  the orbits  only admit regular multi-centre solutions \cite{Bossard:2011kz}.}
\end{figure}

\begin{figure*}[!]
\includegraphics[width=\linewidth]{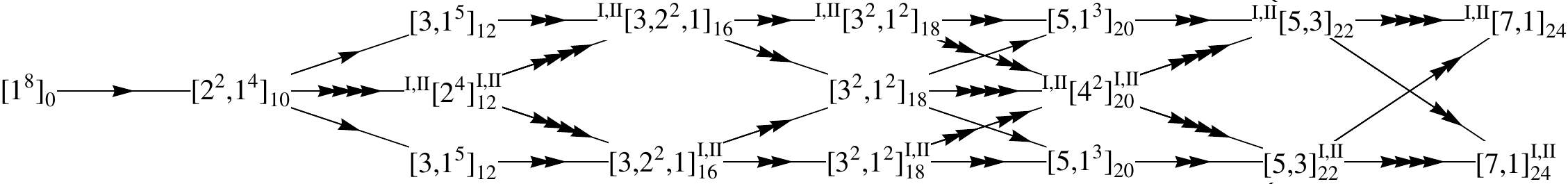}
\caption{$\SO_0(4,4)$ Hasse diagram. The integers inside the bracket indicate the structure of the appropriate Young tableau. The subscript indicates the real dimension of the orbit. The arrows indicate their closure ordering defining a partial order \cite{Djokovic:2000}.\label{fig:hasse}}
\end{figure*}

We are now in a position to summarise the single-centre/multi-centre black hole solutions of the $STU$ model and their corresponding four-qubit entanglement classes. The details of the orbit classification and the black hole solutions may be found in the original references \cite{Collingwood:1993,Behrndt:1996hu,Bellucci:2006xz,Bergshoeff:2008be,Goldstein:2008fq,Bena:2009en,Bena:2009ev,Bossard:2009at,Bossard:2009my,Bossard:2009we,Bossard:2011kz,Borsten:2011ai,Borsten:2011nq}. The explicit mappings to the four-qubit entanglement classes via the Kostant-Sekiguchi correspondence may be found in \cite{Rubens:2011phd}.
The 31 classes reduce to 9 under the permutation of the qubits, as described in \cite{Borsten:2011is}.   These 9 classes of orbits split into two: those for which the orbits have dimension less than 20, which admit regular single-centre solutions, and those for which the orbits have dimension 20 or greater, which only admit regular multi-centre solutions. The basic structure of the orbit classification is presented in \autoref{fig:pres}.

The trivial nilpotent orbit corresponds to the family of purely semi-simple states $G_{abcd}$, which are identically zero for all the four-qubit invariants set to zero. Equally, the non-trivial black hole solutions in this orbit are by definition non-extremal and are not treated here.

The supergravity interpretation of the $\SO_{0}\left( 4,4\right)
$-nilpotent orbits of dim$_{\mathds{R}}\leqslant 18$ (corresponding to single-centre solutions) considered
below is based on the  embedding of the $STU$ model in
$\mathcal{N}=8$, $D=4$ supergravity
discussed \textit{e.g.} in \cite{Ferrara:2006em}. This amounts to identifying the $%
\mathcal{N}=2$ central charge,  introduced in \eqref{central},  and the three associated ``matter'' charges with the
four skew-eigenvalues $Z_{i}$ ($i=1,...,4$ throughout) of the $\mathcal{N}=8$
central charge matrix as follows \cite{Ferrara:2006em,Ceresole:2009vp}
\begin{equation}
Z\equiv Z_{1};~\sqrt{g^{s\overline{s}}}\overline{D}_{\overline{s}}\overline{Z%
}\equiv iZ_{2};~\sqrt{g^{t\overline{t}}}\overline{D}_{\overline{t}}\overline{%
Z}\equiv iZ_{3};~\sqrt{g^{u\overline{u}}}\overline{D}_{\overline{u}}%
\overline{Z}\equiv iZ_{4}.  \label{stu-identifications}
\end{equation}
Thus, the effective BH potential $V_{BH}$, its criticality conditions or, equivalently, the  attractor equations, and the quartic invariant $
I_{4}=-D(\psi) $ of $\mathcal{N}=2$, $D=4$ $STU$ model can be traded for the
ones pertaining to maximal supergravity, respectively reading \cite
{Ferrara:2006em,Kallosh:1996uy,Ferrara:1997ci}:
\begin{eqnarray}
V_{BH} &=&\sum_{i}\left| Z_{i}\right| ^{4};  \label{V_BH} \\
\partial _{\phi }V_{BH} &=&0\Leftrightarrow Z_{i}Z_{j}+\overline{Z_{k}}%
\overline{Z_{l}}=0,~\forall i\neq j\neq k\neq l;  \label{N=8-AEs-gen} \\
{I}_{4} &=&\sum_{i}\left| Z_{i}\right| ^{4}-2\sum_{i<j}\left|
Z_{i}\right| ^{2}\left| Z_{j}\right| ^{2}+4\left( \prod_{i}Z_{i}+\prod_{i}%
\overline{Z_{i}}\right).  \label{I4}
\end{eqnarray}
Note, the central charge $Z_1=Z$ is on a different footing to the matter charges $Z_2, Z_3, Z_4$ for the $STU$ model, while they are all equivalent from the $\mathcal{N}=8$ perspective.

Here we summarise the black hole solutions and their associated entanglement classes. They are grouped into the three broad classes of small, large and multi-centre. Each orbit is labelled $\mathcal{O}_{\dim_\R}$, where $\dim_\R$ is the real dimension of the orbit.

\paragraph*{\textbf{Small black holes and totally/partially separable four-qubit states.}}

\paragraph{$\mathcal{O}_{10}$: $A$-$B$-$C$-$D$ and doubly-critical $\frac{1}{2}$-BPS black holes.}

 There is a unique orbit in this class, given by the unique unlabelled signed Young tableaux
\begin{equation}
 \young(+-,-+,+,-,+,-)
 \end{equation}

The doubly-critical $\frac{1}{2}$-BPS orbit comprises of small single-centre single-charge black holes with vanishing classical entropy \cite{Ferrara:1997uz}. The scalar fields take constant values along the attractor flow. The uniqueness of the doubly-critical $\frac{1}{2}$-BPS orbit was explained in \cite{Borsten:2011ai}. The permutation invariance and uniqueness  may be understood in terms of the conditions on dressed charges defining the solution \cite{Borsten:2011is},
\begin{equation}
\left\{
\begin{array}{l}
\left|Z_{1}\right| ^{2}=\left| Z_{2}\right| ^{2}=\left| Z_{3}\right|
^{2}=\left| Z_{4}\right| ^{2}; \\
\\
Z_{i}Z_{j}-\overline{Z_{k}}\overline{Z_{l}}=0,~\forall i\neq j\neq k\neq l,
\end{array}
\right.  \label{1-charge-constraint}
\end{equation}
which are themselves permutation invariant.

The corresponding four-qubit entanglement class given by Kostant-Sekiguchi is  the totally separable $A$-$B$-$C$-$D$ states with permutation invariant representative state,
\be
\ket{0000}.
\ee   
 This class belongs to the entanglement family $L_{abc_2}$ . The uniqueness of the Young tableau is reflected in the invariance of the $A$-$B$-$C$-$D$ class under the four-qubit permutation group.

\paragraph{$\mathcal{O}_{12}$: $A$-$B$-EPR and critical $\frac{1}{2}$-BPS/non-BPS black holes.}

This class has  six distinct $\SO_0(4,4)$-orbits, given by the six (labelled) signed Young tableaux,
\begin{equation}\label{eq:critab}
\young(-+-,+,-,+,-,+),\quad  \young(+-+,-,+,-,+,-),  \quad\begin{pmatrix}I,II&\young(+-,-+,+-,-+)&I,II\end{pmatrix}.   
 \end{equation}

This class gives the critical $\frac{1}{2}$-BPS/non-BPS orbits which comprises of small single-centre two-charge black holes with vanishing classical entropy.  This set of solutions  may be defined in terms of an $\left[ \SL\left( 2,\mathds{R}%
\right) \right] ^{3}$-invariant set of constraints \cite{Ferrara:1997uz, Cerchiai:2009pi,Borsten:2011is}:
\begin{equation}
2Z_{i}\overline{Z_{i}}^{2}-2%
\overline{Z_{i}}\sum_{j\neq i}\left| Z_{j}\right| ^{2}+4\prod_{j\neq
i}Z_{j}=0,\quad i=1,2,3,4.  \label{2-charge-constraint}
\end{equation}
This set of contraints is manifestly invariant under cyclic permutations of the index $%
j\neq i$. 

The four constraints (\ref{2-charge-constraint}) admit six representative
solutions \cite{Borsten:2011is}:
\begin{equation}
\left\{
\begin{array}{ll}
\mathbf{I}: &
\begin{array}{l}
Z_{1}=0=Z_{2},~Z_{3}=Z_{4}\neq 0;
\end{array}
\\
\mathbf{II}: &
\begin{array}{l}
Z_{1}=0=Z_{3},~Z_{2}=Z_{4}\neq 0;
\end{array}
\\
\mathbf{III}: &
\begin{array}{l}
Z_{1}=0=Z_{4},~Z_{2}=Z_{3}\neq 0;
\end{array}
\\
\mathbf{IV}: &
\begin{array}{l}
Z_{2}=0=Z_{3},~Z_{1}=Z_{4}\neq 0;
\end{array}
\\
\mathbf{V}: &
\begin{array}{l}
Z_{2}=0=Z_{4},~Z_{1}=Z_{3}\neq 0;
\end{array}
\\
\mathbf{VI}: &
\begin{array}{l}
Z_{3}=0=Z_{4},~Z_{1}=Z_{2}\neq 0,
\end{array}
\end{array}
\right.  \label{2-charge-sols-mcs-expl}
\end{equation}
These split into three non-BPS solutions $\{\mathbf{I}, \mathbf{II}, \mathbf{III}\}$ with $Z_1=0$ and three 1/2-BPS solutions   $\{\mathbf{IV}, \mathbf{V}, \mathbf{VI}\}$ with $Z_1\not=0$. Each of the six representative solutions are related under the permutations. To understand the relationship with the signed tableaux in \eqref{eq:critab}  we can define the sets:
\begin{equation}\label{2-charge-sols-mcs}
\{Z_{i}=Z_{i+1}=0, 
Z_{i+2}=Z_{i+3}\not= 0\}; \qquad
\{
Z_{i}=Z_{i+2}=0, Z_{i+1}=Z_{i+3}\not= 0\},
\end{equation}
which correspond to $\{\mathbf{I}, \mathbf{III}, \mathbf{IV}, \mathbf{VI}\}$ and $\{\mathbf{II}, \mathbf{V}\}$ respectively. These two sets are given by the two distinct types of Young tableaux. Embedding the $STU$ model in the $\mathcal{N}=8$ the four $Z_i$ are all put on the same footing and each orbit is identified, yielding a single class of critical two-charge 1/4-BPS black holes.

The corresponding four-qubit entanglement classes given by  Kostant-Sekiguchi  are the six possible $A$-$B$-EPR type configurations, where one pair of qubits forms a Bell state while the remaining two are  separable, 
\be
\ket{00}\otimes(\ket{00}+\ket{11}). 
\ee   
There are six such classes corresponding to the six possible entangled pairs.
 These classes belong to the entanglement families of type $L_{a_2b_{2}}$.  Clearly these six classes collapse into a single class under the four-qubit permutations. 

\paragraph{$\mathcal{O}_{16}$: $A$-$W$ and light-like $\frac{1}{2}$-BPS/non-BPS black holes.}

This class has  four distinct $\SO_0(4,4)$-orbits, given by the four (labelled) signed Young tableaux,
\begin{equation}\label{eq:lltab}
\begin{pmatrix}I,II&\young(+-+,-+,+-,-)\;\end{pmatrix}\quad \begin{pmatrix}\;\young(-+-,+-,-+,+)&I,II\end{pmatrix}.   
 \end{equation}

This class gives the light-like $\frac{1}{2}$-BPS/non-BPS orbits which comprises of small single-centre three-charge black holes with vanishing classical entropy.  This set of solutions  may be defined in terms of an $\left[ \SL\left( 2,\mathds{R}%
\right) \right] ^{3}$-invariant set of constraints on the central/matter charges \cite{Ferrara:1997uz, Cerchiai:2009pi,Borsten:2011is}:
\begin{equation}
\sum_{i}\left| Z_{i}\right| ^{4}-2\sum_{i<j}\left|
Z_{i}\right| ^{2}\left| Z_{j}\right| ^{2}+4\left( \prod_{i}Z_{i}+\prod_{i}%
\overline{Z_{i}}\right) =0.  \label{3-charge-constraint}
\end{equation}
This set of constraints is manifestly invariant under cyclic permutations of $Z_1, Z_2, Z_3, Z_4$. 

The  constraint (\ref{3-charge-constraint}) admits four representative
solutions \cite{Borsten:2011is}, each corresponding to one of the four labelled Young tableaux in \eqref{eq:lltab}.  
These split into three non-BPS solutions  and one 1/2-BPS solution, according as the particular properties of the central/matter charges. See \cite{Borsten:2011is} for details. Each of the four representative solutions are related under the permutations. Again, embedding the $STU$ model in the $\mathcal{N}=8$ theory the four $Z_i$ are all put on the same footing and each orbit is identified, yielding a single class of light-like three-charge 1/8-BPS black holes.

The corresponding four-qubit entanglement classes given by  Kostant-Sekiguchi  are the four possible $A$-$W$ type configurations, where three of the qubits are in the totally entangled $W$ state while the remaining one is  separable,
\be
\ket{0}\otimes(\ket{001}+\ket{010}+\ket{100}). 
\ee   
There are obviously four such classes corresponding to the four possible choices of separable quit. These classes belong to the entanglement families of type $L_{a_{2}0_{3}\oplus \overline{1}}$.   These four classes collapse into a single class under the four-qubit permutations. 
  
 \paragraph*{\textbf{Large black holes and  three-way/totally entangled  four-qubit states.}}

   \paragraph{$\mathcal{O}_{18a}$: $A$-GHZ and time-like $\frac{1}{2}$-BPS/non-BPS black holes.}
   
   This class has  four distinct $\SO_0(4,4)$-orbits, given by the four (labelled) signed Young tableaux,
\begin{equation}\label{eq:tltab}
\begin{pmatrix}I,II&\young(+-+,+-+,-,-)\end{pmatrix}\\\begin{pmatrix}\;\young(-+-,-+-,+,+)&I,II\end{pmatrix}.   
 \end{equation}

This class gives the time-like $\frac{1}{2}$-BPS/non-BPS orbits which comprises of the large single-centre four-charge black holes with non-zero classical entropy\cite{Ferrara:1997uz, Bellucci:2006xz, Cerchiai:2009pi,Borsten:2011is},
\be
S=\pi\sqrt{-D(\psi)},
\ee
where $\psi$ is the three-qubit state built out of the 4-dimensional electromagnetic charges as in \eqref{corres}. Here, Cayley's hyperdeterminant is constrained to be negative $D(\psi)<0$, or equivalently $\mathcal{I}_4>0$.

The four orbits correspond to the four classes of representative solution to the constraints  $Z_{i}Z_{j}+\overline{Z_{k}} \overline{Z_{l}}=0$, $\forall i\neq j\neq k\neq l$ and $\mathcal{I}_4>0$ given by \cite{Ferrara:1997uz, Bellucci:2006xz, Cerchiai:2009pi,Borsten:2011is},
\be
Z_i\not=0, Z_j=Z_k=Z_l=0, \qquad \forall i\not=j\not=k\not=l.
\ee

As explained in \cite{Bellucci:2006xz, Bossard:2009we, Borsten:2011ai} there just is one 1/2-BPS $\SO_0(4,4)$-orbit given by the case $Z_1\not=0$, which is on a different footing from the $STU$ perspective to the remaining three possibilities. These are indeed all non-BPS, but actually fall into two separate classes as explained in \cite{Bellucci:2006xz, Borsten:2011is}. 

Embedding these solutions in the $\mathcal{N}=8$ theory they become equivalent and correspond to a single class of 1/8-BPS large black holes. 

This class also includes the interacting 1/2-BPS multi-centre solutions first discovered by Bates and Denef \cite{Denef:2000nb, Bates:2003vx, Bossard:2011kz}, where every centre is individually 1/2-BPS. Accordingly, this orbit is referred to as the ``Denef'' system in \cite{Bossard:2011kz}.

The corresponding four-qubit entanglement classes given by  Kostant-Sekiguchi  are the four possible $A$-GHZ type configurations, where three of the qubits are in the totally entangled GHZ state, with non-zero three-tangle $\tau_{ABC}\not=0$,  while the remaining one is  separable,
\be
\ket{0}\otimes(-\ket{111}+\ket{001}+\ket{010}+\ket{100}). 
\ee   
There are obviously four such classes corresponding to the four possible choices of separable qubit. These classes belong to the entanglement families of type $L_{0_{3\oplus\bar{1}}0_{3\oplus\bar{1}}}$.   This is a pure single nilpotent orbit in the sense that all four algebraically independent SLOCC invariants are identically zero for the normal form. These four classes collapse into a single class under the four-qubit permutations. 

  \paragraph{$\mathcal{O}_{18b}$: Genuine four-way entanglement and space-like non-BPS black holes.}

 There is a unique orbit in this class, given by the unique unlabelled signed Young tableaux
\begin{equation}\label{eq:nonBPSspacelike}
\young(-+-,+-+,-,+) \end{equation}

This class gives the space-like non-BPS solution which comprises of the large single-centre four-charge black holes with non-zero classical entropy \cite{Ferrara:1997uz, Bellucci:2006xz, Cerchiai:2009pi,Borsten:2011is},
\be
S=\pi\sqrt{D(\psi)},
\ee
where Cayley's hyperdeterminant is now constrained to be positive $D(\psi)>0$, or equivalently $\mathcal{I}_4<0$. 

This solution is determined by the representative solution to the constraints  $Z_{i}Z_{j}+\overline{Z_{k}} \overline{Z_{l}}=0$ and $\mathcal{I}_4<0$ given by \cite{Ferrara:1997uz, Bellucci:2006xz, Cerchiai:2009pi,Borsten:2011is},
\be
Z_{i}=\rho e^{i\frac{\varphi }{4}},\qquad \rho \in \mathds{R}_{0}^{+},~\forall i,\qquad
\varphi =\pi +2k\pi ,~k\in \mathds{Z}, \label{nBPS}
\ee
which is permutation invariant as required for consistency  with our single Young tableau. 
Accordingly, when embedded in the $\mathcal{N}=8$ theory this solution continues to be non-BPS. However, unlike the previous class it does not admit any interacting multi-centre solutions \cite{Gaiotto:2007ag}. 

Interestingly, the corresponding four-qubit entanglement class  given by  Kostant-Sekiguchi  is the first instance of genuine four-way entanglement for a nilpotent state. These states belongs to the class $L_{ab_3}$. A representative state is given by,
\be\label{eq:4waynBPS}
\tfrac{i}{\sqrt{2}}(\ket{0001}+\ket{0010}-\ket{0111}-\ket{1011}). 
\ee   
While the permutation invariance is not manifest in this case, all states obtained from \eqref{eq:4waynBPS} by swapping the qubits are related by $[\SL(2, \C)]^4$.

 \paragraph*{\textbf{Multi-centre black holes and  four-way entangled  four-qubit states.}}

The three remaining classes of $\SO_0(4,4)$-orbits do not admit regular single centre solutions since they all have representative nilpotent elements of degree greater than three. However, they do admit interacting multi-centre solutions for which, although the total charge has degree greater than three, the individual centres have charges of nilpotency degree  three  and belong to the orbits described above \cite{Bossard:2011kz}. Consequently, the three associated nilpotent four-qubit entanglement classes can only be interpreted in terms of interacting multi-centre solutions. What this means for the entanglement properties for these classes remains unclear.

   \paragraph{$\mathcal{O}_{22}$: $L_{0_{5\oplus \bar{3}}}$ and non-BPS composite black holes.} This class has  four distinct $\SO_0(4,4)$-orbits, given by the four (labelled) signed Young tableaux,
\begin{equation}\label{eq:22tab}
\begin{pmatrix}I,II&\young(-+-+-,+-+)\end{pmatrix},\quad \begin{pmatrix}\;\young(+-+-+,-+-)&I,II\end{pmatrix}.   
 \end{equation}
In the mathematical literature it is referred to as the ``subregular'' orbit meaning it is has  second highest dimension and nilpotency degree.      
   
 The two-centre solutions deriving from this orbit were recently obtained in \cite{Bossard:2011kz}.  At the horizon of each centre the scalar momentum $P$ will lie in the single-centre orbit of space-like non-BPS black holes given by the Young tableaux in \eqref{eq:nonBPSspacelike}. Hence, it can be understood as an interacting composite system where each centre is space-like non-BPS.
Note, this does not contradict the fact that there are no interacting multi-centre space-like non-BPS balck holes in the orbit defined by \eqref{eq:nonBPSspacelike}, since away from the horizons $P$ will  lie strictly in the subregular  nilpotent orbit considered here and given by \eqref{eq:22tab}.

This can be understood clearly in terms of a representative nilpotent charge $Q\in \mathfrak{p}$ of the subregular orbit. $Q$ satisfies $Q^5=0$ but not $Q^4=0$ and hence there are no regular single-centre solutions. However, it is simple to show  \cite{Bossard:2011kz} that there exists a finite set of charges $Q_A$, satisfying  $Q_{A}^3=0$ and, in particular, $Q_A\in \mathcal{O}_{18b}$, such that,
\be
Q=\sum_A Q_A,
\ee
for which one can construct an everywhere regular solution. Another interesting feature that we have yet to encounter  is that each black hole in the two-centre case can carry intrinsic angular momentum along the axis of interaction.

Turning our attention to the corresponding class of entangled states given by Kostant-Sekiguchi we obtain the first example of a purely  nilpotent family with genuine  four-way entanglement:
\be
\ket{0000}+\ket{0101}+\ket{1000}+\ket{1110}.
\ee

  \paragraph{$\mathcal{O}_{24}$: $L_{0_{7\oplus \bar{1}}}$    and Almost-BPS composite black holes.}
This class has  four distinct $\SO_0(4,4)$-orbits, given by the four (labelled) signed Young tableaux,
\begin{equation}\label{eq:24tab}
\begin{pmatrix}I,II&\young(+-+-+-+,-)\end{pmatrix},\quad \begin{pmatrix}\;\young(-+-+-+-,+)&I,II\end{pmatrix}.   
 \end{equation}
 In the mathematical literature it is referred to as the ``principle'' orbit meaning it is has  the highest dimension and nilpotency degree.      
   
 The black hole solutions  originating  from this orbit form the  ``almost-BPS'' class, originally obtained by different methods in \cite{Goldstein:2008fq, Bena:2009ev, Bena:2009en}.  In \cite{Bossard:2011kz} Bossard and Ruef showed that all previously known solutions in the almost-BPS class may be recover from their nilpotent orbit analysis. As for the composite non-BPS black holes in $\mathcal{O}_{22}$, described above, there are no regular single-centre solutions falling strictly in this orbit. The almost-BPS class does admit regular multi-centre solutions \cite{Bena:2009ev, Bena:2009en}, but, unlike the composite non-BPS class, there can be no interactions between space-like non-BPS centres. For an interacting  two-centre solution  the scalar momentum $P$ at one of the two horizons will lie in the single-centre orbit $\mathcal{O}_{18a}$ of time-like 1/2-BPS black holes given by the Young tableaux in \eqref{eq:tltab}. Hence, it can be understood as an interacting composite system where  one centre is space-like non-BPS while the other is time-like 1/2-BPS.

Again, this can be understood  in terms of a representative nilpotent charge $Q\in \mathfrak{p}$ of the principle orbit. $Q$ satisfies $Q^7=0$ but not $Q^6=0$ and hence there are no regular single-centre solutions. But, as with the composite non-BPS class, it is not difficult to show  \cite{Bossard:2011kz} that $Q$ may be split $Q=Q_1 + Q_2$, where $Q_1\in\mathcal{O}_{18a}$ and  $Q_1\in\mathcal{O}_{18b}$.

 The corresponding class of entangled states given by Kostant-Sekiguchi is, again in common with the subregular $\mathcal{O}_{22}$ orbit,   a purely  nilpotent family with genuine  four-way entanglement:
\be
 \ket{0000}+\ket{1011}+\ket{1101}+\ket{1110}.  
\ee

  \paragraph{$\mathcal{O}_{20}$: $L_{a_4}$ and composite black holes.} This class has  six distinct $\SO_0(4,4)$-orbits, given by the six (labelled) signed Young tableaux,
\begin{equation}\label{eq:20tab}
 \young(-+-+-,+,-,+),\quad   \young(+-+-+,-,+,-) ,  \quad\begin{pmatrix}I,II&\tyoung(+-+-,-+-+)& I,II\end{pmatrix} .   
 \end{equation}
The system of equations for this class of orbits may be obtained by a straight forward truncation of either the composite non-BPS black holes or the almost-BPS system \cite{Bossard:2011kz}. However, a complete treatment,   clarifying this point, is given  in \cite{Borsten:2012wr}. In fact, this observation is generically true: all systems of equations, and therefore their black hole solutions, can be obtained by suitable truncations from just three of the nilpotent orbits, $\mathcal{O}_{18a}, \mathcal{O}_{22}$ and $\mathcal{O}_{24}$, corresponding to the time-like 1/2-BPS, composite non-BPS and almost-BPS classes respectively \cite{Bossard:2011kz}.
 
One of the four-qubit representatives is given by the four-way entangled state,
  \be
  i\ket{0001}+\ket{0110}-i\ket{1011},
  \ee
however, unlike the previous two cases, the family containing this orbit has a semisimple component as can be seen from its representative state,
\be
 a\left(\ket{0000} + \ket{1111}\right) + \ket{	0011} + 	\ket{0101} + \ket{0110}.
\ee

\paragraph*{General comments.} 
 It is intriguing that the three governing nilpotent orbits are precisely  those which constitute entanglement families in their own right, i.e. their semisimple components are identically zero. 
These same orbits are also interesting from an entanglement perspective. While all the SLOCC invariants are identically zero in these instances, two of the orbits have genuine four-way entanglement. In this sense they are on the same footing as the three-qubit $W$ state. Since the SLOCC invariants (or $n^{th}$-roots thereof) provide good entanglement measures \cite{Verstraete:2003}, it is natural to ask what are the physical properties of such entangled states. 

In the case of three qubits there is  a very precise notion  of the ``degree'' of non-locality (or contexuality)  given by \cite{2012arXiv1203.1352A},
\be
A\text{-EPR} < W < \text{GHZ}.
\ee
While both being three-way entangled the GHZ state is \emph{strongly} contextual while the $W$ state is not \cite{2012arXiv1203.1352A}. This agrees  precisely with the entanglement classification obtained using the SLOCC paradigm. Can this hierarchy be generalised to the four-qubit case (this, at first sight at least, seems difficult since the strong contextually condition satisfied by the three-qubit GHZ state is, in a certain sense,  maximal) and, if so, will it agree with the SLOCC entanglement classification. Of course, the SLOCC entanglement classification is unambiguously ``correct''  in its own terms - two states may be probabilistically interrelated by local operations and classical communication if and only if they are in the same SLOCC entanglement class. What we are asking here rather is whether two distinct SLOCC classes of entanglement can always be distinguished by some physical and observable measure of their (generalised) non-locality properties \footnote{Indeed, failure would call into question the usefulness of the SLOCC entanglement classification.}. The simplicity of the two totally entangled orbits,  $\mathcal{O}_{22}$ and $\mathcal{O}_{24}$, coupled with the fact that the strongly contextual GHZ state is in a different orbit altogether,  makes them a natural testing ground. 
\begingroup
\squeezetable
\newlength\BHCol
\setlength\BHCol{1.4cm}
\newlength\RepCol
\setlength\RepCol{2.4cm}
\begin{table*}[!]
\caption{Each black hole nilpotent $\SO_0(4,4)$ orbit  corresponds to a 4-qubit nilpotent $[\SL(2,\field{C})]^4$ orbit. $z_H$ is the horizon value of the $\SUSY=2$, $D=4$ central charge.\label{tab:realcosets}}
\begin{ruledtabular}
\begin{tabular*}{\textwidth}{@{\extracolsep{\fill}}>{\centering}m{\BHCol}*{4}{M{c}}>{\centering$}m{\RepCol}<{$}cM{c}}
\multicolumn{3}{c}{$STU$ black holes}                                                                                                                                                                                                                                                                          & \multirow{2}{*}{$\dim_\field{R}$} & \multicolumn{4}{c}{Four qubits}                                                                                                                                                                                                   \\
\cline{1-3}\cline{5-8}\\[-6pt]
description                           & \text{Young tableaux}                                                                                                                       & \SO_0(4,4)\text{ coset}                                                                                                  &                                   & [\SL(2,\field{C})]^4\text{ coset}                                                                        & \text{nilpotent rep}                                            &  	     & \text{family}                               \\
\\[-6pt]\hline\hline\\[-6pt]
trivial                               & \text{trivial}                                                                                                                              & \frac{\SO_0(4,4)}{\SO_0(4,4)}                                                                                            & 1                                 & \frac{[\SL(2,\field{C})]^4}{[\SL(2,\field{C})]^4}                                                        & 0                                                               & $\in$  & G_{abcd}                                    \\
\\[-6pt]\hline\\[-6pt]
doubly-critical \half BPS             & \tyoung(+-,-+,+,-,+,-)                                                                                                                      & \frac{\SO_0(4,4)}{[\SL(2,\field{R})\times \SO(2,2,\field{R})]\ltimes[(\rep{2,4})^{(1)} \oplus{\rep1}^{(2)}]}             & 10                                & \frac{[\SL(2,\field{C})]^4}{[\SO(2,\field{C})]^3\ltimes\field{C}^4}                                      & \ket{0110}                                                      & $\in$  & L_{abc_2}                                   \\
\\[-6pt]\hline\\[-6pt]
                                      & \tyoung(-+-,+,-,+,-,+)                                                                                                                      & \frac{\SO_0(4,4)}{\SO(3,2;\field{R})\ltimes[(\rep{5\oplus1})^{(2)}]}                                                     &                                   &                                                                                                          &                                                                 &        &                                             \\
\\[-6pt]
critical, \half BPS and non-BPS       & \tyoung(+-+,-,+,-,+,-)                                                                                                                      & \frac{\SO_0(4,4)}{\SO(2,3;\field{R})\ltimes[(\rep{5\oplus1})^{(2)}]}                                                     & 12                                & \frac{[\SL(2,\field{C})]^4}{[\SO(3,\field{C})\times \field{C}]\times[\SO(2,\field{C})\ltimes \field{C}]} & \ket{0110}+\ket{0011}                                           & $\in$  & L_{a_2b_2}                                  \\
\\[-6pt]
                                      & \begin{pmatrix}I,II&\tyoung(+-,-+,+-,-+)&I,II\end{pmatrix}                                                                                  & \frac{\SO_0(4,4)}{\Sp(4,\field{R})\ltimes[(\rep{5\oplus1})^{(2)}]}                                                       &                                   &                                                                                                          &                                                                 &        &                                             \\
\\[-6pt]\hline\\[-6pt]
lightlike \half BPS and non-BPS       & \begin{array}{c}\begin{pmatrix}I,II&\tyoung(+-+,-+,+-,-)\;\end{pmatrix}\\\begin{pmatrix}\;\tyoung(-+-,+-,-+,+)&I,II\end{pmatrix}\end{array} & \frac{\SO_0(4,4)}{\SL(2,\field{R})\ltimes[(2\times {\rep 2})^{(1)}\oplus(3\times{\rep 1})^{(2)}\oplus{\rep 2}^{(3)}]}    & 16                                & \frac{[\SL(2,\field{C})]^4}{[\SO(2,\field{C})\ltimes\field{C}]\times \field{C}^2}                        & \ket{0110}+\ket{0101}+\ket{0011}                                & $\in$  & L_{a_2 0_{3\oplus \bar{1}}}                 \\
\\[-6pt]\hline\\[-6pt]
large non-BPS $z_H \neq 0$            & \tyoung(-+-,+-+,-,+)                                                                                                                        & \frac{\SO_0(4,4)}{\SO(1,1,\field{R})\times\SO(1,1,\field{R})\ltimes[(\rep{(2,2)\oplus(3,1)})^{(2)}\oplus{\rep 1}^{(4)}]} & 18                                & \frac{[\SL(2,\field{C})]^4}{\field{C}^3}                                                                 & \tfrac{i}{\sqrt{2}}(\ket{0001}+\ket{0010}-\ket{0111}-\ket{1011})& $\in$  & L_{ab_3}                                    \\
\\[-6pt]\hline\\[-6pt]
                                      & \tyoung(-+-+-,+,-,+)                                                                                                                        & \frac{\SO_0(4,4)}{\SO(2,1;\field{R})\ltimes[\rep{1}^{(2)}\oplus\rep{3}^{(4)}\oplus\rep{1}^{(6)}]}                        &                                   &                                                                                                          &                                                                 &        &                                             \\
\\[-6pt]
composite                          & \tyoung(+-+-+,-,+,-)                                                                                                                        & \frac{\SO_0(4,4)}{\SO(1,2;\field{R})\ltimes[\rep{1}^{(2)}\oplus\rep{3}^{(4)}\oplus\rep{1}^{(6)}]}                        & 20                                & \frac{[\SL(2,\field{C})]^4}{\SO(2,\field{C})\times\field{C}}                                             & i\ket{0001}+\ket{0110}-i\ket{1011}                              & $\in$  & L_{a_4}                                     \\
\\[-6pt]
                                      & \begin{pmatrix}I,II&\tyoung(+-+-,-+-+)& I,II\end{pmatrix}                                                                                   & \frac{\SO_0(4,4)}{\Sp(2,\field{R})\ltimes[\rep{1}^{(2)}\oplus\rep{3}^{(4)}\oplus\rep{1}^{(6)} ]}                         &                                   &                                                                                                          &                                                                 &        &                                             \\
\\[-6pt]\hline\hline\\[-6pt]
large \half BPS and non-BPS  $z_H=0$  & \begin{array}{c}\begin{pmatrix}I,II&\tyoung(+-+,+-+,-,-)\end{pmatrix}\\\begin{pmatrix}\;\tyoung(-+-,-+-,+,+)&I,II\end{pmatrix}\end{array}   & \frac{\SO_0(4,4)}{\SO(2,\field{R})\times\SO(2,\field{R})\ltimes[(\rep{(2,2)\oplus(3,1)})^{(2)}\oplus{\rep 1}^{(4)}]}     & 18                                & \frac{[\SL(2,\field{C})]^4}{[\SO(2,\field{C})]^2\times \field{C}}                                        & \ket{0000}+\ket{0111}                                           & $\in$  & L_{0_{3\oplus \bar{1}} 0_{3\oplus \bar{1}}} \\
\\[-6pt]\hline\\[-6pt]
composite non-BPS                          & \begin{array}{c}\begin{pmatrix}I,II&\tyoung(-+-+-,+-+)\end{pmatrix}\\\begin{pmatrix}\;\tyoung(+-+-+,-+-)&I,II\end{pmatrix}\end{array}       & \frac{\SO_0(4,4)}{\field{R}^{3(2)}\oplus\field{R}^{1(4)}\oplus\field{R}^{2(6)}}                                          & 22                                & \frac{[\SL(2,\field{C})]^4}{\field{C}}                                                                   & \ket{0000}+\ket{0101}+\ket{1000}+\ket{1110}                     & $\in$  & L_{0_{5\oplus \bar{3}}}                     \\
\\[-6pt]\hline\\[-6pt]
composite Almost-BPS                          & \begin{array}{c}\begin{pmatrix}I,II&\tyoung(+-+-+-+,-)\end{pmatrix}\\\begin{pmatrix}\;\tyoung(-+-+-+-,+)&I,II\end{pmatrix}\end{array}       & \frac{\SO_0(4,4)}{\field{R}^{(2)}\oplus\field{R}^{2(6)}\oplus\field{R}^{(10)}}                                           & 24                                & \frac{[\SL(2,\field{C})]^4}{\id}                                                                         & \ket{0000}+\ket{1011}+\ket{1101}+\ket{1110}                     & $\in$  & L_{0_{7\oplus \bar{1}}}                     \\
\end{tabular*}
\end{ruledtabular}
\end{table*}
\endgroup

\subsection{The attractor mechanism for $STU$ black holes}
\subsubsection{Attractors}

We have seen that there are static spherically symmetric extremal
black hole solutions in the $STU$ model of three basic types. There
are supersymmetric ($\frac{1}{2}$ BPS) black holes , and
nonsupersymmetric (non-BPS) ones with either vanishing or
nonvanishing central charge. These can be characterized by either
the entanglement properties of the four-qubit state
$\vert\Lambda\rangle$ or by the nilpotent orbits of the associated
Noether charge $Q$. As discussed in the previous subsection the
latter procedure results in a finer classification of black holes
that can be mapped in a one to one manner to a corresponding {\it
complex} SLOCC classification of nilpotent states of four-qubits.

We have also demonstrated how we can elegantly repackage the
information on the charges the moduli, and the warp factor in a
{\it complex} three-qubit state satisfying special reality
conditions. The moduli characterize the geometry of the extra
dimensions, on the other hand the warp factor characterizes the
geometry of the spacetime manifold. From the spacetime
perspective the black hole solutions we are interested in are of
Reissner-Nordstrom type, with asymptotically Minkowski behavior.
As we have noted elsewhere the moduli are really massless scalar
fields without any potential. Their dynamics taken together with
 the warp factor makes it possible to calculate the
macroscopic black hole entropy via the Bekenstein-Hawking area
law. However, fixing the values of these moduli at the asymptotic
region and then solving the dynamical equations governing their
radial flow, gives rise to their horizon values. These
continuously adjustable asymptotic values  would then feature in
the macroscopic
 black hole entropy. This is a dangerous possibility since the
 entropy should depend
 only on quantities that take  discrete values, such as electric
 and magnetic charges. This could be a problem for a microscopic
 reinterpretation of our macroscopic entropy since the number of
 microstates is an integer that should not depend on continuous
 parameters.

Luckily the radial dependence of these moduli fields gives rise to
{\it attractors}  \cite{Ferrara:1995ih,Strominger:1996kf,Ferrara:1996dd}. This means that regardless of
their asymptotic values, the moduli flow to particular horizon
values that can be expressed in terms of the quantized charges.
The existence of such attractors, necessary for a microscopic
reinterpretation of the black hole entropy, is the essence of the
{\it attractor mechanism}. In the following we would like to see
how the attractor mechanism unfolds itself in terms of our three
qubit state of (\ref{ezittaklassz}), as we start from the
asymptotic region and approach the horizon.

More precisely our basic concern will be a study of the radial
behavior of the three-qubit state \beq \vert\chi(\tau)\rangle
=e^{U(\tau)}({ V}\otimes { V} \otimes {
V})(S_3(\tau)\otimes S_2(\tau)\otimes
S_1(\tau))\vert\gamma\rangle.
 \label{ezittaklassz2}
\eeq \noindent where \beq \vert\gamma\rangle=(\sigma_3\otimes
\sigma_3\otimes \sigma_3) \vert\Gamma\rangle \label{naeztkapdki2}
\eeq obtained from (\ref{ezittaklassz}) after setting
$k=p_{\sigma}=0$. Here for the definitions see (\ref{Smatrix})
and for later use we also give the explicit form of the amplitudes
of $\vert\gamma\rangle$
 \beq
\begin{pmatrix}{\gamma}_{000},&{\gamma}_{001},&{\gamma}_{010},&{\gamma}_{100}\\{\gamma}_{111},&{\gamma}_{110},&{\gamma}_{101},&{\gamma}_{011}\end{pmatrix}
\equiv
\frac{1}{\sqrt{2}}\begin{pmatrix}p^0,&p^1,&p^2,&p^3\\-q_0,&q_1,&q_2,&q_3
\label{gammachargeuj}
\end{pmatrix}.
\eeq \noindent We will also use the state $\vert\psi(\tau)\rangle$
defined as \beq \vert\psi(\tau)\rangle := e^{-U(\tau)}\vert
\chi(\tau)\rangle \label{nowrap} \eeq \noindent obtained after
removing the dependence on the warp factor. For the amplitudes of
the state $\vert\psi\rangle$ at the horizon we use the shorthand
notation $\psi_{ABC}\equiv \lim_{\tau\to\infty}\psi_{ABC}(\tau)$
with $A,B,C=0,1$.

Recall, the effective Lagrangian of the $STU$ model 
describes the motion of a fiducial particle on ${\mathds R}\times
{\cal M}_4$ subject to the potential $-e^{2U}V_{BH}$. Here the
four dimensional moduli space is ${\cal M}_4=[\SL(2,{\mathds
R})/\SO(2)]^{\otimes 3}$. The parameter $\tau=1/r$ plays the role of
time. According to (\ref{constraint})   energy is conserved. 
Close to local maxima of
$e^{2U}V_{BH}$ we expect oscillatory motion, close to minima of
$e^{2U}V_{BH}$ we expect unstable solutions {\it except} when the
initial conditions are fine tuned corresponding to extremal black
holes. In this case the particle climbes the hill with just enough
energy to reach the top ending its motion there. Such finite
particle actions correspond to {\it solitons} which are finite
energy solutions of the original field theory. Hence we can regard
our extremal black holes as solitons.
 In  order to classify such   solutions
one has to find  the critical points of $V_{BH}$
at the horizon.
Using the amplitudes of $\lim_{\tau\to\infty}\vert\psi(\tau)\rangle$ 
extremization of the black-hole potential (\ref{potential}) with respect              to the moduli
yields the following classification of attractor equations \cite{Saraikin:2007jc}:

\begin{itemize}
\item {\bf BPS solutions}
\beq
{\psi}_1={\psi}_2={\psi}_4=0, 
\label{BPSattra}
\eeq
\item
{\bf Non-BPS solutions with $Z\not= 0$}
\beq
\label{nBPSsol}
\vert{\psi}_0\vert^2=\vert{\psi}_1\vert^2=\vert{\psi}_2\vert^2=\vert{\psi}_
4\vert^2,
\eeq
Notice that the amplitudes  ${\psi}_0={\psi}_{000}$ and ${\psi}_7={\psi}_{111}$  play a special role as they are related to the central charge and its complex conjugate,
\beq
Z=-{\psi}_{7},\qquad \overline{Z}={\psi}_{0}.
\label{centghz}
\eeq
\noindent
Since $Z\neq 0$ the corresponding amplitudes are non-zero. 

\item  {\bf Non-BPS solutions with $Z= 0$}. \beq
\psi_0=\psi_1=\psi_2=0 \eeq \noindent and two more cases with $12$
replaced by $23$ and $31$ respectively in the last two amplitudes.

\end{itemize}

\subsubsection{BPS attractors}

Let us consider now the case of BPS attractors.
In this case since
$D_iZ=2e^{i\alpha}{\partial}_i\vert Z\vert$
according to (\ref{alternativepot}) critical points of $\vert Z\vert$ are also critical points of $V_{BH}$.
It turns out  \cite{Ferrara:1995ih,Strominger:1996kf,Ferrara:1996dd} that such critical points are also minima.
Recall also from (\ref{bpsexpl}) that for BPS solutions the second order equations of (\ref{Euler}) can be replaced by first order ones.
If we assume $Z_{c}\neq 0$ then from the first of (\ref{bpsexpl}) we get
\beq
\lim_{\tau\to\infty}e^{-U}=\vert Z_c\vert \tau\qquad \text{where}\qquad Z_c:=\lim_{\tau\to\infty}Z(\tau).\eeq
Hence  from (\ref{metricans}) the near horizon geometry of the black hole is $AdS_2\times S^2$
\beq
ds^2=\left(-\frac{r^2}{{\vert Z\vert_c}^2}dt^2+\frac{{\vert Z\vert_c}^2}{r^2}
dr^2\right)+{\vert Z\vert_c}^2\left(d\theta^2+{\sin}^2\theta d\varphi^2\right)
\eeq
Here as usual $\tau=1/r$ where $r$ is the radial distance from the horizon.
The horizon area is
\beq A=4\pi \vert Z_c\vert^2\eeq
hence the thermodynamic Bekenstein-Hawking entropy is
\beq
S_{BH}=\frac{A}{4}=\pi\vert Z_c\vert^2.\label{entro}\eeq

In order to calculate the value of $Z_c$ in terms of the charges let us have a look at the attractor equations (\ref{BPSattra}).
In the case of the $STU$ model these equations determine the horizon values of the moduli in terms of the charges.
These equations also have a very interesting quantum information
theoretic interpretation  \cite{Levay:2006kf}.
We see that though the amplitudes of the state $\vert\psi(\tau)\rangle$ away from the
horizon are generally nonzero, at the horizon 
all of the amplitudes  die out
except for  $\psi_{
000}$ and $\psi_{111}$.
According to (\ref{GHZW}) these are the non-zero amplitudes of the GHZ
states: the BPS attractor flow can be interpreted as a
distillation procedure
for a GHZ state.
Moreover, since at the critical point $V_{BH}=\vert Z\vert^2$ is minimal,
according to (\ref{normsquared}) and (\ref{nowrap})
it
follows that at this point the magnitudes of the GHZ amplitudes are minimized.
(\ref{entro})
 also shows that the black hole entropy is proportional to the horizon  value of the magnitudes of the GHZ amplitudes of our three-qubit state $\vert\psi(\tau)\rangle$.

Let us write
the superpotential of (\ref{central}) as
 \beq
  W(z_3,z_2,z_1)={\gamma}_{CBA}c^Cb^Ba^A=\gamma_{CBA}\varepsilon^{AA^{\prime}}
   \varepsilon^{BB^{\prime}}\varepsilon^{CC^{\prime}}
    c_{C^{\prime}}b_{B^{\prime}}a_{A^{\prime}} \eeq

    \beq
    a_A\leftrightarrow
    \begin{pmatrix}1\\z_1\end{pmatrix},\qquad
    b_B\leftrightarrow
    \begin{pmatrix}1\\z_2\end{pmatrix}\qquad c_C\leftrightarrow
    \begin{pmatrix}1\\z_3\end{pmatrix}.
    \nonumber\eeq Then the (\ref{BPSattra}) BPS attractor equations can be written as
    \beq
    W(\overline{z}_3,z_2,z_1)=0,\qquad
    W(z_3,\overline{z}_2,z_1)=0,\qquad
    W(z_3z_2,\overline{z}_1)=0\eeq
    and their complex conjugates. The amplitudes of $\vert\gamma\rangle$
    are given by (\ref{gammachargeuj}).

Using the fact that $\gamma_{CBA}$ is real these equations taken
together with their complex conjugates are equivalent to the
vanishing of the $2\times 2$ determinants  \cite{Behrndt:1996hu,Levay:2006kf}
\beq {\rm Det}\left(\gamma_{CBA}c^C\right)=0,\qquad {\rm Det}
\left(\gamma_{CBA}b^B\right)=0,\qquad{\rm
Det}\left(\gamma_{CBA}a^A\right)=0\eeq provided the imaginary
parts of the moduli are non vanishing. The result is three
quadratic equations. Assuming
$y_1,y_2$ and $y_3$ positive,   the stabilized  values
for the moduli are given by, \beq
z^a(\infty;p,q)=\frac{(\gamma_0\cdot\gamma_1)^a+i\sqrt{-D}}
{(\gamma_0\cdot\gamma_0)^a},\qquad a=1,2,3.\nonumber\eeq
\beq(\gamma_0\cdot\gamma_1)^1\equiv
\gamma_{CB0}\varepsilon^{CC^{\prime}}\varepsilon^{BB^{\prime}}
\gamma_{C^{\prime}B^{\prime}1},\quad
(\gamma_0\cdot\gamma_1)^2\equiv
\gamma_{C0A}\varepsilon^{CC^{\prime}}\varepsilon^{AA^{\prime}}
\gamma_{C^{\prime}1A^{\prime}} \quad
(\gamma_0\cdot\gamma_1)^3\equiv
\gamma_{0BA}\varepsilon^{BB^{\prime}}\varepsilon^{AA^{\prime}}
\gamma_{1B^{\prime}A^{\prime}} 
\eeq with $D$ being Cayley's hyperdeterminant. In order
to have such solutions $-D$ should be positive and
$(\gamma_0\cdot\gamma_0)^a$ should be negative. The latter means
that the quantities $p^ip^j-p^0q_k>0$ where $i,j,k$ are distinct
elements of the set $\{1,2,3\}$.

Using the stabilized values of the moduli one can calculate the
GHZ amplitudes of $\vert\psi\rangle$. The final form of this
three-qubit state is  \cite{Levay:2006kf} \beq \label{1bpshor}
\lim_{\tau\to\infty}\vert\psi(\tau)\rangle =
(-D)^{1/4}\frac{1}{\sqrt{2}}\left(e^{-i\alpha}\vert 000\rangle
-e^{i\alpha}\vert 111\rangle\right)\eeq \beq
\tan\alpha=\sqrt{-D}\frac{p^0}{2p^1p^2p^3-p^0(p^0q_0+p^1q_1+p^2q_2+p^3q_3
) }\eeq
\begin{eqnarray}
\label{2bpshor}
D&=&(p^0q_0+p^1q_1+p^2q_2+p^3q_3)^2\nonumber\\&-&4((p^1q_1)(p^2q_2)+(p^2q_2)(p^3q_3)+(p^3q_3)(p^1q_1))\\&+&4p^0q_1q_2q_3-4q_0p^1p^2p^3\nonumber
\end{eqnarray}
This state is indeed of the unnormalized GHZ form.
The norm of this state is proportional to the black hole entropy  \cite{Behrndt:1996hu,Duff:2006uz,Levay:2006kf}
\beq
S_{BH}=\pi\sqrt{-D}.\eeq
\noindent
Notice that apart from reproducing the result of (\ref{threetanglebh}) we
also managed to understand the behavior of BPS attractors in terms of a {\it complex} three-qubit state depending on the charges {\it and the moduli}.
It is also interesting that the phase of the GHZ amplitudes, i.e. the phase of the central charge at the horizon, can be expressed in terms of the Freudenthal dual  \cite{Borsten:2009zy} amplitude $\hat{p}^0$ as  \cite{Levay:2006kf}
\beq
\tan\alpha=\frac{p^0}{\hat{p}^0}.
\label{Borsten:2009zy}
\eeq
\noindent

We remark in closing that for non-BPS solutions
with vanishing central charge the attractor states
will be again of the (\ref{1bpshor}) form with
the important difference that now the basis
vectors of the states are changed  \cite{Levay:2010qp}
by flipping one of the bits via the use of the bit flip operator $\sigma_1$.
In this case modifying (\ref{1bpshor})
one can now obtains GHZ states by combining the
basis vectors $\vert 100\rangle$ and
$\vert 011\rangle$ etc.

\subsubsection{The distillation procedure for non-BPS attractors}

In this section we would like to demonstrate how the radial flow
for the moduli gives rise to the distillation of a special
three-qubit (attractor) state at the black hole horizon  \cite{Levay:2010qp}. Here
we demonstrate this process for the non-BPS seed
solutions  \cite{Gimon:2007mh} of
the class of (\ref{nBPSsol}). In this case of the
eight charges only the four  $q_0<0$ and $p^i>0, i=1,2,3$ are
switched on. If we are embedding the $STU$ model into type IIA
string theory the interpretation of these charges is given in
terms of wrapping numbers of $D0$ and $D4$ branes on the $0$ and
$4$ cycles of the six dimensional torus $T^6$. In this case the
basic U-duality invariant is again related to Cayley's
hyperdeterminant calculated for the four charge version of the
state of (\ref{gammachargeuj}). Its value  is
$D=4q_0p^1p^2p^3$.
Using definition (\ref{ujkoord}) the equations to be
solved are given by
(\ref{phidefi})-(\ref{gim2}). The solution of these equations
can be expressed in terms of harmonic functions
$\hat{h}_I, I=0,1,2,3$ as  \cite{Gimon:2007mh} \beq
e^{{\alpha}_i+{\alpha}_j-{\alpha}_k}=\frac{1}{d_k+D^{1/4}\tau}\equiv\frac{1}
{\hat{h}_k} ,\qquad i\neq j\neq k,\quad i,j,k=1,2,3 \label{al}
\eeq \noindent \beq
e^{-\beta}=-\hat{h}_0-\frac{B^2}{\hat{h}_1\hat{h}_2\hat{h}_3},\qquad
h_0\equiv -d_0-D^{1/4}\tau \label{tbeta} \eeq \noindent where \beq
d_i=\frac{D^{1/4}}{\sqrt{2}p^i},\qquad
d_0=-\frac{D^{1/4}}{\sqrt{2}q_0}(1+B^2) \label{boundaryvalues}
\eeq \noindent For this $5$ parameter  solution we have \beq
z^i=R_i\frac{B-ie^{-2U}}{\frac{1}{2}\vert\epsilon_{ijk}\vert\hat{h}_j\hat{h}_k},\qquad
e^{-4U}= -\hat{h}_0\hat{h}_1\hat{h}_2\hat{h}_3-B^2,\qquad
R_i=\sqrt{\frac{-2q_0p^i}{\vert\epsilon_{ijk}\vert p^jp^k}}. \eeq
\noindent
Notice that $B\equiv
x^1(0)=x^2(0)=x^3(0)$ describes the real part for the asymptotic
values of the moduli.

Given the radial
dependence of the moduli and the warp factor one can calculate
the Fourier amplitudes of
the state
$\vert\tilde{\chi}\rangle=(H\otimes H\otimes H)\vert\chi\rangle$.
This state coming from (\ref{ezittaklassz}) can now display how the
attractor mechanism as a distillation procedure unfolds. After
removing the warp factor as in (\ref{nowrap}) one can examine
the behavior of the three-qubit state
$\vert\tilde{\psi}(\tau)\rangle=(H\otimes H\otimes
H)\vert\psi(\tau)\rangle$ in the asymptotic region and on the
horizon. In the asymptotic Minkowski region we get
\begin{eqnarray}
\lim_{\tau\to 0}\vert\tilde{\psi}(\tau)\rangle=\frac{1}{\sqrt{2}}(p^1\vert 001\rangle &+&
p^2\vert 010\rangle +p^3\vert 100\rangle
-iB(p^2+p^3)\vert 110\rangle \nonumber\\
 &-&iB(p^1+p^3)\vert 101\rangle -iB(p^1+p^2)\vert 011\rangle\nonumber\\
 &+&[q_0-B^2(p^1+p^2+p^3)]\vert 111\rangle).
 \label{Minkowski}
 \end{eqnarray}
 \noindent
 On the other hand at the horizon one obtains
 \beq
 \label{at3}
 \lim_{\tau\to\infty}\vert\tilde{\psi}(\tau)\rangle=
 (-4q_0p^1p^2p^3)^{1/4}\frac{1}{2}\left(\vert 001\rangle +
 \vert 010\rangle +\vert 100\rangle - \vert 111\rangle\right).
 \eeq
 \noindent
This result shows that if we ``start" asymptotically with the state
of  (\ref{Minkowski}) with seven nonvanishing generically
different amplitudes we end up with the state of (\ref{at3})
 having merely four nonvanishing amplitudes that
are the same up to a sign. Notice that although the asymptotic
state  features $B$, the asymptotic value for the moduli, the
horizon state is independent of this quantity. Hence, in
accordance with the attractor mechanism, different values for $B$
defining different initial three-qubit states flow to the {\it
same} three-qubit {\it attractor state} determined only by the
charges.

There is an important subtlety here. For BPS black holes no
asymptotic values for the moduli can appear in their horizon
values. Indeed, according to (\ref{1bpshor})-(\ref{2bpshor})
the states on the horizon can be expressed entirely in terms of
the charges. Based on our results as represented by
(\ref{Minkowski})-(\ref{at3}) for the seed solution one
would conclude that a similar result holds also for
non-BPS solutions. However, there are special directions given by particular asymptotic
conditions which give rise to radial flows resulting in
special values of asymptotic moduli at the black hole event
horizon. These directions are referred to as {\it flat}  \cite{Bellucci:2008sv,Gimon:2007mh,Nampuri:2007gv,Ferrara:2007tu}. In other words, the attractor mechanism fixes all the moduli for a BPS charge vector, but may leave flat directions for a non-BPS charge configuration.

An analysis  \cite{Levay:2010qp} based on the most general non-BPS
solution  \cite{Bellucci:2008sv} with nonvanishing central charge, as obtained
from the seed solution, shows that, for the $D0-D4$ configuration
with the most general asymptotic conditions, the radial flow gives
rise to the following three-qubit state on the horizon,
\begin{equation}
\begin{split}
\lim_{\tau\to\infty}\vert\tilde\psi(\tau)\rangle=&\frac{1}{2}
\frac{ (-4q_0p^1p^2p^3)^{\frac1 4} }
{\sqrt{ \cosh(\alpha_3)\cosh(\alpha_2)\cosh(\alpha_1)}}\times\\
&\Bigl[ -\vert111\rangle +\cosh(\alpha_1)\vert001\rangle
+\cosh(\alpha_2)\vert010\rangle
+\cosh(\alpha_3)\vert100\rangle\\
&+i\sinh(\alpha_1)\vert110\rangle +i\sinh(\alpha_2)\vert101\rangle
+i\sinh(\alpha_3)\vert011\rangle\Bigr].
\end{split}
\end{equation}
Here the real constants $\alpha_i, i=1,2,3$ satisfy the
constraint \beq \alpha_1+\alpha_2+\alpha_3=0 \label{sumrule}\eeq
\noindent and, showing up in the asymptotic form of the moduli,
account for the flat directions. In the type IIA picture the
$\alpha_i$ parameterize deformations of the six torus that 
preserve its overall volume. We see that for $\alpha_i=0$ we get
back to the result of (\ref{at3}), but for the most
general case of the non-BPS attractor flow these asymptotic moduli
make an appearance  in the attractor state.

One can consider the dual case of the D2-D6
charge-configuration  \cite{Kallosh:2006ib,Cai:1900ve,Bellucci:2008sv}. In this case
$p^0>0$ and $q_a>0$. The corresponding attractor state is
  \cite{Levay:2010qp}
\begin{equation}
\begin{split}
\lim_{\tau\to\infty}\vert\tilde\psi(\tau)\rangle= &\frac{1}{2} \frac{
(4p^0q_1q_2q_3)^{\frac1 4} }
{\sqrt{ \cosh(\alpha_3)\cosh(\alpha_2)\cosh(\alpha_1)}}\times \\
&\Bigl[ \sinh(\alpha_1)\vert001\rangle
+\sinh(\alpha_2)\vert010\rangle
+\sinh(\alpha_3)\vert100\rangle \\
&-i\vert000\rangle +i\cosh(\alpha_1)\vert110\rangle
+i\cosh(\alpha_2)\vert101\rangle
+i\cosh(\alpha_3)\vert011\rangle\Bigr].
\end{split}
\label{d6d2}
\end{equation}
Notice that for vanishing flat directions the $D0-D4$ and $D2-D6$
cases are dual in the sense that their attractor states are
related by the bit flip operation $\sigma_1\otimes
\sigma_1\otimes\sigma_1$.

It is interesting to analyse the effect of the asymptotic data on
the attractor states at the horizon. In the $D2-D6$ case let us
change the {\it signs} of the charges $q_1,q_2,q_3$ in such a way
that the combination $p^0q_1q_2q_3$ showing up in the entropy
formula is left invariant. Then one can show  \cite{Levay:2007nm} that the
attractor state is \beq
\vert\tilde{\psi}\rangle_{m_1m_2m_3}=i(4p^0q_1q_2q_3)^{1/4}
\frac{1}{2}[m_1\vert 110\rangle +m_2\vert 101\rangle+m_3\vert
011\rangle -\vert 000\rangle] \eeq \noindent where \beq
(m_1,m_2,m_3)\in\{(+,+,+),(+,-,-),(-,+,-),(-,-,+)\}. \eeq
\noindent Hence, although the changes in sign  do not change the
black hole entropy,  they have an effect on the particular
form of the state. The possible sign changes can be implemented
via the action of the phase flip error operators $\sigma_3\otimes
\sigma_3\otimes I$ plus cyclic permutations. Notice that the
Fourier transformed state $\vert\psi\rangle$  features the bit
flip operators $\sigma_1$ in a similar combination.

Let us now fix the signs of the charges of the $D2-D6$ system,
and vary the values for the asymptotic parameters $\alpha_i$
responsible for the flat directions. Let us also refer to the state of
(\ref{d6d2}) as
$\vert\tilde{\psi}\rangle_{\alpha_3\alpha_2\alpha_1}$.
One can then see that by virtue of (\ref{sumrule})  \beq
\vert\tilde{\psi}\rangle_{\alpha_3\alpha_2\alpha_1}= (E_3\otimes
E_2\otimes E_1)\vert\tilde{\psi}\rangle_{+++} \eeq \noindent where
\beq
E_i=\frac{1}{\sqrt{\cosh\alpha_i}}\begin{pmatrix}1&0\\i\sinh\alpha_i&\cosh\alpha_i\end{pmatrix}.
\eeq \noindent Hence the changes on the state originating from the
flat directions have the interpretation of errors of more general
kind depending on continuously changing parameters.
It is amusing to
see that though the error operators $E_i$ act locally,  due
to the constraint of (\ref{sumrule}), they are not independent.
In quantum information
 such constraints usually refer to an agreement between the
 parties affected via the use of classical channels  \cite{Dur:2000,Nielsen:2000}.

These results clearly show the relevance of ideas known from the
theory of quantum error correcting codes. Taken together with our
earlier observations that the structure of the continuous
U-duality group $E_{7(7)}$ can be elegantly described via the use
of the Hamming code shows that these mathematical coincidences are
worth exploring further. Indeed, the tripartite entanglement of
seven qubits interpretation shows that for the most general $56$
charge black hole configurations there are U-duality
transformations of two kind. Either they  transform
within any of the seven possible $STU$ sectors, or  
transform   the sectors among themselves.  U-duality transformations
of the latter form include  elements belonging to the Weyl
group $W(E_7)$. For example, the order seven automorphism
 (\ref{fifth}) that  rotates the $STU$ sectors into each other.
As we have seen such transformations can be represented in the form
featuring CNOT gates. On the other hand U-duality
transformations of the former incorporate naturally bit and phase
flip error operations operating in the solution space of a fixed
$STU$ truncation.
 These considerations might be indications that the
black hole entropy formulae in the semiclassical limit represent
some sort of effective ``area codes" protecting the information
encoded in wrapping configurations of branes from errors arising
from fluctuations of the asymptotic data of moduli. We explore
this idea in some detail in the next section.

\section{Entanglement from extra dimensions}\label{sec:qfed}
In the previous sections we formally employed unnormalized qubits
and qutrits to understand the structure of the semiclassical
Bekenstein-Hawking entropy and the phenomenon of moduli
stabilization in entanglement terms.  In all of our considerations we have used an
effective four or five dimensional picture given by the relevant 
classical supergravities. Hence, 
thus far we have not really made use of the higher dimensional
picture that string theory provides. Indeed superstring (M-theory)
lives in ten (eleven) spacetime dimensions so it would be
desirable to embed our findings in an extra dimensional setting.
Moreover, one might  hope that via this embedding many aspects
of our considerations will be illuminated by the new perspective offered by inclusion of 
the extra dimensions.

Apart from these interesting possibilities, there is a  more
down-to-earth reason to embark upon exploring these ideas. Namely,
 looking at, for example, (\ref{at3}) it is not at all obvious what
  symbols like $\vert 001\rangle$ actually mean in the string context. Of course these
symbols refer to the basis vectors of some sort of Hilbert space.
But how this Hilbert space is defined? Is this Hilbert space
independent of the moduli, or should we rather consider a {\it
family} of such spaces parametrized by them?

In this section we address these important issues. First in the
next section we provide a physical basis for connecting the qubits
and qutrits to the structure of the extra dimensions. The main
idea  \cite{Borsten:2008ur} is to relate them to wrapping
configurations of membranes around the extra dimensions. In later
sections we give a precise meaning to such issues, by identifying
the underlying Hilbert space giving home to the qubits whithin the
cohomology of the extra dimensions  \cite{Levay:2011ph}. We will see
that as an extra bonus these considerations also give a rationale
for our use of charge and moduli dependent three-qubit states in
the previous sections. In the following for simplicity we consider
toroidal compactifications and use the type IIB duality frame,
meaning that we formulate our ideas within the framework of the
corresponding string theory.

\subsection{Wrapped branes as qubits}\label{sec:wrapped}

The microscopic string-theoretic interpretation of the charges is given by configurations of intersecting D3-branes,  wrapping around the six compact dimensions $T^6$. The 3-qubit basis vectors $\ket{ABC}$ are associated   with the corresponding eight wrapping cycles.  In particular, one can relate a well-known fact of quantum information theory, that the most general real three-qubit state up to local unitaries can be parameterised by four real numbers and an angle, to a well-known fact of string theory, that the most general $STU$ black hole can be described by four D3-branes intersecting at an angle.

The microscopic analysis is not unique since there are many ways of embedding the $STU$ model in string/M-theory, but a useful example from our point of view is that of four D3-branes of Type IIB wrapping the $(579)$, $(568)$, $(478)$, $(469)$ cycles of $T^6$ with wrapping numbers $N_0$, $N_1$, $N_2$, $N_3$ and intersecting over a string   \cite{Klebanov:1996mh}. The wrapped circles are denoted by crosses and the unwrapped circles by noughts as shown in \autoref{tab:3QubitIntersect}. This picture is consistent with the interpretation of the 4-charge black hole as bound state at threshold of four 1-charge black holes   \cite{Duff:1994jr,Duff:1996qp,Duff:1995sm}. The fifth parameter $\theta$ is obtained   \cite{Balasubramanian:1997ak,Bertolini:2000ei} by allowing the $N_3$ brane to intersect at an angle which induces additional effective charges on the $(579),(569),(479)$ cycles. 

To make the black hole/qubit correspondence we associate the three $T^2$ with the $\SL(2)_A \times \SL(2)_B \times \SL(2)_C$ of the three qubits Alice, Bob, and Charlie. The eight different cycles then yield eight different basis vectors $\ket{ABC}$ as in the last column of \autoref{tab:3QubitIntersect}, where $\ket{0}$ corresponds to \textsf{xo} and $\ket{1}$ to \textsf{ox}. To wrap or not to wrap; that is the qubit.
We see immediately that we reproduce the five parameter three-qubit state $\ket{\Psi}$ of \eqref{eq:five}:
\begin{equation}
\begin{split}
\ket{\Psi} &= -N_3\cos^2\theta\ket{001}-N_2\ket{010}+N_3\sin\theta\cos\theta\ket{011}\\
&\phantom{=}-N_1\ket{100}-N_3\sin\theta\cos\theta\ket{101}+(N_0+N_3\sin^2\theta)\ket{111}.
\end{split}
\end{equation}
Note  the GHZ state  describes four D3-branes intersecting over a string.
\begin{table}[ht]
\begin{tabular*}{\textwidth}{@{\extracolsep{\fill}}*{11}{c}>{$\lvert}c<{\rangle$}c}
\toprule
& 4    & 5    & & 6    & 7    & & 8    & 9    & macro charges & micro charges              & ABC & \\
\colrule
& \sfx & \sfo & & \sfx & \sfo & & \sfx & \sfo & $p^0$         & 0                          & 000 & \\
& \sfo & \sfx & & \sfo & \sfx & & \sfx & \sfo & $q_1$         & 0                          & 110 & \\
& \sfo & \sfx & & \sfx & \sfo & & \sfo & \sfx & $q_2$         & $-N_3\sin\theta\cos\theta$ & 101 & \\
& \sfx & \sfo & & \sfo & \sfx & & \sfo & \sfx & $q_3$         & $N_3\sin\theta\cos\theta$  & 011 & \\
\colrule
& \sfo & \sfx & & \sfo & \sfx & & \sfo & \sfx & $q_0$         & $N_0+N_3\sin^2\theta$      & 111 & \\
& \sfx & \sfo & & \sfx & \sfo & & \sfo & \sfx & $-p^1$        & $ -N_3\cos^2\theta$        & 001 & \\
& \sfx & \sfo & & \sfo & \sfx & & \sfx & \sfo & $-p^2$        & $-N_2$                     & 010 & \\
& \sfo & \sfx & & \sfx & \sfo & & \sfx & \sfo & $-p^3$        & $-N_1$                     & 100 & \\
\botrule
\end{tabular*}
\caption[Wrapped D3-branes]{Three qubit interpretation of the 8-charge $D=4$ black hole from four D3-branes wrapping around the lower four cycles of $T^6$ with wrapping numbers $N_0,N_1,N_2,N_3$ and then allowing $N_3$ to intersect at an angle $\theta$.}\label{tab:3QubitIntersect}
\end{table}
By embedding this picture in $\mathcal{N}=8$ supergravity we find an exact correspondence between the possible  intersections preserving different degree of supersymmetry and the entanglement classes of 3-qubits as illustrated in \autoref{fig:conventional}.
\captionsetup{justification=raggedright}
\begin{figure}
\caption{Left: Classification of $\mathcal{N}=8$ black holes according to susy. $N$ denotes the number of intersecting D-branes in the microscopic picture. Right: The entanglement classification of 3-qubits. The  arrows represent the removal of a D-brane or a non-invertible SLOCC operation.\label{fig:conventional}}
\centering
{\label{fig:entanglement_bps}\includegraphics[width=\linewidth]{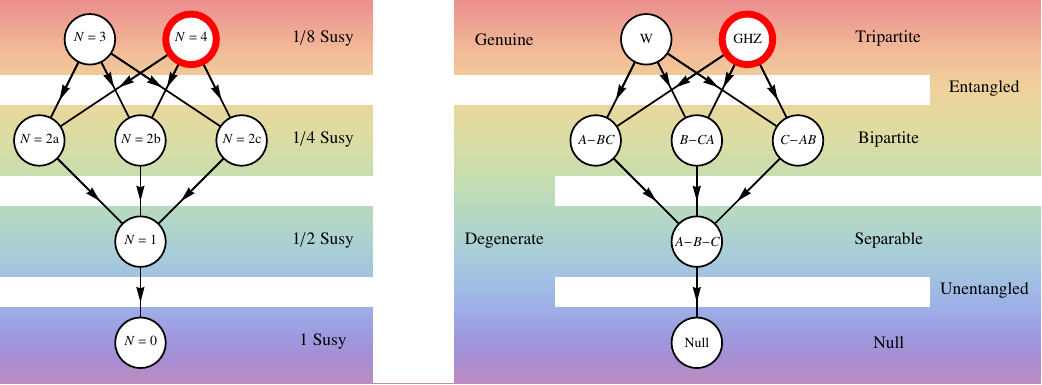}}
\end{figure}

\subsection{Qubits from extra dimensions}

In the previous section we have seen that wrapped branes can be
used to realize qubits, the basic building blocks used in quantum
information. Based on these findings it is natural to expect that
such brane configurations wrapped on different cycles of the
manifold of extra dimensions should be capable of accounting for
many more of the surprising findings of the BHQC.
The aim of the present section is to show that by
simply reinterpreting some of the well-known results of toroidal
compactification of type IIB string theory in a quantum
information theoretic fashion this expectation can indeed be
justified. In particular we identify the Hilbert space giving home
to the qubits inside the cohomology of the extra dimensions,
establishing for the phrase ``to wrap or not to wrap, that is the
qubit" a mathematical meaning. In order to do this let us recall
some basic ideas of Calabi-Yau compactifications of type II string
theories  \cite{2007stmt.book.....B}.

Calabi-Yau compactifications play an important role in string
theory. Calabi-Yau spaces are six dimensional complex K\"ahler
manifolds that are Ricci flat. They are promising candidates for
producing sensible four dimensional low energy phenomenology.
Moreover due to their special structure they are ideal playing
grounds for obtaining results illustrating the interplay between
the geometry of extra dimensions and ordinary four-dimensional
spacetime as tested by extended objects. As we have seen in
sections discussing the attractor mechanism there are scalar
fields called moduli describing the fluctuations of such spaces.
For type II string theories
such moduli can be separated into two characteristic
types. Either they are associated with the deformations of the
complex structure or with the K\"ahler structure of the underlying
Calabi-Yau. For type IIA supergravity the moduli are associated
with the latter and for type IIB with the former. In the following
we merely consider the special case of toroidal compactifications,
hence we assume that the geometry of the extra dimensions is that
of a six-torus $T^6$. In the type IIA picture K\"ahler structure
deformations correspond to deformations of the size, on the
other hand in the type IIB picture complex structure deformations
alter the shape. We adopt the type IIB picture,
hence by moduli in the following we mean the complex structure.
Restricting further to the special case of regarding $T^6$ as
$T^2\times T^2\times T^2$ as in the previous subsection, we can
then associate three-qubits with the three two dimensional tori.

The moduli are now $z^j=x^j-iy^j$ with $y^j>0$ as in
(\ref{moduli}). They correspond to the deformation parameters
for the tori as labelled by $j=1,2,3$. Let us denote the six real
coordinates of $T^6$ by $0\leq u^j\leq 1$ and $0\leq v^j\leq 1$. We can imagine these
coordinates as the ones adapted to the deformed tori. This means
that regarding $T^6$ as a complex manifold the complex coordinates
$w^j$ of $T^6$ should depend on the moduli $z^j$. Notice that
$T^6=T^2\times T^2\times T^2$ is a very special case since now the
number of complex moduli equals the number of complex dimensions
of the manifold. For a general Calabi-Yau this need not be the
case, since the number of complex moduli  \cite{2007stmt.book.....B} is then
given by the Hodge number $h^{2,1}$. Next we write \beq
w^j=u^j+z^jv^j. \label{adapted} \eeq \noindent Hence the real
pairs $(u^j,v^j)$ and $(x^j,y^j)$ are coordinates for $T^6$ and
its moduli space ${\cal M}=[\SL(2,{\mathds R})/\SO(2)]^{\otimes 3}$
respectively. Alternatively $w^j$ and $z^j$ are the corresponding
complex coordinates for these spaces.

For a Calabi-Yau space we have a nowhere vanishing holomorphic
three-form. For our torus  it is \beq \Omega_0=dw^1\wedge dw^2\wedge
dw^3. \label{holthree} \eeq \noindent We have\beq
\int_{T^6}\Omega_0\wedge\overline{\Omega}_0=i(8y^1y^2y^3)=
ie^{-K},\label{holnorm}\eeq\noindent where $K$ is the K\"ahler
potential of (\ref{Kahler}) giving rise to the metric
$G_{i\overline{j}}$ of (\ref{targetmetric}) on the special
K\"ahler manifold ${\cal M}$. Let us now define the nonholomorphic
three-form  $\Omega$ as \beq \Omega\equiv
e^{K/2}\Omega_0.\label{nonhol}\eeq\noindent Define flat covariant
derivatives $D_{\hat{i}}$ acting on $\Omega$ as \beq
D_{\hat{i}}\Omega=(\overline{z}^i-z^i)D_i\Omega=
(\overline{z}^i-z^i)
\left({\partial}_{i}+\frac{1}{2}{\partial}_iK\right)\Omega,
\label{covder3}\eeq\noindent where
${\partial}_i={\partial}/{\partial{z}^i}$. Now one has \beq
\Omega=e^{K/2}dw^1\wedge dw^2\wedge dw^3,\qquad
\overline{\Omega}=e^{K/2}d\overline{w}^1\wedge
d\overline{w}^2\wedge d\overline{w}^3\label{0cov}\eeq\noindent
\beq D_{\hat{1}}\Omega=e^{K/2}d\overline{w}^1\wedge dw^2\wedge
dw^3,\qquad
\overline{D}_{\hat{\overline{1}}}\overline{\Omega}=e^{K/2}dw^1\wedge
d\overline{w}^2\wedge d\overline{w}^3\label{1cov}\eeq\noindent
with the remaining covariant derivatives obtained via cyclic permutation.
Let us consider the action of the Hodge star on our
basis of three-forms as given by (\ref{0cov})-(\ref{1cov}). For
a form of $(p,q)$ type the action of the Hodge star is defined as
\beq
(\varphi,\varphi)\frac{\omega^n}{n!}=\varphi\wedge\star\overline{\varphi}
\label{hodgedef}\eeq\noindent where for our $T^6$ in accord with
our conventions \beq \omega=i(d\overline{w}^1\wedge dw^1+
d\overline{w}^2\wedge dw^2+d\overline{w}^3\wedge
dw^3)\label{omegat6}.\eeq\noindent
Hence we get \beq \star\Omega=i\Omega,\qquad
\star\overline{\Omega}=-i\overline{\Omega}\label{elsohodge}\eeq\noindent
\beq \star D_{\hat{a}}\Omega=-iD_{\hat{a}}\Omega,\qquad
\star\overline{D}_{\hat{\overline{a}}}\overline{\Omega}=i\overline{D}_{\hat{\overline{a}}}
\overline{\Omega}\label{masodikhodge}\eeq\noindent

Now we regard the $8$ complex dimensional untwisted primitive
part  \cite{Blumenhagen:2006ci} of the $20$ dimensional space $H^3(T^6,{\mathds
C})\equiv H^{3,0}\oplus H^{2,1}\oplus H^{1,2}\oplus H^{0,3}$
equipped with the Hermitian inner product \beq
\langle\varphi\vert\eta\rangle\equiv\int_{T^6}
\varphi\wedge\star\overline{\eta}\label{inner3}\eeq\noindent as a
Hilbert space isomorphic to ${\cal H}\equiv ({\mathds
C}^2)^{\otimes 3}\simeq {\mathds C}^8$ of three qubits. In order to
set up the correspondence between the three-forms and the basis
vectors of the three-qubit system we use the {\it negative} of the
basis vectors $\Omega,D_{\hat{1}}\Omega,\ldots$ multiplied by the
imaginary unit $i$. Changing the order of the one-forms we have, for example, 
$ -iD_{\hat{1}}\Omega=ie^{K/2}dz^3\wedge dz^2\wedge
d\overline{z}^1$. Hence, we can take as its representative basis qubit
state $\vert 001\rangle$, which corresponds to the usual binary
labelling provided we label the qubits from the right to the left.

Using  these conventions the basis states of our
computational base are given by, \beq
-i\Omega\leftrightarrow \vert 000\rangle,\qquad
-iD_{\hat{1}}\Omega\leftrightarrow \vert 001\rangle, \qquad
-iD_{\hat{2}}\Omega\leftrightarrow \vert 010\rangle,\qquad
-iD_{\hat{3}}\Omega\leftrightarrow \vert
100\rangle\label{qubitcorrespondence}\eeq\noindent \beq
-i\overline{\Omega}\leftrightarrow \vert 111\rangle,\qquad
-i\overline{D}_{\hat{\overline{1}}}\Omega\leftrightarrow \vert
110\rangle, \qquad
-i\overline{D}_{\hat{\overline{2}}}\Omega\leftrightarrow \vert
101\rangle,\qquad
-i\overline{D}_{\hat{\overline{3}}}\Omega\leftrightarrow \vert
011\rangle.\label{qubitcorrespondence2}\eeq\noindent Now the
locations of the ``1''s correspond to the slots where complex
conjugation is effected. One can check that the states above form
a basis with respect to the inner product of (\ref{inner3})
with the usual set of properties in terms of the three-qubit intepretation.

A further check shows that the action of the flat covariant
derivatives $D_{\hat{a}}, j=1,2,3$ corresponds to the action of
the projective bit flips of the form $I\otimes
I\otimes\sigma_+,\quad I\otimes\sigma_+\otimes I$ and
$\sigma_+\otimes I\otimes I$, where $I$ is the $2\times 2$
identity matrix. For the conjugate flat covariant derivatives
$\sigma_+$ has to be replaced by ${\sigma}_-$. Moreover, the
diagonal action of the Hodge star in the computational base is
represented by the corresponding action of the {\it negative} of
the parity check operator $i\sigma_3\otimes\sigma_3\otimes
\sigma_3$.

Now for a three-form representing the cohomology class of a
wrapped $D3$ brane configuration we take  \beq
\psi=p^I\alpha_I-q_I\beta^I\in H^3(T^6,{\mathds
Z})\label{threebrane}\eeq\noindent 
and \beq \alpha_0=du^1\wedge du^2\wedge du^3,\qquad
\beta^0=-dv^1\wedge dv^2\wedge dv^3\qquad
\alpha_1=dv^1\wedge du^2\wedge du^3,\qquad \beta^1=du^1\wedge
dv^2\wedge dv^3\label{egyesek}\eeq
 with the remaining
four obtained via cyclic permutation. With the canonical choice of
orientation 
we have
$\int_{T^6}{\alpha}_I\wedge {\beta}^J={\delta}_I^J$.

It is well-known  \cite{Denef:2000nb} that in the Hodge diagonal basis we
can express this as \beq \psi=iZ(\psi)\overline{\Omega}
-ig^{j\overline{k}}D_jZ(\psi)
\overline{D}_{\overline{k}}\overline{\Omega}+{\rm c.c.}=
iZ(\psi)\overline{\Omega}
-i\delta^{\hat{j}\hat{\overline{k}}}D_{\hat{j}}Z(\psi)
\overline{D}_{\hat{\overline{k}}}\overline{\Omega}+{\rm
c.c.}\label{expand}\eeq\noindent Here
$Z(\psi)=\int_{T^6}\psi\wedge\Omega$ is the central charge. For
its explicit expression see (\ref{central}).

Employing our basic correspondence between three-forms and
three-qubit states of 
(\ref{qubitcorrespondence})-(\ref{qubitcorrespondence2}) we can
write $\psi \leftrightarrow \vert\psi\rangle$ where \beq
\vert\psi\rangle=\psi_{000}\vert 000\rangle +\psi_{001}\vert
001\rangle +\dots +{\psi}_{110}\vert 110\rangle +{\psi}_{111}\vert
111\rangle\label{qubitform2}\eeq
\beq
{\psi}_{111}=-e^{K/2}W(z^3,z^2,z^1)=-\overline{\psi}_{000},\qquad
{\psi}_{001}=-e^{K/2}W(\overline{z}^3,\overline{z}^2,z^1)=-\overline{\psi}_{110}
\label{ampl1} \eeq\noindent and the remaining amplitudes are given
by cyclic permutation. Now one can check that, using  the definitions in
(\ref{ezittaklassz2}) and (\ref{Smatrix}), the state
$\vert\psi\rangle$ can be written in the form  given by
(\ref{nowrap}).

Hence our state $\vert\psi\rangle$ seems to be exactly the same as
the one that has already appeared in our previous considerations.
It is important to realize however, the basic difference in
interpretation. Untill now the state $\vert\psi\rangle$ was merely
a {\it charge} and {\it moduli} dependent state connected to the
$4$ dimensional setting of the $STU$ model. Moreover, in that
setting the basis states $\vert ABC\rangle$ with $A,B,C=0,1$ had
no obvious physical meaning. They merely served as basis vectors
providing a suitable frame for a three-qubit reformulation.

 Now
$\vert\psi\rangle$  depends on the {\it
charges}, the {\it moduli} and the {\it coordinates of the extra
dimensions}. 
Hence, this state is connected to a $10$-dimensional setting of the
$STU$ model in the type IIB duality frame. Now the basis vectors
$\vert ABC\rangle$ have an obvious physical meaning: they are the
Hodge diagonal complex basis vectors of the untwisted primitive
part of the third cohomology group of the extra dimensions, i.e. of
$H^3(T^6,{\mathds C})$. They are also basis vectors of a genuine
Hilbert space equipped with a natural Hermitian inner product 
(\ref{inner3}), isomorphic to the usual  space of three qubits.
The basis vectors $\vert ABC\rangle$  not only depend on
the coordinates of the tori, but also  on the
{\it moduli}. Hence the notation should reflect that $\vert
ABC\rangle $ refers to a {\it parametrized family} of basis
vectors. Since the notation  $\vert ABC ;
z^1,z^2,z^3\rangle $  is rather awkward
we omit the  $z^i$ and tacitly assume an implicit
dependence on them.

In the new formalism the state $\vert\psi\rangle$ also has a nice
physical meaning.  It is the Poincar\'e dual of the homology cycle
representing wrapped $D3$-brane configurations. Moreover
$\vert\psi\rangle$ can be represented in two different forms.
Namely as in (\ref{qubitform2}) which is an expansion in a
Hodge-diagonal moduli dependent complex base, or in an equivalent
way based on the qubit version of (\ref{threebrane}) which is a
Hodge-non-diagonal but moduli independent real base.

These results also shed some light on the phenomenon that the
fluctuations of the moduli are related to phase flip and bit flip
errors we encountered at the end of Section 5.6.3. Indeed, now we
see that on the basis vectors $\vert ABC\rangle$ the flat
covariant derivatives with respect to the moduli act naturally as
elementary error operations. As a byproduct of this the attractor
equations can be given an alternative explanation. For the
attractor flow we have a three-qubit state with amplitudes
associated with all of the {\it moduli dependent} basis vectors
$\vert ABC\rangle$. We have seen that at the horizon only special
combinations of the amplitudes associated with special basis
vectors survive. For example according to (\ref{1bpshor}) for
the BPS flow we have only amplitudes multiplied by the basis
states $\vert 000\rangle$ and $\vert 111\rangle$. This means that
at the horizon bit flip errors of the form $\sigma_1\otimes
I\otimes I$ and their cyclic permutations acting on these basis
vectors are suppressed   \cite{Levay:2007nm}. These conditions can be
expressed precisely in the form of the BPS attractor equations
i.e. (\ref{BPSattra}) and their conjugates. This result is a
quantum information theoretic reinterpretation of the well-known
property of supersymmetric attractors in the type IIB picture
namely that in this case only the $H^{3,0}$ and $H^{0,3}$ parts of
the cohomology survive  \cite{2007stmt.book.....B}.

\subsection{Fermionic entanglement from extra dimensions}

As a generalization of our considerations giving rise to qubits
we consider the problem of obtaining
entangled systems of a more general kind from toroidal
compactification  \cite{Levay:2011ph}. The trick is to embed our
simple systems featuring few qubits into larger ones. Here we
discuss the natural generalization of embedding qubits (based on
entangled systems with distinguishable constituents) into
fermionic systems (based on entangled systems with
indistinguishable ones  \cite{2002AnPhy.299...88E,2004PhRvA..70a2109G}). In the quantum
information theoretic context this possibility has already been
elaborated  \cite{levay-2008}, see  \autoref{sec:ferFTS}. Here we show that toroidal
compactifications also incorporate this idea quite naturally.

As in the special case of the $STU$ model we choose analytic
coordinates for the complex torus such that the holomorphic
one-forms are defined as $dw^i=du^i+z^{ij}dv^j$ where now
$z^{ij},\quad 1\leq i,j\leq 3$ is the period matrix of the torus
with the convention \beq
z^{ij}=x^{ij}-iy^{ij}.\label{matrconvention}\eeq\noindent For
principally polarized Abelian varieties we have the additional
constraints \beq z^{ij}=z^{ji},\qquad
{y}^{ij}>0.\label{matrconstraints}\eeq\noindent
 We choose as usual ${\Omega}_0=dw^1\wedge dw^2\wedge dw^3$, and
 the canonical orientation.

Now we exploit the
full $20$ dimensional space of $H^3(T^6,{\mathds C})$. We expand
$\psi\in H^3(T^6,{\mathds C})$ in the basis 

\beq \alpha_0=du^1\wedge du^2\wedge du^3,\qquad
\alpha_{ij}=\frac{1}{2}{\varepsilon}_{ii^{\prime}j^{\prime}}
du^{i^{\prime}}\wedge du^{j^{\prime}}\wedge dv^j \eeq \noindent
\beq \beta^0=-dv^1\wedge dv^2\wedge dv^3,\qquad
\beta^{ij}=\frac{1}{2}{\varepsilon}_{ji^{\prime}j^{\prime}}du^i\wedge
dv^{i^{\prime}}\wedge dv^{j^{\prime}}. \eeq\noindent  One can then
show that \beq
\Omega_0=\alpha_0+z^{ij}\alpha_{ij}+{z^{\sharp}}_{ij}\beta^{ji}-({\rm
Det}z)\beta^0\label{kifejtes2}\eeq\noindent where ${z}^{\sharp}$
is the transposed cofactor matrix satisfying $zz^{\sharp}={\rm
Det}(z) \mathds{1}$. One can
check that the generalization of the identity of
(\ref{holnorm}) holds and $e^{-K}=8{\rm Det}y$.

An element $\psi$ of $H^3(T^6,{\mathds Z})$ can be expanded as
\beq
\psi=p^0\alpha_0+P^{ij}\alpha_{ij}-Q_{ij}\beta^{ij}-q_0\beta^0.\eeq\noindent
We can rewrite this as \beq \psi=\frac{1}{3!}\gamma_{ABC}f^A\wedge
f^B\wedge f^C \label{fermistate}\eeq\noindent where \beq
(f^1,f^2,f^3,f^4,f^5,f^6)\equiv
(f^1,f^2,f^3,f^{\overline{1}},f^{\overline{2}},f^{\overline{3}})=(du^1,du^2,du^3,dv^1,dv^2,dv^3).\eeq\noindent
Here ${\psi}_{ABC}$ has been generalized to a completely
antisymmetric tensor of rank three with $20$ independent
components. Clearly the independent components of $\gamma_{ABC}$
are identified with the $20$ quantized charges $(p^0,P^{ij}
,Q_{ij},q_0)$ related to wrapping three-branes on the
corresponding homology cycles. The explicit identification is
given by  \beq p^0=\gamma_{123},\qquad
\begin{pmatrix}P^{11}&P^{12}&P^{13}\\
               P^{21}&P^{22}&P^{23}\\
                      P^{31}&P^{32}&P^{33}\end{pmatrix}
              =\begin{pmatrix}\gamma_{23\overline{1}}&
              \gamma_{23\overline{2}}&\gamma_{23\overline{3}}\\
              \gamma_{31\overline{1}}&\gamma_{31\overline{2}}
              &\gamma_{31\overline{3}}\\
              \gamma_{12\overline{1}}&\gamma_{12\overline{2}}&
              \gamma_{12\overline{3}}\end{pmatrix}
              \label{pma}\eeq\noindent

\beq q^0=\gamma_{\overline{1}\overline{2}\overline{3}},\qquad
\begin{pmatrix}Q_{11}&Q_{12}&Q_{13}\\
               Q_{21}&Q_{22}&Q_{23}\\
                          Q_{31}&Q_{32}&Q_{33}\end{pmatrix}
                  =-\begin{pmatrix}
                  \gamma_{1\overline{2}\overline{3}}&
                  \gamma_{1\overline{3}\overline{1}}&
                  \gamma_{1\overline{1}\overline{2}}\\
                  \gamma_{2\overline{2}\overline{3}}&
                  \gamma_{2\overline{3}\overline{1}}&
                  \gamma_{2\overline{1}\overline{2}}\\
                  \gamma_{3\overline{2}\overline{3}}&
                  \gamma_{3\overline{3}\overline{1}}&
                  \gamma_{3\overline{1}\overline{2}}
                  \end{pmatrix}.\label{qma}\eeq\noindent
Using the language of fermionic entanglement  \cite{2002AnPhy.299...88E,2004PhRvA..70a2109G}
$\psi$ can also be regarded as an unnormalized three fermion
state with six single particle states  \cite{levay-2008}, as described in \autoref{sec:ferFTS}.

Now we introduce the new moduli dependent basis vectors \beq
e^A=f^{A^{\prime}}{S_{A^{\prime}}}^A, \qquad
{S_{A^{\prime}}}^A=\begin{pmatrix}I&I\\z&\overline{z}\end{pmatrix}.
\eeq\noindent One can then write \beq
\psi=\frac{1}{3!}\psi_{A^{\prime}B^{\prime}C^{\prime}}
\left(-ie^{K/2}e^{A^{\prime}} \wedge e^{B^{\prime}}\wedge
e^{C^{\prime}}\right) \label{indish} \eeq \noindent where \beq
\psi_{A^{\prime}B^{\prime}C^{\prime}}={{\cal S}_{A^{\prime}}}^A
{{\cal S}_{B^{\prime}}}^B{{\cal S}_{C^{\prime}}}^C\gamma_{ABC}
\label{kisgamma2} \eeq \noindent and \beq {\cal S}\equiv
-ie^{-K/6} S^{-1}=-ie^{-K/6}(z-\overline{z})^{-1}\begin{pmatrix}-
\overline{z}&I\\z&-I\end{pmatrix}.\eeq\noindent In this new form
the amplitudes $\psi_{ABC}$ depend on the charges and the
moduli. Notice also that now we have the {\it same} matrix ${\cal
S}\in \GL(6,{\mathds C})$ acting on all indices of $\gamma_{ABC}$.
This reflects the fact known from the theory of quantum
entanglement that the SLOCC group  \cite{Dur:2000,2002AnPhy.299...88E} for a quantum
system consisting of {\it indistinguishable} subsystems (now with
six single particle states  \cite{levay-2008}) is represented by the
same $\GL(6,{\mathds C})$ matrices acting on each entry of a tensor
representing the set of amplitudes (now of a tripartite system).
The basis states $-ie^{K/2}e^A\wedge e^B\wedge
e^C$ for $1\leq A<B<C\leq 6$ form an orthonormal basis with
respect to the inner product of (\ref{inner3}).

It is instructive to see how one recovers the $STU$ case of the
previous section. In particular one would like to see how the {\it
indistinguishable} character of the subsystems represented by
$\psi$ boils down to the {\it distinguishable} one of the
corresponding three-qubit system. In order to see this just notice
that in the $STU$ case we merely have $8$ nonzero amplitudes to be
used in  (\ref{kisgamma2}). Namely  we have $\gamma_{ABC}$ with
labels ${123},{12\overline{3}},\dots,{\overline{1}\overline{2}3},
{\overline{1}\overline{2}\overline{3}}$. Moreover, the $3\times 3$
matrix $z$ is now diagonal, hence the explicit form of ${\cal S}$
is \beq {\cal S}=\frac{1}{2}e^{-K/6}\begin{pmatrix}-
\overline{z}^1/y^1&0&0&1/y^1&0&0\\0&-\overline{z}^2/y^2&0&0&1/y^2&0\\                                         0&0&-\overline{z}^3/y^3&0&0&1/y^3\\
z^1/y^1&0&0&-1/y^1&0&0\\                                                     0&
z^2/y^2&0&0&-1/y^2&0\\
0&0&z^3/y^3&0&0&-1/y^3\end{pmatrix}                                          \label{kiirva}
\eeq                                                                            \noindent
After switching to our usual ordering convention let us make the
correspondence ${321}\leftrightarrow {000}, {32\overline{1}}\leftrightarrow {001}$ etc. meaning
 that the labels $1$ and $\overline{1}$, $2$ and $\overline{2}$, $3$ and $\overline{3}$
 refer to the labels $0$ and $1$ of the {\it first}, {\it second} and {\it third} qubit.
 Looking at the structure of the tensor product
 ${\cal S}\otimes {\cal S}\otimes {\cal S}$ and recalling that $e^{-K/2}=\sqrt{8y^1y^2y^3}$
 we quickly recover the
  structure of the three qubit state of (\ref{qubitform2}). (See also the definitions of
(\ref{nowrap})
(\ref{ezittaklassz2}) and (\ref{Smatrix}).)

Let us finally comment on the structure of BPS
attractors  \cite{Moore:1998pn} in our entanglement based approach. As in
 the $STU$ case the attractor equations  require that only
the $H^{3,0}$ and $H^{0,3}$ parts of the cohomology classes are
non-vanishing. This implies that for our ``state" of fermionic
entanglement at the horizon we have \beq
\psi=\psi_{321}(-ie^{K/2}e^3\wedge e^2\wedge e^1)-
\psi_{\overline{321}}(-ie^{K/2}e^{\overline{3}}\wedge
e^{\overline{2}}\wedge e^{\overline{1}}), \eeq\noindent where
\beq\overline{\psi}_{321}=
-\psi_{\overline{3}\overline{2}\overline{1}}= Z(z_{\rm
fix},p^0,q_0,P,Q). \label{attractorfermi} \eeq \noindent According
to the general theory  \cite{levay-2008} for classifying the SLOCC
entanglement types  \cite{Dur:2000} for tripartite fermionic systems with
six single particle states, such attractor states belong to the
fermionic generalization of the usual GHZ state well-known for
three qubits. Hence our result on the reinterpretation of the
attractor mechanism as a quantum information theoretic
distillation procedure in this fermionic context still holds.

In order to see this one has to solve the attractor
equations  \cite{Moore:1998pn,Levay:2011ph} with the result $z=x-iy$ where
\beq y=\frac{1}{2}\sqrt{\cal
D}(p^0Q+P^{\sharp})^{-1},\qquad
x=\frac{1}{2}(2PQ-[p^0q_0+{\rm
Tr}(PQ)]I)(p^0Q+P^{\sharp})^{-1}.\label{valoska}\eeq\noindent 
The $STU$ case is recovered when using diagonal
matrices for $z=x-iy$, $P$ and $Q$. In these expressions the
definition of ${\cal D}$ is
 \beq {\cal D}=-(p^0q_0+{\rm Tr}(PQ))^2+4{\rm
Tr}(P^{\sharp}Q^{\sharp})+4p^0{\rm Det}Q-4q_0{\rm
Det}P\label{cartanka}\eeq\noindent which is minus half of the
usual quartic invariant of Freudenthal triple
systems  \cite{Krutelevich:2004}. Here the Freudenthal system is the one
based on the cubic Jordan algebra of $3\times 3$ matrices with
complex entries (see also \autoref{sec:ferFTS}).

One can then show that 
as the result of a distillation procedure the GHZ-like state at the horizon is of the form
as given by (\ref{1bpshor}) with suitable replacements. First
Cayley's hyperdeterminant $D$ has to be replaced by its
generalization ${\cal D}$ as given by (\ref{cartanka}).
Moreover, the phase $\alpha$ of the central charge is determined
by the equation \beq
\tan\alpha=\frac{p^0}{\hat{p}^0}\label{ujfazis}\eeq\noindent where
$\hat{p}^0$ is the Freudenthal dual  \cite{Borsten:2009zy} of $p^0$. Its
explicit expression is \beq \hat{p^0}=\frac{2{\rm Det}P+p^0{\rm
Tr}(PQ)+p^0q_0}{\sqrt{-{\cal D}}}. \eeq \noindent The stabilized
states $\vert 000\rangle$ and $\vert 111\rangle$ of
(\ref{1bpshor}) should be replaced by their ``fermionic"
counterparts $(-ie^{K/2} e^3\wedge e^2\wedge e^1)$ and
$(-ie^{K/2}e^{\overline{3}}\wedge e^{\overline{2}}\wedge
e^{\overline{1}})$. These basis vectors should be evaluated at the
stabilized values of the moduli as given by
(\ref{valoska}). Finally the formula for
the entropy as expected is \beq S_{BH}=\pi \sqrt{{\cal
D}}.\label{entropyfermionic}\eeq\noindent

Based on our experience with the $STU$ case where according to
(\ref{threetanglebh}) the entropy formula was given in terms of
a genuine tripartite measure, it is tempting to interpret ${\cal
T}_{123}\equiv 4\vert{\cal D}\vert$ as an entanglement measure for
three fermions with six single particle states as represented by
the {\it normalized} state $\psi\in\wedge^3 V^{*}$ where
$V={\mathds C}^6$. According to Ref.  \cite{levay-2008} within the
realm of quantum information the quantity ${\cal T}_{123}$ indeed
works well as a basic quantity to characterize the entanglement
types under the SLOCC group $\GL(6,{\mathds C})$. Within the
context of black hole solutions  the SLOCC group  should be
restricted to its real subgroup $\GL(6,{\mathds R})$. It is not
difficult to see then that the different types of black holes
correspond to the different entanglement types of fermionic
entanglement. This correspondence runs in parallel with the
original observation of Kallosh and Linde  \cite{Kallosh:2006zs} that
the entanglement types of three qubit states correspond to
different types of $STU$ black holes.

\subsection{Other entangled systems from extra dimensions}

Finally let us comment on possible generalizations. One can
show  \cite{Levay:2011ph} that the idea of the attractor mechanism as
a distillation procedure works nicely also for flux attractors
  \cite{Dasgupta:1999ss,2007stmt.book.....B}. One can study this mechanism within the
context of F-theory compactifications on elliptically fibered
Calabi-Yau four-folds  \cite{2007stmt.book.....B}.
 Here the flux attractor equations are just a
rephrasing of the imaginary self duality condition  \cite{Dasgupta:1999ss}
(ISD) $\star_6 G=iG$ for the complex flux form defined as \beq
G_3=F_3-\tau H_3\label{G3}\eeq\noindent where $G$ is a combination
of the type IIB NS and RR three-forms $H_3$ and $F_3$ into a new
three-form $G_3$ which has also a dependence on a special type of
new moduli \beq\tau=a+ie^{-\Phi} \eeq \noindent the axion-dilaton
field. The extra moduli can be incorporated into the formalism via
an additional torus  \cite{Kallosh:2005ax,Denef:2004cf,Levay:2011ph}.

 In the special case of choosing the Calabi-Yau space as the orbifold
  $T^6/{\mathds Z}_2\times{\mathds Z}_2$ one finds
{\it four-qubit systems}  \cite{Levay:2011ph}. Indeed, it is
convenient to incorporate the three-forms $H$ and $F$ into a
complex four-form $G$.  Then applying a correspondence similar to
(\ref{qubitcorrespondence})-(\ref{qubitcorrespondence2})
between the Hodge diagonal basis vectors for complex
four-forms  \cite{Denef:2004cf} and the four-qubit basis vectors one
can define a four-qubit state. This state  depends on the
quantized fluxes and the four moduli and it is of the usual form
\beq \vert G\rangle={\cal S}_4\otimes{\cal S}_3\otimes {\cal
S}_2\otimes {\cal S}_1\vert g\rangle.\label{fluxstate}\eeq\noindent
Here the extra ${\cal S}_4$ is featuring the axion-dilaton as
$z^4\equiv \overline{\tau}$ and the explicit form of the $16$
amplitudes of $\vert g\rangle$ is labelled as
 \beq
\begin{pmatrix}g_{0000}&g_{0001}&g_{0010}&g_{0100}\\g_{0111}&g_{0110}&g_{0101}&g_{0011}
\end{pmatrix}
=\begin{pmatrix}-p^0&-p^1&-p^2&-p^3\\-q_0&q_1&q_2&q_3\end{pmatrix}
\eeq\noindent \beq
\begin{pmatrix}g_{1000}&g_{1001}&g_{1010}&g_{1100}\\g_{1111}&g_{1110}&g_{1101}&g_{1011}
\end{pmatrix}
=\begin{pmatrix}-P^0&-P^1&-P^2&-P^3\\-Q_0&Q_1&Q_2&Q_3\end{pmatrix}.
\label{flux2charge}\eeq\noindent

 As an explicit example
for this system one can revisit and reinterpret the solution found
by Larsen and O'Connell  \cite{Larsen:2009fw} in the language of
four-qubit entangled systems. In this special case only $8$ fluxes
are switched on. They are  $(p^0,q_1,q_2,q_3)$ and
$(Q_0,P^1,P^2,P^3)$. Analysing the attractor equations, i.e. the ISD
condition,  one can show  \cite{Larsen:2009fw} that this configuration
with $8$ fluxes has a purely imaginary solution for the four
moduli $z^i, i=1,2,3,4$ of the form \beq
z^1=-i\left(-\frac{Q_0P^1q_2q_3}{P^2P^3p^0q_1}\right)^{1/4},\qquad
z^2=-i\left(-\frac{Q_0P^2q_1q_3}{P^1P^3p^0q_2}\right)^{1/4}\label{12mod}
 \eeq\noindent
\beq
z^3=-i\left(-\frac{Q_0P^3q_1q_2}{P^1P^2p^0q_3}\right)^{1/4},\qquad
z^4=-i\left(-\frac{p^0q_1q_2q_3}{Q_0P^1P^2P^3}\right)^{1/4}\label{34mod}
 \eeq\noindent
\beq -{\rm sgn}(Q_0p^0)={\rm sgn}(P^1q_1)={\rm sgn}(P^2q_2)={\rm
sgn}(P^3q_3)=+1. \label{signconve}\eeq\noindent Recall that
$z^4=\overline{\tau}$ where $\tau=a+ie^{-\phi}$ is the
axion-dilaton. Now $a=0$ hence $-z^4$ gives the stabilized value of
the dilaton. These results can be rephrased in the language of the
state $\vert G\rangle$ as the distillation of the attractor state
\begin{eqnarray}\vert G\rangle_{fix}&=&\frac{1}{2}(a+d)\left(\vert
0000\rangle-\vert 1111\rangle\right)+\frac{1}{2}(a-d)\left(\vert
0011\rangle - \vert 1100\rangle\right)\\\nonumber
&+&\frac{1}{2}(b+c)\left(\vert 0101\rangle-\vert
1010\rangle\right)+\frac{1}{2}(b-c)\left(\vert 0110\rangle -\vert
1001\rangle \right).\end{eqnarray}\noindent Here \beq
a=i(t+z),\qquad b=i(t-z),\qquad c=i(y-x),\qquad
d=i(y+x)\label{szotarka}\eeq\noindent and
 \beq x={\rm
sgn}(q_1)\sqrt{P^1q_1},\quad y={\rm sgn}(q_2)\sqrt{P^2q_2},\quad
z={\rm sgn}(q_3)\sqrt{P^3q_3},\quad t=-{\rm
sgn}(-Q_0)\sqrt{-Q_0p^0}.\label{mennyisegek}\eeq\noindent This
state up to some phase conventions is of the same form as the
normal form representative of the generic class of four-qubit entangled
states  \cite{Verstraete:2002}. This is the only entanglement class with non-vanishing hyperdeterminant, which, in this case, is an order 24 polynomial built out of the four algebraically independent invariants \cite{Gelfand:1994, Luque:2002}.
 The state $\vert G\rangle$ above is the result of a distillation
procedure similar in character to the one discussed in the black
hole context. 

Note that one half the norm squared of $\vert G\rangle$ at the
attractor point is  analogous to  the black hole
potential in the three-qubit case. Depending
merely on the fluxes at the attractor point it should be an
$[\SL(2)]^{\otimes 4}$  four-qubit invariant. For our example this
quantity is also related to the sum of the a gravitino and chiral
fermion mass squares  \cite{Kallosh:2005ax}. A quick calculation shows
that \beq \frac{1}{2}\vert\vert G\vert\vert^2_{fix}= 2I_1=\int
F_3\wedge H_3 ,\eeq\noindent where $I_1$ is the quadratic
four-qubit invariant  \cite{Luque:2002}.

Another interesting quantity to look at is the four-qubit
generalization of Cayley's hyperdeterminant.
 For the definition of this $[\SL(2)]^{\otimes 4}$ and permutation
invariant polynomial of order $24$ we refer to the
literature  \cite{Luque:2002,2006JPhA...39.9533L}. Here we just give its explicit
form for our example \beq D_{4}=
(-Q_0P^1P^2P^3)(p^0q_1q_2q_3)\prod_{{lkji}\in({\mathds
Z}_2)^{\times
4}}\left((-1)^lt+(-1)^kz+(-1)^jy+(-1)^ix\right).\label{hyper4}\eeq\noindent
It is easy to check that $D_4>0$ due to our sign conventions of
(\ref{signconve}). A necessary condition for $D_4\neq 0$ for
this example of $8$ nonvanishing fluxes is the nonvanishing of the
$4$ independent amplitudes of $\vert G\rangle_{fix}$ showing up in
the $16$ terms of the product. The four-qubit hyperdeterminant
$D_4$ also plays an important role in the structure of two-center
black hole solutions of the $STU$ model  \cite{Levay:2011bq}.

Why only tori?
Clearly one should be able to remove the rather disturbing
restriction to toroidal compactifications by embarking on the rich
field of Calabi-Yau compactifications. Notice in this respect that
the decomposition of (\ref{elsohodge})-(\ref{masodikhodge}) in
the Hodge diagonal basis can then be used to reinterpret such
formulae as
    {\it qudits} i.e. $d$-level systems with $d=h^{2,1}+1$.
F-theoretical flux compactifications for elliptically fibered
Calabi-Yau fourfolds can then be associated with entangled systems
comprising a qubit (a $T^2$ accounting for the axion-dilaton) and
a qudit coming from a Calabi-Yau three-fold ($CY_3$).
Alternatively, using instead of $CY_3$ the combination
$T^2\times K3$ we can have tripartite systems consisting of two
qubits and a qudit. The idea that separable states geometrically
should correspond to product manifolds and entangled ones to
fibered ones was already discussed in the literature, for the
simplest cases of two and three qubits.
We also emphasize that though our main motivation was to account for the occurrence of qubits
in these exotic scenarios we have revealed that
in the string theoretical  context
 entangled systems of more general kind than qubits should rather be considered.
 In particular for toroidal models we have 
 seen that the natural arena where these
systems  live is
the realm of fermionic entanglement  \cite{2002AnPhy.299...88E,2004PhRvA..70a2109G} of
subsystems with {\it indistinguishable} parts. Of course the notion ``fermionic"
entanglement is simply associated with the structure of the
cohomology of $p$-forms related to $p$-branes.
It would be interesting to
explore further consequences of these ideas in connection with the
black-hole/qubit correspondence.

\section{The Freudenthal triple system classification of entanglement classes}
\label{FTSthreequbits}

There is a remarkable connection between 4-dimensional supergravity theories and what have come to be known as ``groups of type $E_7$'' \cite{Brown:1969,Gunaydin:1983bi, Gunaydin:1983rk, Borsten:2009zy,Ferrara:2011gv,Borsten:2011ai,Borsten:2011nq,Ferrara:2011dz,Ferrara:2011xb}. These groups may be characterized by the FTS. In particular, the $\mathcal{N}=8$ theory is related to the FTS defined over the Jordan algebra of $3\times 3$ Hermitian matrices, which has an $E_{7(7)}$ automorphism symmetry. Since the $STU$ model may be embedded in the $\mathcal{N}=8$ theory there should be a corresponding $STU$ Jordan algebra and FTS. There is and, as we shall see, it is particularly simple. 

Consequently, we would expect the  elegant mathematics  of Jordan algebras and Freudenthal triple systems to naturally capture the entanglement classification of three qubits. Indeed, the FTS ranks,  in a succinct algebraic manner, do indeed yield the correct classification.   The entanglement classes correspond to FTS ranks 0, 1, 2a, 2b, 2c, 3 and 4,  or, for SLOCC* (SLOCC plus permutations), simply 0, 1, 2, 3, 4. In fact, we would argue that this is perhaps the most natural classification scheme. This is not only a matter of aesthetics. The classification of \cite{Dur:2000} based on the local entropies $S_{A,B,C}$  did not make the $\SL(2,\C)^{\otimes 3}$ symmetry manifest. While the  hyperdeteminant is SLOCC-covariant, the local entropies are not; they are natural objects for a classification based on local unitaries not SLOCC. This observation is important from the perspective of generalising to $n$ qubits. Three qubits is the only non-trivial data point we have for a full  SLOCC classification.  If we seek to generalise this result, we should first formulate it in terms of the objects that will generalise, i.e. SLOCC-covariants. The FTS formulation is manifestly SLOCC-covariant since the automorphism group coincides with the SLOCC-equivalence group.

More speculatively, by studying the FTS classification, one might hope to identify those algebraic features which would usefully carry over to  an $n$-qubit generalisation. There is in fact an $n$-qubit generalization of the FTS, but it is not yet clear how it captures the entanglement classification \cite{Borsten:2010ths}.

\subsection{Three qubit entanglement}\label{sec:conventional}

The first SLOCC classification of three qubit entanglement was performed in \cite{Dur:2000} using a subset of the algebraically independent local unitary invariants and the  3-tangle.

The number of parameters needed to describe unnormalised LOCC (local unitary) inequivalent states   is  given by the dimension of the space of orbits \cite{Linden:1997qd},
\begin{equation}
\frac{\mathds{C}^2\times\mathds{C}^2\times\mathds{C}^2}{\U(1) \times \SU(2) \times \SU(2) \times \SU(2)},
\end{equation}
namely $16-10=6$.
This is also equivalent to the number of algebraically independent LU invariants \cite{Sudbery:2001},

\begin{enumerate}
\item  The norm squared:
\begin{equation}
|\Psi|^2=\braket{\Psi}{\Psi}.
\end{equation}
\item  The local entropies:
 \begin{equation}
S_{A}=4\det\rho_A, \qquad
S_{B}=4\det\rho_B, \qquad
S_{C}=4\det\rho_C,
\end{equation}
where $\rho_A, \rho_B, \rho_C$ are the doubly  reduced density matrices:
\begin{equation}
\rho_A=\Tr_{BC}\ket{\Psi}\bra{\Psi}, \qquad
\rho_B=\Tr_{CA}\ket{\Psi}\bra{\Psi}, \qquad
\rho_C=\Tr_{AB}\ket{\Psi}\bra{\Psi}.
\end{equation}
\item The Kempe invariant \cite{Kempe:1999vk}: 
\begin{equation}
\begin{split}
K&=\tr(\rho_A\otimes\rho_B\rho_{AB})-\tr(\rho_{A}^{3})-\tr(\rho_{B}^{3}) \\
&=\tr(\rho_B\otimes\rho_C\rho_{BC})-\tr(\rho_{B}^{3})-\tr(\rho_{C}^{3}) \\
&=\tr(\rho_C\otimes\rho_A\rho_{CA})-\tr(\rho_{C}^{3})-\tr(\rho_{A}^{3}),
\end{split}
\end{equation}
where $\rho_{AB}, \rho_{BC}, \rho_{CA}$ are the singly  reduced density matrices:
\begin{equation}
\rho_{AB}=\Tr_{C}\ket{\Psi}\bra{\Psi}, \qquad
\rho_{BC}=\Tr_{A}\ket{\Psi}\bra{\Psi}, \qquad
\rho_{CA}=\Tr_{B}\ket{\Psi}\bra{\Psi}.
\end{equation}

\item The 3-tangle \cite{Coffman:1999jd}:
\begin{equation}
\tau_{ABC}=4|\Det a_{ABC}|
\end{equation}
where for the explicit expression\cite{Cayley:1845,Miyake:2002}
see Eq.(\ref{Cayley}).
\end{enumerate}
Hence, under local unitary operations the most general state may be written as a six real parameter generating solution \cite{Acin:2001}. For subsequent comparison with the $STU$ black hole we restrict our attention to states with \emph{real} coefficients $a_{ABC}$. In this case, one can show that there are five algebraically independent LU invariants \cite{Acin:2001}: $\Det a$, $S_A$, $S_B$, $S_C$ and the norm $\braket{\Psi}{\Psi}$, corresponding to the dimension of
\begin{equation}
\frac{{\mathds R}^2 \times {\mathds R}^2 \times {\mathds R}^2}{\SO(2) \times \SO(2) \times \SO(2)},
\end{equation}
namely $8-3=5$.  Hence, the most general real three-qubit state can be described by just five parameters \cite{Acin:2001}, conveniently taken as four real numbers $N_0,N_1,N_2,N_3$ and an angle $\theta$:
\begin{equation}\label{eq:five}
\begin{split}
\ket{\Psi} &= -N_3\cos^2\theta\ket{001}-N_2\ket{010}+N_3\sin\theta\cos\theta\ket{011}\\
&\phantom{=}-N_1\ket{100}-N_3\sin\theta\cos\theta\ket{101}+(N_0+N_3\sin^2\theta)\ket{111}.
\end{split}
\end{equation}

\subsubsection{Entanglement classification}\label{sec:3qubitEnt}

 D\"ur et al. \cite{Dur:2000} used simple arguments concerning the conservation of ranks of reduced density matrices to show that there are only six types of 3-qubit equivalence classes (or seven if we count the null state); only two of which show
\emph{genuine} tripartite entanglement.  They are as follows:
\captionsetup{justification=raggedright}
\begin{figure}
\centering
\subfloat[Onion structure]{\label{fig:orbits}\includegraphics[width=.4\linewidth]{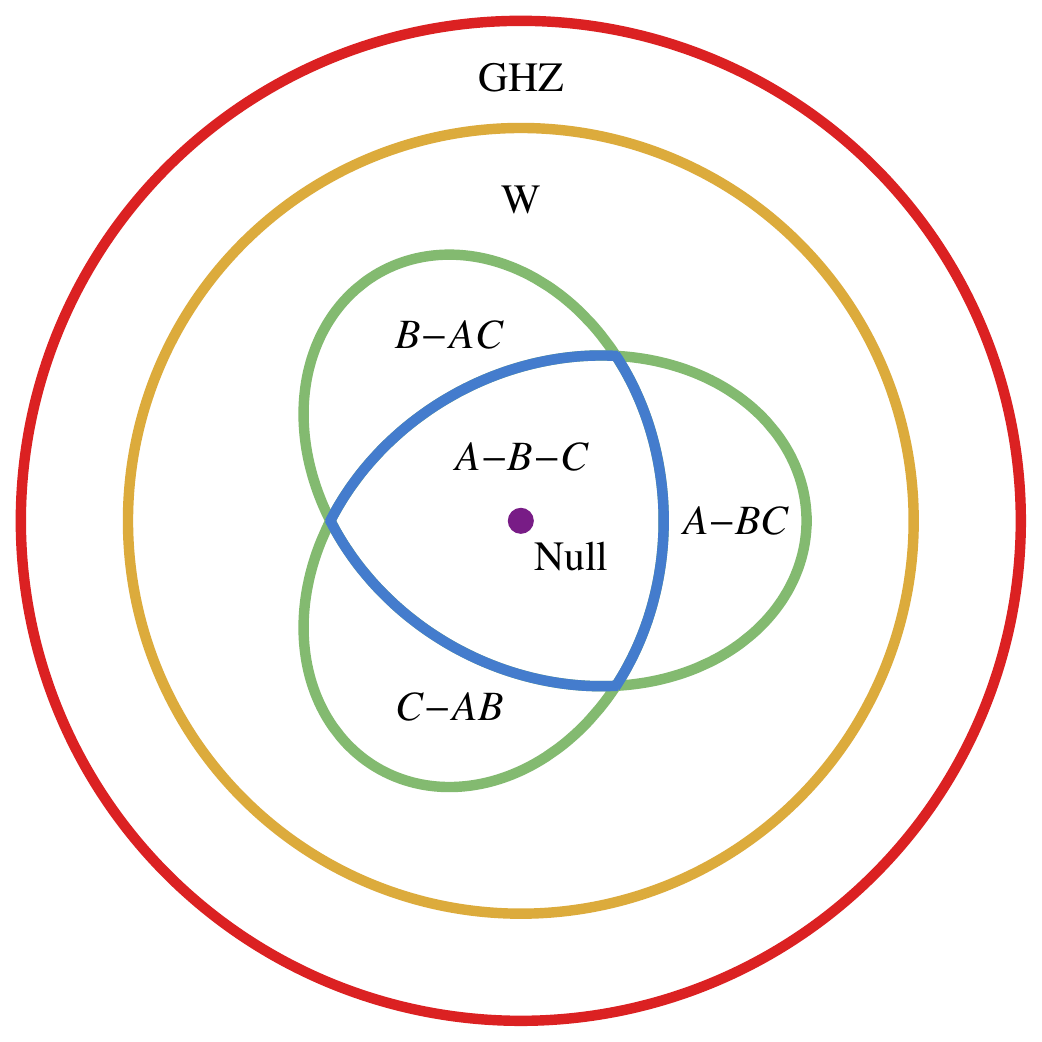}}
\subfloat[Hierarchy]{\label{fig:hierarchy}\includegraphics[width=.5\linewidth]{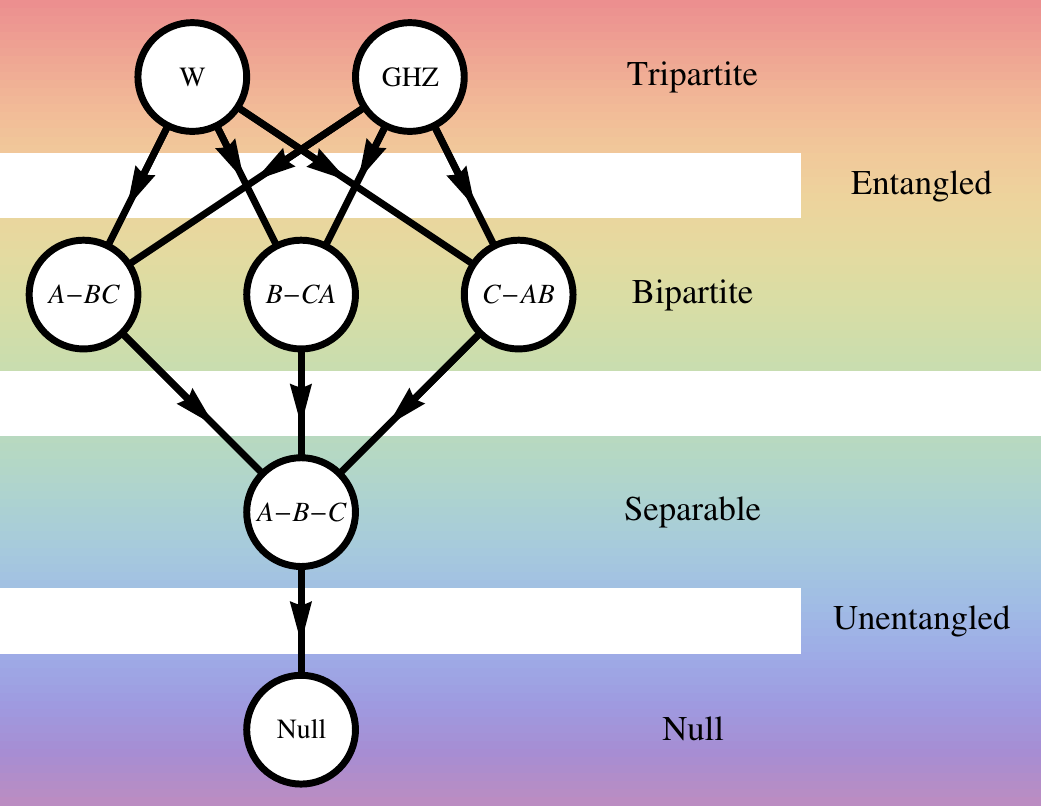}}
\caption{(a) Onion-like classification of SLOCC orbits. (b) Stratification. The arrows are non-invertible SLOCC transformations between classes that generate the entanglement hierarchy. The partial order defined by the arrows is transitive, so we may omit e.g. GHZ $\to$ $A$-$B$-$C$ and $A$-$BC$ $\to$ Null arrows for clarity.\label{fig:conventional}}
\end{figure}
\begin{table}
\caption{The values of the local entropies $S_A, S_B$, and $S_C$ and the hyperdeterminant $\Det a$ are used to partition three-qubit states into entanglement classes.\label{tab:conventional}}
\begin{tabular*}{\textwidth}{@{\extracolsep{\fill}}cc*{8}{M{c}}c}
\toprule
& \multirow{2}{*}{Class} & \multirow{2}{*}{Representative} & & \multicolumn{5}{c}{Condition}          & & \\
\cline{4-10}
&                        &                                 & & \Psi  & S_A   & S_B   & S_C   & \Det a & & \\
\colrule
& Null                   & 0                               & & =0    & =0    & =0    & =0    & =0     & & \\
& $A$-$B$-$C$            & \ket{000}                       & & \neq0 & =0    & =0    & =0    & =0     & & \\
& $A$-$BC$               & \ket{010}+\ket{001}             & & \neq0 & =0    & \neq0 & \neq0 & =0     & & \\
& $B$-$CA$               & \ket{100}+\ket{001}             & & \neq0 & \neq0 & =0    & \neq0 & =0     & & \\
& $C$-$AB$               & \ket{010}+\ket{100}             & & \neq0 & \neq0 & \neq0 & =0    & =0     & & \\
& W                      & \ket{100}+\ket{010}+\ket{001}   & & \neq0 & \neq0 & \neq0 & \neq0 & =0     & & \\
& GHZ                    & \ket{000}+\ket{111}             & & \neq0 & \neq0 & \neq0 & \neq0 & \neq0  & &\\
\botrule
\end{tabular*}
\end{table}
\begin{description}
  \item[Null:] The trivial zero entanglement orbit corresponding to vanishing states,
      \begin{equation}\text{Null}:\quad0.\end{equation}
  \item[Separable:] Another zero entanglement orbit for completely factorisable product states,
      \begin{equation}A\text{-}B\text{-}C:\quad\ket{000}.\end{equation}
  \item[Biseparable:] Three classes of bipartite entanglement,
      \begin{equation}
      \begin{split}
      A\text{-}BC:\quad\ket{010}+\ket{001},\\
      B\text{-}CA:\quad\ket{100}+\ket{001},\\
      C\text{-}AB:\quad\ket{010}+\ket{100}.
      \end{split}
      \end{equation}
      Note, these three classes are identified under SLOCC*.
  \item[W:] Three-way entangled states that do not maximally violate Bell-type inequalities in the same way as the GHZ class. However, they are robust in the sense that tracing out a subsystem generically results in a bipartite mixed state that is maximally entangled under a number of criteria \cite{Dur:2000},
      \begin{equation}\text{W}:\quad\ket{100}+\ket{010}+\ket{001}.\end{equation}
  \item[GHZ:] Genuinely tripartite entangled Greenberger-Horne-Zeilinger \cite{Greenberger:1989} states. These maximally violate Bell-type inequalities but, in contrast to class W, are fragile under the tracing out of a subsystem since the resultant state is completely unentangled,
      \begin{equation}\text{GHZ}:\quad\ket{000}+\ket{111}.\end{equation}
\end{description}
These classes and their  representative states are summarised in \autoref{tab:conventional}. They are characterised \cite{Dur:2000} by the vanishing or not of the invariants listed in the table. Note that the Kempe invariant is redundant in this SLOCC classification. A visual representation of these SLOCC orbits is provided by the onion-like classification \cite{Miyake:2002} of \hyperref[fig:orbits]{Figure 1a}.

These SLOCC equivalence classes are then stratified by \emph{non-invertible} SLOCC operations into an entanglement hierarchy \cite{Dur:2000} as depicted in \hyperref[fig:hierarchy]{Figure 1b}. Note that no SLOCC  operations (invertible or not) relate the GHZ and W classes; they are genuinely distinct classes of tripartite entanglement. However, from either the GHZ class or W class one may use non-invertible SLOCC transformations to descend to one of the biseparable or separable classes and hence we have a hierarchical entanglement structure. For more on three qubit entanglement see \cite{Gelfand:1994,Gibbs:2001,Fernando:2006,Brody:2007,Lee:2005,Lee:2007}.

\subsection{Cubic Jordan algebras}\label{sec:J}

A Jordan algebra $\mathfrak{J}$ is vector space defined over a ground field $\mathds{F}$ equipped with a bilinear product satisfying
\begin{equation}\label{eq:Jid}
X\circ Y =Y\circ X,\quad X^2\circ (X\circ Y)=X\circ (X^2\circ Y), \quad\forall\ X, Y \in \mathfrak{J}.
\end{equation}
The class of \emph{cubic} Jordan algebras are constructed as follows \cite{McCrimmon:2004}. Let $V$ be a vector space equipped with a cubic norm, i.e. a homogeneous map of degree three, $N:V\to \mathds{F}$ s.t. $N(\lambda X)=\lambda^3N(X)$, $\forall \lambda \in \mathds{F}, X\in V$, with trilinear linearisation
\begin{equation}
\begin{split}
N(X, Y, Z):=
&\frac{1}{6}[N(X+ Y+ Z)-N(X+Y)-N(X+ Z)-N(Y+
Z)\\&+N(X)+N(Y)+N(Z)].
\end{split}
\end{equation}
If $V$ further contains a base point $N(c)=1, c\in V$  one may define the following three maps,
    \begin{equation}\label{eq:cubicdefs}
    \begin{split}
    \Tr:V\to\mathds{F};\quad& X    \mapsto3N(c, c, X),\\
 		 S: V\times V\to\mathds{F};\quad& (X,Y)       \mapsto6N(X, Y, c),\\
    \Tr:V\times V\to\mathds{F};\quad& (X,Y)      \mapsto\Tr(X)\Tr(Y)-S(X, Y).
    \end{split}
    \end{equation}

A cubic Jordan algebra $\mathfrak{J}$, with multiplicative identity $\mathds{1}=c$, may be derived from any such vector space if $N$ is \emph{Jordan cubic}. That is: (i) the trace bilinear form \eqref{eq:cubicdefs} is non-degenerate (ii) the quadratic adjoint map, $\sharp\colon\mathfrak{J}\to\mathfrak{J}$, uniquely defined by $\Tr(X^\sharp, Y) = 3N(X, X, Y)$, satisfies
$(X^{\sharp})^\sharp=N(X)X$, $\forall X\in \mathfrak{J}$. The Jordan product is then given by 
\be
X\circ Y = \half\big(X\times Y+\Tr(X)Y+\Tr(Y)X-S(X, Y)\mathds{1}\big),
\ee
where, $X\times Y$ is the linearisation of the quadratic adjoint, $X\times Y = (X+Y)^\sharp-X^\sharp-Y^\sharp$.

\subsection{The Freudenthal triple system}\label{sec:FTS}

Given a cubic Jordan algebra $\mathfrak{J}$ defined over a field $\mathds{F}$, one is able to construct an FTS by defining the vector space $\mathfrak{F(J)}$,
\begin{equation}
\mathfrak{F(J)}=\mathds{F\oplus F}\oplus \mathfrak{J\oplus J}.
\end{equation}
An arbitrary element $x\in \mathfrak{F(J)}$ may be written as a ``$2\times 2$ matrix'',
\begin{equation}
x=\begin{pmatrix}\alpha&A\\B&\beta\end{pmatrix} \quad\text{where} ~\alpha, \beta\in\mathds{F}\quad\text{and}\quad A, B\in \mathfrak{J}.
\end{equation}
The FTS comes equipped with a non-degenerate bilinear antisymmetric quadratic form, a quartic form and a trilinear triple product \cite{Freudenthal:1954,Brown:1969,Faulkner:1971, Ferrar:1972, Krutelevich:2004}:
\begin{subequations}
\begin{enumerate}
\item Quadratic form $ \{x, y\}$: $\mathfrak{F}\times \mathfrak{F}\to \mathds{F}$
    \begin{equation}\label{eq:bilinearform}
    \{x, y\}=\alpha\delta-\beta\gamma+\Tr(A,D)-\Tr(B,C),\quad
\ee    
where
\be
\quad x=\begin{pmatrix}\alpha&A\\B&\beta\end{pmatrix},\quad y=\begin{pmatrix}\gamma&C\\D&\delta\end{pmatrix}.
      \end{equation}
\item Quartic form $q:\mathfrak{F}\to \mathds{F}$
    \begin{equation}\label{eq:quarticnorm}
        q(x)=-2[\alpha\beta-\Tr(A, B)]^2 
         -8[\alpha N(A)+\beta N(B)-\Tr(A^\sharp, B^\sharp)].
        \end{equation}
\item Triple product $T:\mathfrak{F\times F\times F\to F}$ which is uniquely defined by
	\begin{equation}\label{eq:tripleproduct}
	\{T(x, y, w), z\}=q(x, y, w, z)
	\end{equation}
	where $q(x, y, w, z)$ is the full linearisation of $q(x)$ such that $q(x, x, x, x)=q(x)$.
\end{enumerate}
\end{subequations}
The \emph{automorphism} group is given by the set of invertible $\F$-linear transformations preserving the quadratic and quartic forms \cite{Freudenthal:1954,Brown:1969},   
\be
\AutF{\mathfrak{F}}=\{\sigma\in \Iso_{\F}(\mathfrak{F})| q(\sigma x)=q(x), \{\sigma x, \sigma y\}=\{x,y\}, \;\forall x,y\in \mathfrak{F}\}.
\ee
The automorphism group corresponds to the U-duality group of a variety 4-dimensional supergravities (see for example \cite{Bellucci:2006xz,Borsten:2010aa} and the references therein). 
The conventional concept of matrix rank may be generalised to Freudenthal triple systems in a natural and $\Aut(\mathfrak{F})$ invariant manner. The rank of an arbitrary element $x\in\mathfrak{F}$ is uniquely defined by \cite{Ferrar:1972, Krutelevich:2004}:
\be\label{eq:FTSrank}
\begin{split}
\textrm{Rank} x=1&\Leftrightarrow 3T(x,x,y)+x\{x,y\}x=0\;\forall y;\\
\textrm{Rank} x=2&\Leftrightarrow \exists y\;\textrm{s.t.}\;3T(x,x,y)+x\{x,y\}x\not=0,\;T(x,x,x)=0;\\
\textrm{Rank} x=3&\Leftrightarrow T(x,x,x)\not=0,\;q(x)=0;\\
\textrm{Rank} x=4&\Leftrightarrow q(x)\not=0.\\
\end{split}
\ee

\subsection{The 3-qubit  Freudenthal triple system}\label{sec:3quibitFTS}

\begin{definition}[$STU$ cubic Jordan algebra] We define the $STU$ cubic Jordan algebra, denoted $\J_{STU}$, as the real vector space $\R\oplus\R\oplus\R$ with elements,
\be
A=(A_1, A_2, A_3),
\ee
and cubic norm,
\be
N_3(A)=A_1A_2A_3.
\ee
\end{definition}
Using the cubic Jordan algebra construction \eqref{eq:cubicdefs}, one finds
\begin{equation}
\Tr(A,B)= A_1B_1+A_2B_2+A_3B_3,
\end{equation}
Then, using $\Tr(A^\sharp,B)=3N(A,A,B)$, the quadratic adjoint is given by
\begin{gather}
A^\sharp=(A_2A_3, A_1A_3, A_1A_2),\\
\shortintertext{and therefore}
\begin{split}
(A^\sharp)^\sharp&=(A_1A_2A_3A_1,A_1A_2A_3A_2,A_1A_2A_3A_3)\\
                 &=N(A)A.
\end{split}
\end{gather}
It is not hard to check $\Tr(A,B)$ is non-degenerate and so $N_3$ is Jordan cubic. Hence, we have a \emph{bona fide} cubic Jordan algebra $\mathfrak{J}_{STU}=\mathds{R\oplus R\oplus R}$ with product given by
\begin{equation}
\begin{split}
A\circ B&=(A_1B_1, A_2B_2, A_3B_3).
\end{split}
\end{equation}
The structure and reduced structure groups are given by $[\SO(2, \mathds{R})]^3$ and $[\SO(2, \mathds{R})]^2$ respectively.

The 3-qubit cubic Jordan algebra is defined by simply promoting $\R$ to $\C$,
\begin{definition}[3-qubit cubic Jordan algebra] We define the 3-qubit cubic Jordan algebra, denoted as $\J_{ABC}$ (or more generally as $\J_{3\C}$), as the complex vector space $\C\oplus\C\oplus\C$ with elements:
\be
A=(A_1, A_2, A_3),
\ee
and cubic norm,
\be
N_3(A)=A_1A_2A_3.
\ee
\end{definition}

\begin{definition}[3-qubit Freudenthal triple system] We define the 3-qubit Freudenthal triple system, denoted $\FTS_{ABC}$, as the complex vector space,
\be
\FTS_{ABC}:=\C\oplus\C\oplus\J_{ABC}\oplus\J_{ABC},
\ee 
with elements
\begin{equation}\label{eq:BasicFts}
        \begin{pmatrix} \alpha & A=(A_1, A_2, A_3) \\
                        B=(B_1,B_2,B_3)   &  \beta
        \end{pmatrix}.
\end{equation}
\end{definition}
We identify the  eight complex components of $\FTS_{ABC}$ with the  three qubit wavefunction $\ket{\psi} = a_{ABC}\ket{ABC}$,
\begin{equation}
        \begin{pmatrix} \alpha & (A_1, A_2, A_3) \\
                        (B_1,B_2,B_3)   &  \beta
        \end{pmatrix}\leftrightarrow
        \begin{pmatrix} a_{111} & (a_{001}, a_{010}, a_{100}) \\
                       (a_{110}, a_{101}, a_{011})&a_{000}
        \end{pmatrix}
\end{equation}
so that 
\begin{equation}
\ket{\Psi}=a_{ABC}\ket{ABC}\quad\leftrightarrow\quad\Psi:=\begin{pmatrix}a_{111}&(a_{001}, a_{010}, a_{100})\\(a_{110}, a_{101}, a_{011})&a_{000}\end{pmatrix}.
\label{eq:3bit}
\end{equation}
Using \eqref{eq:quarticnorm} one finds that the quartic norm $\Delta(\Psi)$ is related to Cayley's hyperdeterminant by
\begin{equation}\label{eq:q}
\begin{split}
\Delta(\Psi)&=\{T(\Psi,\Psi,\Psi),\Psi\}\\
       &=2\det\gamma^{A}=2\det\gamma^{B}=2\det\gamma^{C}\\
       &=-2\Det a_{ABC},
\end{split}
\end{equation}
The triple product maps a state $\Psi$, which transforms as a $\rep{(2,2,2)}$ of  $[\SL(2, \mathds{C})]^3$, to another state $T(\Psi, \Psi, \Psi)$, cubic in the state vector coefficients, also transforming as a $\rep{(2,2,2)}$. Explicitly, $T(\Psi, \Psi, \Psi)$ may be written as
\begin{equation}\label{eq:Tdef}
T(\Psi, \Psi, \Psi)=T_{ABC}\ket{ABC}
\end{equation}
where $T_{ABC}$ takes one of three equivalent forms
\begin{equation}\label{eq:Tofgamma}
\begin{split}
T_{A_3B_1C_1}=\epsilon^{A_1A_2}a_{A_1B_1C_1}(\gamma^{A})_{A_{2}A_{3}}\\
T_{A_1B_3C_1}=\epsilon^{B_1B_2}a_{A_1B_1C_1}(\gamma^{B})_{B_{2}B_{3}}\\
T_{A_1B_1C_3}=\epsilon^{C_1C_2}a_{A_1B_1C_1}(\gamma^{C})_{C_{2}C_{3}}.
\end{split}
\end{equation}
The $\gamma$'s are related to the local entropies of \autoref{sec:conventional} by
\begin{equation}\label{eq:S_i(gamma)}
S_A = 4\Big[\tr\gamma^{B\dag}\gamma^B+\tr\gamma^{C\dag}\gamma^C\Big], \quad
\tr\gamma^{A\dag}\gamma^A = \tfrac{1}{8}\big[S_B+S_C-S_A\big]
\end{equation}
and their cyclic permutations.
This  permits us to link $T$ to the norm, local entropies and  the Kempe invariant  of \autoref{sec:conventional}:
\begin{equation}
\braket{T}{T}=\tfrac{2}{3}(K-|\psi|^6)+\tfrac{1}{16}|\psi|^2(S_A+S_B+S_C).
\end{equation}
Having couched the 3-qubit system within the FTS framework we may assign an abstract FTS rank \eqref{eq:FTSrank} to an arbitrary state $\Psi$:
\be
\begin{split}
\textrm{Rank} \Psi =1& \Leftrightarrow \Upsilon(\Psi, \Psi, \Phi)=0,\; \Psi\not=0;\\
\textrm{Rank} \Psi =2& \Leftrightarrow T(\Psi)=0,\;\Upsilon(\Psi, \Psi, \Phi)\not=0;\\
\textrm{Rank} \Psi =3& \Leftrightarrow \Delta(\Psi)=0,\;T(\Psi)\not=0;\\
\textrm{Rank} \Psi =4& \Leftrightarrow \Delta(\Psi)\not=0.\\
\end{split}
\ee

Strictly speaking, the automorphism group $\Aut(\FTS_{ABC})$ is not simply  $\SL(2,\mathds{C}) \times \SL(2,\mathds{C}) \times \SL(2,\mathds{C})$ but includes a semi-direct product with the interchange triality $A \leftrightarrow B   \leftrightarrow C$.  The rank conditions are invariant under this triality.  Hence the ranks naturally provide an SLOCC* classification. However, as we shall demonstrate, the set of rank 2 states may be subdivided into three distinct classes  which are inter-related by this triality. In the next section we show that these rank conditions give the correct entanglement classification of three qubits as in \autoref{tab:merge}.
\begin{table}
\begin{tabular*}{\textwidth}{@{\extracolsep{\fill}}ccccM{c}M{c}cc}
\toprule
& \multirow{2}{*}{Class} & \multirow{2}{*}{Rank} & & \multicolumn{2}{c}{\text{FTS rank condition}}               & & \\
\cline{4-7}
&                        &                       & & \text{vanishing}                     & \text{non-vanishing} & & \\
\colrule
& Null                   & 0  		             & & \Psi  				                  & -                    & & \\
& $A$-$B$-$C$            & 1  		             & & 3T(\Psi,\Psi,\Phi)+\{\Psi,\Phi\}\Psi & \Psi 		         & & \\
& $A$-$BC$               & 2a 		             & & T(\Psi,\Psi,\Psi) 					  & \gamma^A             & & \\
& $B$-$CA$               & 2b 		             & & T(\Psi,\Psi,\Psi) 					  & \gamma^B             & & \\
& $C$-$AB$               & 2c 		             & & T(\Psi,\Psi,\Psi)                    & \gamma^C	         & & \\
& W                      & 3  		             & & \Delta(\Psi)							  & T(\Psi,\Psi,\Psi) 	 & & \\
& GHZ                    & 4  		             & & -                                    & \Delta(\Psi)              & &\\
\botrule
\end{tabular*}
\caption{The entanglement classification of three qubits as according to the FTS rank system.\label{tab:merge}}
\end{table}

\subsubsection{The FTS rank entanglement classes}
\label{sec:rank}

\paragraph{Rank 1 and the class of separable states}
\label{sec:rank1}

A non-zero state $\Psi$  is rank 1 if and only if
\begin{equation}\label{eq:FTSrank1}
\Upsilon(\Psi, \Psi, \Phi) :=3T(\Psi, \Psi, \Phi) + \{\Psi , \Phi\}\Psi=0, \quad \forall\ \Phi.
\end{equation}
The weaker condition $T(\Psi,\Psi,\Psi)=0$ implies that there is at most one non-vanishing $\gamma$ since, \begin{equation}
\begin{split}
(\gamma^A)_{A_{1}A_{2}}(\gamma^C)_{C_{1}C_{2}}&=\phantom{\times}\epsilon^{B_1B_2}\epsilon^{Z_1Z_2} a_{A_1B_1Z_1}a_{A_2B_2Z_2}(\gamma^C)_{C_{1}C_{2}}\\
&=\phantom{+}\epsilon^{B_2B_1}a_{A_1B_1C_1}T_{A_2B_2C_2}+\epsilon^{B_1B_2}a_{A_2B_2C_1}T_{A_1B_1C_2},
\end{split}
\end{equation}
and similarly for $(\gamma^B)_{B_{1}B_{2}}(\gamma^A)_{A_{1}A_{2}}$ and $(\gamma^C)_{C_{1}C_{2}}(\gamma^B)_{B_{1}B_{2}}$.
In component form $\Upsilon$ is given by,
\be
\begin{split}
-\Upsilon_{A_1B_1C_1}=&\epsilon^{A_2A_3}b_{A_3B_1C_1}(\gamma^A)_{A_1A_2}\\
&+\epsilon^{B_2B_3}b_{A_1B_3C_1}(\gamma^B)_{B_1B_2}\\
&+\epsilon^{C_2C_3}b_{A_1B_1C_3}(\gamma^C)_{C_1C_2}
\end{split}
\ee
where
\be
\ket{\phi}=b_{ABC}\ket{ABC}\quad\leftrightarrow\quad\Phi=\begin{pmatrix}b_{111}&(b_{001}, b_{010}, b_{100})\\(b_{110}, b_{101}, b_{011})&b_{000}\end{pmatrix}.
\ee
Hence, \eqref{eq:FTSrank1} implies all three gammas must vanish. Using \eqref{eq:S_i(gamma)} it is then clear that all three local entropies vanish.

Conversely,  $S_A=S_B=S_C=0$ implies that each of the three $\gamma$'s vanish and the rank 1 condition is satisfied. Hence FTS rank 1 is equivalent to the class of separable states as in \autoref{tab:merge}.

\paragraph{Rank 2 and the class of biseparable states}

 A  nonzero state $\Psi$ is rank 2 or less if and only if $T(\Psi,\Psi,\Psi)=0$, which implies there is at most one non-vanishing $\gamma$. To  be rank $>1$ there must exist some $\Phi$ such that $3T(\Psi, \Psi, \Phi) + \{\Psi , \Phi\}\Psi\not=0$, which implies there is at least one non-vanishing $\gamma$. 
 Hence,  rank  2 states have precisely one nonzero $\gamma$.

Using \eqref{eq:S_i(gamma)} it is clear that the choices $\gamma^A\not=0$ or $\gamma^B\not=0$ or $\gamma^C\not=0$  give $S_A=0, S_{B,C}\neq0$ or $S_B=0, S_{C,A}\neq0$ or $S_C=0, S_{A,B,}\neq0$, respectively. These are precisely the conditions for the biseparable class $A$-$BC$ or $B$-$CA$ or $C$-$AB$ presented in \autoref{tab:conventional}.

Conversely, using \eqref{eq:S_i(gamma)}  and the fact that the local entropies and $\tr(\gamma^{\dag}\gamma)$ are positive semidefinite, we find that all states in the biseparable class are rank 2, the particular subdivision being given by the corresponding non-zero $\gamma$. Hence FTS rank 2  is equivalent to the class of biseparable states as in \autoref{tab:merge}.

\paragraph{Rank 3 and the class of W-states}

A  non-zero state $\Psi$ is rank 3 if $\Delta(\Psi)=-2\Det a=0$ but $T(\Psi, \Psi, \Psi)\neq 0$.
From \eqref{eq:Tofgamma} all three $\gamma$'s are then non-zero but from \eqref{eq:q} all have vanishing determinant.   In this case \eqref{eq:S_i(gamma)} implies that all three local entropies are non-zero but $\Det a=0$.  So all rank 3 $\Psi$ belong to the W-class.

Conversely, from \eqref{eq:S_i(gamma)} it is clear that no two $\gamma$'s may simultaneously vanish when all three $S$'s $>0$. We saw in \autoref{sec:rank1} that $T(\Psi, \Psi, \Psi)=0$ implied at least two of the $\gamma$'s vanish. Consequently, for all W-states $T(\Psi, \Psi, \Psi)\not=0$ and, therefore, all W-states are rank 3.  Hence FTS rank 3 is equivalent to the class of W-states as in \autoref{tab:merge}.

\paragraph{Rank 4 and the class of GHZ-states}

The rank 4 condition is given by $\Delta(\Psi)\neq0$ and, since for the 3-qubit FTS $\Delta(\Psi)=-2 \Det a$, we immediately see that the set of rank 4 states is equivalent to the GHZ class of genuine tripartite entanglement  as in \autoref{tab:merge}.

Note, $\Aut(\mathfrak{F}_{ABC})$ acts transitively only on rank 4 states with the same value of $\Delta(\Psi)$ as in the standard treatment. The GHZ class really corresponds to a 1-dimensional space of orbits parametrised by $\Delta$.

In summary, we have demonstrated that each rank corresponds to one of the entanglement classes described in   \autoref{sec:conventional}.

\subsubsection{SLOCC orbits}
\label{sec:orbits}

\begin{table}
\begin{tabular*}{\textwidth}{@{\extracolsep{\fill}}ccD{c}cD{c}cc}
\toprule
& Class  & \text{Orbits} & dim &	\text{Projective orbits} & dim & \\
\colrule
& Separable    & \frac{[\SL(2,\mathds{C})]^3}{[\SO(2,\mathds{C})]^2\ltimes\mathds{C}^3} & 4 & \frac{[\SL(2,\mathds{C})]^3}{[\SO(2,\mathds{C})\ltimes\mathds{C}]^3}                    & 3 & \\
& Biseparable & \frac{[\SL(2,\mathds{C})]^3}{\Orth(3,\mathds{C})\times\mathds{C}}         & 5 & \frac{[\SL(2,\mathds{C})]^3}{\Orth(3,\mathds{C})\times(\SO(2,\mathds{C})\ltimes\mathds{C})} & 4 & \\
& W            & \frac{[\SL(2,\mathds{C})]^3}{\mathds{C}^2}                            & 7 & \frac{[\SL(2,\mathds{C})]^3}{\SO(2,\mathds{C})\ltimes\mathds{C}^2}                      & 6 & \\
& GHZ          & \frac{[\SL(2,\mathds{C})]^3}{[\SO(2,\mathds{C})]^2}                    & 7 & \frac{[\SL(2,\mathds{C})]^3}{[\SO(2,\mathds{C})]^2}                                     & 7 &\\
\botrule
\end{tabular*}
\caption{Coset spaces of the orbits of the 3-qubit state space $\mathds{C}^2\otimes\mathds{C}^2\otimes\mathds{C}^2$ under the action of the SLOCC group $[\SL(2, \mathds{C})]^3$.\label{tab:3qubitscosets}}
\end{table}
\begin{table}
\begin{tabular*}{\textwidth}{@{\extracolsep{\fill}}cccM{c}D{c}cc}
\toprule
& Class       & FTS Rank & \Delta(\Psi) & \text{Orbits}                                               & dim & \\
\colrule
& Separable   & 1        & =0      & \frac{[\SL(2,\mathds{R})]^3}{[\SO(1,1)]^2\ltimes\mathds{R}^3} & 4   & \\
& Biseparable & 2        & =0      & \frac{[\SL(2,\mathds{R})]^3}{\Orth(2,1)\times\mathds{R}}         & 5   & \\
& W           & 3        & =0      & \frac{[\SL(2,\mathds{R})]^3}{\mathds{R}^2}                   & 7   & \\
& GHZ         & 4        & <0      & \frac{[\SL(2,\mathds{R})]^3}{[\SO(1,1)]^2}                    & 7   & \\
& GHZ         & 4        & >0      & \frac{[\SL(2,\mathds{R})]^3}{[\U(1)]^2}                       & 7   & \\
& GHZ         & 4        & >0      & \frac{[\SL(2,\mathds{R})]^3}{[\U(1)]^2}                       & 7   & \\
\botrule
\end{tabular*}
\caption{Coset spaces of the orbits of the real case $\mathfrak{J}_{3\mathds{R}}=\mathds{R\oplus R\oplus R}$ under $[\SL(2, \mathds{R})]^3$.\label{tab:realcosets}}
\end{table}
We now turn our attention to the  coset parametrisation of the entanglement classes. The coset space of each orbit $(i=1,2,3,4)$ is given by $G/H_i$ where $G=[\SL(2, \mathds{C})]^3$ is the SLOCC group and $H_i \subset [\SL(2, \mathds{C})]^3$ is the stability subgroup leaving the representative state of the $i $th orbit invariant. We proceed by considering the infinitesimal action of $\Aut(\FTS)$ on the representative states of each class. The subalgebra annihilating the representative state gives, upon exponentiation, the stability group $H$. Note, $\mathfrak{der}(\mathfrak{J}_{ABC})$ is empty due to the associativity of $\mathfrak{J}_{ABC}$. Consequently, $\mathfrak{Str}(\mathfrak{J}_{ABC})=L_\mathds{1}\mathds{C}\oplus \mathfrak{Str}_0(\mathfrak{J}_{ABC})$  has complex dimension 3, while $\mathfrak{Str}_0(\mathfrak{J}_{ABC})$ is now simply $L_{\mathfrak{J}'}$ and has complex dimension 2. Recall, $\mathfrak{Str}(\mathfrak{J}_{ABC})$   and $\mathfrak{Str}_0(\mathfrak{J}_{ABC})$ generate $[\SO(2, \mathds{C})]^3$ and $[\SO(2, \mathds{C})]^2$,  respectively the structure and reduced structure groups of $\mathfrak{J}_\mathds{C}$. 

The results are summarised in \autoref{tab:3qubitscosets}.  To be clear, in the preceding analysis we have regarded the three-qubit state as a  point in $\mathds{C}^2\otimes\mathds{C}^2\otimes\mathds{C}^2$, the philosophy adopted in, for example, \cite{Linden:1997qd, Carteret:2000-1, Sudbery:2001}. We could have equally well considered the projective Hilbert space regarding states as rays in $\mathds{C}^2\otimes\mathds{C}^2\otimes\mathds{C}^2$, that is, identifying states related by a global complex scalar factor, as was done in \cite{Miyake:2002, Miyake:2003, Brody:2007}. The coset spaces obtained in this case are also presented in \autoref{tab:3qubitscosets}, the dimensions of which agree with the results of \cite{Miyake:2002, Gelfand:1994}. Note that the three-qubit separable projective coset  is just a direct product of  three individual qubit cosets
$\SL(2,\mathds{C})/\SO(2,\mathds{C})\ltimes \mathds{C}$.	Furthermore, the biseparable projective coset is just the direct product of  the two entangled qubits coset $[\SL(2,\mathds{C})]^2/\Orth(3,\mathds{C})$ and an individual qubit coset.

As noted in \cite{Acin:2000,Levay:2006kf}, the case of real qubits or ``rebits'' is qualitatively different from the complex case. An interesting observation is that on restricting to real states the GHZ class actually has two distinct orbits, characterised by the sign of $\Delta(\Psi)$. This difference shows up in the cosets in the different possible real forms of $[\SO(2, \mathds{C})]^2$. For positive $\Delta(\Psi)$ there are two disconnected orbits, both with $[\SL(2,\mathds{R})]^3/[\U(1)]^2$ cosets, while for negative $\Delta(\Psi)$ there is one orbit $[\SL(2,\mathds{R})]^3/[\SO(1,1, \mathds{R})]^2$. In which of the two positive   $\Delta(\Psi)$ orbits a given state lies is determined by the sign of the eigenvalues of the three $\gamma$'s, as shown in \autoref{tab:realcosets}. This phenomenon also has its counterpart in the black-hole context \cite{Ferrara:1997ci, Ferrara:1997uz,Bellucci:2006xz, Bellucci:2007gb, Bellucci:2008sv, Borsten:2008wd}, where the two disconnected $\Delta(\Psi)>0$ orbits are given by $1/2$-BPS black holes and non-BPS black holes with vanishing central charge respectively \cite{Bellucci:2006xz}.

\subsection{Fermionic entanglement classification from the  Freudenthal triple system}\label{sec:ferFTS}

The 3-qubit Freudenthal triple system can be generalised to classify the entanglement of several other multipartite systems simply by considering different Jordan algebras. In particular, the FTS naturally captures the structure of entanglement for various multipartite systems of both distinguishable and \emph{indistinguishable} constituents \cite{levay-2008,LevayVrana}.  While there is a significant body of literature on the entanglement classification of distinguishable multipartite systems, much less is known for indistinguishable constituents. For bipartite fermionic and bosonic systems a number of useful results exist \cite{Eckert200288,PhysRevA.64.054302,PhysRevA.64.042310,PhysRevA.70.012109,PhysRevA.64.022303,PhysRevB.63.085311}. For example, a two-fermion systems with $2k$ Òsingle-particleÓ states admits a Schmidt-like decomposition \cite{PhysRevA.64.022303,PhysRevB.63.085311}.

To differentiate the various possible distinguishable and indistinguishable (bosonic/fermionic) systems we will  denote the basis vectors for a $k$-level distinguishable, indistinguishable-bosonic and indistinguishable-fermionic system by, $e_i$, $b_i$ and $f_i$, where $i=1,\ldots, k$, respectively.  The $n$-partite basis vectors are then given by,
\be
e_{i_1}\otimes e_{i_2}\otimes\cdots\otimes e_{i_n}, \quad b_{(i_1}\otimes b_{i_2}\otimes\cdots\otimes b_{i_n)}, \quad f_{i_1}\wedge f_{i_2}\wedge\cdots\wedge f_{i_n}.
\ee
 \begin{table}
\begin{tabular*}{\textwidth}{@{\extracolsep{\fill}}cccl}
\toprule
Space $\FTS$ ($\mathcal{H}$)&$\Aut(\FTS)$ & Class  &  Representatives  \\
\hline
\multirow{4}{*}{$\bigwedge^3\mathds{C}^6$}  & \multirow{4}{*}{$\SL(6, \C)$} & GHZ&$\frac{1}{\sqrt{2}}(f_{1}\wedge f_{2}\wedge f_{3}+f_{4}\wedge f_{5}\wedge f_{6})$  \\

 &&   $W$&$\frac{1}{\sqrt{3}}(f_{4}\wedge f_{2}\wedge f_{3}+f_{1}\wedge f_{5}\wedge f_{3}+ f_{1}\wedge f_{2}\wedge f_{6})$  \\
 &&   Biseparable&$\frac{1}{\sqrt{2}}(f_{1}\wedge f_{2}\wedge f_{3}+f_{1}\wedge f_{5}\wedge f_{6})$  \\
 &&   Separable&$f_{1}\wedge f_{2}\wedge f_{3}$  \\
\hline
 \multirow{5}{*}{$\mathds{C}^2\otimes\bigwedge^2\mathds{C}^4$}  & \multirow{5}{*}{$\SL(2, \C)\times\SL(4, \C)$}&  GHZ&$\frac{1}{\sqrt{2}}(e_{0}\otimes(f_{0}\wedge f_{1})+e_{1}\otimes(f_{2}\wedge f_{3}))$  \\

  &&   $W$&$\frac{1}{\sqrt{3}}(e_{0}\otimes(f_{2}\wedge f_{3})+e_{1}\otimes(f_{0}\wedge f_{3})+e_{1}\otimes(f_{2}\wedge f_{1}))$  \\
  &&  Biseparable (a) &$\frac{1}{\sqrt{2}}e_{0}\otimes(f_{0}\wedge f_{1}+f_{2}\wedge f_{3})$  \\
  &&   Biseparable (b)&$\frac{1}{\sqrt{2}}(e_{0}\otimes(f_{0}\wedge f_{1})+e_{1}\otimes(f_{0}\wedge f_{3}))$  \\
  &&   Separable &$e_{0}\otimes(f_{0}\wedge f_{1})$  \\
\hline
 \multirow{6}{*}{$\mathds{C}^2\otimes\mathds{C}^2\otimes\mathds{C}^2$
}  & \multirow{6}{*}{$\SL(2, \C)\times\SL(2, \C)\times\SL(2, \C)$}&  GHZ&$\frac{1}{\sqrt{2}}(e_{0}\otimes e_{0}\otimes e_{0}+e_{1}\otimes e_{1}\otimes e_{1})$  \\
  &&  $W$&$\frac{1}{\sqrt{3}}(e_{1}\otimes e_{0}\otimes e_{0}+e_{0}\otimes e_{1}\otimes e_{0}+e_{0}\otimes e_{0}\otimes e_{1})$  \\
  &&  Biseparable (a)&$\frac{1}{\sqrt{2}}(e_{0}\otimes (e_{0}\otimes e_{0}+e_{1}\otimes e_{1}))$  \\
  &&  Biseparable (b)&$\frac{1}{\sqrt{2}}(e_{0}\otimes e_{0}\otimes e_{0}+e_{1}\otimes e_{0}\otimes e_{1})$  \\
  &&  Biseparable (c)&$\frac{1}{\sqrt{2}}((e_{0}\otimes e_{0}+e_{1}\otimes e_{1})\otimes e_{0})$  \\
  &&  Separable&$e_{0}\otimes e_{0}\otimes e_{0}$  \\
\hline
 \multirow{4}{*}{$\mathds{C}^2\otimes \Sym^2(\mathds{C}^2)$} & \multirow{4}{*}{$\SL(2, \C)\times\SL(2, \C)$} &  GHZ&$\frac{1}{\sqrt{2}}(e_{0}\otimes(b_{0}\otimes b_{0})+e_{1}\otimes(b_{1}\otimes b_{1}))$  \\

  &&  $W$&$\frac{1}{\sqrt{3}}(e_{1}\otimes(b_{0}\otimes b_{0})+e_{0}\otimes(b_{1}\otimes b_{0}+b_{0}\otimes b_{1}))$  \\
  &&  Biseparable (a) &$\frac{1}{\sqrt{2}}(e_{0}\otimes (b_{0}\otimes b_{0}+b_{1}\otimes b_{1}))$  \\
  && Separable&$e_{0}\otimes(b_{0}\otimes b_{0})$  \\
\hline
 \multirow{3}{*}{$\Sym^3(\mathds{C}^2)$} & \multirow{3}{*}{$\SL(2, \C)$} &  GHZ&$\frac{1}{\sqrt{2}}(b_{0}\otimes b_{0}\otimes b_{0}+b_{1}\otimes b_{1}\otimes b_{1})$  \\

  &&  $W$&$\frac{1}{\sqrt{3}}(b_{1}\otimes b_{0}\otimes b_{0}+b_{0}\otimes b_{1}\otimes b_{0}+b_{0}\otimes b_{0}\otimes b_{1})$  \\
  &&  Separable&$b_{0}\otimes b_{0}\otimes b_{0}$  \\
\botrule
\end{tabular*}
\caption{\label{tab:reps}Representatives of SLOCC orbits of quantum mechanical systems classified via Freudenthal's construction \cite{LevayVrana}.}
\end{table}

The two simplest examples we consider are essentially degenerations of the 3-qubit system $\FTS(\J_{3\C})$, given by replacing $\J_{3\C}$ with $\J_{2\C}=\C\oplus\C$ and $\J_{\C}=\C$. The cubic norms of $\J_{2\C}$ and $\J_{\C}$ are given by setting $A_2=A_3$ and $A_1=A_2=A_3$, respectively,
\be
N_{\J_{2\C}}(A)=A_1A_{2}^{2}, \qquad N_{\J_{3\C}}(A)=A_{1}^{3}.
\ee
Since $\J_{\C}\subset\J_{2\C}\subset\J_{3\C}$,  clearly $\FTS(\J_{\C})\subset\FTS(\J_{2\C})\subset\FTS(\J_{3\C})$. As vector spaces $\FTS(\J_{\C})$ and $\FTS(\J_{2\C})$ are given by $\Sym^3(\C^2)$ and $\C^2\otimes\Sym^2(\C^2)$, respectively. The former is the Hilbert space of three indistinguishable bosonic quibits, while the later corresponds to the Hilbert space of a single distinguishable qubit and two indistinguishable bosonic quibits. 
Using the identification  between $\Psi\in\FTS(\J_{3\C})$ and $a_{ABC}\ket{ABC}$ given in  \eqref{eq:3bit}, elements in
  $\FTS(\J_{2\C})$ and $\FTS(\J_{\C})$ can be written in terms of the appropriate symmetrizations, i.e. $a_{A(BC)}$ and $a_{(ABC)}$, respectively.  The antisymmetric bilinear form, quartic norm and triple product are then simply given by symmetrizing those of  $\FTS(\J_{3\C})$. See \cite{LevayVrana, Borsten:2011nq} for details. 
  
The automorphism (SLOCC) groups are given by $\SL(2,\C)_{A}\times \SL(2,\C)_{(BC)}$ and $\SL(2,\C)_{(ABC)}$, under which $a_{A(BC)}$ and $a_{(ABC)}$ transform as a $\mathbf{(2,3)}$ and a $\mathbf{4}$, respectively. Just as in the conventional 3-qubit case, the orbit (``entanglement'') classification is given  by the FTS ranks defined in \eqref{eq:FTSrank}. The classes/orbits and their representative states are given in \autoref{tab:reps} \cite{LevayVrana}.
These essentially have the same structure as the 3-qubit case. One subtlety is that the biseparable class is absent for  $\FTS(\J_{\C})$. This is a consequence of $N_{\J_{3\C}}(A)=0\Rightarrow A=0$. The corresponding  orbit classification over the reals is given in \cite{Borsten:2011nq}. The real case underlies the structure of the $ST^2$  and $T^3$ models of $\mathcal{N}=2$ supergravity coupled to two and one vector multiplets respectively. The analysis of the single-centre extremal black hole solutions in these models, via the orbits, is given in \cite{Bellucci:2006xz, Bellucci:2007zi,Gaiotto:2007ag,Bellucci:2008sv,Kim:2010bf,Borsten:2011ai}.

These bosonic systems were embedded in the  3-qubit FTS. Going the other way, i.e. embedding the 3-qubit system in larger FTSs, we can accommodate  a 3-fermion system. As pointed out in \cite{levay-2008} the first non-trivial 3-fermion system has 6-level constituents, since the the 5-level and 4-level cases may be mapped to 2- and 1-fermion systems, respectively, using $\bigwedge^3 \C^5\cong\bigwedge^2\C^5$  and  $\bigwedge^3 \C^4\cong\bigwedge^1\C^4$. The system of three 6-level fermions has Hilbert space $\bigwedge^3 \C^6$. 

Let $\{f_{1},f_{2},f_{3}, f_{\bar{1}}\equiv f_{4}, f_{\bar{2}}\equiv f_{5}, f_{\bar{3}}\equiv f_{6}\}$
be an orthonormal basis of $\mathds{C}^6$, and let $x\wedge y\wedge z$ denote
the normalized wedge product of the vectors $x,y,z\in\mathds{C}^6$:
\be
x\wedge y\wedge z =  \frac{1}{\sqrt{6}}(x\otimes y\otimes z+y\otimes z\otimes x+z\otimes x\otimes y -x\otimes z\otimes y-z\otimes y\otimes x-y\otimes x\otimes z)
\ee
Then a generic (unnormalised) 3-fermion state may be written as
\begin{equation}
\ket{P}=\sum_{1\le a<b<c\le \bar{3}}P_{abc}f_{a}\wedge f_{b}\wedge f_{c}.
\end{equation}

The state space coincides with the FTS defined over the Jordan algebra $H_3(\tilde{\C}_\C)$ of $3\times 3$ Hermitian matrices over the complexified split-complexes \cite{Krutelevich:2004, Rios:2010br}, or, more simply, the Jordan algebra $M_3(\C)$ of $3\times 3$ complex matrices \cite{levay-2008},
\be
\FTS_{\text{3-fermion}}=\C\oplus\C\oplus M_3(\C)\oplus M_3(\C).
\ee
The cubic norm on $M_3(\C)$ is simply given by the determinant and the remaining Jordan/FTS structures then follow from the definitions given in \autoref{sec:J} and \autoref{sec:FTS}.

Explicitly, 
for $x=(\alpha,\beta,A,B)\in \FTS_{\text{3-fermion}}$ we have
\begin{equation}
\alpha=P_{123}\quad\beta=P_{\bar{1}\bar{2}\bar{3}}\quad
A=\left[\begin{array}{ccc}
P_{1\bar{2}\bar{3}}  &  P_{1\bar{3}\bar{1}}  &  P_{1\bar{1}\bar{2}}  \\
P_{2\bar{2}\bar{3}}  &  P_{2\bar{3}\bar{1}}  &  P_{2\bar{1}\bar{2}}  \\
P_{3\bar{2}\bar{3}}  &  P_{3\bar{3}\bar{1}}  &  P_{3\bar{1}\bar{2}}
\end{array}\right]\quad
B=\left[\begin{array}{ccc}
P_{\bar{1}23}  &  P_{\bar{1}31}  &  P_{\bar{1}12}  \\
P_{\bar{2}23}  &  P_{\bar{2}31}  &  P_{\bar{2}12}  \\
P_{\bar{3}23}  &  P_{\bar{3}31}  &  P_{\bar{3}12}
\end{array}\right].
\end{equation}
Note, the conventional 3-qubit system is given by the sub Jordan algebra of diagonal matrices in  $M_3(\C)$, in which case the quartic norm reduces to Cayley's hyperdeterminant \cite{levay-2008}. 

The automorphism group is given by $\SL(6,\C)$, which partitions the state space into four orbits  given precisely by the ranks, as in the 3-qubit case. Of course, the GHZ class is actually a 1-dimensional space of orbits parametrised by the quartic norm, which collapses to a single orbit under $\GL(6,\C)$. The representative states are given \autoref{tab:reps}. Finally, there is  an intermediate FTS sitting between $\FTS_{3\C}$ and $\FTS_{\text{3-fermion}}$, given by the Jordan algebra $\C\oplus M_2(\C)$, with cubic norm $a\det(A)$ for $a\in\C, A\in M_2(\C)$, which describes a single distinguishable qubit with two 4-level fermions \cite{LevayVrana}. The entanglement classes are given in \autoref{tab:reps} and further details may be found in  \cite{LevayVrana}.

\section{Conclusions}

Our purpose in this review has been to highlight the numerous (and still growing) useful ways in which black holes in string theory can inform qubit entanglement in QIT and vice-versa, whether or not there is physical connection between the two. These include not only the originally discovered relations between the U-duality invariant entropies of the black holes and the corresponding entanglement measures in QIT, and the more recent work on the attractor mechanism and concurrence and classifying four qubit entanglement and Freudenthal triple systems, but also less familiar results on the relation to Hamming codes, geometric hyperplanes and Mermin Squares. All this suggests that there is till much more to be uncovered.

\section*{Acknowledgement} We would like to thank Samson Abramsky, Anna Ceresole, Alessio Marrani, Philip Gibbs and William Rubens for helpful contributions.  The work of L. B. is
supported by an Imperial College Junior Research Fellowship.
The work of M. J. D. is
supported by the STFC under rolling grant ST/G000743/1. P. L.  would like to acknowledge the support provided by
the New Hungary Development Plan (Project ID:
T\'AMOP-4.2.1/B-09/1/KMR-2010-002).

%\bibliography{arxiv,notarxiv}

\begin{thebibliography}{190}
\expandafter\ifx\csname natexlab\endcsname\relax\def\natexlab#1{#1}\fi
\expandafter\ifx\csname bibnamefont\endcsname\relax
  \def\bibnamefont#1{#1}\fi
\expandafter\ifx\csname bibfnamefont\endcsname\relax
  \def\bibfnamefont#1{#1}\fi
\expandafter\ifx\csname citenamefont\endcsname\relax
  \def\citenamefont#1{#1}\fi
\expandafter\ifx\csname url\endcsname\relax
  \def\url#1{\texttt{#1}}\fi
\expandafter\ifx\csname urlprefix\endcsname\relax\def\urlprefix{URL }\fi
\providecommand{\bibinfo}[2]{#2}
\providecommand{\eprint}[2][]{\url{#2}}

\bibitem[{\citenamefont{Duff}(2007)}]{Duff:2006uz}
\bibinfo{author}{\bibfnamefont{M.~J.} \bibnamefont{Duff}},
  \bibinfo{journal}{Phys. Rev.} \textbf{\bibinfo{volume}{D76}},
  \bibinfo{pages}{025017} (\bibinfo{year}{2007}), \eprint{hep-th/0601134}.

\bibitem[{\citenamefont{Kallosh and Linde}(2006)}]{Kallosh:2006zs}
\bibinfo{author}{\bibfnamefont{R.}~\bibnamefont{Kallosh}} \bibnamefont{and}
  \bibinfo{author}{\bibfnamefont{A.}~\bibnamefont{Linde}},
  \bibinfo{journal}{Phys. Rev.} \textbf{\bibinfo{volume}{D73}},
  \bibinfo{pages}{104033} (\bibinfo{year}{2006}), \eprint{hep-th/0602061}.

\bibitem[{\citenamefont{L\'evay}(2006)}]{Levay:2006kf}
\bibinfo{author}{\bibfnamefont{P.}~\bibnamefont{L\'evay}},
  \bibinfo{journal}{Phys. Rev.} \textbf{\bibinfo{volume}{D74}},
  \bibinfo{pages}{024030} (\bibinfo{year}{2006}), \eprint{hep-th/0603136}.

\bibitem[{\citenamefont{Nielsen and Chuang}(2000)}]{Nielsen:2000}
\bibinfo{author}{\bibfnamefont{M.~A.} \bibnamefont{Nielsen}} \bibnamefont{and}
  \bibinfo{author}{\bibfnamefont{I.~L.} \bibnamefont{Chuang}},
  \emph{\bibinfo{title}{Quantum Computation and Quantum Information}}
  (\bibinfo{publisher}{Cambridge University Press}, \bibinfo{address}{New York,
  NY, USA}, \bibinfo{year}{2000}), ISBN \bibinfo{isbn}{0-521-63503-9}.

\bibitem[{\citenamefont{Bekenstein}(1973)}]{Bekenstein:1973ur}
\bibinfo{author}{\bibfnamefont{J.~D.} \bibnamefont{Bekenstein}},
  \bibinfo{journal}{Phys. Rev.} \textbf{\bibinfo{volume}{D7}},
  \bibinfo{pages}{2333} (\bibinfo{year}{1973}).

\bibitem[{\citenamefont{Hawking}(1975)}]{Hawking:1974sw}
\bibinfo{author}{\bibfnamefont{S.~W.} \bibnamefont{Hawking}},
  \bibinfo{journal}{Commun. Math. Phys.} \textbf{\bibinfo{volume}{43}},
  \bibinfo{pages}{199} (\bibinfo{year}{1975}).

\bibitem[{\citenamefont{Plenio and Virmani}(2007)}]{Plenio:2007}
\bibinfo{author}{\bibfnamefont{M.~B.} \bibnamefont{Plenio}} \bibnamefont{and}
  \bibinfo{author}{\bibfnamefont{S.}~\bibnamefont{Virmani}},
  \bibinfo{journal}{Quant. Inf. Comp.} \textbf{\bibinfo{volume}{7}},
  \bibinfo{pages}{1} (\bibinfo{year}{2007}), \eprint{quant-ph/0504163}.

\bibitem[{\citenamefont{Cremmer and Julia}(1978)}]{Cremmer:1978ds}
\bibinfo{author}{\bibfnamefont{E.}~\bibnamefont{Cremmer}} \bibnamefont{and}
  \bibinfo{author}{\bibfnamefont{B.}~\bibnamefont{Julia}},
  \bibinfo{journal}{Phys.Lett.} \textbf{\bibinfo{volume}{B80}},
  \bibinfo{pages}{48} (\bibinfo{year}{1978}).

\bibitem[{\citenamefont{Cremmer and Julia}(1979)}]{Cremmer:1979up}
\bibinfo{author}{\bibfnamefont{E.}~\bibnamefont{Cremmer}} \bibnamefont{and}
  \bibinfo{author}{\bibfnamefont{B.}~\bibnamefont{Julia}},
  \bibinfo{journal}{Nucl. Phys.} \textbf{\bibinfo{volume}{B159}},
  \bibinfo{pages}{141} (\bibinfo{year}{1979}).

\bibitem[{\citenamefont{Duff and Lu}(1990)}]{Duff:1990hn}
\bibinfo{author}{\bibfnamefont{M.~J.} \bibnamefont{Duff}} \bibnamefont{and}
  \bibinfo{author}{\bibfnamefont{J.~X.} \bibnamefont{Lu}},
  \bibinfo{journal}{Nucl. Phys.} \textbf{\bibinfo{volume}{B347}},
  \bibinfo{pages}{394} (\bibinfo{year}{1990}).

\bibitem[{\citenamefont{Hull and Townsend}(1995)}]{Hull:1994ys}
\bibinfo{author}{\bibfnamefont{C.~M.} \bibnamefont{Hull}} \bibnamefont{and}
  \bibinfo{author}{\bibfnamefont{P.~K.} \bibnamefont{Townsend}},
  \bibinfo{journal}{Nucl. Phys.} \textbf{\bibinfo{volume}{B438}},
  \bibinfo{pages}{109} (\bibinfo{year}{1995}), \eprint{hep-th/9410167}.

\bibitem[{\citenamefont{Obers and Pioline}(1999)}]{Obers:1998fb}
\bibinfo{author}{\bibfnamefont{N.}~\bibnamefont{Obers}} \bibnamefont{and}
  \bibinfo{author}{\bibfnamefont{B.}~\bibnamefont{Pioline}},
  \bibinfo{journal}{Phys.Rept.} \textbf{\bibinfo{volume}{318}},
  \bibinfo{pages}{113} (\bibinfo{year}{1999}), \eprint{hep-th/9809039}.

\bibitem[{\citenamefont{Bennett et~al.}(2000)\citenamefont{Bennett, Popescu,
  Rohrlich, Smolin, and Thapliyal}}]{Bennett:1999}
\bibinfo{author}{\bibfnamefont{C.~H.} \bibnamefont{Bennett}},
  \bibinfo{author}{\bibfnamefont{S.}~\bibnamefont{Popescu}},
  \bibinfo{author}{\bibfnamefont{D.}~\bibnamefont{Rohrlich}},
  \bibinfo{author}{\bibfnamefont{J.~A.} \bibnamefont{Smolin}},
  \bibnamefont{and} \bibinfo{author}{\bibfnamefont{A.~V.}
  \bibnamefont{Thapliyal}}, \bibinfo{journal}{Phys. Rev.}
  \textbf{\bibinfo{volume}{A63}}, \bibinfo{pages}{012307}
  (\bibinfo{year}{2000}), \eprint{quant-ph/9908073}.

\bibitem[{\citenamefont{{D\"ur} et~al.}(2000)\citenamefont{{D\"ur}, Vidal, and
  Cirac}}]{Dur:2000}
\bibinfo{author}{\bibfnamefont{W.}~\bibnamefont{{D\"ur}}},
  \bibinfo{author}{\bibfnamefont{G.}~\bibnamefont{Vidal}}, \bibnamefont{and}
  \bibinfo{author}{\bibfnamefont{J.~I.} \bibnamefont{Cirac}},
  \bibinfo{journal}{Phys. Rev.} \textbf{\bibinfo{volume}{A62}},
  \bibinfo{pages}{062314} (\bibinfo{year}{2000}), \eprint{quant-ph/0005115}.

\bibitem[{\citenamefont{Duff and Ferrara}(2007{\natexlab{a}})}]{Duff:2006ue}
\bibinfo{author}{\bibfnamefont{M.~J.} \bibnamefont{Duff}} \bibnamefont{and}
  \bibinfo{author}{\bibfnamefont{S.}~\bibnamefont{Ferrara}},
  \bibinfo{journal}{Phys. Rev.} \textbf{\bibinfo{volume}{D76}},
  \bibinfo{pages}{025018} (\bibinfo{year}{2007}{\natexlab{a}}),
  \eprint{quant-ph/0609227}.

\bibitem[{\citenamefont{L\'evay}(2007{\natexlab{a}})}]{Levay:2006pt}
\bibinfo{author}{\bibfnamefont{P.}~\bibnamefont{L\'evay}},
  \bibinfo{journal}{Phys. Rev.} \textbf{\bibinfo{volume}{D75}},
  \bibinfo{pages}{024024} (\bibinfo{year}{2007}{\natexlab{a}}),
  \eprint{hep-th/0610314}.

\bibitem[{\citenamefont{Duff and Ferrara}(2007{\natexlab{b}})}]{Duff:2007wa}
\bibinfo{author}{\bibfnamefont{M.~J.} \bibnamefont{Duff}} \bibnamefont{and}
  \bibinfo{author}{\bibfnamefont{S.}~\bibnamefont{Ferrara}},
  \bibinfo{journal}{Phys. Rev.} \textbf{\bibinfo{volume}{D76}},
  \bibinfo{pages}{124023} (\bibinfo{year}{2007}{\natexlab{b}}),
  \eprint{0704.0507}.

\bibitem[{\citenamefont{L\'{e}vay et~al.}(2008)\citenamefont{L\'{e}vay, Saniga,
  and Vrana}}]{Levay:2008mi}
\bibinfo{author}{\bibfnamefont{P.}~\bibnamefont{L\'{e}vay}},
  \bibinfo{author}{\bibfnamefont{M.}~\bibnamefont{Saniga}}, \bibnamefont{and}
  \bibinfo{author}{\bibfnamefont{P.}~\bibnamefont{Vrana}},
  \bibinfo{journal}{Phys. Rev.} \textbf{\bibinfo{volume}{D78}},
  \bibinfo{pages}{124022} (\bibinfo{year}{2008}), \eprint{0808.3849}.

\bibitem[{\citenamefont{Levay et~al.}(2009)\citenamefont{Levay, Saniga, Vrana,
  and Pracna}}]{Levay:2009bp}
\bibinfo{author}{\bibfnamefont{P.}~\bibnamefont{Levay}},
  \bibinfo{author}{\bibfnamefont{M.}~\bibnamefont{Saniga}},
  \bibinfo{author}{\bibfnamefont{P.}~\bibnamefont{Vrana}}, \bibnamefont{and}
  \bibinfo{author}{\bibfnamefont{P.}~\bibnamefont{Pracna}},
  \bibinfo{journal}{Phys. Rev.} \textbf{\bibinfo{volume}{D79}},
  \bibinfo{pages}{084036} (\bibinfo{year}{2009}), \eprint{0903.0541}.

\bibitem[{\citenamefont{L\'{e}vay and Vrana}(2008)}]{levay-2008}
\bibinfo{author}{\bibfnamefont{P.}~\bibnamefont{L\'{e}vay}} \bibnamefont{and}
  \bibinfo{author}{\bibfnamefont{P.}~\bibnamefont{Vrana}},
  \bibinfo{journal}{Phys. Rev.} \textbf{\bibinfo{volume}{A78}},
  \bibinfo{pages}{022329} (\bibinfo{year}{2008}), \eprint{0806.4076}.

\bibitem[{\citenamefont{Vrana and L{\'e}vay}(2009)}]{Levay:2009}
\bibinfo{author}{\bibfnamefont{P.}~\bibnamefont{Vrana}} \bibnamefont{and}
  \bibinfo{author}{\bibfnamefont{P.}~\bibnamefont{L{\'e}vay}},
  \bibinfo{journal}{Journal of Physics A: Mathematical and Theoretical}
  \textbf{\bibinfo{volume}{42}}, \bibinfo{pages}{285303}
  (\bibinfo{year}{2009}), \eprint{0902.2269},
  \urlprefix\url{http://stacks.iop.org/1751-8121/42/i=28/a=285303}.

\bibitem[{\citenamefont{Borsten
  et~al.}(2010{\natexlab{a}})\citenamefont{Borsten, Dahanayake, Duff, Marrani,
  and Rubens}}]{Borsten:2010db}
\bibinfo{author}{\bibfnamefont{L.}~\bibnamefont{Borsten}},
  \bibinfo{author}{\bibfnamefont{D.}~\bibnamefont{Dahanayake}},
  \bibinfo{author}{\bibfnamefont{M.~J.} \bibnamefont{Duff}},
  \bibinfo{author}{\bibfnamefont{A.}~\bibnamefont{Marrani}}, \bibnamefont{and}
  \bibinfo{author}{\bibfnamefont{W.}~\bibnamefont{Rubens}},
  \bibinfo{journal}{Phys. Rev. Lett.} \textbf{\bibinfo{volume}{105}},
  \bibinfo{pages}{100507} (\bibinfo{year}{2010}{\natexlab{a}}),
  \eprint{1005.4915}.

\bibitem[{\citenamefont{Borsten
  et~al.}(2011{\natexlab{a}})\citenamefont{Borsten, Duff, Marrani, and
  Rubens}}]{Borsten:2011is}
\bibinfo{author}{\bibfnamefont{L.}~\bibnamefont{Borsten}},
  \bibinfo{author}{\bibfnamefont{M.}~\bibnamefont{Duff}},
  \bibinfo{author}{\bibfnamefont{A.}~\bibnamefont{Marrani}}, \bibnamefont{and}
  \bibinfo{author}{\bibfnamefont{W.}~\bibnamefont{Rubens}},
  \bibinfo{journal}{Eur.Phys.J.Plus} \textbf{\bibinfo{volume}{126}},
  \bibinfo{pages}{37} (\bibinfo{year}{2011}{\natexlab{a}}), \eprint{1101.3559}.

\bibitem[{\citenamefont{Ferrara et~al.}(1995)\citenamefont{Ferrara, Kallosh,
  and Strominger}}]{Ferrara:1995ih}
\bibinfo{author}{\bibfnamefont{S.}~\bibnamefont{Ferrara}},
  \bibinfo{author}{\bibfnamefont{R.}~\bibnamefont{Kallosh}}, \bibnamefont{and}
  \bibinfo{author}{\bibfnamefont{A.}~\bibnamefont{Strominger}},
  \bibinfo{journal}{Phys. Rev.} \textbf{\bibinfo{volume}{D52}},
  \bibinfo{pages}{5412} (\bibinfo{year}{1995}), \eprint{hep-th/9508072}.

\bibitem[{\citenamefont{Strominger}(1996)}]{Strominger:1996kf}
\bibinfo{author}{\bibfnamefont{A.}~\bibnamefont{Strominger}},
  \bibinfo{journal}{Phys. Lett.} \textbf{\bibinfo{volume}{B383}},
  \bibinfo{pages}{39} (\bibinfo{year}{1996}), \eprint{hep-th/9602111}.

\bibitem[{\citenamefont{Ferrara and
  Kallosh}(1996{\natexlab{a}})}]{Ferrara:1996dd}
\bibinfo{author}{\bibfnamefont{S.}~\bibnamefont{Ferrara}} \bibnamefont{and}
  \bibinfo{author}{\bibfnamefont{R.}~\bibnamefont{Kallosh}},
  \bibinfo{journal}{Phys. Rev.} \textbf{\bibinfo{volume}{D54}},
  \bibinfo{pages}{1514} (\bibinfo{year}{1996}{\natexlab{a}}),
  \eprint{hep-th/9602136}.

\bibitem[{\citenamefont{Duff et~al.}(1996)\citenamefont{Duff, Liu, and
  Rahmfeld}}]{Duff:1995sm}
\bibinfo{author}{\bibfnamefont{M.~J.} \bibnamefont{Duff}},
  \bibinfo{author}{\bibfnamefont{J.~T.} \bibnamefont{Liu}}, \bibnamefont{and}
  \bibinfo{author}{\bibfnamefont{J.}~\bibnamefont{Rahmfeld}},
  \bibinfo{journal}{Nucl. Phys.} \textbf{\bibinfo{volume}{B459}},
  \bibinfo{pages}{125} (\bibinfo{year}{1996}), \eprint{hep-th/9508094}.

\bibitem[{\citenamefont{Bellucci et~al.}(2008)\citenamefont{Bellucci, Ferrara,
  Marrani, and Yeranyan}}]{Bellucci:2008sv}
\bibinfo{author}{\bibfnamefont{S.}~\bibnamefont{Bellucci}},
  \bibinfo{author}{\bibfnamefont{S.}~\bibnamefont{Ferrara}},
  \bibinfo{author}{\bibfnamefont{A.}~\bibnamefont{Marrani}}, \bibnamefont{and}
  \bibinfo{author}{\bibfnamefont{A.}~\bibnamefont{Yeranyan}},
  \bibinfo{journal}{Entropy} \textbf{\bibinfo{volume}{10}},
  \bibinfo{pages}{507} (\bibinfo{year}{2008}), \eprint{0807.3503}.

\bibitem[{\citenamefont{L\'evay}(2007{\natexlab{b}})}]{Levay:2007nm}
\bibinfo{author}{\bibfnamefont{P.}~\bibnamefont{L\'evay}},
  \bibinfo{journal}{Phys. Rev.} \textbf{\bibinfo{volume}{D76}},
  \bibinfo{pages}{106011} (\bibinfo{year}{2007}{\natexlab{b}}),
  \eprint{0708.2799}.

\bibitem[{\citenamefont{Levay and Szalay}(2010)}]{Levay:2010qp}
\bibinfo{author}{\bibfnamefont{P.}~\bibnamefont{Levay}} \bibnamefont{and}
  \bibinfo{author}{\bibfnamefont{S.}~\bibnamefont{Szalay}},
  \bibinfo{journal}{Phys.Rev.} \textbf{\bibinfo{volume}{D82}},
  \bibinfo{pages}{026002} (\bibinfo{year}{2010}), \eprint{1004.2346}.

\bibitem[{\citenamefont{Levay and Szalay}(2011)}]{Levay:2010yh}
\bibinfo{author}{\bibfnamefont{P.}~\bibnamefont{Levay}} \bibnamefont{and}
  \bibinfo{author}{\bibfnamefont{S.}~\bibnamefont{Szalay}},
  \bibinfo{journal}{Phys.Rev.} \textbf{\bibinfo{volume}{D83}},
  \bibinfo{pages}{045005} (\bibinfo{year}{2011}), \eprint{1011.4180}.

\bibitem[{\citenamefont{Levay}(2010)}]{Levay:2010ua}
\bibinfo{author}{\bibfnamefont{P.}~\bibnamefont{Levay}},
  \bibinfo{journal}{Phys.Rev.} \textbf{\bibinfo{volume}{D82}},
  \bibinfo{pages}{026003} (\bibinfo{year}{2010}), \eprint{1004.3639}.

\bibitem[{\citenamefont{Borsten
  et~al.}(2009{\natexlab{a}})\citenamefont{Borsten, Dahanayake, Duff, Rubens,
  and Ebrahim}}]{Borsten:2009yb}
\bibinfo{author}{\bibfnamefont{L.}~\bibnamefont{Borsten}},
  \bibinfo{author}{\bibfnamefont{D.}~\bibnamefont{Dahanayake}},
  \bibinfo{author}{\bibfnamefont{M.~J.} \bibnamefont{Duff}},
  \bibinfo{author}{\bibfnamefont{W.}~\bibnamefont{Rubens}}, \bibnamefont{and}
  \bibinfo{author}{\bibfnamefont{H.}~\bibnamefont{Ebrahim}},
  \bibinfo{journal}{Phys. Rev.} \textbf{\bibinfo{volume}{A80}},
  \bibinfo{pages}{032326} (\bibinfo{year}{2009}{\natexlab{a}}),
  \eprint{0812.3322}.

\bibitem[{\citenamefont{Freudenthal}(1954)}]{Freudenthal:1954}
\bibinfo{author}{\bibfnamefont{H.}~\bibnamefont{Freudenthal}},
  \bibinfo{journal}{Nederl. Akad. Wetensch. Proc. Ser.}
  \textbf{\bibinfo{volume}{57}}, \bibinfo{pages}{218} (\bibinfo{year}{1954}).

\bibitem[{\citenamefont{McCrimmon}(2004)}]{McCrimmon:2004}
\bibinfo{author}{\bibfnamefont{K.}~\bibnamefont{McCrimmon}},
  \emph{\bibinfo{title}{{A Taste of Jordan Algebras}}}
  (\bibinfo{publisher}{Springer-Verlag New York Inc.}, \bibinfo{address}{New
  York}, \bibinfo{year}{2004}), ISBN \bibinfo{isbn}{0-387-95447-3}.

\bibitem[{\citenamefont{Krutelevich}(2007)}]{Krutelevich:2004}
\bibinfo{author}{\bibfnamefont{S.}~\bibnamefont{Krutelevich}},
  \bibinfo{journal}{J. Algebra} \textbf{\bibinfo{volume}{314}},
  \bibinfo{pages}{924} (\bibinfo{year}{2007}), \eprint{math/0411104}.

\bibitem[{\citenamefont{G{\"u}naydin et~al.}(1984)\citenamefont{G{\"u}naydin,
  Sierra, and Townsend}}]{Gunaydin:1983bi}
\bibinfo{author}{\bibfnamefont{M.}~\bibnamefont{G{\"u}naydin}},
  \bibinfo{author}{\bibfnamefont{G.}~\bibnamefont{Sierra}}, \bibnamefont{and}
  \bibinfo{author}{\bibfnamefont{P.~K.} \bibnamefont{Townsend}},
  \bibinfo{journal}{Nucl. Phys.} \textbf{\bibinfo{volume}{B242}},
  \bibinfo{pages}{244} (\bibinfo{year}{1984}).

\bibitem[{\citenamefont{G{\"u}naydin et~al.}(1983)\citenamefont{G{\"u}naydin,
  Sierra, and Townsend}}]{Gunaydin:1983rk}
\bibinfo{author}{\bibfnamefont{M.}~\bibnamefont{G{\"u}naydin}},
  \bibinfo{author}{\bibfnamefont{G.}~\bibnamefont{Sierra}}, \bibnamefont{and}
  \bibinfo{author}{\bibfnamefont{P.~K.} \bibnamefont{Townsend}},
  \bibinfo{journal}{Phys. Lett.} \textbf{\bibinfo{volume}{B133}},
  \bibinfo{pages}{72} (\bibinfo{year}{1983}).

\bibitem[{\citenamefont{G{\"u}naydin et~al.}(1985)\citenamefont{G{\"u}naydin,
  Sierra, and Townsend}}]{Gunaydin:1984ak}
\bibinfo{author}{\bibfnamefont{M.}~\bibnamefont{G{\"u}naydin}},
  \bibinfo{author}{\bibfnamefont{G.}~\bibnamefont{Sierra}}, \bibnamefont{and}
  \bibinfo{author}{\bibfnamefont{P.~K.} \bibnamefont{Townsend}},
  \bibinfo{journal}{Nucl. Phys.} \textbf{\bibinfo{volume}{B253}},
  \bibinfo{pages}{573} (\bibinfo{year}{1985}).

\bibitem[{\citenamefont{Borsten}(2008)}]{Borsten:2008}
\bibinfo{author}{\bibfnamefont{L.}~\bibnamefont{Borsten}},
  \bibinfo{journal}{Fortschr. Phys.} \textbf{\bibinfo{volume}{56}},
  \bibinfo{pages}{842} (\bibinfo{year}{2008}).

\bibitem[{\citenamefont{Borsten
  et~al.}(2009{\natexlab{b}})\citenamefont{Borsten, Dahanayake, Duff, and
  Rubens}}]{Borsten:2009zy}
\bibinfo{author}{\bibfnamefont{L.}~\bibnamefont{Borsten}},
  \bibinfo{author}{\bibfnamefont{D.}~\bibnamefont{Dahanayake}},
  \bibinfo{author}{\bibfnamefont{M.~J.} \bibnamefont{Duff}}, \bibnamefont{and}
  \bibinfo{author}{\bibfnamefont{W.}~\bibnamefont{Rubens}},
  \bibinfo{journal}{Phys. Rev.} \textbf{\bibinfo{volume}{D80}},
  \bibinfo{pages}{026003} (\bibinfo{year}{2009}{\natexlab{b}}),
  \eprint{0903.5517}.

\bibitem[{\citenamefont{Ferrara
  et~al.}(2011{\natexlab{a}})\citenamefont{Ferrara, Marrani, and
  Yeranyan}}]{Ferrara:2011gv}
\bibinfo{author}{\bibfnamefont{S.}~\bibnamefont{Ferrara}},
  \bibinfo{author}{\bibfnamefont{A.}~\bibnamefont{Marrani}}, \bibnamefont{and}
  \bibinfo{author}{\bibfnamefont{A.}~\bibnamefont{Yeranyan}},
  \bibinfo{journal}{Phys.Lett.} \textbf{\bibinfo{volume}{B701}},
  \bibinfo{pages}{640} (\bibinfo{year}{2011}{\natexlab{a}}),
  \eprint{1102.4857}.

\bibitem[{\citenamefont{Ferrara
  et~al.}(2011{\natexlab{b}})\citenamefont{Ferrara, Marrani, Orazi, Stora, and
  Yeranyan}}]{Ferrara:2010ug}
\bibinfo{author}{\bibfnamefont{S.}~\bibnamefont{Ferrara}},
  \bibinfo{author}{\bibfnamefont{A.}~\bibnamefont{Marrani}},
  \bibinfo{author}{\bibfnamefont{E.}~\bibnamefont{Orazi}},
  \bibinfo{author}{\bibfnamefont{R.}~\bibnamefont{Stora}}, \bibnamefont{and}
  \bibinfo{author}{\bibfnamefont{A.}~\bibnamefont{Yeranyan}},
  \bibinfo{journal}{J.Math.Phys.} \textbf{\bibinfo{volume}{52}},
  \bibinfo{pages}{062302} (\bibinfo{year}{2011}{\natexlab{b}}),
  \bibinfo{note}{1+31 pages/ v2: amendments in Sec. 9, App. C added, other
  minor refinements, Refs. added/ v3: Ref. added, typos fixed. To appear on
  J.Math.Phys}, \eprint{1011.5864}.

\bibitem[{\citenamefont{Levay}(2011{\natexlab{a}})}]{Levay:2011bq}
\bibinfo{author}{\bibfnamefont{P.}~\bibnamefont{Levay}},
  \bibinfo{journal}{Phys.Rev.} \textbf{\bibinfo{volume}{D84}},
  \bibinfo{pages}{025023} (\bibinfo{year}{2011}{\natexlab{a}}),
  \eprint{1104.0144}.

\bibitem[{\citenamefont{Luque and Thibon}(2003)}]{Luque:2002}
\bibinfo{author}{\bibfnamefont{J.-G.} \bibnamefont{Luque}} \bibnamefont{and}
  \bibinfo{author}{\bibfnamefont{J.-Y.} \bibnamefont{Thibon}},
  \bibinfo{journal}{Phys. Rev.} p. \bibinfo{pages}{042303}
  (\bibinfo{year}{2003}), \eprint{quant-ph/0212069}.

\bibitem[{\citenamefont{{L{\'e}vay}}(2006)}]{2006JPhA...39.9533L}
\bibinfo{author}{\bibfnamefont{P.}~\bibnamefont{{L{\'e}vay}}},
  \bibinfo{journal}{Journal of Physics A Mathematical General}
  \textbf{\bibinfo{volume}{39}}, \bibinfo{pages}{9533} (\bibinfo{year}{2006}),
  \eprint{arXiv:quant-ph/0605151}.

\bibitem[{\citenamefont{Gibbs}(2010)}]{Gibbs:2010uz}
\bibinfo{author}{\bibfnamefont{P.}~\bibnamefont{Gibbs}} (\bibinfo{year}{2010}),
  \eprint{1010.4219}.

\bibitem[{\citenamefont{Bellucci et~al.}(2011)\citenamefont{Bellucci, Marrani,
  and Roychowdhury}}]{Bellucci:2010zd}
\bibinfo{author}{\bibfnamefont{S.}~\bibnamefont{Bellucci}},
  \bibinfo{author}{\bibfnamefont{A.}~\bibnamefont{Marrani}}, \bibnamefont{and}
  \bibinfo{author}{\bibfnamefont{R.}~\bibnamefont{Roychowdhury}},
  \bibinfo{journal}{J.Math.Phys.} \textbf{\bibinfo{volume}{52}},
  \bibinfo{pages}{082302} (\bibinfo{year}{2011}), \bibinfo{note}{1+39 pages},
  \eprint{1011.0705}.

\bibitem[{\citenamefont{{Ribeiro} and {Mosseri}}(2011)}]{2011PhRvL.106r0502R}
\bibinfo{author}{\bibfnamefont{P.}~\bibnamefont{{Ribeiro}}} \bibnamefont{and}
  \bibinfo{author}{\bibfnamefont{R.}~\bibnamefont{{Mosseri}}},
  \bibinfo{journal}{Physical Review Letters} \textbf{\bibinfo{volume}{106}},
  \bibinfo{eid}{180502} (\bibinfo{year}{2011}), \eprint{1101.2828}.

\bibitem[{\citenamefont{Borsten et~al.}(2008)\citenamefont{Borsten, Dahanayake,
  Duff, Rubens, and Ebrahim}}]{Borsten:2008ur}
\bibinfo{author}{\bibfnamefont{L.}~\bibnamefont{Borsten}},
  \bibinfo{author}{\bibfnamefont{D.}~\bibnamefont{Dahanayake}},
  \bibinfo{author}{\bibfnamefont{M.~J.} \bibnamefont{Duff}},
  \bibinfo{author}{\bibfnamefont{W.}~\bibnamefont{Rubens}}, \bibnamefont{and}
  \bibinfo{author}{\bibfnamefont{H.}~\bibnamefont{Ebrahim}},
  \bibinfo{journal}{Phys. Rev. Lett.} \textbf{\bibinfo{volume}{100}},
  \bibinfo{pages}{251602} (\bibinfo{year}{2008}), \eprint{0802.0840}.

\bibitem[{\citenamefont{Levay}(2011{\natexlab{b}})}]{Levay:2011ph}
\bibinfo{author}{\bibfnamefont{P.}~\bibnamefont{Levay}}
  (\bibinfo{year}{2011}{\natexlab{b}}), \eprint{1109.0361}.

\bibitem[{\citenamefont{Gukov et~al.}(2000)\citenamefont{Gukov, Vafa, and
  Witten}}]{Gukov:1999ya}
\bibinfo{author}{\bibfnamefont{S.}~\bibnamefont{Gukov}},
  \bibinfo{author}{\bibfnamefont{C.}~\bibnamefont{Vafa}}, \bibnamefont{and}
  \bibinfo{author}{\bibfnamefont{E.}~\bibnamefont{Witten}},
  \bibinfo{journal}{Nucl.Phys.} \textbf{\bibinfo{volume}{B584}},
  \bibinfo{pages}{69} (\bibinfo{year}{2000}), \eprint{hep-th/9906070}.

\bibitem[{\citenamefont{Ronan}(1987)}]{Ronan:1987:EHD:38247.38256}
\bibinfo{author}{\bibfnamefont{M.~A.} \bibnamefont{Ronan}},
  \bibinfo{journal}{Eur. J. Comb.} \textbf{\bibinfo{volume}{8}},
  \bibinfo{pages}{179} (\bibinfo{year}{1987}), ISSN \bibinfo{issn}{0195-6698},
  \urlprefix\url{http://dl.acm.org/citation.cfm?id=38247.38256}.

\bibitem[{\citenamefont{{Gottesman}}(1996)}]{1996PhRvA..54.1862G}
\bibinfo{author}{\bibfnamefont{D.}~\bibnamefont{{Gottesman}}},
  \bibinfo{journal}{\pra} \textbf{\bibinfo{volume}{54}}, \bibinfo{pages}{1862}
  (\bibinfo{year}{1996}), \eprint{arXiv:quant-ph/9604038}.

\bibitem[{\citenamefont{{L{\'e}vay}}(2010)}]{2010SPPhy.134...85L}
\bibinfo{author}{\bibfnamefont{P.}~\bibnamefont{{L{\'e}vay}}}, in
  \emph{\bibinfo{booktitle}{Springer Proceedings in Physics}}, edited by
  \bibinfo{editor}{\bibnamefont{{S.~Bellucci}}} (\bibinfo{year}{2010}), vol.
  \bibinfo{volume}{134} of \emph{\bibinfo{series}{Springer Proceedings in
  Physics}}, p.~\bibinfo{pages}{85}.

\bibitem[{\citenamefont{{Mermin}}(1990)}]{Mermin}
\bibinfo{author}{\bibfnamefont{N.~D.} \bibnamefont{{Mermin}}},
  \bibinfo{journal}{Physical Review Letters} \textbf{\bibinfo{volume}{65}},
  \bibinfo{pages}{1838} (\bibinfo{year}{1990}).

\bibitem[{\citenamefont{Vrana and Levay}(2009)}]{Vrana:2009ph}
\bibinfo{author}{\bibfnamefont{P.}~\bibnamefont{Vrana}} \bibnamefont{and}
  \bibinfo{author}{\bibfnamefont{P.}~\bibnamefont{Levay}}
  (\bibinfo{year}{2009}), \eprint{0906.3655}.

\bibitem[{\citenamefont{{Cerchiai} and {van
  Geemen}}(2010)}]{2010JMP....51l2203C}
\bibinfo{author}{\bibfnamefont{B.~L.} \bibnamefont{{Cerchiai}}}
  \bibnamefont{and} \bibinfo{author}{\bibfnamefont{B.}~\bibnamefont{{van
  Geemen}}}, \bibinfo{journal}{Journal of Mathematical Physics}
  \textbf{\bibinfo{volume}{51}}, \bibinfo{pages}{122203}
  (\bibinfo{year}{2010}), \eprint{1003.4255}.

\bibitem[{\citenamefont{Borsten
  et~al.}(2009{\natexlab{c}})\citenamefont{Borsten, Dahanayake, Duff, Ebrahim,
  and Rubens}}]{Borsten:2008wd}
\bibinfo{author}{\bibfnamefont{L.}~\bibnamefont{Borsten}},
  \bibinfo{author}{\bibfnamefont{D.}~\bibnamefont{Dahanayake}},
  \bibinfo{author}{\bibfnamefont{M.~J.} \bibnamefont{Duff}},
  \bibinfo{author}{\bibfnamefont{H.}~\bibnamefont{Ebrahim}}, \bibnamefont{and}
  \bibinfo{author}{\bibfnamefont{W.}~\bibnamefont{Rubens}},
  \bibinfo{journal}{Phys. Rep.} \textbf{\bibinfo{volume}{471}},
  \bibinfo{pages}{113} (\bibinfo{year}{2009}{\natexlab{c}}),
  \eprint{0809.4685}.

\bibitem[{\citenamefont{Borsten
  et~al.}(2010{\natexlab{b}})\citenamefont{Borsten, Dahanayake, Duff, and
  Rubens}}]{Borsten:2009ae}
\bibinfo{author}{\bibfnamefont{L.}~\bibnamefont{Borsten}},
  \bibinfo{author}{\bibfnamefont{D.}~\bibnamefont{Dahanayake}},
  \bibinfo{author}{\bibfnamefont{M.~J.} \bibnamefont{Duff}}, \bibnamefont{and}
  \bibinfo{author}{\bibfnamefont{W.}~\bibnamefont{Rubens}},
  \bibinfo{journal}{Phys. Rev.} \textbf{\bibinfo{volume}{D81}},
  \bibinfo{pages}{105023} (\bibinfo{year}{2010}{\natexlab{b}}),
  \eprint{0908.0706}.

\bibitem[{\citenamefont{Rios}(2011)}]{Rios:2011fa}
\bibinfo{author}{\bibfnamefont{M.}~\bibnamefont{Rios}} (\bibinfo{year}{2011}),
  \eprint{1102.1193}.

\bibitem[{\citenamefont{Verstraete et~al.}(2003)\citenamefont{Verstraete,
  Dehaene, and {De Moor}}}]{Verstraete:2003}
\bibinfo{author}{\bibfnamefont{F.}~\bibnamefont{Verstraete}},
  \bibinfo{author}{\bibfnamefont{J.}~\bibnamefont{Dehaene}}, \bibnamefont{and}
  \bibinfo{author}{\bibfnamefont{B.}~\bibnamefont{{De Moor}}},
  \bibinfo{journal}{Phys. Rev.} \textbf{\bibinfo{volume}{A68}},
  \bibinfo{pages}{012103} (\bibinfo{year}{2003}), \eprint{quant-ph/0105090}.

\bibitem[{\citenamefont{Cayley}()}]{Cayley:1845}
\bibinfo{author}{\bibfnamefont{A.}~\bibnamefont{Cayley}},
  \bibinfo{note}{\href{http://www.archive.org/download/collmathpapers01caylrich/collmathpapers01caylrich.pdf}{\textit{Camb.
  Math. J.} \textbf{4} (1845) 193--209}}.

\bibitem[{\citenamefont{Gelfand et~al.}(1994)\citenamefont{Gelfand, Kapranov,
  and Zelevinsky}}]{Gelfand:1994}
\bibinfo{author}{\bibfnamefont{I.~M.} \bibnamefont{Gelfand}},
  \bibinfo{author}{\bibfnamefont{M.~M.} \bibnamefont{Kapranov}},
  \bibnamefont{and} \bibinfo{author}{\bibfnamefont{A.~V.}
  \bibnamefont{Zelevinsky}}, \emph{\bibinfo{title}{{Discriminants, Resultants
  and Multidimensional Determinants}}} (\bibinfo{publisher}{Birkh\"{a}user},
  \bibinfo{address}{Boston}, \bibinfo{year}{1994}), ISBN
  \bibinfo{isbn}{0-8176-3660-9}.

\bibitem[{\citenamefont{Coffman et~al.}(2000)\citenamefont{Coffman, Kundu, and
  Wootters}}]{Coffman:1999jd}
\bibinfo{author}{\bibfnamefont{V.}~\bibnamefont{Coffman}},
  \bibinfo{author}{\bibfnamefont{J.}~\bibnamefont{Kundu}}, \bibnamefont{and}
  \bibinfo{author}{\bibfnamefont{W.~K.} \bibnamefont{Wootters}},
  \bibinfo{journal}{Phys. Rev.} \textbf{\bibinfo{volume}{A61}},
  \bibinfo{pages}{052306} (\bibinfo{year}{2000}), \eprint{quant-ph/9907047}.

\bibitem[{\citenamefont{L\'evay}(2005)}]{Levay:2004}
\bibinfo{author}{\bibfnamefont{P.}~\bibnamefont{L\'evay}},
  \bibinfo{journal}{Phys. Rev.} \textbf{\bibinfo{volume}{A71}},
  \bibinfo{pages}{012334} (\bibinfo{year}{2005}), \eprint{quant-ph/0403060}.

\bibitem[{\citenamefont{Brody et~al.}(2007)\citenamefont{Brody, Gustavsson, and
  Hughston}}]{Brody:2007}
\bibinfo{author}{\bibfnamefont{D.~C.} \bibnamefont{Brody}},
  \bibinfo{author}{\bibfnamefont{A.~C.~T.} \bibnamefont{Gustavsson}},
  \bibnamefont{and} \bibinfo{author}{\bibfnamefont{L.~P.}
  \bibnamefont{Hughston}}, \bibinfo{journal}{J. Phys.: Conf. Ser.}
  \textbf{\bibinfo{volume}{67}}, \bibinfo{pages}{012044}
  (\bibinfo{year}{2007}), \eprint{quant-ph/0612117}.

\bibitem[{\citenamefont{Sen and Vafa}(1995)}]{Sen:1995ff}
\bibinfo{author}{\bibfnamefont{A.}~\bibnamefont{Sen}} \bibnamefont{and}
  \bibinfo{author}{\bibfnamefont{C.}~\bibnamefont{Vafa}},
  \bibinfo{journal}{Nucl. Phys.} \textbf{\bibinfo{volume}{B455}},
  \bibinfo{pages}{165} (\bibinfo{year}{1995}), \eprint{hep-th/9508064}.

\bibitem[{\citenamefont{Gregori et~al.}(1999)\citenamefont{Gregori, Kounnas,
  and Petropoulos}}]{Gregori:1999ns}
\bibinfo{author}{\bibfnamefont{A.}~\bibnamefont{Gregori}},
  \bibinfo{author}{\bibfnamefont{C.}~\bibnamefont{Kounnas}}, \bibnamefont{and}
  \bibinfo{author}{\bibfnamefont{P.~M.} \bibnamefont{Petropoulos}},
  \bibinfo{journal}{Nucl. Phys.} \textbf{\bibinfo{volume}{B553}},
  \bibinfo{pages}{108} (\bibinfo{year}{1999}), \eprint{hep-th/9901117}.

\bibitem[{\citenamefont{Behrndt et~al.}(1996)\citenamefont{Behrndt, Kallosh,
  Rahmfeld, Shmakova, and Wong}}]{Behrndt:1996hu}
\bibinfo{author}{\bibfnamefont{K.}~\bibnamefont{Behrndt}},
  \bibinfo{author}{\bibfnamefont{R.}~\bibnamefont{Kallosh}},
  \bibinfo{author}{\bibfnamefont{J.}~\bibnamefont{Rahmfeld}},
  \bibinfo{author}{\bibfnamefont{M.}~\bibnamefont{Shmakova}}, \bibnamefont{and}
  \bibinfo{author}{\bibfnamefont{W.~K.} \bibnamefont{Wong}},
  \bibinfo{journal}{Phys. Rev.} \textbf{\bibinfo{volume}{D54}},
  \bibinfo{pages}{6293} (\bibinfo{year}{1996}), \eprint{hep-th/9608059}.

\bibitem[{\citenamefont{Caves et~al.}(2001)\citenamefont{Caves, Fuchs, and
  Rungta}}]{Caves:2000}
\bibinfo{author}{\bibfnamefont{C.~M.} \bibnamefont{Caves}},
  \bibinfo{author}{\bibfnamefont{C.~A.} \bibnamefont{Fuchs}}, \bibnamefont{and}
  \bibinfo{author}{\bibfnamefont{C.~A.} \bibnamefont{Rungta}},
  \bibinfo{journal}{Found. Phys. Lett.} \textbf{\bibinfo{volume}{14}},
  \bibinfo{pages}{199} (\bibinfo{year}{2001}), \eprint{quant-ph/0009063}.

\bibitem[{\citenamefont{Tripathy and Trivedi}(2006)}]{Tripathy:2005qp}
\bibinfo{author}{\bibfnamefont{P.~K.} \bibnamefont{Tripathy}} \bibnamefont{and}
  \bibinfo{author}{\bibfnamefont{S.~P.} \bibnamefont{Trivedi}},
  \bibinfo{journal}{JHEP} \textbf{\bibinfo{volume}{03}}, \bibinfo{pages}{022}
  (\bibinfo{year}{2006}), \eprint{hep-th/0511117}.

\bibitem[{\citenamefont{Kallosh
  et~al.}(2006{\natexlab{a}})\citenamefont{Kallosh, Sivanandam, and
  Soroush}}]{Kallosh:2006bt}
\bibinfo{author}{\bibfnamefont{R.}~\bibnamefont{Kallosh}},
  \bibinfo{author}{\bibfnamefont{N.}~\bibnamefont{Sivanandam}},
  \bibnamefont{and} \bibinfo{author}{\bibfnamefont{M.}~\bibnamefont{Soroush}},
  \bibinfo{journal}{JHEP} \textbf{\bibinfo{volume}{03}}, \bibinfo{pages}{060}
  (\bibinfo{year}{2006}{\natexlab{a}}), \eprint{hep-th/0602005}.

\bibitem[{\citenamefont{Sen}(1995)}]{Sen:1995in}
\bibinfo{author}{\bibfnamefont{A.}~\bibnamefont{Sen}}, \bibinfo{journal}{Mod.
  Phys. Lett.} \textbf{\bibinfo{volume}{A10}}, \bibinfo{pages}{2081}
  (\bibinfo{year}{1995}), \eprint{hep-th/9504147}.

\bibitem[{\citenamefont{Acin et~al.}(2001)\citenamefont{Acin, Andrianov, Jane,
  and Tarrach}}]{Acin:2001}
\bibinfo{author}{\bibfnamefont{A.}~\bibnamefont{Acin}},
  \bibinfo{author}{\bibfnamefont{A.}~\bibnamefont{Andrianov}},
  \bibinfo{author}{\bibfnamefont{E.}~\bibnamefont{Jane}}, \bibnamefont{and}
  \bibinfo{author}{\bibfnamefont{R.}~\bibnamefont{Tarrach}},
  \bibinfo{journal}{J. Phys.} \textbf{\bibinfo{volume}{A34}},
  \bibinfo{pages}{6725} (\bibinfo{year}{2001}), \eprint{quant-ph/0009107}.

\bibitem[{\citenamefont{Cartan}(1984)}]{Cartan}
\bibinfo{author}{\bibfnamefont{E.}~\bibnamefont{Cartan}},
  \emph{\bibinfo{title}{{{\OE}euvres compl\`{e}tes}}},
  \bibinfo{howpublished}{Editions du Centre National de la Recherche
  Scientifique} (\bibinfo{year}{1984}).

\bibitem[{\citenamefont{Kallosh and Kol}(1996)}]{Kallosh:1996uy}
\bibinfo{author}{\bibfnamefont{R.}~\bibnamefont{Kallosh}} \bibnamefont{and}
  \bibinfo{author}{\bibfnamefont{B.}~\bibnamefont{Kol}},
  \bibinfo{journal}{Phys. Rev.} \textbf{\bibinfo{volume}{D53}},
  \bibinfo{pages}{5344} (\bibinfo{year}{1996}), \eprint{hep-th/9602014}.

\bibitem[{\citenamefont{{Becker} et~al.}(2007)\citenamefont{{Becker}, {Becker},
  and {Schwarz}}}]{2007stmt.book.....B}
\bibinfo{author}{\bibfnamefont{K.}~\bibnamefont{{Becker}}},
  \bibinfo{author}{\bibfnamefont{M.}~\bibnamefont{{Becker}}}, \bibnamefont{and}
  \bibinfo{author}{\bibfnamefont{J.~H.} \bibnamefont{{Schwarz}}},
  \emph{\bibinfo{title}{{String Theory and M-Theory}}} (\bibinfo{year}{2007}).

\bibitem[{\citenamefont{Ferrara and Kallosh}(2006)}]{Ferrara:2006em}
\bibinfo{author}{\bibfnamefont{S.}~\bibnamefont{Ferrara}} \bibnamefont{and}
  \bibinfo{author}{\bibfnamefont{R.}~\bibnamefont{Kallosh}},
  \bibinfo{journal}{Phys. Rev.} \textbf{\bibinfo{volume}{D73}},
  \bibinfo{pages}{125005} (\bibinfo{year}{2006}), \eprint{hep-th/0603247}.

\bibitem[{\citenamefont{Lu et~al.}(1996)\citenamefont{Lu, Pope, and
  Stelle}}]{Lu:1996ge}
\bibinfo{author}{\bibfnamefont{H.}~\bibnamefont{Lu}},
  \bibinfo{author}{\bibfnamefont{C.~N.} \bibnamefont{Pope}}, \bibnamefont{and}
  \bibinfo{author}{\bibfnamefont{K.~S.} \bibnamefont{Stelle}},
  \bibinfo{journal}{Nucl. Phys.} \textbf{\bibinfo{volume}{B476}},
  \bibinfo{pages}{89} (\bibinfo{year}{1996}), \eprint{hep-th/9602140}.

\bibitem[{\citenamefont{{Planat} and {Kibler}}(2008)}]{2008arXiv0807.3650P}
\bibinfo{author}{\bibfnamefont{M.}~\bibnamefont{{Planat}}} \bibnamefont{and}
  \bibinfo{author}{\bibfnamefont{M.~R.} \bibnamefont{{Kibler}}},
  \bibinfo{journal}{ArXiv e-prints}  (\bibinfo{year}{2008}),
  \eprint{0807.3650}.

\bibitem[{\citenamefont{Manivel}(2006)}]{Manivel:2005}
\bibinfo{author}{\bibfnamefont{L.}~\bibnamefont{Manivel}}, \bibinfo{journal}{J.
  Algebra} \textbf{\bibinfo{volume}{304}}, \bibinfo{pages}{457}
  (\bibinfo{year}{2006}), \eprint{math/0507118}.

\bibitem[{\citenamefont{{Alberto Elduque}}(2007)}]{Elduque:2005}
\bibinfo{author}{\bibnamefont{{Alberto Elduque}}}, \bibinfo{journal}{Rev. Mat.
  Iberoamericana} \textbf{\bibinfo{volume}{23}}, \bibinfo{pages}{57}
  (\bibinfo{year}{2007}), \eprint{math/0507282}.

\bibitem[{\citenamefont{Cvetic and Tseytlin}(1996)}]{Cvetic:1995bj}
\bibinfo{author}{\bibfnamefont{M.}~\bibnamefont{Cvetic}} \bibnamefont{and}
  \bibinfo{author}{\bibfnamefont{A.~A.} \bibnamefont{Tseytlin}},
  \bibinfo{journal}{Phys. Rev.} \textbf{\bibinfo{volume}{D53}},
  \bibinfo{pages}{5619} (\bibinfo{year}{1996}), \eprint{hep-th/9512031}.

\bibitem[{\citenamefont{Ferrara et~al.}(2006)\citenamefont{Ferrara, Gimon, and
  Kallosh}}]{Ferrara:2006yb}
\bibinfo{author}{\bibfnamefont{S.}~\bibnamefont{Ferrara}},
  \bibinfo{author}{\bibfnamefont{E.~G.} \bibnamefont{Gimon}}, \bibnamefont{and}
  \bibinfo{author}{\bibfnamefont{R.}~\bibnamefont{Kallosh}},
  \bibinfo{journal}{Phys. Rev.} \textbf{\bibinfo{volume}{D74}},
  \bibinfo{pages}{125018} (\bibinfo{year}{2006}), \eprint{hep-th/0606211}.

\bibitem[{\citenamefont{Maldacena et~al.}(1999)\citenamefont{Maldacena, Moore,
  and Strominger}}]{Maldacena:1999bp}
\bibinfo{author}{\bibfnamefont{J.~M.} \bibnamefont{Maldacena}},
  \bibinfo{author}{\bibfnamefont{G.~W.} \bibnamefont{Moore}}, \bibnamefont{and}
  \bibinfo{author}{\bibfnamefont{A.}~\bibnamefont{Strominger}}
  (\bibinfo{year}{1999}), \eprint{hep-th/9903163}.

\bibitem[{\citenamefont{Ferrara and
  Kallosh}(1996{\natexlab{b}})}]{Ferrara:1996um}
\bibinfo{author}{\bibfnamefont{S.}~\bibnamefont{Ferrara}} \bibnamefont{and}
  \bibinfo{author}{\bibfnamefont{R.}~\bibnamefont{Kallosh}},
  \bibinfo{journal}{Phys. Rev.} \textbf{\bibinfo{volume}{D54}},
  \bibinfo{pages}{1525} (\bibinfo{year}{1996}{\natexlab{b}}),
  \eprint{hep-th/9603090}.

\bibitem[{\citenamefont{Andrianopoli et~al.}(1997)\citenamefont{Andrianopoli,
  D'Auria, and Ferrara}}]{Andrianopoli:1997hb}
\bibinfo{author}{\bibfnamefont{L.}~\bibnamefont{Andrianopoli}},
  \bibinfo{author}{\bibfnamefont{R.}~\bibnamefont{D'Auria}}, \bibnamefont{and}
  \bibinfo{author}{\bibfnamefont{S.}~\bibnamefont{Ferrara}},
  \bibinfo{journal}{Phys. Lett.} \textbf{\bibinfo{volume}{B411}},
  \bibinfo{pages}{39} (\bibinfo{year}{1997}), \eprint{hep-th/9705024}.

\bibitem[{\citenamefont{Ferrara and G{\"u}naydin}(1998)}]{Ferrara:1997uz}
\bibinfo{author}{\bibfnamefont{S.}~\bibnamefont{Ferrara}} \bibnamefont{and}
  \bibinfo{author}{\bibfnamefont{M.}~\bibnamefont{G{\"u}naydin}},
  \bibinfo{journal}{Int. J. Mod. Phys.} \textbf{\bibinfo{volume}{A13}},
  \bibinfo{pages}{2075} (\bibinfo{year}{1998}), \eprint{hep-th/9708025}.

\bibitem[{\citenamefont{Payne and Thas}(2009)}]{678649}
\bibinfo{author}{\bibfnamefont{S.~E.} \bibnamefont{Payne}} \bibnamefont{and}
  \bibinfo{author}{\bibfnamefont{J.}~\bibnamefont{Thas}},
  \emph{\bibinfo{title}{Finite Generalized Quadrangles}}, EMS Series of
  Lectures in Mathematics (\bibinfo{publisher}{European Mathematical Society},
  \bibinfo{year}{2009}), ISBN \bibinfo{isbn}{978-3-03719-066-1}.

\bibitem[{\citenamefont{Polster}(1991)}]{Polster}
\bibinfo{author}{\bibfnamefont{B.}~\bibnamefont{Polster}},
  \emph{\bibinfo{title}{A Geometrical Picture Book}}
  (\bibinfo{publisher}{Springer}, \bibinfo{address}{New York},
  \bibinfo{year}{1991}).

\bibitem[{\citenamefont{Cartan}(1986)}]{Cartan2}
\bibinfo{author}{\bibfnamefont{E.}~\bibnamefont{Cartan}},
  \bibinfo{journal}{Amer. J. Math.} \textbf{\bibinfo{volume}{18}}
  (\bibinfo{year}{1986}).

\bibitem[{\citenamefont{Saniga et~al.}(2009)\citenamefont{Saniga, Levay,
  Pracna, and Vrana}}]{Saniga:2009ik}
\bibinfo{author}{\bibfnamefont{M.}~\bibnamefont{Saniga}},
  \bibinfo{author}{\bibfnamefont{P.}~\bibnamefont{Levay}},
  \bibinfo{author}{\bibfnamefont{P.}~\bibnamefont{Pracna}}, \bibnamefont{and}
  \bibinfo{author}{\bibfnamefont{P.}~\bibnamefont{Vrana}}
  (\bibinfo{year}{2009}), \eprint{0903.0715}.

\bibitem[{\citenamefont{Baez}(2001)}]{Baez:2001dm}
\bibinfo{author}{\bibfnamefont{J.~C.} \bibnamefont{Baez}},
  \bibinfo{journal}{Bull. Amer. Math. Soc.} \textbf{\bibinfo{volume}{39}},
  \bibinfo{pages}{145} (\bibinfo{year}{2001}), \eprint{math/0105155}.

\bibitem[{\citenamefont{Cvetic and Hull}(1996)}]{Cvetic:1996zq}
\bibinfo{author}{\bibfnamefont{M.}~\bibnamefont{Cvetic}} \bibnamefont{and}
  \bibinfo{author}{\bibfnamefont{C.~M.} \bibnamefont{Hull}},
  \bibinfo{journal}{Nucl. Phys.} \textbf{\bibinfo{volume}{B480}},
  \bibinfo{pages}{296} (\bibinfo{year}{1996}), \eprint{hep-th/9606193}.

\bibitem[{\citenamefont{Balasubramanian
  et~al.}(1998)\citenamefont{Balasubramanian, Larsen, and
  Leigh}}]{Balasubramanian:1997az}
\bibinfo{author}{\bibfnamefont{V.}~\bibnamefont{Balasubramanian}},
  \bibinfo{author}{\bibfnamefont{F.}~\bibnamefont{Larsen}}, \bibnamefont{and}
  \bibinfo{author}{\bibfnamefont{R.~G.} \bibnamefont{Leigh}},
  \bibinfo{journal}{Phys. Rev.} \textbf{\bibinfo{volume}{D57}},
  \bibinfo{pages}{3509} (\bibinfo{year}{1998}), \eprint{hep-th/9704143}.

\bibitem[{\citenamefont{Bertolini and Trigiante}(2000)}]{Bertolini:2000ei}
\bibinfo{author}{\bibfnamefont{M.}~\bibnamefont{Bertolini}} \bibnamefont{and}
  \bibinfo{author}{\bibfnamefont{M.}~\bibnamefont{Trigiante}},
  \bibinfo{journal}{Nucl. Phys.} \textbf{\bibinfo{volume}{B582}},
  \bibinfo{pages}{393} (\bibinfo{year}{2000}), \eprint{hep-th/0002191}.

\bibitem[{\citenamefont{Andrianopoli
  et~al.}(1998{\natexlab{a}})\citenamefont{Andrianopoli, D'Auria, Ferrara, Fre,
  and Trigiante}}]{Andrianopoli:1997wi}
\bibinfo{author}{\bibfnamefont{L.}~\bibnamefont{Andrianopoli}},
  \bibinfo{author}{\bibfnamefont{R.}~\bibnamefont{D'Auria}},
  \bibinfo{author}{\bibfnamefont{S.}~\bibnamefont{Ferrara}},
  \bibinfo{author}{\bibfnamefont{P.}~\bibnamefont{Fre}}, \bibnamefont{and}
  \bibinfo{author}{\bibfnamefont{M.}~\bibnamefont{Trigiante}},
  \bibinfo{journal}{Nucl.Phys.} \textbf{\bibinfo{volume}{B509}},
  \bibinfo{pages}{463} (\bibinfo{year}{1998}{\natexlab{a}}), \bibinfo{note}{60
  pages, 1 LaTeX file, 3 eps figures included}, \eprint{hep-th/9707087}.

\bibitem[{\citenamefont{{Mermin}}(1993)}]{Mermin2}
\bibinfo{author}{\bibfnamefont{N.~D.} \bibnamefont{{Mermin}}},
  \bibinfo{journal}{Reviews of Modern Physics} \textbf{\bibinfo{volume}{65}},
  \bibinfo{pages}{803} (\bibinfo{year}{1993}).

\bibitem[{\citenamefont{{Peres}}(1991)}]{1991JPhA...24L.175P}
\bibinfo{author}{\bibfnamefont{A.}~\bibnamefont{{Peres}}},
  \bibinfo{journal}{Journal of Physics A Mathematical General}
  \textbf{\bibinfo{volume}{24}}, \bibinfo{pages}{L175} (\bibinfo{year}{1991}).

\bibitem[{\citenamefont{{Saniga} and {L{\'e}vay}}(2012)}]{2012EL.....9750006S}
\bibinfo{author}{\bibfnamefont{M.}~\bibnamefont{{Saniga}}} \bibnamefont{and}
  \bibinfo{author}{\bibfnamefont{P.}~\bibnamefont{{L{\'e}vay}}},
  \bibinfo{journal}{EPL (Europhysics Letters)} \textbf{\bibinfo{volume}{97}},
  \bibinfo{pages}{50006} (\bibinfo{year}{2012}), \eprint{1111.5923}.

\bibitem[{\citenamefont{Breitenlohner et~al.}(1988)\citenamefont{Breitenlohner,
  Maison, and Gibbons}}]{Breitenlohner:1987dg}
\bibinfo{author}{\bibfnamefont{P.}~\bibnamefont{Breitenlohner}},
  \bibinfo{author}{\bibfnamefont{D.}~\bibnamefont{Maison}}, \bibnamefont{and}
  \bibinfo{author}{\bibfnamefont{G.~W.} \bibnamefont{Gibbons}},
  \bibinfo{journal}{Commun. Math. Phys.} \textbf{\bibinfo{volume}{120}},
  \bibinfo{pages}{295} (\bibinfo{year}{1988}).

\bibitem[{\citenamefont{Gaiotto et~al.}(2007)\citenamefont{Gaiotto, Li, and
  Padi}}]{Gaiotto:2007ag}
\bibinfo{author}{\bibfnamefont{D.}~\bibnamefont{Gaiotto}},
  \bibinfo{author}{\bibfnamefont{W.}~\bibnamefont{Li}}, \bibnamefont{and}
  \bibinfo{author}{\bibfnamefont{M.}~\bibnamefont{Padi}},
  \bibinfo{journal}{JHEP} \textbf{\bibinfo{volume}{0712}}, \bibinfo{pages}{093}
  (\bibinfo{year}{2007}), \eprint{0710.1638}.

\bibitem[{\citenamefont{Bossard et~al.}(2010)\citenamefont{Bossard, Michel, and
  Pioline}}]{Bossard:2009we}
\bibinfo{author}{\bibfnamefont{G.}~\bibnamefont{Bossard}},
  \bibinfo{author}{\bibfnamefont{Y.}~\bibnamefont{Michel}}, \bibnamefont{and}
  \bibinfo{author}{\bibfnamefont{B.}~\bibnamefont{Pioline}},
  \bibinfo{journal}{JHEP} \textbf{\bibinfo{volume}{01}}, \bibinfo{pages}{038}
  (\bibinfo{year}{2010}), \eprint{0908.1742}.

\bibitem[{\citenamefont{{Newman} et~al.}(1963)\citenamefont{{Newman},
  {Tamburino}, and {Unti}}}]{1963JMP.....4..915N}
\bibinfo{author}{\bibfnamefont{E.}~\bibnamefont{{Newman}}},
  \bibinfo{author}{\bibfnamefont{L.}~\bibnamefont{{Tamburino}}},
  \bibnamefont{and} \bibinfo{author}{\bibfnamefont{T.}~\bibnamefont{{Unti}}},
  \bibinfo{journal}{Journal of Mathematical Physics}
  \textbf{\bibinfo{volume}{4}}, \bibinfo{pages}{915} (\bibinfo{year}{1963}).

\bibitem[{\citenamefont{Craps et~al.}(1997)\citenamefont{Craps, Roose, Troost,
  and Proeyen}}]{Craps1997565}
\bibinfo{author}{\bibfnamefont{B.}~\bibnamefont{Craps}},
  \bibinfo{author}{\bibfnamefont{F.}~\bibnamefont{Roose}},
  \bibinfo{author}{\bibfnamefont{W.}~\bibnamefont{Troost}}, \bibnamefont{and}
  \bibinfo{author}{\bibfnamefont{A.~V.} \bibnamefont{Proeyen}},
  \bibinfo{journal}{Nuclear Physics B} \textbf{\bibinfo{volume}{503}},
  \bibinfo{pages}{565 } (\bibinfo{year}{1997}), ISSN \bibinfo{issn}{0550-3213},
  \urlprefix\url{http://www.sciencedirect.com/science/article/pii/S0550321397004082}.

\bibitem[{\citenamefont{{Ehlers}}(1955)}]{1955ZPhy..143..239E}
\bibinfo{author}{\bibfnamefont{J.}~\bibnamefont{{Ehlers}}},
  \bibinfo{journal}{Zeitschrift fur Physik} \textbf{\bibinfo{volume}{143}},
  \bibinfo{pages}{239} (\bibinfo{year}{1955}).

\bibitem[{\citenamefont{Bergshoeff et~al.}(2009)\citenamefont{Bergshoeff,
  Chemissany, Ploegh, Trigiante, and Van~Riet}}]{Bergshoeff:2008be}
\bibinfo{author}{\bibfnamefont{E.}~\bibnamefont{Bergshoeff}},
  \bibinfo{author}{\bibfnamefont{W.}~\bibnamefont{Chemissany}},
  \bibinfo{author}{\bibfnamefont{A.}~\bibnamefont{Ploegh}},
  \bibinfo{author}{\bibfnamefont{M.}~\bibnamefont{Trigiante}},
  \bibnamefont{and} \bibinfo{author}{\bibfnamefont{T.}~\bibnamefont{Van~Riet}},
  \bibinfo{journal}{Nucl. Phys.} \textbf{\bibinfo{volume}{B812}},
  \bibinfo{pages}{343} (\bibinfo{year}{2009}), \eprint{0806.2310}.

\bibitem[{\citenamefont{Bossard et~al.}(2009)\citenamefont{Bossard, Nicolai,
  and Stelle}}]{Bossard:2009at}
\bibinfo{author}{\bibfnamefont{G.}~\bibnamefont{Bossard}},
  \bibinfo{author}{\bibfnamefont{H.}~\bibnamefont{Nicolai}}, \bibnamefont{and}
  \bibinfo{author}{\bibfnamefont{K.}~\bibnamefont{Stelle}},
  \bibinfo{journal}{JHEP} \textbf{\bibinfo{volume}{0907}}, \bibinfo{pages}{003}
  (\bibinfo{year}{2009}), \eprint{0902.4438}.

\bibitem[{\citenamefont{Gibbons et~al.}(1996)\citenamefont{Gibbons, Kallosh,
  and Kol}}]{Gibbons:1996af}
\bibinfo{author}{\bibfnamefont{G.~W.} \bibnamefont{Gibbons}},
  \bibinfo{author}{\bibfnamefont{R.}~\bibnamefont{Kallosh}}, \bibnamefont{and}
  \bibinfo{author}{\bibfnamefont{B.}~\bibnamefont{Kol}},
  \bibinfo{journal}{Phys.Rev.Lett.} \textbf{\bibinfo{volume}{77}},
  \bibinfo{pages}{4992} (\bibinfo{year}{1996}), \eprint{hep-th/9607108}.

\bibitem[{\citenamefont{Andrianopoli
  et~al.}(1998{\natexlab{b}})\citenamefont{Andrianopoli, D'Auria, and
  Ferrara}}]{Andrianopoli:1996ve}
\bibinfo{author}{\bibfnamefont{L.}~\bibnamefont{Andrianopoli}},
  \bibinfo{author}{\bibfnamefont{R.}~\bibnamefont{D'Auria}}, \bibnamefont{and}
  \bibinfo{author}{\bibfnamefont{S.}~\bibnamefont{Ferrara}},
  \bibinfo{journal}{Int. J. Mod. Phys.} \textbf{\bibinfo{volume}{A13}},
  \bibinfo{pages}{431} (\bibinfo{year}{1998}{\natexlab{b}}),
  \eprint{hep-th/9612105}.

\bibitem[{\citenamefont{Ferrara et~al.}(1997)\citenamefont{Ferrara, Gibbons,
  and Kallosh}}]{Ferrara:1997tw}
\bibinfo{author}{\bibfnamefont{S.}~\bibnamefont{Ferrara}},
  \bibinfo{author}{\bibfnamefont{G.~W.} \bibnamefont{Gibbons}},
  \bibnamefont{and} \bibinfo{author}{\bibfnamefont{R.}~\bibnamefont{Kallosh}},
  \bibinfo{journal}{Nucl. Phys.} \textbf{\bibinfo{volume}{B500}},
  \bibinfo{pages}{75} (\bibinfo{year}{1997}), \eprint{hep-th/9702103}.

\bibitem[{\citenamefont{Moore}(1998)}]{Moore:1998pn}
\bibinfo{author}{\bibfnamefont{G.~W.} \bibnamefont{Moore}}
  (\bibinfo{year}{1998}), \bibinfo{note}{107pp. harvmac b-mode, 4 figures:
  minor mistakes, typos corrected. references added:v3: typo fixed, reference
  added Report-no: YCTP-P17-98}, \eprint{hep-th/9807087}.

\bibitem[{\citenamefont{Bellucci
  et~al.}(2007{\natexlab{a}})\citenamefont{Bellucci, Marrani, Orazi, and
  Shcherbakov}}]{Bellucci:2007zi}
\bibinfo{author}{\bibfnamefont{S.}~\bibnamefont{Bellucci}},
  \bibinfo{author}{\bibfnamefont{A.}~\bibnamefont{Marrani}},
  \bibinfo{author}{\bibfnamefont{E.}~\bibnamefont{Orazi}}, \bibnamefont{and}
  \bibinfo{author}{\bibfnamefont{A.}~\bibnamefont{Shcherbakov}},
  \bibinfo{journal}{Phys. Lett.} \textbf{\bibinfo{volume}{B655}},
  \bibinfo{pages}{185} (\bibinfo{year}{2007}{\natexlab{a}}),
  \eprint{0707.2730}.

\bibitem[{\citenamefont{Gimon et~al.}(2008)\citenamefont{Gimon, Larsen, and
  Simon}}]{Gimon:2007mh}
\bibinfo{author}{\bibfnamefont{E.~G.} \bibnamefont{Gimon}},
  \bibinfo{author}{\bibfnamefont{F.}~\bibnamefont{Larsen}}, \bibnamefont{and}
  \bibinfo{author}{\bibfnamefont{J.}~\bibnamefont{Simon}},
  \bibinfo{journal}{JHEP} \textbf{\bibinfo{volume}{01}}, \bibinfo{pages}{040}
  (\bibinfo{year}{2008}), \eprint{0710.4967}.

\bibitem[{\citenamefont{Bossard and Nicolai}(2010)}]{Bossard:2009my}
\bibinfo{author}{\bibfnamefont{G.}~\bibnamefont{Bossard}} \bibnamefont{and}
  \bibinfo{author}{\bibfnamefont{H.}~\bibnamefont{Nicolai}},
  \bibinfo{journal}{Gen. Rel. Grav.} \textbf{\bibinfo{volume}{42}},
  \bibinfo{pages}{509} (\bibinfo{year}{2010}), \eprint{0906.1987}.

\bibitem[{\citenamefont{Bossard and Ruef}(2012)}]{Bossard:2011kz}
\bibinfo{author}{\bibfnamefont{G.}~\bibnamefont{Bossard}} \bibnamefont{and}
  \bibinfo{author}{\bibfnamefont{C.}~\bibnamefont{Ruef}},
  \bibinfo{journal}{Gen.Rel.Grav.} \textbf{\bibinfo{volume}{44}},
  \bibinfo{pages}{21} (\bibinfo{year}{2012}), \eprint{1106.5806}.

\bibitem[{\citenamefont{Sekiguchi}(1987)}]{Sekiguchi:1987}
\bibinfo{author}{\bibfnamefont{J.}~\bibnamefont{Sekiguchi}},
  \bibinfo{journal}{J. Math. Soc. Japan} \textbf{\bibinfo{volume}{39}},
  \bibinfo{pages}{127} (\bibinfo{year}{1987}), ISSN \bibinfo{issn}{0025-5645},
  \urlprefix\url{http://dx.doi.org/10.2969/jmsj/03910127}.

\bibitem[{\citenamefont{Collingwood and McGovern}(1993)}]{Collingwood:1993}
\bibinfo{author}{\bibfnamefont{D.~H.} \bibnamefont{Collingwood}}
  \bibnamefont{and} \bibinfo{author}{\bibfnamefont{W.~M.}
  \bibnamefont{McGovern}}, \emph{\bibinfo{title}{Nilpotent orbits in semisimple
  Lie algebras}}, Van Nostrand Reinhold mathematics series
  (\bibinfo{publisher}{CRC Press}, \bibinfo{year}{1993}), ISBN
  \bibinfo{isbn}{0-5341-8834-6}.

\bibitem[{\citenamefont{Verstraete et~al.}(2002)\citenamefont{Verstraete,
  Dehaene, {De Moor}, and Verschelde}}]{Verstraete:2002}
\bibinfo{author}{\bibfnamefont{F.}~\bibnamefont{Verstraete}},
  \bibinfo{author}{\bibfnamefont{J.}~\bibnamefont{Dehaene}},
  \bibinfo{author}{\bibfnamefont{B.}~\bibnamefont{{De Moor}}},
  \bibnamefont{and}
  \bibinfo{author}{\bibfnamefont{H.}~\bibnamefont{Verschelde}},
  \bibinfo{journal}{Phys. Rev.} \textbf{\bibinfo{volume}{A65}},
  \bibinfo{pages}{052112} (\bibinfo{year}{2002}), \eprint{quant-ph/0109033}.

\bibitem[{\citenamefont{Chterental and Djokovi{\'{c}}}(2007)}]{Chterental:2007}
\bibinfo{author}{\bibfnamefont{O.}~\bibnamefont{Chterental}} \bibnamefont{and}
  \bibinfo{author}{\bibfnamefont{D.~{\v{Z}}.} \bibnamefont{Djokovi{\'{c}}}}, in
  \emph{\bibinfo{booktitle}{{Linear Algebra Research Advances}}}, edited by
  \bibinfo{editor}{\bibfnamefont{G.~D.} \bibnamefont{Ling}}
  (\bibinfo{publisher}{Nova Science Publishers Inc}, \bibinfo{year}{2007}),
  chap.~\bibinfo{chapter}{4}, pp. \bibinfo{pages}{133--167},
  \eprint{quant-ph/0612184}.

\bibitem[{\citenamefont{Rubens}(2011)}]{Rubens:2011phd}
\bibinfo{author}{\bibfnamefont{W.}~\bibnamefont{Rubens}}, Ph.D. thesis,
  \bibinfo{school}{Imperial College, London} (\bibinfo{year}{2011}).

\bibitem[{\citenamefont{Borsten and Rubens}(2012)}]{Borsten:2012wr}
\bibinfo{author}{\bibfnamefont{L.}~\bibnamefont{Borsten}} \bibnamefont{and}
  \bibinfo{author}{\bibfnamefont{W.}~\bibnamefont{Rubens}}
  (\bibinfo{year}{2012}), \bibinfo{note}{to appear}.

\bibitem[{\citenamefont{Wallach}(2008)}]{Wallach:2008}
\bibinfo{author}{\bibfnamefont{N.}~\bibnamefont{Wallach}}, in
  \emph{\bibinfo{booktitle}{Representation Theory and Complex Analysis}}
  (\bibinfo{publisher}{Springer Berlin / Heidelberg}, \bibinfo{year}{2008}),
  vol. \bibinfo{volume}{1931} of \emph{\bibinfo{series}{Lecture Notes in
  Mathematics}}, pp. \bibinfo{pages}{345--376},
  \urlprefix\url{http://dx.doi.org/10.1007/978-3-540-76892-0_6}.

\bibitem[{\citenamefont{Lamata et~al.}(2007)\citenamefont{Lamata, Le\'{o}n,
  Salgado, and Solano}}]{Lamata:2006b}
\bibinfo{author}{\bibfnamefont{L.}~\bibnamefont{Lamata}},
  \bibinfo{author}{\bibfnamefont{J.}~\bibnamefont{Le\'{o}n}},
  \bibinfo{author}{\bibfnamefont{D.}~\bibnamefont{Salgado}}, \bibnamefont{and}
  \bibinfo{author}{\bibfnamefont{E.}~\bibnamefont{Solano}},
  \bibinfo{journal}{Phys. Rev.} \textbf{\bibinfo{volume}{A75}},
  \bibinfo{pages}{022318} (\bibinfo{year}{2007}), \eprint{quant-ph/0610233}.

\bibitem[{\citenamefont{Cao and Wang}(2007)}]{Cao:2007}
\bibinfo{author}{\bibfnamefont{Y.}~\bibnamefont{Cao}} \bibnamefont{and}
  \bibinfo{author}{\bibfnamefont{A.~M.} \bibnamefont{Wang}},
  \bibinfo{journal}{Eur. Phys. J.} \textbf{\bibinfo{volume}{D44}},
  \bibinfo{pages}{159} (\bibinfo{year}{2007}).

\bibitem[{\citenamefont{Li et~al.}(2007)\citenamefont{Li, Li, Huang, and
  Li}}]{Li:2007c}
\bibinfo{author}{\bibfnamefont{D.}~\bibnamefont{Li}},
  \bibinfo{author}{\bibfnamefont{X.}~\bibnamefont{Li}},
  \bibinfo{author}{\bibfnamefont{H.}~\bibnamefont{Huang}}, \bibnamefont{and}
  \bibinfo{author}{\bibfnamefont{X.}~\bibnamefont{Li}},
  \bibinfo{journal}{Quant. Info. Comp.} \textbf{\bibinfo{volume}{9, 10}},
  \bibinfo{pages}{0778} (\bibinfo{year}{2007}), \eprint{0712.1876}.

\bibitem[{\citenamefont{Akhtarshenas and Ghahi}(2010)}]{Akhtarshenas:2010}
\bibinfo{author}{\bibfnamefont{S.~J.} \bibnamefont{Akhtarshenas}}
  \bibnamefont{and} \bibinfo{author}{\bibfnamefont{M.~G.} \bibnamefont{Ghahi}}
  (\bibinfo{year}{2010}), \eprint{1003.2762}.

\bibitem[{\citenamefont{Buniy and Kephart}(2010)}]{Buniy:2010a}
\bibinfo{author}{\bibfnamefont{R.~V.} \bibnamefont{Buniy}} \bibnamefont{and}
  \bibinfo{author}{\bibfnamefont{T.~W.} \bibnamefont{Kephart}}
  (\bibinfo{year}{2010}), \eprint{1012.2630}.

\bibitem[{\citenamefont{Briand et~al.}(2003)\citenamefont{Briand, Luque, and
  Thibon}}]{Briand:2003a}
\bibinfo{author}{\bibfnamefont{E.}~\bibnamefont{Briand}},
  \bibinfo{author}{\bibfnamefont{J.-G.} \bibnamefont{Luque}}, \bibnamefont{and}
  \bibinfo{author}{\bibfnamefont{J.-Y.} \bibnamefont{Thibon}},
  \bibinfo{journal}{J. Phys.} \textbf{\bibinfo{volume}{A36}},
  \bibinfo{pages}{9915} (\bibinfo{year}{2003}), \eprint{quant-ph/0304026}.

\bibitem[{\citenamefont{Gunaydin et~al.}(2006)\citenamefont{Gunaydin, Neitzke,
  Pioline, and Waldron}}]{Gunaydin:2005mx}
\bibinfo{author}{\bibfnamefont{M.}~\bibnamefont{Gunaydin}},
  \bibinfo{author}{\bibfnamefont{A.}~\bibnamefont{Neitzke}},
  \bibinfo{author}{\bibfnamefont{B.}~\bibnamefont{Pioline}}, \bibnamefont{and}
  \bibinfo{author}{\bibfnamefont{A.}~\bibnamefont{Waldron}},
  \bibinfo{journal}{Phys. Rev.} \textbf{\bibinfo{volume}{D73}},
  \bibinfo{pages}{084019} (\bibinfo{year}{2006}), \eprint{hep-th/0512296}.

\bibitem[{\citenamefont{Bossard}(2012)}]{Bossard:2012ge}
\bibinfo{author}{\bibfnamefont{G.}~\bibnamefont{Bossard}}
  (\bibinfo{year}{2012}), \bibinfo{note}{89 pages}, \eprint{1203.0530}.

\bibitem[{\citenamefont{Djokovic et~al.}(2000)\citenamefont{Djokovic, Lemire,
  and Sekiguchi}}]{Djokovic:2000}
\bibinfo{author}{\bibfnamefont{Z.~D.} \bibnamefont{Djokovic}},
  \bibinfo{author}{\bibfnamefont{N.}~\bibnamefont{Lemire}}, \bibnamefont{and}
  \bibinfo{author}{\bibfnamefont{J.}~\bibnamefont{Sekiguchi}},
  \bibinfo{journal}{Tohoku. Math J.} \textbf{\bibinfo{volume}{53}},
  \bibinfo{pages}{395} (\bibinfo{year}{2000}).

\bibitem[{\citenamefont{Bellucci et~al.}(2006)\citenamefont{Bellucci, Ferrara,
  G{\"u}naydin, and Marrani}}]{Bellucci:2006xz}
\bibinfo{author}{\bibfnamefont{S.}~\bibnamefont{Bellucci}},
  \bibinfo{author}{\bibfnamefont{S.}~\bibnamefont{Ferrara}},
  \bibinfo{author}{\bibfnamefont{M.}~\bibnamefont{G{\"u}naydin}},
  \bibnamefont{and} \bibinfo{author}{\bibfnamefont{A.}~\bibnamefont{Marrani}},
  \bibinfo{journal}{Int. J. Mod. Phys.} \textbf{\bibinfo{volume}{A21}},
  \bibinfo{pages}{5043} (\bibinfo{year}{2006}), \eprint{hep-th/0606209}.

\bibitem[{\citenamefont{Goldstein and Katmadas}(2009)}]{Goldstein:2008fq}
\bibinfo{author}{\bibfnamefont{K.}~\bibnamefont{Goldstein}} \bibnamefont{and}
  \bibinfo{author}{\bibfnamefont{S.}~\bibnamefont{Katmadas}},
  \bibinfo{journal}{JHEP} \textbf{\bibinfo{volume}{0905}}, \bibinfo{pages}{058}
  (\bibinfo{year}{2009}), \eprint{0812.4183}.

\bibitem[{\citenamefont{Bena et~al.}(2009{\natexlab{a}})\citenamefont{Bena,
  Giusto, Ruef, and Warner}}]{Bena:2009en}
\bibinfo{author}{\bibfnamefont{I.}~\bibnamefont{Bena}},
  \bibinfo{author}{\bibfnamefont{S.}~\bibnamefont{Giusto}},
  \bibinfo{author}{\bibfnamefont{C.}~\bibnamefont{Ruef}}, \bibnamefont{and}
  \bibinfo{author}{\bibfnamefont{N.~P.} \bibnamefont{Warner}},
  \bibinfo{journal}{JHEP} \textbf{\bibinfo{volume}{0911}}, \bibinfo{pages}{032}
  (\bibinfo{year}{2009}{\natexlab{a}}), \eprint{0908.2121}.

\bibitem[{\citenamefont{Bena et~al.}(2009{\natexlab{b}})\citenamefont{Bena,
  Dall'Agata, Giusto, Ruef, and Warner}}]{Bena:2009ev}
\bibinfo{author}{\bibfnamefont{I.}~\bibnamefont{Bena}},
  \bibinfo{author}{\bibfnamefont{G.}~\bibnamefont{Dall'Agata}},
  \bibinfo{author}{\bibfnamefont{S.}~\bibnamefont{Giusto}},
  \bibinfo{author}{\bibfnamefont{C.}~\bibnamefont{Ruef}}, \bibnamefont{and}
  \bibinfo{author}{\bibfnamefont{N.~P.} \bibnamefont{Warner}},
  \bibinfo{journal}{JHEP} \textbf{\bibinfo{volume}{0906}}, \bibinfo{pages}{015}
  (\bibinfo{year}{2009}{\natexlab{b}}), \eprint{0902.4526}.

\bibitem[{\citenamefont{Borsten
  et~al.}(2011{\natexlab{b}})\citenamefont{Borsten, Duff, Ferrara, Marrani, and
  Rubens}}]{Borsten:2011ai}
\bibinfo{author}{\bibfnamefont{L.}~\bibnamefont{Borsten}},
  \bibinfo{author}{\bibfnamefont{M.}~\bibnamefont{Duff}},
  \bibinfo{author}{\bibfnamefont{S.}~\bibnamefont{Ferrara}},
  \bibinfo{author}{\bibfnamefont{A.}~\bibnamefont{Marrani}}, \bibnamefont{and}
  \bibinfo{author}{\bibfnamefont{W.}~\bibnamefont{Rubens}},
  \bibinfo{journal}{Phys.Rev. D (to appear)}
  (\bibinfo{year}{2011}{\natexlab{b}}), \eprint{1108.0424}.

\bibitem[{\citenamefont{Borsten
  et~al.}(2011{\natexlab{c}})\citenamefont{Borsten, Duff, Ferrara, Marrani, and
  Rubens}}]{Borsten:2011nq}
\bibinfo{author}{\bibfnamefont{L.}~\bibnamefont{Borsten}},
  \bibinfo{author}{\bibfnamefont{M.}~\bibnamefont{Duff}},
  \bibinfo{author}{\bibfnamefont{S.}~\bibnamefont{Ferrara}},
  \bibinfo{author}{\bibfnamefont{A.}~\bibnamefont{Marrani}}, \bibnamefont{and}
  \bibinfo{author}{\bibfnamefont{W.}~\bibnamefont{Rubens}}
  (\bibinfo{year}{2011}{\natexlab{c}}), \eprint{1108.0908}.

\bibitem[{\citenamefont{Ceresole et~al.}(2010)\citenamefont{Ceresole,
  Dall'Agata, Ferrara, and Yeranyan}}]{Ceresole:2009vp}
\bibinfo{author}{\bibfnamefont{A.}~\bibnamefont{Ceresole}},
  \bibinfo{author}{\bibfnamefont{G.}~\bibnamefont{Dall'Agata}},
  \bibinfo{author}{\bibfnamefont{S.}~\bibnamefont{Ferrara}}, \bibnamefont{and}
  \bibinfo{author}{\bibfnamefont{A.}~\bibnamefont{Yeranyan}},
  \bibinfo{journal}{Nucl. Phys.} \textbf{\bibinfo{volume}{B832}},
  \bibinfo{pages}{358} (\bibinfo{year}{2010}), \eprint{0910.2697}.

\bibitem[{\citenamefont{Ferrara and Maldacena}(1998)}]{Ferrara:1997ci}
\bibinfo{author}{\bibfnamefont{S.}~\bibnamefont{Ferrara}} \bibnamefont{and}
  \bibinfo{author}{\bibfnamefont{J.~M.} \bibnamefont{Maldacena}},
  \bibinfo{journal}{Class. Quant. Grav.} \textbf{\bibinfo{volume}{15}},
  \bibinfo{pages}{749} (\bibinfo{year}{1998}), \eprint{hep-th/9706097}.

\bibitem[{\citenamefont{Cerchiai et~al.}(2009)\citenamefont{Cerchiai, Ferrara,
  Marrani, and Zumino}}]{Cerchiai:2009pi}
\bibinfo{author}{\bibfnamefont{B.~L.} \bibnamefont{Cerchiai}},
  \bibinfo{author}{\bibfnamefont{S.}~\bibnamefont{Ferrara}},
  \bibinfo{author}{\bibfnamefont{A.}~\bibnamefont{Marrani}}, \bibnamefont{and}
  \bibinfo{author}{\bibfnamefont{B.}~\bibnamefont{Zumino}},
  \bibinfo{journal}{Phys. Rev.} \textbf{\bibinfo{volume}{D79}},
  \bibinfo{pages}{125010} (\bibinfo{year}{2009}), \eprint{0902.3973}.

\bibitem[{\citenamefont{Denef}(2000)}]{Denef:2000nb}
\bibinfo{author}{\bibfnamefont{F.}~\bibnamefont{Denef}},
  \bibinfo{journal}{JHEP} \textbf{\bibinfo{volume}{0008}}, \bibinfo{pages}{050}
  (\bibinfo{year}{2000}), \bibinfo{note}{40 pages, 13 figures, reference
  added}, \eprint{hep-th/0005049}.

\bibitem[{\citenamefont{Bates and Denef}(2011)}]{Bates:2003vx}
\bibinfo{author}{\bibfnamefont{B.}~\bibnamefont{Bates}} \bibnamefont{and}
  \bibinfo{author}{\bibfnamefont{F.}~\bibnamefont{Denef}},
  \bibinfo{journal}{JHEP} \textbf{\bibinfo{volume}{1111}}, \bibinfo{pages}{127}
  (\bibinfo{year}{2011}), \bibinfo{note}{13 pages, 1 figure Report-no:
  RUNHETC-2003-10}, \eprint{hep-th/0304094}.

\bibitem[{\citenamefont{{Abramsky} and {Hardy}}(2012)}]{2012arXiv1203.1352A}
\bibinfo{author}{\bibfnamefont{S.}~\bibnamefont{{Abramsky}}} \bibnamefont{and}
  \bibinfo{author}{\bibfnamefont{L.}~\bibnamefont{{Hardy}}},
  \bibinfo{journal}{ArXiv e-prints}  (\bibinfo{year}{2012}),
  \eprint{1203.1352}.

\bibitem[{\citenamefont{Saraikin and Vafa}(2008)}]{Saraikin:2007jc}
\bibinfo{author}{\bibfnamefont{K.}~\bibnamefont{Saraikin}} \bibnamefont{and}
  \bibinfo{author}{\bibfnamefont{C.}~\bibnamefont{Vafa}},
  \bibinfo{journal}{Class.Quant.Grav.} \textbf{\bibinfo{volume}{25}},
  \bibinfo{pages}{095007} (\bibinfo{year}{2008}), \eprint{hep-th/0703214}.

\bibitem[{\citenamefont{Nampuri et~al.}(2007)\citenamefont{Nampuri, Tripathy,
  and Trivedi}}]{Nampuri:2007gv}
\bibinfo{author}{\bibfnamefont{S.}~\bibnamefont{Nampuri}},
  \bibinfo{author}{\bibfnamefont{P.~K.} \bibnamefont{Tripathy}},
  \bibnamefont{and} \bibinfo{author}{\bibfnamefont{S.~P.}
  \bibnamefont{Trivedi}}, \bibinfo{journal}{JHEP}
  \textbf{\bibinfo{volume}{0708}}, \bibinfo{pages}{054} (\bibinfo{year}{2007}),
  \eprint{0705.4554}.

\bibitem[{\citenamefont{Ferrara and Marrani}(2007)}]{Ferrara:2007tu}
\bibinfo{author}{\bibfnamefont{S.}~\bibnamefont{Ferrara}} \bibnamefont{and}
  \bibinfo{author}{\bibfnamefont{A.}~\bibnamefont{Marrani}},
  \bibinfo{journal}{Phys. Lett.} \textbf{\bibinfo{volume}{B652}},
  \bibinfo{pages}{111} (\bibinfo{year}{2007}), \eprint{0706.1667}.

\bibitem[{\citenamefont{Kallosh
  et~al.}(2006{\natexlab{b}})\citenamefont{Kallosh, Sivanandam, and
  Soroush}}]{Kallosh:2006ib}
\bibinfo{author}{\bibfnamefont{R.}~\bibnamefont{Kallosh}},
  \bibinfo{author}{\bibfnamefont{N.}~\bibnamefont{Sivanandam}},
  \bibnamefont{and} \bibinfo{author}{\bibfnamefont{M.}~\bibnamefont{Soroush}},
  \bibinfo{journal}{Phys.Rev.} \textbf{\bibinfo{volume}{D74}},
  \bibinfo{pages}{065008} (\bibinfo{year}{2006}{\natexlab{b}}),
  \eprint{hep-th/0606263}.

\bibitem[{\citenamefont{Cai and Pang}(2008)}]{Cai:1900ve}
\bibinfo{author}{\bibfnamefont{R.-G.} \bibnamefont{Cai}} \bibnamefont{and}
  \bibinfo{author}{\bibfnamefont{D.-W.} \bibnamefont{Pang}},
  \bibinfo{journal}{JHEP} \textbf{\bibinfo{volume}{0801}}, \bibinfo{pages}{046}
  (\bibinfo{year}{2008}), \eprint{0712.0217}.

\bibitem[{\citenamefont{Klebanov and Tseytlin}(1996)}]{Klebanov:1996mh}
\bibinfo{author}{\bibfnamefont{I.~R.} \bibnamefont{Klebanov}} \bibnamefont{and}
  \bibinfo{author}{\bibfnamefont{A.~A.} \bibnamefont{Tseytlin}},
  \bibinfo{journal}{Nucl. Phys.} \textbf{\bibinfo{volume}{B475}},
  \bibinfo{pages}{179} (\bibinfo{year}{1996}), \eprint{hep-th/9604166}.

\bibitem[{\citenamefont{Duff and Rahmfeld}(1995)}]{Duff:1994jr}
\bibinfo{author}{\bibfnamefont{M.~J.} \bibnamefont{Duff}} \bibnamefont{and}
  \bibinfo{author}{\bibfnamefont{J.}~\bibnamefont{Rahmfeld}},
  \bibinfo{journal}{Phys. Lett.} \textbf{\bibinfo{volume}{B345}},
  \bibinfo{pages}{441} (\bibinfo{year}{1995}), \eprint{hep-th/9406105}.

\bibitem[{\citenamefont{Duff and Rahmfeld}(1996)}]{Duff:1996qp}
\bibinfo{author}{\bibfnamefont{M.~J.} \bibnamefont{Duff}} \bibnamefont{and}
  \bibinfo{author}{\bibfnamefont{J.}~\bibnamefont{Rahmfeld}},
  \bibinfo{journal}{Nucl. Phys.} \textbf{\bibinfo{volume}{B481}},
  \bibinfo{pages}{332} (\bibinfo{year}{1996}), \eprint{hep-th/9605085}.

\bibitem[{\citenamefont{Balasubramanian}(1997)}]{Balasubramanian:1997ak}
\bibinfo{author}{\bibfnamefont{V.}~\bibnamefont{Balasubramanian}}, in
  \emph{\bibinfo{booktitle}{Cargese 1997, Strings, branes and dualities}}
  (\bibinfo{year}{1997}), pp. \bibinfo{pages}{399--410},
  \bibinfo{note}{published in the proceedings of NATO Advanced Study Institute
  on Strings, Branes and Dualities, Cargese, France, 26 May - 14 Jun 1997},
  \eprint{hep-th/9712215}.

\bibitem[{\citenamefont{Blumenhagen et~al.}(2007)\citenamefont{Blumenhagen,
  Kors, Lust, and Stieberger}}]{Blumenhagen:2006ci}
\bibinfo{author}{\bibfnamefont{R.}~\bibnamefont{Blumenhagen}},
  \bibinfo{author}{\bibfnamefont{B.}~\bibnamefont{Kors}},
  \bibinfo{author}{\bibfnamefont{D.}~\bibnamefont{Lust}}, \bibnamefont{and}
  \bibinfo{author}{\bibfnamefont{S.}~\bibnamefont{Stieberger}},
  \bibinfo{journal}{Phys.Rept.} \textbf{\bibinfo{volume}{445}},
  \bibinfo{pages}{1} (\bibinfo{year}{2007}), \eprint{hep-th/0610327}.

\bibitem[{\citenamefont{{Eckert} et~al.}(2002)\citenamefont{{Eckert},
  {Schliemann}, {Bru{\ss}}, and {Lewenstein}}}]{2002AnPhy.299...88E}
\bibinfo{author}{\bibfnamefont{K.}~\bibnamefont{{Eckert}}},
  \bibinfo{author}{\bibfnamefont{J.}~\bibnamefont{{Schliemann}}},
  \bibinfo{author}{\bibfnamefont{D.}~\bibnamefont{{Bru{\ss}}}},
  \bibnamefont{and}
  \bibinfo{author}{\bibfnamefont{M.}~\bibnamefont{{Lewenstein}}},
  \bibinfo{journal}{Annals of Physics} \textbf{\bibinfo{volume}{299}},
  \bibinfo{pages}{88} (\bibinfo{year}{2002}), \eprint{arXiv:quant-ph/0203060}.

\bibitem[{\citenamefont{{Ghirardi} and
  {Marinatto}}(2004)}]{2004PhRvA..70a2109G}
\bibinfo{author}{\bibfnamefont{G.}~\bibnamefont{{Ghirardi}}} \bibnamefont{and}
  \bibinfo{author}{\bibfnamefont{L.}~\bibnamefont{{Marinatto}}},
  \bibinfo{journal}{\pra} \textbf{\bibinfo{volume}{70}}, \bibinfo{eid}{012109}
  (\bibinfo{year}{2004}), \eprint{arXiv:quant-ph/0401065}.

\bibitem[{\citenamefont{Dasgupta et~al.}(1999)\citenamefont{Dasgupta, Rajesh,
  and Sethi}}]{Dasgupta:1999ss}
\bibinfo{author}{\bibfnamefont{K.}~\bibnamefont{Dasgupta}},
  \bibinfo{author}{\bibfnamefont{G.}~\bibnamefont{Rajesh}}, \bibnamefont{and}
  \bibinfo{author}{\bibfnamefont{S.}~\bibnamefont{Sethi}},
  \bibinfo{journal}{JHEP} \textbf{\bibinfo{volume}{9908}}, \bibinfo{pages}{023}
  (\bibinfo{year}{1999}), \eprint{hep-th/9908088}.

\bibitem[{\citenamefont{Kallosh}(2005)}]{Kallosh:2005ax}
\bibinfo{author}{\bibfnamefont{R.}~\bibnamefont{Kallosh}},
  \bibinfo{journal}{JHEP} \textbf{\bibinfo{volume}{0512}}, \bibinfo{pages}{022}
  (\bibinfo{year}{2005}), \eprint{hep-th/0510024}.

\bibitem[{\citenamefont{Denef and Douglas}(2005)}]{Denef:2004cf}
\bibinfo{author}{\bibfnamefont{F.}~\bibnamefont{Denef}} \bibnamefont{and}
  \bibinfo{author}{\bibfnamefont{M.~R.} \bibnamefont{Douglas}},
  \bibinfo{journal}{JHEP} \textbf{\bibinfo{volume}{0503}}, \bibinfo{pages}{061}
  (\bibinfo{year}{2005}), \bibinfo{note}{28 pages, JHEP Latex format. v2: a
  correction further favoring high scale, v3: minor clarifications},
  \eprint{hep-th/0411183}.

\bibitem[{\citenamefont{Larsen and O'Connell}(2009)}]{Larsen:2009fw}
\bibinfo{author}{\bibfnamefont{F.}~\bibnamefont{Larsen}} \bibnamefont{and}
  \bibinfo{author}{\bibfnamefont{R.}~\bibnamefont{O'Connell}},
  \bibinfo{journal}{JHEP} \textbf{\bibinfo{volume}{0907}}, \bibinfo{pages}{049}
  (\bibinfo{year}{2009}), \eprint{0905.2130}.

\bibitem[{\citenamefont{Brown}(1969)}]{Brown:1969}
\bibinfo{author}{\bibfnamefont{R.~B.} \bibnamefont{Brown}},
  \bibinfo{journal}{J. Reine Angew. Math.} \textbf{\bibinfo{volume}{236}},
  \bibinfo{pages}{79} (\bibinfo{year}{1969}).

\bibitem[{\citenamefont{Ferrara and Kallosh}(2011)}]{Ferrara:2011dz}
\bibinfo{author}{\bibfnamefont{S.}~\bibnamefont{Ferrara}} \bibnamefont{and}
  \bibinfo{author}{\bibfnamefont{R.}~\bibnamefont{Kallosh}}
  (\bibinfo{year}{2011}), \eprint{1110.4048}.

\bibitem[{\citenamefont{Ferrara and Marrani}(2011)}]{Ferrara:2011xb}
\bibinfo{author}{\bibfnamefont{S.}~\bibnamefont{Ferrara}} \bibnamefont{and}
  \bibinfo{author}{\bibfnamefont{A.}~\bibnamefont{Marrani}}
  (\bibinfo{year}{2011}), \eprint{1112.2664}.

\bibitem[{\citenamefont{Borsten}(2010)}]{Borsten:2010ths}
\bibinfo{author}{\bibfnamefont{L.}~\bibnamefont{Borsten}}
  (\bibinfo{year}{2010}), \bibinfo{note}{ph.D. thesis, Imperial College}.

\bibitem[{\citenamefont{Linden and Popescu}(1998)}]{Linden:1997qd}
\bibinfo{author}{\bibfnamefont{N.}~\bibnamefont{Linden}} \bibnamefont{and}
  \bibinfo{author}{\bibfnamefont{S.}~\bibnamefont{Popescu}},
  \bibinfo{journal}{Fortschr. Phys.} \textbf{\bibinfo{volume}{46}},
  \bibinfo{pages}{567} (\bibinfo{year}{1998}), \eprint{quant-ph/9711016}.

\bibitem[{\citenamefont{Sudbery}(2001)}]{Sudbery:2001}
\bibinfo{author}{\bibfnamefont{A.}~\bibnamefont{Sudbery}}, \bibinfo{journal}{J.
  Phys.} \textbf{\bibinfo{volume}{A34}}, \bibinfo{pages}{643}
  (\bibinfo{year}{2001}), \eprint{quant-ph/0001116}.

\bibitem[{\citenamefont{Kempe}(1999)}]{Kempe:1999vk}
\bibinfo{author}{\bibfnamefont{J.}~\bibnamefont{Kempe}},
  \bibinfo{journal}{Phys. Rev.} \textbf{\bibinfo{volume}{A60}},
  \bibinfo{pages}{910} (\bibinfo{year}{1999}), \eprint{quant-ph/9902036}.

\bibitem[{\citenamefont{Miyake and Wadati}(2002)}]{Miyake:2002}
\bibinfo{author}{\bibfnamefont{A.}~\bibnamefont{Miyake}} \bibnamefont{and}
  \bibinfo{author}{\bibfnamefont{M.}~\bibnamefont{Wadati}},
  \bibinfo{journal}{Quant. Info. Comp.} \textbf{\bibinfo{volume}{2 (Special)}},
  \bibinfo{pages}{540} (\bibinfo{year}{2002}), \eprint{quant-ph/0212146}.

\bibitem[{\citenamefont{Greenberger et~al.}(1989)\citenamefont{Greenberger,
  Horne, and Zeilinger}}]{Greenberger:1989}
\bibinfo{author}{\bibfnamefont{D.~M.} \bibnamefont{Greenberger}},
  \bibinfo{author}{\bibfnamefont{M.}~\bibnamefont{Horne}}, \bibnamefont{and}
  \bibinfo{author}{\bibfnamefont{A.}~\bibnamefont{Zeilinger}},
  \emph{\bibinfo{title}{{Bell's Theorem, Quantum Theory and Conceptions of the
  Universe}}} (\bibinfo{publisher}{Kluwer Academic},
  \bibinfo{address}{Dordrecht}, \bibinfo{year}{1989}), ISBN
  \bibinfo{isbn}{0-7923-0496-9}.

\bibitem[{\citenamefont{Gibbs}(2001)}]{Gibbs:2001}
\bibinfo{author}{\bibfnamefont{P.}~\bibnamefont{Gibbs}} (\bibinfo{year}{2001}),
  \eprint{math/0107203}.

\bibitem[{\citenamefont{Fernando}(2006)}]{Fernando:2006}
\bibinfo{author}{\bibfnamefont{K.~V.} \bibnamefont{Fernando}}
  (\bibinfo{year}{2006}), \bibinfo{note}{oxford preprint}.

\bibitem[{\citenamefont{Lee et~al.}(2005)\citenamefont{Lee, Joo, and
  Kim}}]{Lee:2005}
\bibinfo{author}{\bibfnamefont{S.}~\bibnamefont{Lee}},
  \bibinfo{author}{\bibfnamefont{J.}~\bibnamefont{Joo}}, \bibnamefont{and}
  \bibinfo{author}{\bibfnamefont{J.}~\bibnamefont{Kim}},
  \bibinfo{journal}{Phys. Rev.} \textbf{\bibinfo{volume}{A72}},
  \bibinfo{pages}{024302} (\bibinfo{year}{2005}), \eprint{quant-ph/0502157}.

\bibitem[{\citenamefont{Lee et~al.}(2007)\citenamefont{Lee, Joo, and
  Kim}}]{Lee:2007}
\bibinfo{author}{\bibfnamefont{S.}~\bibnamefont{Lee}},
  \bibinfo{author}{\bibfnamefont{J.}~\bibnamefont{Joo}}, \bibnamefont{and}
  \bibinfo{author}{\bibfnamefont{J.}~\bibnamefont{Kim}},
  \bibinfo{journal}{Phys. Rev.} \textbf{\bibinfo{volume}{A76}},
  \bibinfo{pages}{012311} (\bibinfo{year}{2007}), \eprint{quant-ph/0702247}.

\bibitem[{\citenamefont{Faulkner}()}]{Faulkner:1971}
\bibinfo{author}{\bibfnamefont{J.~R.} \bibnamefont{Faulkner}},
  \bibinfo{note}{\href{http://www.jstor.org/stable/pdfplus/1995694.pdf}{\textit{Trans.
  Amer. Math. Soc.} \textbf{155} (1971) no. 2, 397--408}}.

\bibitem[{\citenamefont{Ferrar}()}]{Ferrar:1972}
\bibinfo{author}{\bibfnamefont{C.~J.} \bibnamefont{Ferrar}},
  \bibinfo{note}{\href{http://www.jstor.org/stable/pdfplus/1996111.pdf}{\textit{Trans.
  Amer. Math. Soc.} \textbf{174} (1972) 313--331}}.

\bibitem[{\citenamefont{Borsten
  et~al.}(2010{\natexlab{c}})\citenamefont{Borsten, Dahanayake, Duff, Ferrara,
  Marrani et~al.}}]{Borsten:2010aa}
\bibinfo{author}{\bibfnamefont{L.}~\bibnamefont{Borsten}},
  \bibinfo{author}{\bibfnamefont{D.}~\bibnamefont{Dahanayake}},
  \bibinfo{author}{\bibfnamefont{M.}~\bibnamefont{Duff}},
  \bibinfo{author}{\bibfnamefont{S.}~\bibnamefont{Ferrara}},
  \bibinfo{author}{\bibfnamefont{A.}~\bibnamefont{Marrani}},
  \bibnamefont{et~al.}, \bibinfo{journal}{Class.Quant.Grav.}
  \textbf{\bibinfo{volume}{27}}, \bibinfo{pages}{185003}
  (\bibinfo{year}{2010}{\natexlab{c}}), \eprint{1002.4223}.

\bibitem[{\citenamefont{Carteret and Sudbery}(2000)}]{Carteret:2000-1}
\bibinfo{author}{\bibfnamefont{H.}~\bibnamefont{Carteret}} \bibnamefont{and}
  \bibinfo{author}{\bibfnamefont{A.}~\bibnamefont{Sudbery}},
  \bibinfo{journal}{J. Phys.} \textbf{\bibinfo{volume}{A33}},
  \bibinfo{pages}{4981} (\bibinfo{year}{2000}), \eprint{quant-ph/0001091}.

\bibitem[{\citenamefont{Miyake}(2003)}]{Miyake:2003}
\bibinfo{author}{\bibfnamefont{A.}~\bibnamefont{Miyake}},
  \bibinfo{journal}{Phys. Rev.} \textbf{\bibinfo{volume}{A67}},
  \bibinfo{pages}{012108} (\bibinfo{year}{2003}), \eprint{quant-ph/0206111}.

\bibitem[{\citenamefont{Acin et~al.}(2000)\citenamefont{Acin, Andrianov, Costa,
  Jan\'e, Latorre, and Tarrach}}]{Acin:2000}
\bibinfo{author}{\bibfnamefont{A.}~\bibnamefont{Acin}},
  \bibinfo{author}{\bibfnamefont{A.}~\bibnamefont{Andrianov}},
  \bibinfo{author}{\bibfnamefont{L.}~\bibnamefont{Costa}},
  \bibinfo{author}{\bibfnamefont{E.}~\bibnamefont{Jan\'e}},
  \bibinfo{author}{\bibfnamefont{J.~I.} \bibnamefont{Latorre}},
  \bibnamefont{and} \bibinfo{author}{\bibfnamefont{R.}~\bibnamefont{Tarrach}},
  \bibinfo{journal}{Phys. Rev. Lett.} \textbf{\bibinfo{volume}{85}},
  \bibinfo{pages}{1560} (\bibinfo{year}{2000}), \eprint{quant-ph/0003050}.

\bibitem[{\citenamefont{Bellucci
  et~al.}(2007{\natexlab{b}})\citenamefont{Bellucci, Ferrara, and
  Marrani}}]{Bellucci:2007gb}
\bibinfo{author}{\bibfnamefont{S.}~\bibnamefont{Bellucci}},
  \bibinfo{author}{\bibfnamefont{S.}~\bibnamefont{Ferrara}}, \bibnamefont{and}
  \bibinfo{author}{\bibfnamefont{A.}~\bibnamefont{Marrani}}
  (\bibinfo{year}{2007}{\natexlab{b}}), \bibinfo{note}{contribution to the
  Proceedings of the XVII SIGRAV Conference, Turin, Italy, 4--7 Sep 2006},
  \eprint{hep-th/0702019}.

\bibitem[{\citenamefont{{Vrana} and {L{\'e}vay}}(2009)}]{LevayVrana}
\bibinfo{author}{\bibfnamefont{P.}~\bibnamefont{{Vrana}}} \bibnamefont{and}
  \bibinfo{author}{\bibfnamefont{P.}~\bibnamefont{{L{\'e}vay}}},
  \bibinfo{journal}{Journal of Physics A Mathematical General}
  \textbf{\bibinfo{volume}{42}}, \bibinfo{pages}{B5303} (\bibinfo{year}{2009}),
  \eprint{0902.2269}.

\bibitem[{Eck(2002)}]{Eckert200288}
\bibinfo{journal}{Annals of Physics} \textbf{\bibinfo{volume}{299}},
  \bibinfo{pages}{88 } (\bibinfo{year}{2002}), ISSN \bibinfo{issn}{0003-4916},
  \urlprefix\url{http://www.sciencedirect.com/science/article/pii/S0003491602962688}.

\bibitem[{\citenamefont{Li et~al.}(2001)\citenamefont{Li, Zeng, Liu, and
  Long}}]{PhysRevA.64.054302}
\bibinfo{author}{\bibfnamefont{Y.~S.} \bibnamefont{Li}},
  \bibinfo{author}{\bibfnamefont{B.}~\bibnamefont{Zeng}},
  \bibinfo{author}{\bibfnamefont{X.~S.} \bibnamefont{Liu}}, \bibnamefont{and}
  \bibinfo{author}{\bibfnamefont{G.~L.} \bibnamefont{Long}},
  \bibinfo{journal}{Phys. Rev. A} \textbf{\bibinfo{volume}{64}},
  \bibinfo{pages}{054302} (\bibinfo{year}{2001}),
  \urlprefix\url{http://link.aps.org/doi/10.1103/PhysRevA.64.054302}.

\bibitem[{\citenamefont{Pa\ifmmode~\check{s}\else \v{s}\fi{}kauskas and
  You}(2001)}]{PhysRevA.64.042310}
\bibinfo{author}{\bibfnamefont{R.}~\bibnamefont{Pa\ifmmode~\check{s}\else
  \v{s}\fi{}kauskas}} \bibnamefont{and}
  \bibinfo{author}{\bibfnamefont{L.}~\bibnamefont{You}},
  \bibinfo{journal}{Phys. Rev. A} \textbf{\bibinfo{volume}{64}},
  \bibinfo{pages}{042310} (\bibinfo{year}{2001}),
  \urlprefix\url{http://link.aps.org/doi/10.1103/PhysRevA.64.042310}.

\bibitem[{\citenamefont{Ghirardi and Marinatto}(2004)}]{PhysRevA.70.012109}
\bibinfo{author}{\bibfnamefont{G.}~\bibnamefont{Ghirardi}} \bibnamefont{and}
  \bibinfo{author}{\bibfnamefont{L.}~\bibnamefont{Marinatto}},
  \bibinfo{journal}{Phys. Rev. A} \textbf{\bibinfo{volume}{70}},
  \bibinfo{pages}{012109} (\bibinfo{year}{2004}),
  \urlprefix\url{http://link.aps.org/doi/10.1103/PhysRevA.70.012109}.

\bibitem[{\citenamefont{Schliemann
  et~al.}(2001{\natexlab{a}})\citenamefont{Schliemann, Cirac,
  Ku\ifmmode~\acute{s}\else \'{s}\fi{}, Lewenstein, and
  Loss}}]{PhysRevA.64.022303}
\bibinfo{author}{\bibfnamefont{J.}~\bibnamefont{Schliemann}},
  \bibinfo{author}{\bibfnamefont{J.~I.} \bibnamefont{Cirac}},
  \bibinfo{author}{\bibfnamefont{M.}~\bibnamefont{Ku\ifmmode~\acute{s}\else
  \'{s}\fi{}}}, \bibinfo{author}{\bibfnamefont{M.}~\bibnamefont{Lewenstein}},
  \bibnamefont{and} \bibinfo{author}{\bibfnamefont{D.}~\bibnamefont{Loss}},
  \bibinfo{journal}{Phys. Rev. A} \textbf{\bibinfo{volume}{64}},
  \bibinfo{pages}{022303} (\bibinfo{year}{2001}{\natexlab{a}}),
  \urlprefix\url{http://link.aps.org/doi/10.1103/PhysRevA.64.022303}.

\bibitem[{\citenamefont{Schliemann
  et~al.}(2001{\natexlab{b}})\citenamefont{Schliemann, Loss, and
  MacDonald}}]{PhysRevB.63.085311}
\bibinfo{author}{\bibfnamefont{J.}~\bibnamefont{Schliemann}},
  \bibinfo{author}{\bibfnamefont{D.}~\bibnamefont{Loss}}, \bibnamefont{and}
  \bibinfo{author}{\bibfnamefont{A.~H.} \bibnamefont{MacDonald}},
  \bibinfo{journal}{Phys. Rev. B} \textbf{\bibinfo{volume}{63}},
  \bibinfo{pages}{085311} (\bibinfo{year}{2001}{\natexlab{b}}),
  \urlprefix\url{http://link.aps.org/doi/10.1103/PhysRevB.63.085311}.

\bibitem[{\citenamefont{Kim et~al.}(2010)\citenamefont{Kim, Hornlund,
  Palmkvist, and Virmani}}]{Kim:2010bf}
\bibinfo{author}{\bibfnamefont{S.-S.} \bibnamefont{Kim}},
  \bibinfo{author}{\bibfnamefont{J.~L.} \bibnamefont{Hornlund}},
  \bibinfo{author}{\bibfnamefont{J.}~\bibnamefont{Palmkvist}},
  \bibnamefont{and} \bibinfo{author}{\bibfnamefont{A.}~\bibnamefont{Virmani}},
  \bibinfo{journal}{JHEP} \textbf{\bibinfo{volume}{08}}, \bibinfo{pages}{072}
  (\bibinfo{year}{2010}), \eprint{1004.5242}.

\bibitem[{\citenamefont{Rios}(2010)}]{Rios:2010br}
\bibinfo{author}{\bibfnamefont{M.}~\bibnamefont{Rios}} (\bibinfo{year}{2010}),
  \eprint{1005.3514}.

\end{thebibliography}
%\bibliographystyle{utphys}

\end{document}